\documentclass[12pt]{article}%
\usepackage{afterpage}
\usepackage{amsmath}
\usepackage{amsfonts}
\usepackage{amssymb}
\usepackage{array}
\usepackage{blkarray}
\usepackage{caption}
\usepackage{enumerate}
\usepackage{float}
\usepackage[bottom, multiple,hang]{footmisc}
\usepackage[margin=1in]{geometry}
\usepackage{graphicx}
\usepackage{hhline}
\usepackage{lscape}
\usepackage{natbib}
\usepackage{rotating}
\usepackage{setspace}
\usepackage{subfigure}
\usepackage{subfloat}
\usepackage{tabularx}
\usepackage{titletoc}
\usepackage{url}
\usepackage{version}
\usepackage{dcolumn}
\usepackage{inputenc}
\usepackage{longtable}
\usepackage{accents}
\usepackage{placeins}
\usepackage{flafter}
\usepackage{afterpage}
\usepackage{titlesec}
\usepackage{mathptmx}
\usepackage{endnotes}%
\providecommand{\U}[1]{\protect\rule{.1in}{.1in}}

\makeatletter
\g@addto@macro\normalsize{
\setlength\abovedisplayskip{7pt}
\setlength\belowdisplayskip{7pt}
\setlength\abovedisplayshortskip{7pt}
\setlength\belowdisplayshortskip{7pt}
}
\newcolumntype{P}[1]{>{\centering\arraybackslash}p{#1}}
\newcolumntype{M}[1]{>{\centering\arraybackslash}m{#1}}
\renewcommand\@makefnmark{
\hbox{\hspace{-.3em}\@textsuperscript{\normalfont\@thefnmark}}}
\makeatother
\titlespacing*{\section}{0pt}{15pt}{6pt}
\titlespacing*{\subsection}{0pt}{9pt}{3pt}
\titlespacing*{\subsubsection}{0pt}{3pt}{1pt}
\renewcommand{\footnotesize}{\fontsize{9pt}{11pt}\selectfont}
\renewcommand{\small}{\fontsize{10pt}{11pt}\selectfont}
\begin{document}

\title{Identifying the Effects of Sanctions on the Iranian Economy using Newspaper
Coverage\thanks{We would like to thank Nick Bloom, Jeff Nugent, Adrian Pagan,
Alessandro Rebucci, and Ron Smith for helpful comments.}}
\author{Dario Laudati\\{\small {}{}University of Southern California, USA }\\M. Hashem Pesaran\\{\small {}{}{}{}{}{}University of Southern California, USA, and Trinity
College, Cambridge, UK}%

\vspace*{-0.2cm}%

}
\date{August 2021}
\maketitle

\begin{abstract}
This paper considers how sanctions affected the Iranian economy using a novel
measure of sanctions intensity based on daily newspaper coverage. It finds
sanctions to have significant effects on exchange rates, inflation, and output
growth, with the Iranian rial over-reacting to sanctions, followed up with a
rise in inflation and a fall in output. In absence of sanctions, Iran's
average annual growth could have been around $4-5$ per cent, as compared to
the $3$ per cent realized. Sanctions are also found to have adverse effects on
employment, labor force participation, secondary and high-school education,
with such effects amplified for females.

\bigskip



\end{abstract}

\pagenumbering{gobble}

\newpage\pagenumbering{arabic} \setcounter{page}{1}%

\doublespacing

\section{Introduction \label{Sec: introduction}}

Over the past forty years Iran has been subject to varying degrees of economic
and financial sanctions, and asset freezes, which began in November 1979 when
the U.S. placed an embargo on Iranian oil trade and froze \$12 billion of
Iranian assets held outside Iran with the aim of securing the release of U.S.
hostages. Although this particular sanction episode was successfully
negotiated in January 1981, U.S. policy towards Iran became increasingly
entrenched, aimed at curtailing the economic and political influence of Iran
in the Middle East region and beyond; a process which escalated over Iran's
nuclear program. As a result, the Iranian economy has been operating for a
prolonged period under severe and often quite harsh international
restrictions, perhaps unique for a sizeable economy with deep historical roots
in the global economy. Given the uncertainty and durability of sanction
regimes, it is also important to bear in mind that, besides the direct effects
of sanctions (arising from loss of oil export revenues, loss of access to
currency reserves and other trade-related losses), sanctions also result in
important and lasting indirect effects, such as rent-seeking, resource
allocation distortions, and general costs associated with efforts involved in
mitigating and circumventing the sanctions regimes. These indirect effects are
likely to be more serious the longer the sanctions are in place, particularly
when the prospect of a sanctions free outcome seems very remote.

The focus of the present paper is on the identification and quantitative
evaluation of the direct and indirect effects of sanctions on the Iranian
economy over the period 1989--2020, since the end of the Iran-Iraq War and the
start of the reconstruction period under President Rafsanjani. We exclude the
period 1979--1988 due to the special circumstances of the 1979 Revolution, the
hostage crises and the ensuing eight year war with Iraq, which ended in August
1988. We also do not address the wider issue of the effectiveness of sanctions
in achieving the foreign policy goals of the U.S. and its Western allies. Nor
do we discuss Iran's ability to respond to sanctions in the rapidly changing
geographical conditions of the region.

We try to capture the intensity of the sanctions and the international
political pressure on Iran to comply by constructing a time series index based
on daily newspaper coverage of the sanctions, their imposition, the intensity
of their use, as well as their occasional removal. The idea of a newspaper
coverage index was developed by \cite{baker2016} for measurement of economic
uncertainty, but to our knowledge it has not been utilized in the analysis of
sanctions. As we shall see, the evolution of our proposed sanctions intensity
index closely tracks the main sanctions time points, such as the U.S. Iran and
Libya Sanctions Act of 1996, the U.S. export ban in 1997, the U.S. investment
bans and asset freezes in 2006 and 2007 ("Iran Freedom and Support Act", and
Executive Order 13438), the United Nations nuclear Resolutions (1737, 1747)
during 2006 and 2007, the U.S. Comprehensive Iran Sanctions, Accountability,
and Divestment Act of 2010, the U.S. National Defence Authorization Act of
2012, the partial lifting of U.N. sanctions under the Joint Comprehensive Plan
of Action (JCPOA) in 2015 and its subsequent implementation in January 2016,
and finally President Trump's unilateral withdrawal from the JCPOA agreement
in 2018. See Figure \ref{fig: s_graph}. The sanctions intensity measure also
correlates with the U.S. Treasury "Specially Designated Nationals And Blocked
Persons List"\ (SDN) for Iran\ which has been publicly available since 2006.

We use the sanctions intensity variable in reduced form equations as well as
in structural vector autoregressive (SVAR) models to identify short run and
long run effects of sanctions (direct and indirect effects) on Iran's rial/USD
exchange rate, money supply growth, inflation and output growth, whilst
controlling for oil price changes, foreign output growth, and other global
factors such as equity market volatility. By making use of the sanctions
intensity index we also avoid some of the limitations of comparative
approaches used in the literature for policy evaluations, such as the
synthetic control method (SCM) proposed by \cite{abadie_gardezabal2003}, and
the panel data approach proposed by \cite{hsiao_et_al2012}. These techniques
require pre-policy intervention outcomes to estimate weighted averages of post
policy outcomes for a "pre-selected" control group to be used as the basis of
comparisons. In the case of Iran there are no clear cut "sanctions on" and
"sanctions off" periods, and it is unclear which countries should be included
in the control group given the continued importance of the Iranian economy in
the region.

The main drawback of our approach is its inability to distinguish between the
direct and indirect effects of sanctions, since we have not been able to come
up with satisfactory measures to control for rent-seeking, and other economic
distortions that might have been caused by the sanctions. There is no doubt
that the Iranian economy would have been subject to distortions and economic
mismanagement even in the absence of any sanctions, and it would seem unlikely
to be able to separate sanctions-induced distortions from all other
distortions either. For these reasons, our results should be approached with
caution. What we estimate can be viewed as measuring the combined effects of
sanctions and sanctions-induced distortions, broadly defined. Seen from this
perspective, we find that the sanctions intensity variable has highly
statistically significant effects on exchange rates, inflation and output
growth, but not on money supply growth. These estimates proved to be robust to
alternative specifications and after allowing for a host of control variables.

Our results also show that strong currency depreciations (with substantial
overshooting), and high inflation rates are important channels through which
sanctions affect the real economy. On the other hand, the overexpansion of the
money supply used to compensate underdeveloped capital and money markets does
not seem to affect the path of other domestic variables, once we control for
inflation and variations in exchange rates.

Using impulse response analysis and forecast error variance decomposition
techniques, we also find a significant over-reaction of the rial to sanctions,
with a subsequent rise in inflation and a fall in output shortly after. The
economy adapts reasonably quickly to sanction shocks, a property that has
already been documented by \cite{esfahani_etal2013}, who consider the effects
of oil revenue shocks on output growth and inflation. The forecast error
variance decompositions also show that, despite the inclusion of the sanctions
intensity variable in the VAR, around 80 per cent of variations in foreign
exchange and 83 per cent of variations in output growth remain unexplained,
and most likely relate to many other latent factors that drive the Iranian
economy. We also estimate that in the absence of sanctions Iran's output
growth could have been around $4-5$ per cent, as compared to the $3$ per cent realized.

We are also able to identify negative effects of sanctions on the labor
market. The employment rate with respect to other countries in the Middle East
and North Africa (\emph{MENA}) region has systematically decreased after
sanctions were imposed, and women seem to have paid the higher price, with
significant declines in female labor force participation. From a
socio-demographic standpoint, we also find that sanctions have negatively
affected secondary school education, with the number of schools and teachers
both sharply down in response to new sanctions waves. Again, gender effects
seem to be at play here. We find changes in sanctions intensity to have
statistically significant negative effects on the ratio of female to male
students. These outcomes could be due to government responses to
sanctions-induced reductions in oil income, with associated budgetary
allocation away from education and female participation.

Sanctions have also had a number of positive unintended consequences.
Interestingly, the Iranian economy at the onset of sanctions was as heavily
dependent on oil exports as countries such as Saudi Arabia. Restricting oil
exports over a relatively long time period has led to important structural
transformations of the Iranian economy, with significant increases in non-oil
exports, most notably petrochemicals, light manufacturing products and
agricultural goods. There have also been significant successes in internet
access and the associated rise of high-tech and knowledge-based companies in
Iran, such as Digikala, Snapp, and Cafe Bazaar, to name just a few. It is
likely that U.S. sanctions have been partly responsible for the rapid rise of
high-tech companies in Iran over the past decade.

Our research does not encompass health outcomes because of the lack of
sufficiently long time series on the healthcare system. A number of recent
on-field reports suggest that medicines (especially quality drugs) were
increasingly harder to find in Iran even before the Covid-19
pandemic.\footnote{To our knowledge there are no U.S. sanctions on the export
of essential drugs to Iran. But large pharmaceutical and medical companies
tend to avoid transactions with Iran because of the highly complex financial
regulatory structure which is in place as a part of U.S. sanctions. Different
laws overlap making some transactions appear legal on the assistance side but
potentially illegal from a financial perspective (\cite{peksen2009},
\cite{kokabisaghi2018}).}

Overall, there seems little doubt that sanctions have harmed the Iranian
economy and weakened its socio-economic infrastructure. But removal of
sanctions on their own is unlikely to ensure a period of sustained growth and
low and stable inflation, and many policy reforms are needed to address
sanctions-induced price distortions as well as other distortions due to
general economic mismanagement, poor governance, and the ambiguities that
surround the relative roles of semi-government agencies and the private sector
in the economy. Subsidies on essential food items and energy (fuel as well as
electricity) have created inefficiencies, smuggling, and damaging unintended
consequences. Subsidies on electricity, for example, have led to excessive
ground water withdrawals from electricity-powered irrigation wells, and more
recently for mining crypto-currencies, one of the sources of Iran's worsening
water shortages, and frequent electricity blackouts.

\noindent\textbf{Related literature}

As noted already, in this paper we are primarily interested in the economic
implications of sanctions with a focus on the case of Iran, which is arguably
the most sanctioned country in the world. Yet there exists limited systematic
knowledge of the effects of a prolonged period of sanctions on a major economy
such as Iran \citep{jhu2020}. We do not address the wider issues discussed in the literature about the
efficacy of sanctions as a foreign policy tool, although it is hoped that the
economic analysis provided in this paper could help in this
regard.\footnote{The effectiveness of sanctions in achieving foreign policy
goals has been studied extensively in the literature. \cite{hufbauer1990}
examine 116 case studies covering the period from the economic blockade of
Germany during World War I through the U.N.-U.S. embargo of Iraq in 1990.
Further overviews are provided in \cite{morgan2014}.and \cite{doxey1996}.
Critical assessments of sanctions as a policy tool are provided by
\cite{cortright1997}, \cite{pape1997, pape1998}, \cite{andreas2005}, and \cite{peksen2010}. These studies highlight possible
counterproductive effects of economic sanctions. \cite{naghavi2015} provide a
game-theoretic analysis of sanctions and its application to Iran.} Studies
that are more directly related to our paper either consider a specific
sub-period or use shocks to oil export revenues as representing a sanction
shock. \cite{gharehgozli2017} considers the effects of sanctions just before
JCPOA, which we discuss in further detail in Section \ref{Sec: methodology}
below. \cite{dizaji2013} study the impact of economic sanctions via changes in
oil revenues using a VAR model. They show that sanctions are effective in the
short-run but their relevance fades with time. Similar results are reported by
\cite{esfahani_etal2013}, who find that shocks to foreign output and oil
exports are rather short-lived for Iran. This is an important feature of the
Iranian economy which is also confirmed by our analysis using the new
sanctions intensity variable. \cite{haidar2017} uses micro-data over the
period 2006--2011 to find that two-thirds of Iran's sanctioned non-oil exports
were redirected to other non-sanctioning countries. It is also found that
large exporters appear to suffer less from export sanctions. \cite{popova2016}
find a similar geographical redirection of Iran's non-oil exports over the
period 2006--2013 of trading partners away from Western economies to countries
in the region (notably Iraq), China and other Asian economies.
\cite{farzanegan_hayo2019} analyze the effect of sanctions to expand the
shadow economy. Although not strictly quantitative in nature, a number of
studies maintained that the burden of economic sanctions for Iranian growth
was high but not decisive to bring about political change in Iran.
(\cite{carswell1981}, \cite{amuzegar1997b}, \cite{amuzegar1997a},
\cite{dadkhah1998}, \cite{downs2011} and \cite{borszik2016}).

Sanctions have also played an important role in shaping Iran's monetary and
financial system. \cite{mazarei2019} analyzes the current state of the Iranian
financial system and its fragility. \cite{farzanegan_markwardt2009} focus on
the extent to which Iran suffers from a form of "Dutch disease", thus
advocating for a sovereign oil fund to mitigate inflationary pressures and
risks of currency crises. \cite{mazarei2020} highlights the danger of
inflation for Iran in the wake of sanctions and the pandemic. There are also
several studies on the determinants of inflation in Iran (not related to
sanctions), which could be of interest. See, for example, the studies by
\ \cite{liu_adedeji2000}, \cite{celasun_goswami2002}, and \cite{bonato2008}.

Sanctions have often led to the establishment of multiple exchange rate
markets with important rent-seeking opportunities and related economic
distortions. Currently, there are three different exchange rates for the
rial.\footnote{The three exchange rates are: (\textit{i}) The official
exchange rate to import essential items such as medicine, grain and sugar;
(\textit{ii}) The \textit{Nima rate}, officially set at 2 per cent above the
official rate by Bank Markazi daily, but in practice it is subject to much
higher premiums and is reserved for non-oil exporters; (\textit{iii}) The free
market rate used for all other transactions.} \cite{bahmani1996} provides an
earlier account of the gains obtainable in Iran from the black market premium,
and the need to consider the free market rate rather than the official one
when the Iranian money demand is to be assessed; we follow this approach when
conducting our analyses and disregard the official rate. The economic
implications of multiple exchange rates in Iran are discussed in
\cite{pesaran1992}, \cite{farzanegan2013} and \cite{majidpour2013}.

The role of gender inequality in Iran as compared to other Middle Eastern
countries both in terms of labor force participation and education has also
been studied by \cite{esfahani_etal2012} and \cite{majbouri2015}, among
others, and \cite{alizadeh2017} provides a review.\footnote{We do not address
the effects of economic sanctions on government and military expenditures.
Interested readers may wish to consult the papers by \cite{dizaji2014},
\cite{dizaji_et_al_2014}, \cite{farzanegan2011oil}, \cite{farzanegan2019scm},
\cite{dizaji2021}, \cite{farzanegan2014military} and \cite{farzanegan2016}.}

The rest of the paper is organized as follows. Section
\ref{Sec: sanctions issues} offers an overview of the Iranian economy under
sanctions. Section \ref{Sec: methodology} discusses alternative approaches to
the analysis of policy interventions, and develops a structural vector
autoregressive framework with latent factors used to identify the effects of
sanctions on the Iranian economy. Dynamics of sanctions and the channels
through which sanctions affect the Iranian economy are discussed in Section
\ref{Sec: Channels and dynamics of sanctions}. Section
\ref{Sec: sanctions measure}\ explains how we construct the sanctions
intensity index from six leading newspapers, and its co-movements with
historical events. Section \ref{Sec: Structural model results} presents the
estimates of the structural model and reports the related impulse responses
and forecast error variance decompositions of sanction and domestic shocks, as
well as providing an estimate of sanctions-induced output losses.\ Additional
economic and socio-demographic results are obtained from the reduced form
analyses in Section \ref{Sec: reduced form results}. Section
\ref{Sec: conclusion} ends with some concluding remarks. An online supplement
provides details on the construction of our sanctions intensity variable, all
data sources, further methodological notes and empirical analyses, and a
comprehensive list of sanctions imposed against Iran over the past forty years.

\section{Sanctions and the Iranian economy: an overview
\label{Sec: sanctions issues}}

The evolution of the Iranian economy over the past forty years has been
largely shaped by the Revolution and the eight-years war with Iraq
(1979-1988), prolonged episodes of economic and financial sanctions, and often
very different policy responses to sanctions and economic management under the
four presidents that have held office since 1989. Initially, U.S. sanctions
were much more clearly targeted. The goal of the 1980--81 sanctions was to
negotiate the release of U.S. hostages, and the 1987 sanctions to end
hostilities in the Persian Gulf and bring about an end to the war with Iraq.
These sanctions aimed at limiting Iran's access to foreign exchange earnings
through asset freezes and, more importantly, by reducing Iran's capacity and
ability to produce and export oil.\footnote{For an overview of U.S. sanctions
against Iran see also Chapter 9 of \cite{maloney2015}.}

\begin{figure}%
\small\caption
{Relevant Iran's and World macroeconomic and financial time series over the period 1979--2020}%
\label{fig: intro descriptives}%

\vspace{-0.2cm}%
%

\[%
\begin{array}
[c]{ccc}%
\text{\emph{Panel} \emph{A}} &  & \text{\emph{Panel B}}\\
\text{Iranian oil exports}^{\text{\emph{1}}} &  & \text{Shares of oil and gas
vs. non-oil and gas exports revenues}^{\text{\emph{2}}}\\%
\vspace{-0.3cm}%

&  & \\%
{\includegraphics[
height=2.2506in,
width=3.0426in
]%
{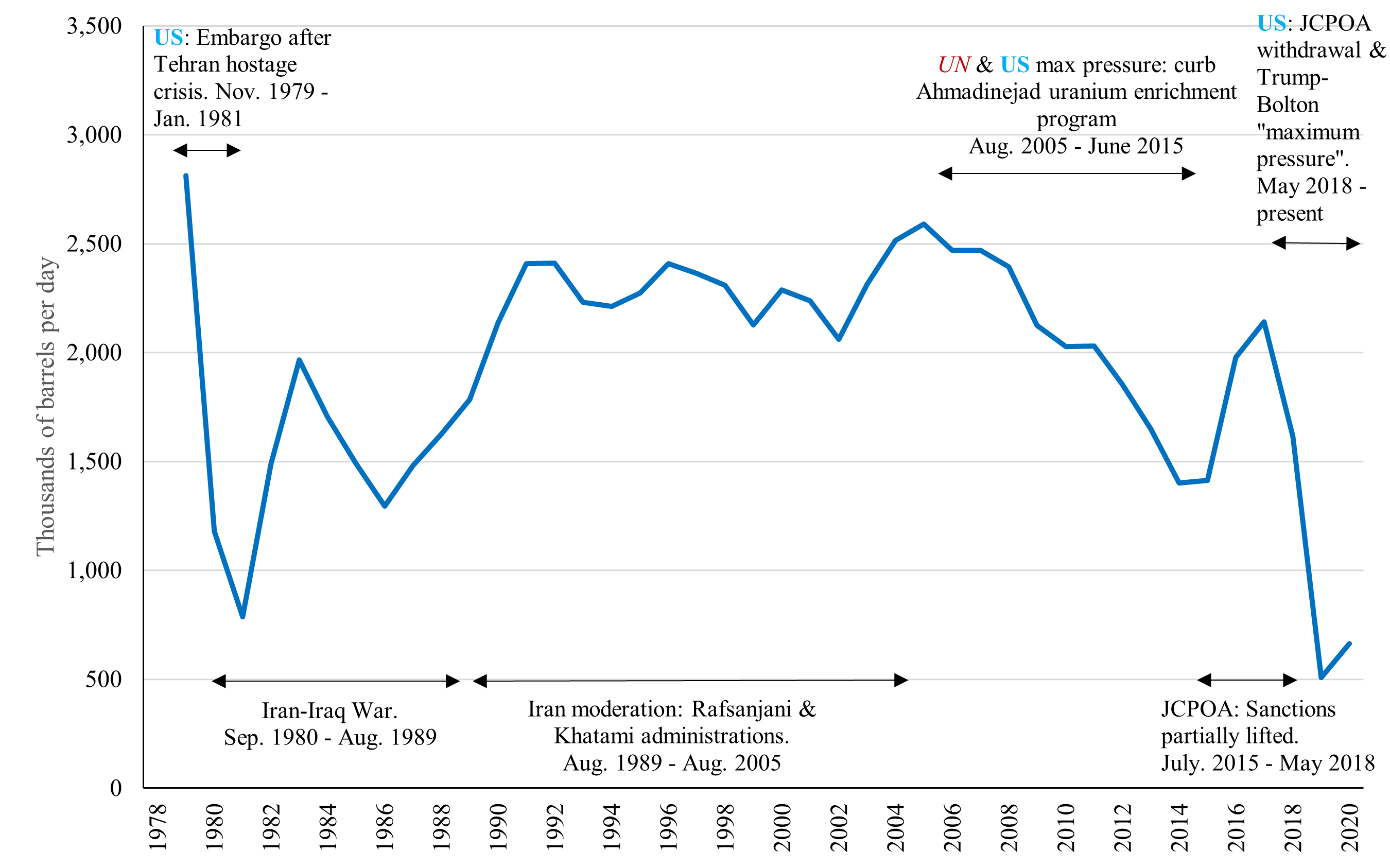}%
}%
&  &
{\includegraphics[
height=2.2506in,
width=3.0426in
]%
{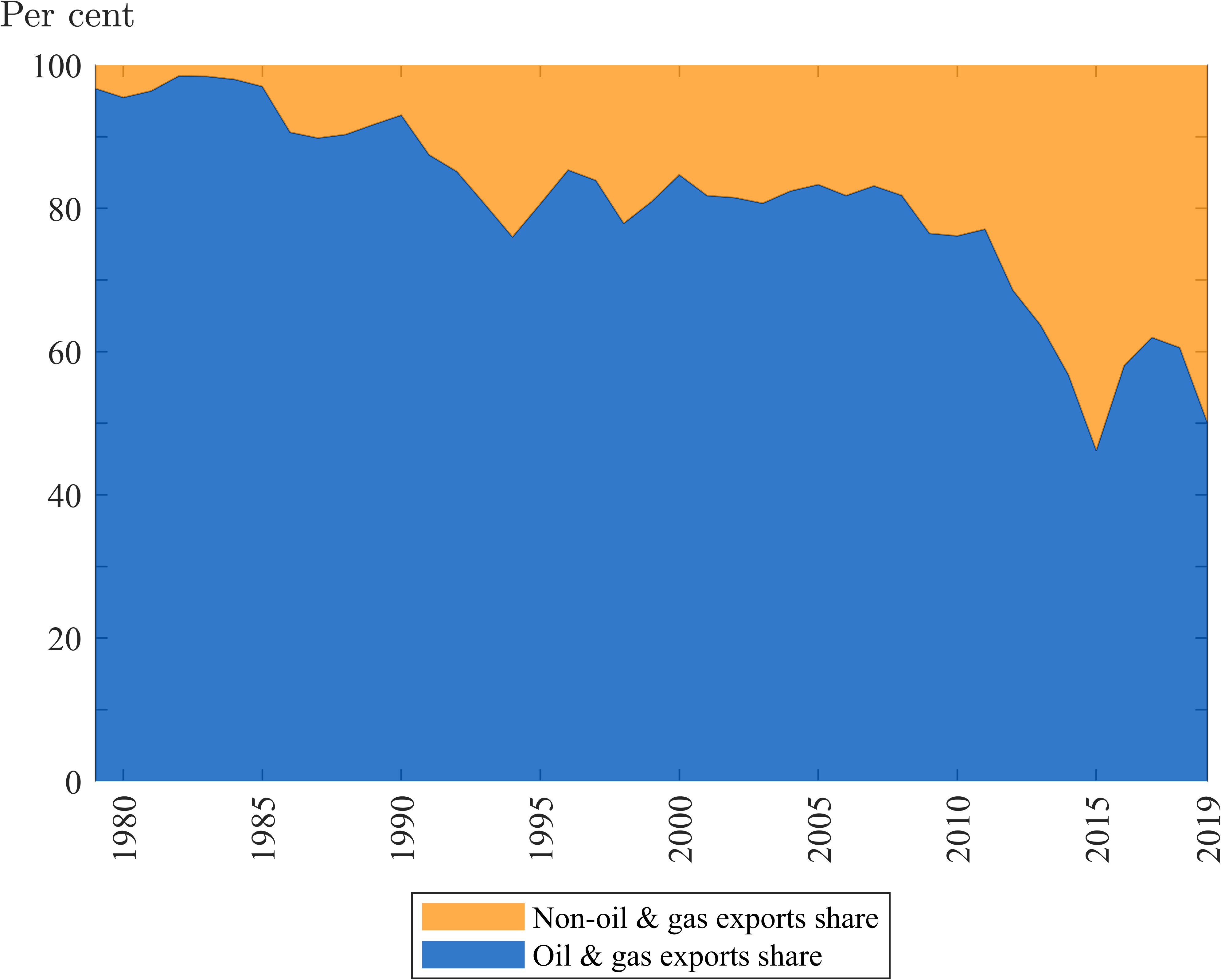}%
}%

\\
&  & \\
\text{\emph{Panel C}} &  & \text{\emph{Panel D}}\\
\text{Free market and official FX rates in logs}^{\text{\emph{3}}} &  &
\text{Free market FX rate and CPI in logs}^{\text{\emph{4}}}\\%
{\includegraphics[
height=2.2505in,
width=3.0398in
]%
{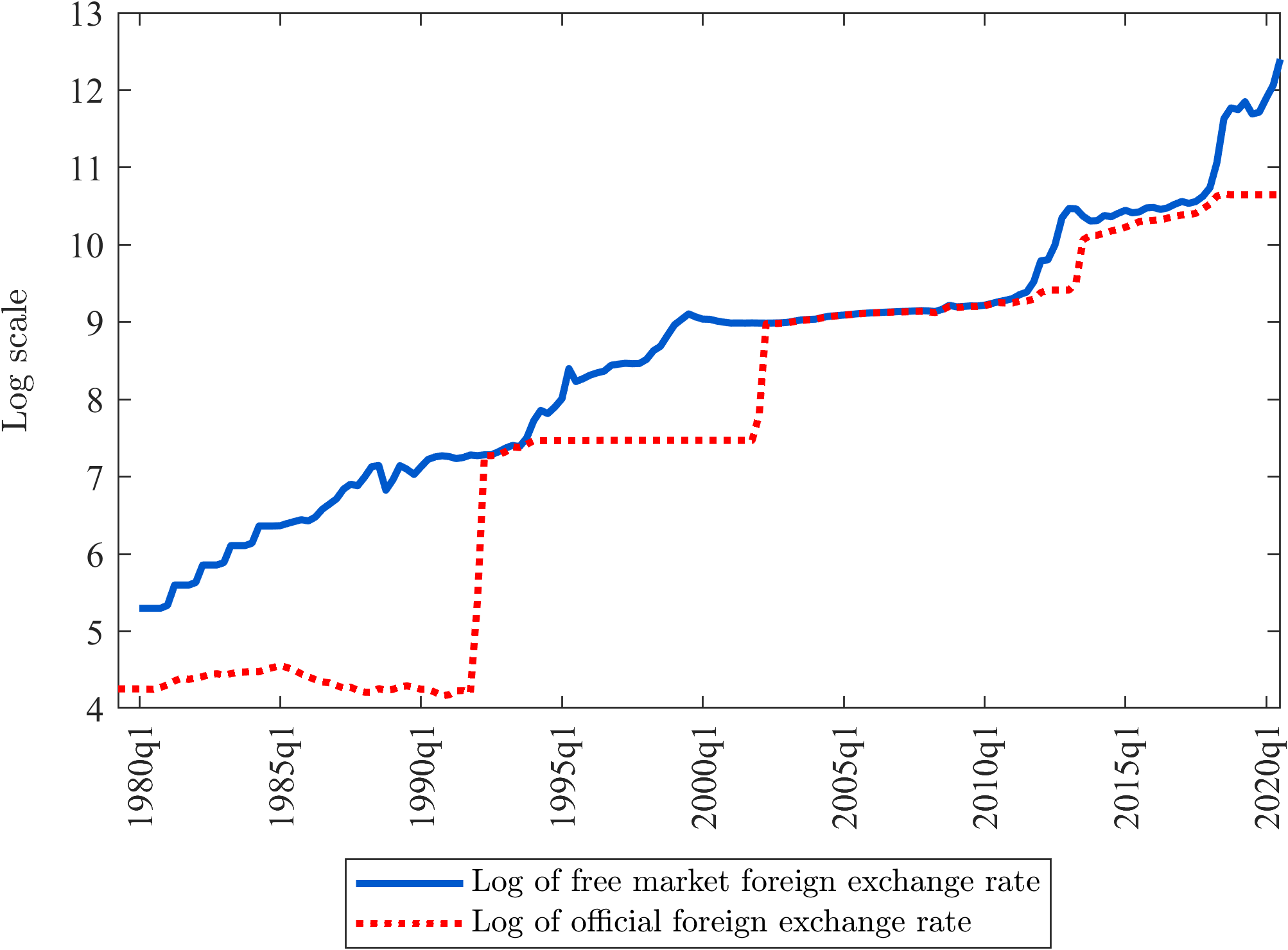}%
}%

&  &
{\includegraphics[
height=2.2505in,
width=3.0398in
]%
{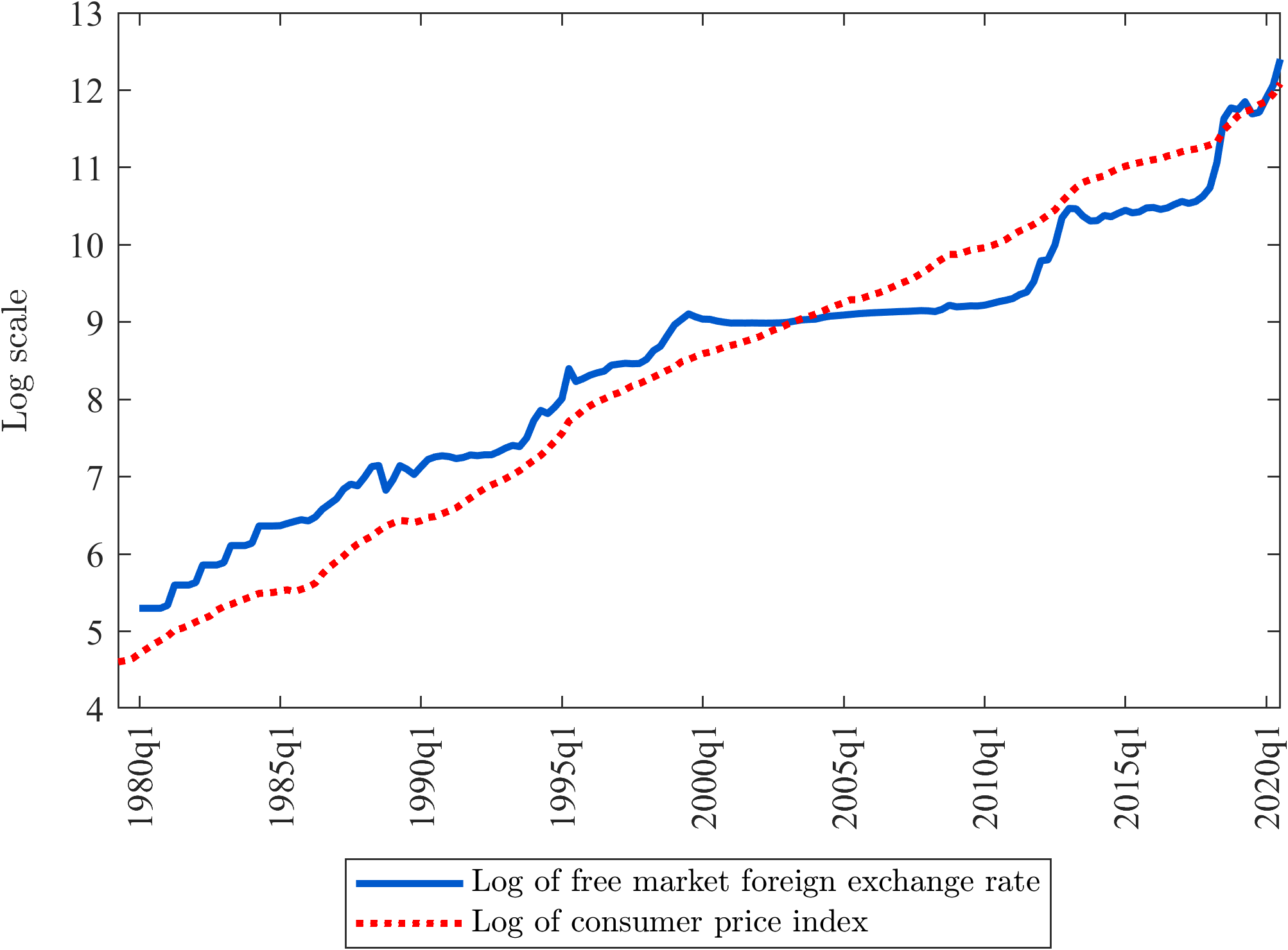}%
}%

\end{array}
\]

\[%
\begin{array}
[c]{c}%
\text{\emph{Panel E}}\\
\text{Iran and World real output in logs}^{\text{\emph{5}}}\\%
{\includegraphics[
height=2.2489in,
width=4.5328in
]%
{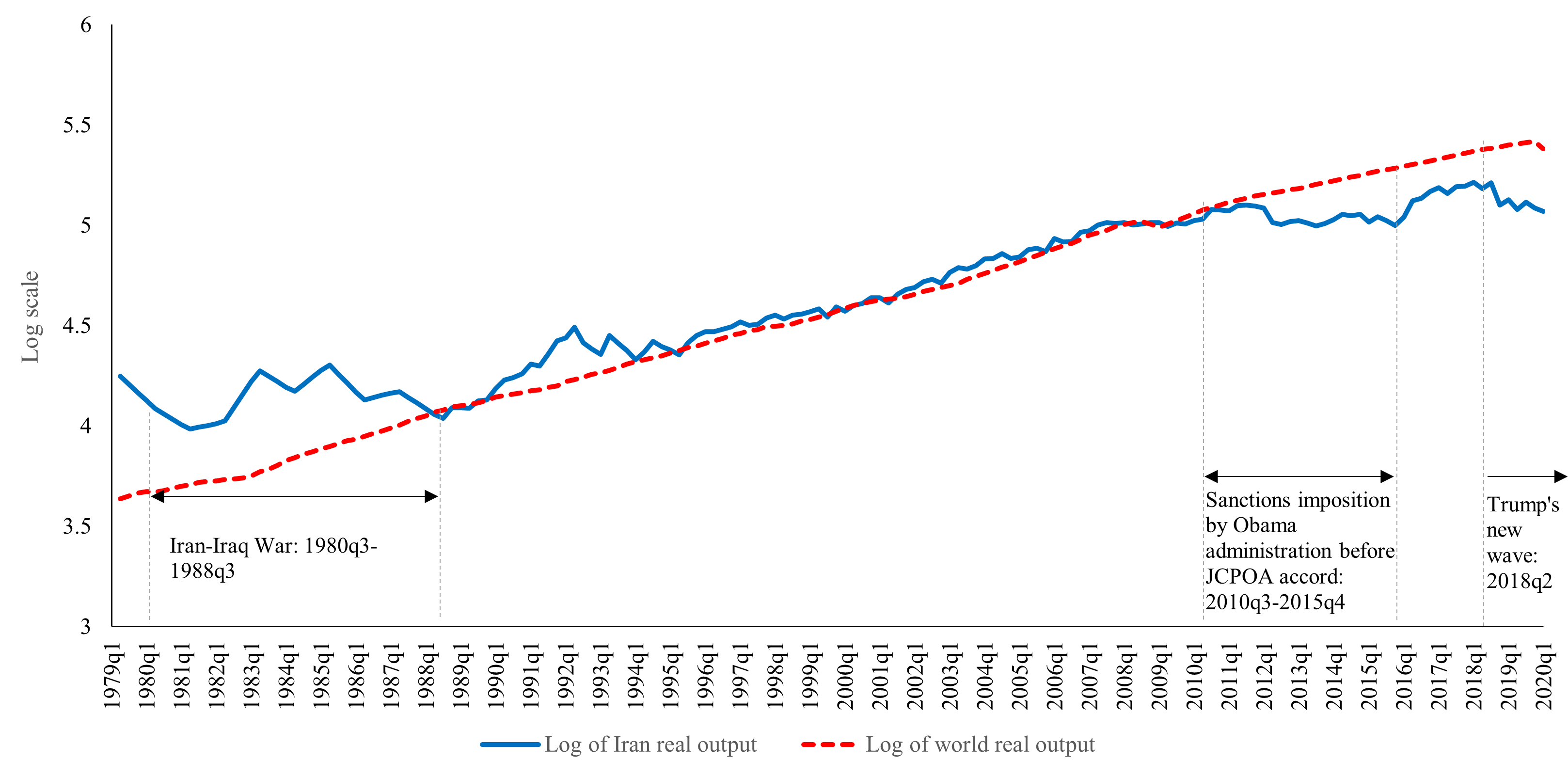}%
}%
\end{array}
\]

\footnotesize
{}\textbf{Notes}: \emph{1.} Annual data over the period 1979--2020. \emph{2.}
Annual data over the period 1979--2019. \emph{3.} Quarterly data over the
period 1979q2--2020q3. Foreign exchange rates are expressed as number of
Iran's rials per U.S. dollars. \emph{4.} Quarterly data over the period
1979q2--2020q3. CPI stands for Consumer Price Index, and it is equal to 100 in
1979q2. \emph{5. }Quarterly data over the period 1979q2--2020q1. The world
real output is a weighted average of the natural logarithm of real output for
33 major economies.

See Sections \ref{Sec. calendar conversion} and
\ref{Sec. socio-economic vars construction} in the data appendix of the online
supplement for details on calendar conversions, and sources of the data.%

\end{figure}

The evolution of Iran's oil exports since the Revolution is shown in
\emph{Panel A} of Figure \ref{fig: intro descriptives}. Iran's oil exports had
been already cut by half from the pre-Revolution peak of 6 millions barrels
per day (mb/d). The first U.S. sanctions drove Iran's oil exports to the low
of 700,000 b/d before recovering somewhat after the sanctions were lifted in
January 1981. However, since the lifting of the sanctions coincided with the
intensification of the war with Iraq, oil exports did not recover fully till
after the war ended in 1988. From 1989 to 2005 with improvements in the
diplomatic relationships between Iran and the U.S. and other Western
countries, oil exports started to rise and stabilized to around 2.5 mb/d under
the presidencies of Rafsanjani (1989q3--1997q2) and Khatami (1997q3--2005q2).
Oil exports began to decline again from 2007 after the imposition of U.S. and
U.N. sanctions in December 2006 aimed at halting Iran's uranium enrichment
program which had gathered pace under the newly elected President Ahmadinejad.
Initially, the sanctions targeted investments in oil, gas and petrochemicals,
and exports of refined products, but then were later expanded to include
banking, insurance and shipping. Sanctions also targeted the Islamic
Revolutionary Guard Corps (IRGC) and its vast commercial and industrial empire
whose activities had been expanded under President Ahmadinejad's
administration (2005q3--2013q2). Sanctions against Iran were intensified
further under President Obama and their coverage expanded to imports of rugs,
pistachios, and caviar, representing some of Iran's major non-oil exports.
Additional financial sanctions were imposed on Iran from\ July 2013, almost
immediately before President Rouhani (2013q3--2021q2) took office. The
coverage of U.N. and U.S. sanctions increased well beyond the oil and gas
sectors and affected all aspects of Iranian foreign trade and international
finance, and even the international payment system of Bank Markazi (Iran's
Central Bank). The extensive coverage of the sanctions, their multilateral
nature, coupled with the start of Rouhani's administration in 2013 paved the
way for the 2015 nuclear agreement, Joint Comprehensive Plan of Action
(\emph{JCPOA}), implemented in January 2016 which led to the easing of some of
the U.S. sanctions and the lifting of U.N. and European Union sanctions
against Iran. But the benefits of the JCPOA to Iran were limited, as many
non-US global companies and banks were hesitant to deal with Iran because of
the remaining U.S. sanctions, as well as concerns over money laundering,
opacity of ownership, and the fragility of the Iranian banking system. As it
turned out, JCPOA was also short lived, with oil exports sharply declining
after May 2018, when U.S. President Trump unilaterally withdrew from JCPOA,
and adopted the policy of "maximum pressure" against Iran. With the election
of President Biden in November 2020, there are negotiations for the U.S. to
return to the 2015 nuclear agreement, although our analysis will be pre-dating
these negotiations.%

\begin{table}%
\caption
{Free market and official foreign exchange rate depreciation, inflation, real output growth, and sanctions intensity
over the period 1979q3--2021q1}
\renewcommand{\arraystretch}{1.05}
\small
\label{table: descriptives by presidency}%

\vspace{-0.53cm}%

\begin{center}
	\begin{tabular}{l cccc M{1.1cm} M{1.1cm}}
		&       &       &       &  \multicolumn{3}{>{\raggedleft}p{5cm}}%
{\textit{Per cent per annum}} \\
		\hline\hline Periods &       \multicolumn{1}{M{2.1cm}}{Free FX depreciation}
& \multicolumn{1}{M{2.1cm}}{Official FX depreciation}
& Inflation & \multicolumn{1}{M{1.5cm}}{Output growth} & \multicolumn
{2}{>{\centering}M{3.1cm}}{Sanctions intensity (0,1)} \\
\cline{6-7}          &       &       &       &      & Mean  & Median \\
		\hline
Revolution and Iran-Iraq War$^1$ & 19.94 & 0.28  & 18.29 & -1.60 & 0.20  & 0.11 \\
		(1979q3--1989q2) &       &       &       &       &       &  \\
		Rafsanjani presidency & 16.55 & 39.83 & 21.17 & 5.16  & 0.11  & 0.10 \\
		(1989q3--1997q2) &       &       &       &       &       &  \\
		Khatami presidency & 7.90  & 20.34 & 14.53 & 4.72  & 0.15  & 0.13 \\
		(1997q3--2005q2) &       &       &       &       &       &  \\
		Ahmadinejad presidency & 17.08 & 5.16  & 18.15 & 1.68  & 0.38  & 0.39 \\
		(2005q3--2013q2) &       &       &       &       &       &  \\
		Rouhani presidency$^2$ & 25.34 & 14.66 & 19.61 & 0.83  & 0.34  & 0.27 \\
		(2013q3--2021q2) &       &       &       &       &       &  \\
		\hline\textit{Post-revolution full sample}%
$^2$ & 17.39 & 15.30 & 18.34 & 2.01  & 0.23  & 0.15 \\
		(1979q3--2021q1) &       &       &       &       &       &  \\
		\textit{Post-War full sample}%
$^2$ & 17.38 & 19.88 & 18.30 & 3.13  & 0.24  & 0.16 \\
		(1989q1--2021q1) &       &       &       &       &       &  \\
		\hline\hline
	\end{tabular}
\end{center}

\vspace{-0.3cm}%
\footnotesize
\textbf{Notes}: \emph{1}. Data on free market foreign exchange rate start in
1980q2. \emph{2}. Data on foreign exchange rates (free market and official
rate), and inflation end in 2021q1, data on output growth end in 2020q1, data
on sanctions intensity end in 2020q3. See Section \ref{Sec: sanctions measure}
of the paper for the sanctions intensity variable definition over the range
(0,1). See Sections \ref{Sec. sanction index construction},
\ref{Sec. calendar conversion}, and
\ref{Sec. socio-economic vars construction} in the data appendix of the online
supplement for details on the construction of the sanctions intensity
variable, calendar conversions, and sources of the data used.%

\label{tab:addlabel}
\end{table}%

The U.S. sanctions against Iran were mainly of extra-territorial nature.
Iran-U.S. trade had already been cut drastically after the Revolution and did
not recover after the resolution of the hostage crisis. In response to
sanctions, the Iranian government made concerted efforts to re-direct Iran's
foreign trade from the West to the East (primarily China) and to neighboring
countries, focussing on trading companies that operate outside the reach of
the U.S. Treasury. The sources of foreign exchange were also diversified from
oil to non-oil exports of goods and services. In this regard, ironically, Iraq
became one of the most important importers of Iranian products and services
(electricity, agricultural products and light manufacturing) after the fall of
Saddam in 2003. Turkey and Qatar have also played important roles in
facilitating Iran's foreign trade and international payments. \emph{Panel B}
of Figure \ref{fig: intro descriptives} shows that the share of oil and gas
exports declined steadily from 96 per cent of total exports in 1979 to around
60 per cent in 2018, before the full impact of the U.S. withdrawal from Iran's
oil exports. Over the same period non-oil exports have increased from 753
million dollars to 37 billion dollars.

Nevertheless, the Iranian financial system has found it difficult to adapt to
the new sanctions sufficiently quickly, resulting in large depreciations of
the free market rate of the rial against the U.S. dollar, with the official
rate lagging behind for a number of years, thus creating opportunities for
rent-seeking and often corrupt business practices.\footnote{The development of
the free market exchange rate, also known as the `black' market rate during
the 1979-1988 period, is discussed in \cite{pesaran1992}.} \emph{Panel C} of
Figure \ref{fig: intro descriptives} shows the evolution of the free and
official rates (in log scales), and clearly shows the episodic nature of the
jumps in the free market rate, followed by the sluggish catch up of the
official rate. Given the relevance of imports in the Iranian economy, and the
role of the U.S. dollar as the store of value and as a hedge against inflation
for many Iranians, the fall in value of the rial quite rapidly translates into
higher consumer prices, with the rise in prices somewhat moderated due to
government imports of food and medicine at official rates. But as the gap
between the official and market rates closes over time, consumer prices end up
reflecting the full extent of depreciation of the rial on the free market. The
time profiles of free market rate and consumer prices (in log scales) are
depicted in \emph{Panel D} of Figure \ref{fig: intro descriptives}, and show
the very close association that exists between the two series. As can be seen
from Table \ref{table: descriptives by presidency}, over the period
1989q1--2021q1 the free market rate has depreciated around 17.4 per cent per
annum as compared to the average annual rate of inflation of around 18.3 per
cent over the same period, representing a gap of around 1 per cent between the
two series. But according to the Purchasing Power Parity (PPP), the difference
between inflation and exchange rate depreciation should match the average
annual U.S. inflation, which is estimated to be around 2.5 per cent over the
same period.\footnote{The PPP is a long-run relationship that relates the
exchange rate between two currencies to their relative price of goods:
$P_{t}=E_{t}P_{t}^{\ast},$ with $E_{t}$ being the exchange rate representing
the number of domestic currency units that can be bought with one unit of
foreign currency, $P_{t}$ and $P_{t}^{\ast}$ denote the domestic and foreign
price levels.} One reason for the discrepancy (of 1.5 per cent) could be due
to the official rate being kept systematically below the free market rate, and
the lower than expected long run inflation should be set against the cost of
price distortion and corruption that multiple exchange rate regimes entail.

It is also important to note that not all foreign exchange crises can be
traced to the intensification of sanctions. Iran has witnessed major currency
crises during all the four presidencies since 1989, whilst only the last two
currency crises can be directly attributed to intensification of the
sanctions. The currency crises during Rafsanjani and Khatami presidencies have
domestic roots resulting from the rapid expansion of imports and low oil
prices, coupled with accommodating fiscal and monetary
policies.\footnote{During the reconstruction period under President Rafsanjani
imports of goods and services doubled over the period 1989--1991 (rising from
13.5 to 25 billion dollars, and Iran's foreign debt rose to 23.2 billion
dollars by the end of 1993. For further details of the developments that led
to the currency crisis under President Rafsanjani, see Section 3 of
\cite{pesaran2000}.} As shown in Table \ref{table: descriptives by presidency}%
, the average rate of inflation has been systematically high across all the
four presidencies, and does not seem to correlate with changes in sanctions
intensity. Even under Khatami's Presidency the average annual inflation still
amounted to 14.5 per cent, despite his conciliatory foreign policy and a much
lower rate of currency depreciation (7.9 per cent as compared to 17.4 per cent
over the full sample).

Turning to the effects of sanctions on real output, \emph{Panel E} of Figure
\ref{fig: intro descriptives} shows the time profile of Iran's real Gross
Domestic Product (GDP) and world GDP in log scales. World output is computed
as a weighted average of some of the largest 33 economies with details
provided in the online supplement. Iran's average output growth rates over
various sub-periods starting with the period of the Revolution and the
Iran-Iraq War, and followed by the four presidencies since 1989 are also
summarized in Table \ref{table: descriptives by presidency}. After the 1.6 per
cent contraction during 1979q3--1989q2, Iran's economy recovered strongly
during President Rafsanjani's administration and grew by an average of 5.2 per
cent per annum, an upward trend which continued over the period of Khatami's
presidency.\footnote{Detailed analyses of reconstruction of the Iranian
economy under President Rafsanjani are provided in
\cite{karshenas_pesaran1995} and Chapter 5 of \cite{maloney2015}.} Iran's
output growth was also in line with the growth of world output during the
1989--2005 period. This period also coincided with the lowest levels of
sanctions intensity that Iran experienced over the past forty years, and is in
stark contrast with the growth performance under Ahmadinejad and Rouhani
presidencies that coincided with new U.S. and U.N. sanctions during the period
before JCPOA in 2015, and the subsequent "maximum pressure" strategy initiated
by Trump in 2018. The results in Table \ref{table: descriptives by presidency}
clearly provide a \textit{prima facie} case for sanctions having important
adverse effects on Iran's output growth. It is also worth noting that Turkey,
a country often compared to Iran being of a similar size and in a similar
region, grew by an average annual rate of 4 per cent over the 2005--2020
period as compared to Iran's 1.3 per cent average growth rate over the same period.

Comparing Iran's output growth with that of world output growth over the
1989--2019 period also suggests an output growth shortfall of around 1 per
cent per annum, which could be contributed to the sanctions, although such a
comparison does not take account of Iran's potential as an emerging economy.
Even if we exclude the war periods, we also observe a much larger output
growth volatility in Iran as compared to the volatility of world output growth
volatility or a number of emerging economies of similar size to Iran, such as
Turkey or Saudi Arabia. Iran's output growth volatility (as measured by
standard deviations of output growth) was almost five times as large as the
global output growth volatility (12.61 versus 2.69 per
cent).\footnote{\cite{mohaddes_pesaran2013} document the high volatility of
Iran's oil export revenues as one of the factors behind Iran's low growth and
high volatility. A large part of the volatility of Iran's oil export revenues
is traced to high volatility of barrels of oil exported, largely due to
vagaries of sanctions. By comparison the volatility of oil prices is of
secondary importance. This contrasts to the volatility of Saudi Arabia oil
revenues which is largely governed by changes in international oil prices.}
Comparing Iran and Turkey over the same period we also find that Turkey grew
at an average annual rate of 4 per cent with volatility of 10.8 per cent, a
country also known for high inflation and repeated currency
crises.\footnote{The average annual output growth of Saudi Arabia over the
2005-2019 period was also similar to Turkey and amounted to 4.3 per cent.}
Finally, sanctions have most likely also contributed to the de-coupling of the
Iranian economy from the rest of the world. Again comparing Iran and Turkey,
we note that the correlation of Iran's output growth with the world output
growth is around 0.12, barely statistically significant, as compared with a
correlation of 0.33 for Turkey.

There seems little doubt that sanctions have adversely affected the Iranian
economy, contributing to low growth, high inflation and increased volatility.
What is less clear is how to carry out a quantitative evaluation and
identification of channels through which sanctions have affected the Iranian
economy over time, in particular the dynamics of responses and the time
horizon over which the effects of sanctions filter out throughout the economy.
To this end, a formal model is required where conditions under which the
effects of sanctions can be identified are made explicit, and their dynamic
implications are estimated and evaluated. It is to this task that we now turn
in the rest of this paper.

\section{Identification of sanctions effects: methodological
issues\label{Sec: methodology}}

Identifying the effects of sanctions on the Iranian economy is challenging
even if a reliable measure of sanctions intensity is available. As with all
macro policy interventions, when identifying the effects of sanctions we also
need to take account of confounding factors that are correlated with changes
in sanctions intensity, and which at the same time have a causal influence on
target variable(s) of interest such as output growth, inflation, health and
education outcomes. In situations where a policy intervention has differential
effects over time and across many different units such as households or firms,
difference-in-difference techniques are used whereby changes in outcomes
during policy on and policy off periods for those affected by the intervention
are compared to corresponding changes in outcomes for a control group that is
not directly affected by the intervention. This method is clearly not
applicable to the analysis of policy interventions that target a particular
entity such as a region or country, and a different approach is needed.
Currently, there are two such approaches: the Synthetic Control Method
(\textit{SCM}) advanced by \cite{abadie_gardezabal2003} and the Panel Data
Approach (\textit{PDA}) proposed by \cite{hsiao_et_al2012}.\footnote{Further
details and extensions of SCM are discussed in \cite{abadie2010} and
\cite{doudchenko_imbens2016}.} Both approaches compare outcomes for the
country (region) subject to the intervention with a weighted average of
outcomes from a control group. The former was originally applied to quantify
the economic costs of political instability in the Basque Country in Spain,
and the latter to evaluate the economic effects of the hand-over of Hong Kong
to China in 1997. Both studies consider discrete policy interventions and do
not allow for the policy intensity to vary over time. Perhaps most importantly
they both use pre-policy outcomes to estimate the weights applied to the
countries included in the control group. The main difference between the two
approaches lies in way the weights are
estimated.\footnote{\cite{gardeazabal_et_al2017} provide a comparative
simulation analysis of SCM and PDA, with a follow up critique by
\cite{wan_et_al2018}.}

The application of these approaches to the case of Iran is complicated by the
fact that imposition of sanctions coincided with the onset of the Revolution
which renders the pre-sanctions period of limited relevance. Also, as noted
earlier, the scope and intensity of sanctions against Iran have undergone
considerable changes over the past forty years and there are no clear cut
periods that one could identify as "sanctions on" periods to be compared to
"sanctions off" periods, in which all sanctions were levied. There is also the
additional challenge of identifying countries for inclusion in the control group.

To our knowledge, the only study that applies SCM to Iran is by
\cite{gharehgozli2017}, who considers the effects of the intensification of
sanctions just before the JCPOA agreement in July 2015 on Iran's real GDP,
treating the years 2011--2014 as the "sanctions on" period as compared to the
preceding years 1995--2010 as the "sanctions off" period. She then selects 13
countries worldwide to mimic a "synthetic" Iran as a weighted average of GDP
of these economies with their respective weights determined using the SCM
based on seven different macroeconomic indicators. She concludes that the
2011--2014 sanctions resulted in Iran's real GDP to fall by as much as 17 per
cent, as compared to the synthetic sanctions free Iran, with all the output
short fall attributed to sanctions.

We depart from the mainstream literature reviewed above and consider the
following model for Iran quarterly output growth%
\begin{equation}
\Delta y_{t}=\alpha+\lambda\Delta y_{t-1}+\psi_{0}s_{t}+\psi_{1}%
s_{t-1}+\boldsymbol{\beta}^{\prime}\mathbf{x}_{t}+\boldsymbol{\gamma}^{\prime
}\mathbf{f}_{t}+u_{t}, \label{Dy}%
\end{equation}
where $\Delta y_{t}$ is the output growth, $s_{t}$ measures the intensity of
sanctions against Iran, $\mathbf{x}_{t}$ and $\mathbf{f}_{t}$ are respectively
observed and unobserved control variables, and $u_{t}$ is an idiosyncratic
error term, distributed independently of $(s_{t},$\thinspace$\mathbf{x}%
_{t},\mathbf{f}_{t})$. The measurement of the sanctions intensity variable,
$s_{t}$, will be discussed in Section \ref{Sec: sanctions measure}, and will
be treated as given for now. It is assumed that part of the change in the
intensity of sanctions affects Iran's output growth with a lag, thus
distinguishing between short term, $\psi_{0},$ and long term, $\theta
=(\psi_{0}+\psi_{1})/(1-\lambda)$, effects of sanctions.

Despite the introduction of $s_{t}$, identification of the sanctions
coefficients, $\psi_{0}$ and $\psi_{1}$, depends on the confounding observed,
$\mathbf{x}_{t}$, and unobserved, $\mathbf{f}_{t}$, factors. In the present
reduced-form setting, for $\mathbf{x}_{t}$ we consider changes in
international oil prices. The effects of sanctions on output growth that
operate through exchange rate, liquidity, and inflation will be addressed in
Section \ref{Sec: Channels and dynamics of sanctions}. Here we are concerned
with both direct and indirect effects of sanctions on output growth and for
this reason we will not be including any of the domestic variables in
$\mathbf{x}_{t}$. The main challenge is how to identify and estimate
$\mathbf{f}_{t}$. In this regard our approach is closely related to the
\textit{PDA }(\cite{hsiao_et_al2012}). To this end, we consider the following
equations for output growth for the rest of the world%
\begin{equation}
\Delta y_{it}=\alpha_{iy}+\boldsymbol{\beta}_{iy}^{\prime}\mathbf{x}%
_{it}+\boldsymbol{\gamma}_{iy}^{\prime}\mathbf{f}_{t}+u_{y,it}\text{,
}i=1,2,...,n, \label{Dyi}%
\end{equation}
where $\Delta y_{it}\,$\ denotes output growth in country $i$ (excluding
Iran), $\mathbf{x}_{it}$ is a $k\times1$ vector of control variables specific
to country $i$, and $\mathbf{f}_{t}$ is the $m\times1$ vector of unobserved
common factors, and $u_{y,it}$ are idiosyncratic shocks to output growth that
are serially uncorrelated but could be weakly cross correlated.\footnote{A set
of random variables, $\left\{  u_{it},i=1,2,...,n\right\}  $ is said to be
weakly cross correlated if sup$_{j}\sum_{i=1}^{n}\left\vert Cov(u_{it}%
,u_{jt})\right\vert <C<\infty$. It then follows that $\sum_{i=1}^{n}%
w_{i}u_{it}=O_{p}(n^{-1/2})$, for any granular weights, $w_{i}$, such that
$w_{i}=O(n^{-1})$ and $\sum_{i=1}^{n}w_{i}^{2}=O(n^{-1})$. An obvious example
is the simple weights $w_{i}=1/n$. For further details see
\cite{chudik_etal2011}.} By allowing the factor loadings, $\boldsymbol{\gamma
}_{i}$, to differ across countries, we do not assume that all economies are
equally affected by the same factors, an assumption that underlies the DiD
approach. We also depart from SCM and PDA and, unlike these approaches, we do
not require a "donor pool" of countries to be selected for comparative
analysis. Instead, we assume that $\mathbf{x}_{it}$ also follows similar
multi-factor structures, and impose a rank condition which allows us to
identify $\mathbf{f}_{t}$ as weighted averages of $\Delta y_{it}$ and
$\mathbf{x}_{it}$ over $i$ (excluding Iran). Any granular weights can be used
to construct these averages, such as simple averages. But in cases where $n$
is not sufficiently large and there are dominant economies such as the U.S.,
it is advisable to use output shares as weights. Accordingly, suppose that%
\begin{equation}
\mathbf{x}_{it}=\boldsymbol{\alpha}_{ix}+\boldsymbol{\Gamma}_{ix}^{\prime
}\mathbf{f}_{t}+\mathbf{u}_{x,it}\text{, }i=1,2,...,n, \label{xi}%
\end{equation}
where $\boldsymbol{\Gamma}_{ix}$ is a $k\times m\,$matrix of factor loadings,
and $\mathbf{u}_{x,it}$ is a $k\times1$ vector that follows general stationary
processes that are weakly cross-sectionally correlated. Combining (\ref{Dyi})
and (\ref{xi}) we have%

\[
\left(
\begin{array}
[c]{cc}%
1 & -\boldsymbol{\beta}_{iy}^{\prime}\\
\mathbf{0} & \mathbf{I}_{k}%
\end{array}
\right)  \mathbf{z}_{it}=\left(
\begin{array}
[c]{c}%
\alpha_{iy}\\
\boldsymbol{\alpha}_{ix}%
\end{array}
\right)  \mathbf{+}\left(
\begin{array}
[c]{c}%
\boldsymbol{\gamma}_{iy}^{\prime}\\
\boldsymbol{\Gamma}_{ix}^{\prime}%
\end{array}
\right)  \mathbf{f}_{t}+\left(
\begin{array}
[c]{c}%
u_{y,it}\\
\mathbf{u}_{x,it}%
\end{array}
\right)  ,
\]
which yields $\mathbf{z}_{it}=\mathbf{c}_{i}+\mathbf{A}_{i}\mathbf{f}%
_{t}+\mathbf{B}_{i}\mathbf{u}_{it}$, where
\[
\mathbf{c}_{i}=\left(
\begin{array}
[c]{c}%
\alpha_{iy}+\boldsymbol{\beta}_{i}^{\prime}\\
\boldsymbol{\alpha}_{ix}%
\end{array}
\right)  \text{, }\mathbf{A}_{i}=\left(
\begin{array}
[c]{c}%
\boldsymbol{\gamma}_{iy}^{\prime}+\boldsymbol{\beta}_{iy}^{\prime
}\boldsymbol{\Gamma}_{ix}^{\prime}\\
\boldsymbol{\Gamma}_{ix}^{\prime}%
\end{array}
\right)  \mathbf{f}_{t}\text{, and }\mathbf{B}_{i}=\left(
\begin{array}
[c]{cc}%
1 & \boldsymbol{\beta}_{iy}^{\prime}\\
\mathbf{0} & \mathbf{I}_{k}%
\end{array}
\right)  .
\]
Averaging $\mathbf{z}_{it}$ over $i$ using the weights $w_{i}$ we now have
$\mathbf{\bar{z}}_{wt}=\mathbf{\bar{c}}_{w}+\mathbf{\bar{A}}_{w}\mathbf{f}%
_{t}+\sum_{i=1}^{n}w_{i}\mathbf{B}_{i}\mathbf{u}_{it},$where $\ \mathbf{\bar
{z}}_{wt}=\sum_{i=1}^{n}w_{i}\mathbf{z}_{it}$, $\mathbf{\bar{c}}_{w}%
=\sum_{i=1}^{n}w_{i}\mathbf{c}_{i}$, $\mathbf{\bar{A}}_{w}=\sum_{i=1}^{n}%
w_{i}\mathbf{A}_{i}$. Suppose now that the $(k+1)\times m$ matrix
$\mathbf{\bar{A}}_{w}$ is full column rank (that requires $m\leq k+1$) then we
can solve for $\mathbf{f}_{t}$ as\footnote{See \cite{pesaran2006} for further
details in a related context.}%
\[
\mathbf{f}_{t}=\left(  \mathbf{\bar{A}}_{w}^{\prime}\mathbf{\bar{A}}%
_{w}\right)  ^{-1}\overline{\mathbf{A}}_{w}^{\prime}\mathbf{\bar{c}}%
_{w}+\left(  \mathbf{\bar{A}}_{w}^{\prime}\mathbf{\bar{A}}_{w}\right)
^{-1}\mathbf{\bar{A}}_{w}^{\prime}\mathbf{\bar{z}}_{wt}-\left(  \mathbf{\bar
{A}}_{w}^{\prime}\mathbf{\bar{A}}_{w}\right)  ^{-1}\mathbf{\bar{A}}%
_{w}^{\prime}\sum_{i=1}^{n}w_{i}\mathbf{B}_{i}\mathbf{u}_{it}.
\]
Under the assumptions that $\mathbf{u}_{it}$ are weakly cross correlated,
$\mathbf{\bar{A}}_{w}^{\prime}\mathbf{\bar{A}}_{w}\rightarrow_{p}>0$, as
$n\rightarrow\infty$, then for \textit{any} choice of weights $w_{i}$ that are
granular it is possible to consistently estimate $\mathbf{f}_{t}$ up to
intercepts and an $m\times m$ rotation matrix using $\mathbf{\bar{z}}%
_{wt}=\left(  \Delta\bar{y}_{wt},\mathbf{x}_{wt}^{\prime}\right)  ^{\prime}$.
More specifically, it is easily established that $\sum_{i=1}^{n}%
w_{i}\mathbf{B}_{i}\mathbf{u}_{it}=O_{p}(n^{-1/2})$, and we have
$\mathbf{f}_{t}=\mathbf{\bar{a}}_{wf}+\mathbf{\bar{H}}_{w}\mathbf{\bar{z}%
}_{wt}+O_{p}(n^{-1/2})$, which can be used to eliminate the unobserved
factors, $\mathbf{f}_{t}$, from Iran's output growth equation. Specifically,
we obtain
\begin{equation}
\Delta y_{t}=\alpha_{yw}+\lambda\Delta y_{t-1}+\psi_{0}s_{t}+\psi_{1}%
s_{t-1}+\boldsymbol{\beta}^{\prime}\mathbf{x}_{t}+\theta_{yw}\Delta\bar
{y}_{wt}+\mathbf{\theta}_{xw}^{\prime}\mathbf{\bar{x}}_{wt}+u_{t}%
+O_{p}(n^{-1/2}). \label{Dy2}%
\end{equation}
Hence, for $n$ sufficiently large, and considering that the Iranian economy is
quite small relative to the rest of the world, the sanctions coefficients
$\psi_{0}$, and $\psi_{1}$ can be identified by augmenting the output growth
equations with the rest of the world average output growth, $\Delta\bar
{y}_{wt}$, and the cross section weighted averages of the observed drivers of
the rest of the world output growth, $\mathbf{\bar{x}}_{wt}$.

It is interesting to note that our approach does not favor selecting a pool of
countries that are close to Iran, but recommends including all countries,
possibly weighted for their relative importance in the world economy.
Selecting specific countries could bias the results by restricting the number
included in the construction of cross section averages. The rank condition,
$rank\left(  \mathbf{\bar{A}}_{w}^{\prime}\mathbf{\bar{A}}_{w}\right)  =m\,,$
for a given $n,$ and as $n\rightarrow\infty$, ensures that $\mathbf{f}_{t}$
has a reasonably pervasive effect on most economies which in turn allows us to
use $\Delta\bar{y}_{wt}$, and $\mathbf{\bar{x}}_{wt}$ as a reliable proxy for
$\mathbf{f}_{t}$.

The analysis of sanctions effects can also be extended to other macro
variables such as inflation and unemployment, and even to some key
socio-economic indicators such as life expectancy, death rate or educational
achievement, which has been made possible due to the unusually long duration
of sanctions in the case of Iran.

\section{Channels and dynamics of
sanctions\label{Sec: Channels and dynamics of sanctions}}

It is also important to consider the main channels through which sanctions
affect the economy, and to learn about the time profile of their effects. In
the case of Iran, new sanctions, or even their announcement, have invariably
led to a significant depreciation of the Iranian rial, reduced oil exports and
foreign exchange revenues, followed by a sharp rise in price inflation and
output losses within 3--6 months after the imposition of the new sanctions.
The dynamic inter-relationships of exchange rate, money supply, inflation and
output growth can be modelled using a structural vector autoregressive
(\emph{SVAR }for short) model augmented with the sanctions intensity variable,
oil price changes and other control variables, denoted by $\mathbf{\bar{z}%
}_{wt}$ above.

We denote by $\mathbf{q}_{t}=\left(  \Delta e_{ft},\Delta m_{t},\Delta
p_{t},\Delta y_{t}\right)  ^{\prime}$ an $m\times1$ (with $m=4$) vector of
endogenous domestic variables, where $\Delta e_{ft}$ represents the rate of
change of free market foreign exchange rate,\footnote{We also tried a weighted
average of free market and official rate, and we found that the free market
rate provides a more accurate and timely measure of the exchange rate
movements for Iran given its higher responsiveness to sanctions. Our variable
is expressed as Iran's rials per U.S. dollar.} $\Delta m_{t}$ is the growth
rate of money supply, $\Delta p_{t}$ is the rate of inflation, and $\Delta
y_{t}$ is real output growth.

To distinguish between different types of shocks and their implications for
the Iranian economy, in our SVAR\textit{ }we assume the direction of causality
goes from $\Delta e_{ft}$ to money supply growth, to inflation, and then to
output growth, as represented by the ordering of the four endogenous variables
in $\mathbf{q}_{t}$. Under this causal ordering, we are able to distinguish
changes in $\mathbf{q}_{t}$ that are due to variations in the intensity of
sanctions from those that are the result of domestic policy
shocks.\footnote{It is also possible to use non-recursive identification
schemes such as sign restrictions, or the more recently developed Bayesian
approach by \cite{baumeister_hamilton2015} to point identify and estimate
contemporaneous effects in the SVAR model and associated impulse responses
using priors. We do not expect that the main results of our paper will be much
affected by such alternative identification schemes.} The assumed causal
ordering can be justified in terms of relative speed with which the Iranian
economy responds to crises. It is the value of the rial in free market that
weakens first, followed by a potential expansion of liquidity, a rise in the
price of imported commodities, before the real economy starts to adjust to
higher prices and interest rates. Due to the relatively underdeveloped nature
of money and capital markets, monetary policy tends to be accommodating by
allowing a commensurate rise in liquidity.

In Equation (\ref{A0qt}), we look at the explanatory variables for
$\mathbf{q}_{t}=\left(  \Delta e_{ft},\Delta m_{t},\Delta p_{t},\Delta
y_{t}\right)  ^{\prime}$%
\begin{equation}
\mathbf{A}_{0}\mathbf{q}_{t}=\mathbf{a}_{q}+\mathbf{A}_{1}\mathbf{q}%
_{t-1}+\mathbf{A}_{2}\mathbf{q}_{t-2}+\boldsymbol{\gamma}_{0s}s_{t}%
+\boldsymbol{\gamma}_{1s}s_{t-1}+\mathbf{D}_{w}\text{ }\mathbf{\bar{z}}%
_{wt}+\boldsymbol{\varepsilon}_{t}, \label{A0qt}%
\end{equation}
where $s_{t}$ is the sanctions intensity variable, and $\mathbf{\bar{z}}%
_{wt}=(\Delta p_{t}^{0},\Delta\bar{y}_{wt},\Delta\overline{req}_{wt}%
,\Delta\bar{r}_{wt},grv_{t},\Delta\overline{e}_{wt})^{\prime}$ is a $k\times1$
($k=6$) vector of control variables that includes: the rate of change of oil
price, $\Delta p_{t}^{0},$ global output growth, $\Delta\bar{y}_{wt}$, global
equity returns, $\Delta\overline{req}_{wt}$, global long term interest rates
(in first difference), $\Delta\bar{r}_{wt}$, global realized volatility,
$grv_{t},$ and the rate of change of the global real exchange rate,
$\Delta\overline{e}_{wt}$.\footnote{Details on data sources and the
computation of the global variables are given in Section
\ref{Sec: data appendix} of the online supplement.} To reflect the assumed
causal ordering, we restrict $\mathbf{A}_{0}$ to be the following lower
triangular matrix%
\[
\mathbf{A}_{0}=\left(
\begin{array}
[c]{cccc}%
1 & 0 & \ldots & 0\\
-a_{\Delta m,\Delta e}^{0} & 1 & 0 & \vdots\\
-a_{\Delta p,\Delta e}^{0} & -a_{\Delta p,\Delta m}^{0} & 1 & 0\\
-a_{\Delta y,\Delta e}^{0} & -a_{\Delta y,\Delta m}^{0} & -a_{\Delta y,\Delta
p}^{0} & 1
\end{array}
\right)  ,
\]
where we expect $a_{\Delta p,\Delta e}^{0}\geq0$, with inflation responding
positively to a contemporaneous rise in $e_{ft}$ (rial depreciation). The
contemporaneous impacts of $\Delta e_{ft},$ $\Delta m_{t}$ and $\Delta p_{t}$
on output growth are less clear cut, considering that all four structural
equations are also conditioned on $s_{t}$, the variable representing sanctions
intensity. The structural shocks, $\boldsymbol{\varepsilon}_{t}=(\varepsilon
_{\Delta e,t},\varepsilon_{\Delta m,t},\varepsilon_{\Delta p,t},\varepsilon
_{\Delta y,t})^{\prime},$ are assumed to be serially uncorrelated with zero
means, $\mathbb{E}(\boldsymbol{\varepsilon}_{t})=0$, and the diagonal
covariance matrix $\mathbb{E}(\boldsymbol{\varepsilon}_{t}%
\boldsymbol{\varepsilon}_{t}^{\prime})=\boldsymbol{\Sigma}=Diag\left(
\sigma_{\Delta e,\Delta e},\sigma_{\Delta m,\Delta m},\sigma_{\Delta p,\Delta
p},\sigma_{\Delta y,\Delta y}\right)  $. Since we condition on sanctions
intensity and global indicators, the structural shocks can be viewed as
"domestic" shocks attributed to policy changes that are unrelated to
sanctions. Specifically, it is assumed that $\boldsymbol{\varepsilon}_{t}$ are
uncorrelated with $s_{t}$ and $\mathbf{\bar{z}}_{wt}$. Under these assumptions
it is now possible to distinguish between the effects of a unit change in the
sanctions variable, from domestic policy changes initiated by a unit standard
error change to the domestic shocks, $\boldsymbol{\varepsilon}_{t}$.
Specifically, for contemporaneous effects we have $\partial\mathbf{q}%
_{t}/\partial s_{t}=\mathbf{A}_{0}^{-1}\boldsymbol{\gamma}_{0s}$, and
$\partial\mathbf{q}_{t}/\partial\varepsilon_{jt}=\sqrt{\sigma_{jj}}%
\mathbf{A}_{0}^{-1}\mathbf{e}_{j}$ where $\mathbf{e}_{j}$ ($j=\Delta
e_{f},\Delta m,\Delta p,\Delta y$) are the vectors of zeros except for their
$j$-th component, which is one.

For the purpose of computing impulse responses and forecast error variance
decompositions, we model $s_{t}$ and $\mathbf{\bar{z}}_{wt}$ as autoregressive
processes:%
\begin{align}
s_{t}  &  =a_{s}+\rho_{s}s_{t-1}+\eta_{t},\label{s_AR1}\\
\mathbf{\bar{z}}_{wt}  &  =\mathbf{a}_{zw}+\mathbf{A}_{zw}\mathbf{\bar{z}%
}_{w,t-1}+\mathbf{v}_{wt}, \label{zw}%
\end{align}

where the sanctions and global shocks, $\eta_{t}$ and $\mathbf{v}_{wt}$, are
serially uncorrelated with zero means, and variances $\omega_{s}^{2}$ and
$\boldsymbol{\Omega}_{w}$. Combine equations (\ref{A0qt}), (\ref{s_AR1}), and
(\ref{zw}), we obtain the following\ SVAR model%
\begin{equation}
\boldsymbol{\Psi}_{0}\mathbf{z}_{t}=\mathbf{a+}\boldsymbol{\Psi}_{1}%
\mathbf{z}_{t-1}+\boldsymbol{\Psi}_{2}\mathbf{z}_{t-2}+\mathbf{u}_{t},
\label{psi0zt}%
\end{equation}
where $\mathbf{z}_{t}=\left(  \mathbf{q}_{t}^{\prime},s_{t},\mathbf{\bar{z}%
}_{wt}^{\prime}\right)  ^{\prime},$ $\mathbf{a}=\left(  \mathbf{a}_{q}%
^{\prime},a_{s},\mathbf{a}_{zw}^{\prime}\right)  ^{\prime},$ $\mathbf{u}%
_{t}=\left(  \boldsymbol{\varepsilon}_{t}^{\prime},\eta_{t},\mathbf{v}%
_{wt}^{\prime}\right)  ^{\prime},$ are $(m+k+1)\times1$ vectors, and%
\[
\boldsymbol{\Psi}_{0}=\left(
\begin{array}
[c]{ccc}%
\mathbf{A}_{0} & -\boldsymbol{\gamma}_{0s} & -\mathbf{D}_{w}\\
\mathbf{0} & 1 & \mathbf{0}\\
\mathbf{0} & \mathbf{0} & \mathbf{I}_{k}%
\end{array}
\right)  ,\text{ }\boldsymbol{\Psi}_{1}=\left(
\begin{array}
[c]{ccc}%
\mathbf{A}_{1} & \boldsymbol{\gamma}_{1s} & \mathbf{0}\\
\mathbf{0} & \rho_{s} & \mathbf{0}\\
\mathbf{0} & \mathbf{0} & \mathbf{A}_{zw}%
\end{array}
\right)  ,\text{ }\boldsymbol{\Psi}_{2}=\left(
\begin{array}
[c]{ccc}%
\mathbf{A}_{2} & \mathbf{0} & \mathbf{0}\\
\mathbf{0} & 0 & \mathbf{0}\\
\mathbf{0} & \mathbf{0} & \mathbf{0}%
\end{array}
\right)  ,
\]
are $(m+k+1)\times(m+k+1)$ matrices. Standard techniques can now be applied to
Equation (\ref{psi0zt}) to obtain impulse response functions and error
variance decompositions assuming the global shocks, $\mathbf{v}_{wt}$, are
uncorrelated with domestic and sanctions shocks (namely
$\boldsymbol{\varepsilon}_{t},$ and $\eta_{t}$). This is a standard small open
economy assumption which applies to the Iranian economy in particular since
its relative size in the world economy is small and has been declining over
the past forty years.

\subsection{Impulse response analysis for SVAR model of the Iranian
economy\label{Sec: irf description}}

The SVAR model can also be used to compute the time profile of the responses
of the economy to shocks (sanction, domestic and foreign) using impulse
response functions (IRFs). For the purpose of computing IRFs, we drop money
supply growth and foreign variables except the world output growth, as none of
these variables will prove to be statistically significant.

\subsubsection{IRFs for domestic shocks}

Our starting point is Equation (\ref{A0qt}) where $\mathbf{q}_{t}=\left(
\Delta e_{ft},\Delta m_{t},\Delta p_{t},\Delta y_{t}\right)  ^{\prime}$, and
there are four domestic shocks $\boldsymbol{\varepsilon}_{t}=(\varepsilon
_{\Delta e,t},\varepsilon_{\Delta m,t},\varepsilon_{\Delta p,t},\varepsilon
_{\Delta y,t})^{\prime}$. The IRFs of one standard error shock to domestic
shocks are given by%
\[
IRF_{\mathbf{q}}\mathcal{(}h,\sqrt{\sigma_{jj}})=E\left(  \mathbf{q}%
_{t+h}\left\vert I_{t-1},\varepsilon_{t,j}=\sqrt{\sigma_{jj}}\right.  \right)
-E\left(  \mathbf{q}_{t+h}\left\vert I_{t-1}\right.  \right)  ,\text{ for
}j=\Delta e_{f},\Delta m,\Delta p,\Delta y,
\]

\noindent where $h$\thinspace$=0,1,2...,H$, is the horizon of the IRFs,
$\sigma_{jj}=Var\left(  \varepsilon_{jt}\right)  $, and $I_{t-1}$ is the
information set at time $t-1$. The IRFs compare the expected outcome of the
shock (intervention) to an alternative counterfactual in the absence of the
shock. Using the reduced form version of (\ref{A0qt}), we have
$IRF_{\mathbf{q}}\mathcal{(}h,\sqrt{\sigma_{jj}})=\sqrt{\sigma_{jj}%
}(\mathbf{G}_{h}\mathbf{A}_{0}^{-1}\mathbf{e}_{j}),$ where
\begin{equation}
\mathbf{G}_{\ell}=\boldsymbol{\Phi}_{1}\mathbf{G}_{\ell-1}+\boldsymbol{\Phi
}_{2}\mathbf{G}_{\ell-2},\text{ for }\ell=1,2,\dots, \label{Gla}%
\end{equation}
with $\mathbf{G}_{-1}=\mathbf{0},$ and $\mathbf{G}_{0}=\mathbf{I}_{m},$
$\boldsymbol{\Phi}_{j}=\mathbf{A}_{0}^{-1}\mathbf{A}_{j}$, for $j=1,2$, and
$\mathbf{e}_{j}$ is a $m\times1$ selection\ vector of zeros except for its
$j^{th}$ element which is unity. See Chapter\ 24 of \cite{pesaran2015}. More
specifically, the impulse response effects of a positive one standard error
shock to the $j^{th}$ domestic variable, $\sqrt{\sigma_{jj}}$, on the $i^{th}$
variable at horizon $h=0,1,...,H$, are given by $IRF_{ij}\mathcal{(}%
h,\sqrt{\sigma_{jj}})=\sqrt{\sigma_{jj}}(\mathbf{e}_{i}^{\prime}\mathbf{G}%
_{h}\mathbf{A}_{0}^{-1}\mathbf{e}_{j})$, for $i,j=\Delta e_{f},\Delta m,\Delta
p,\Delta y$.

\subsubsection{IRFs for a shock to the sanctions intensity variable}

Since global factors are assumed to be strictly exogenous to the Iranian
economy and unrelated to sanctions, then without loss of generality the IRFs
of sanction shocks can be obtained abstracting from the global shocks.
Accordingly, using (\ref{A0qt}) and (\ref{s_AR1}), the moving average ($MA$)
representation of the domestic variables can be written as%
\begin{equation}
\mathbf{q}_{t}=\mathbf{G}(1)\mathbf{A}_{0}^{-1}\left(  \mathbf{a}_{q}%
+\frac{a_{s}}{1-\rho_{s}}\boldsymbol{\gamma}_{s}\right)  +\mathbf{b(}%
L\mathbf{)}\eta_{t}+\mathbf{G}(L)\mathbf{A}_{0}^{-1}\boldsymbol{\varepsilon
}_{t}, \label{qt}%
\end{equation}
where $\boldsymbol{\gamma}_{s}=\boldsymbol{\gamma}_{0s}+\boldsymbol{\gamma
}_{1s},$ $\mathbf{b(}L\mathbf{)}=\mathbf{G}(L)\mathbf{A}_{0}^{-1}(1-\rho
_{s}L)^{-1}\left(  \boldsymbol{\gamma}_{0s}+\boldsymbol{\gamma}_{1s}L\right)
$, $\mathbf{G}(L)=\sum_{\ell=0}^{\infty}\mathbf{G}_{\ell}L^{\ell}$, and
$\mathbf{G}_{\ell}$ is defined by the recursions in (\ref{Gla}). Therefore,
the responses of the $i^{th}$ domestic variable (the $i^{th}$ element of
$\mathbf{q}_{t}$) to a positive one standard error shock to the sanctions
intensity variable, $\omega_{s}$, are given by%
\begin{equation}
IRF_{i}\mathcal{(}h,\omega_{s})=\omega_{s}(\mathbf{e}_{i}^{\prime}%
\mathbf{b}_{h}),\text{ }h=0,1,...,H\text{, }i=\Delta e_{f},\Delta m,\Delta
p,\Delta y. \label{IRF_s}%
\end{equation}

\subsubsection{IRFs for a global factor shock}

As noted earlier, we only consider the shock to the world output growth,
$\Delta\overline{y}_{wt}$, as the global factor in our analysis, and consider
the following general linear process for $\Delta\overline{y}_{wt}$%
\begin{equation}
\Delta\overline{y}_{wt}=g_{0}+c(L)v_{\Delta\overline{y},t}. \label{zbar_wt}%
\end{equation}
Since the sanctions intensity variable and the world output growth are assumed
to be uncorrelated, abstracting from the sanctions intensity variable we can
re-write (\ref{A0qt}),\
\[
\mathbf{A}_{0}\mathbf{q}_{t}=\mathbf{a}_{q}+\mathbf{A}_{1}\mathbf{q}%
_{t-1}+\mathbf{A}_{2}\mathbf{q}_{t-2}+\boldsymbol{\delta}_{w}\Delta
\overline{y}_{wt}+\boldsymbol{\varepsilon}_{t}.
\]
By combining (\ref{zbar_wt}) with the moving average representation of the
above equation we have%
\begin{equation}
\mathbf{q}_{t}=\mathbf{G}(1)\mathbf{A}_{0}^{-1}(\mathbf{a}_{q}%
+\boldsymbol{\delta}_{w}g_{0})+\boldsymbol{\kappa}\mathbf{(}L\mathbf{)}%
v_{\Delta\overline{y},t}+\mathbf{G}(L)\mathbf{A}_{0}^{-1}%
\boldsymbol{\varepsilon}_{t}, \label{qt_bis}%
\end{equation}
where $\boldsymbol{\kappa}\mathbf{(}L\mathbf{)=\sum_{\ell=0}^{\infty
}\mathbf{\kappa}_{\ell}}L\mathbf{^{\ell}=G}(L)\mathbf{A}_{0}^{-1}%
\boldsymbol{\delta}_{w}c(L),$ and $\mathbf{G}(L)$ is as defined above. Hence,
the impulse responses of the $i^{th}$ element of $\mathbf{q}_{t}$ to a single
period shock to world output growth is then given by%
\begin{equation}
IRF_{i}\mathcal{(}h,\omega_{\Delta\overline{y}})=\omega_{\Delta\overline{y}%
}(\mathbf{e}_{i}^{\prime}\boldsymbol{\kappa}_{h}),\text{ }h=0,1,...,H\text{,
}i=\Delta e_{f},\Delta m,\Delta p,\Delta y, \label{IRF yw}%
\end{equation}
where $\omega_{\Delta\overline{y}}^{2}$ is the variance of $v_{\Delta
\overline{y},t}$.

\subsection{Forecast error variance decompositions\label{FEVD}}

Another useful measure of dynamic propagation of shocks is forecast error
variance decompositions (FEVDs), which measure the proportion of forecast
error variance of variable $q_{it}$ (say, output growth) which is accounted
for by a particular domestic shock, $\varepsilon_{jt}$, at different horizons.
We are particularly interested in estimating the relative importance of
domestic shocks \textit{vis-\`{a}-vis} sanctions or world output shocks in
explaining output growth at different horizons. To obtain the FEVDs of both
types of shocks, we first note that, by building on (\ref{qt}) and
(\ref{qt_bis}), the $n$-step ahead forecast errors for the vector of domestic
variables, $\mathbf{q}_{t}$, is given by%
\[
\boldsymbol{\xi}_{t}(n)=\sum_{\ell=0}^{n}\mathbf{b}_{\ell}\eta_{t+n-\ell}%
+\sum_{\ell=0}^{n}\boldsymbol{\kappa}_{\ell}v_{\Delta\overline{y},t+n-\ell
}+\sum_{\ell=0}^{n}\mathbf{G}_{\ell}\mathbf{A}_{0}^{-1}\boldsymbol{\varepsilon
}_{t+n-\ell},
\]
where, as before, $\boldsymbol{\varepsilon}_{t}$ is a $m\times1$ (with $m=4$)
vector of domestic shocks. Using standard results from the literature, the
$h$-step ahead FEVD of the $i^{th}$ variable in $\mathbf{q}_{t}$ which is
accounted by the domestic shock $\varepsilon_{jt}$ is given by%
\begin{equation}
\theta_{ij}(h)=\frac{\sigma_{jj}\sum_{\ell=0}^{h}\left(  \mathbf{e}%
_{i}^{\prime}\mathbf{G}_{\ell}\mathbf{A}_{0}^{-1}\mathbf{e}_{j}\right)  ^{2}%
}{\sum_{\ell=0}^{h}\mathbf{e}_{i}^{\prime}\mathbf{G}_{\ell}\mathbf{A}_{0}%
^{-1}\boldsymbol{\Sigma}\mathbf{\mathbf{A}_{0}^{\prime-1}G}_{\ell}^{\prime
}\mathbf{e}_{i}+\omega_{s}^{2}\sum_{\ell=0}^{h}\mathbf{e}_{i}^{\prime
}\mathbf{b}_{\ell}\mathbf{b}_{\ell}^{\prime}\mathbf{e}_{i}+\omega
_{\Delta\overline{y}}^{2}\sum_{\ell=0}^{h}\mathbf{e}_{i}^{\prime
}\boldsymbol{\kappa}_{\ell}\boldsymbol{\kappa}_{\ell}^{\prime}\mathbf{e}_{i}},
\label{FEVDij}%
\end{equation}
for $i,j=\Delta e_{f},\Delta m,\Delta p,\Delta y,$ and $\boldsymbol{\Sigma
}=Diag(\sigma_{\Delta e\Delta e},\sigma_{\Delta m\Delta m},\sigma_{\Delta
p\Delta p},\sigma_{\Delta y\Delta y})$. Similarly, the proportion of the
forecast error variance of the $i^{th}$ variable due to sanctions intensity
and world output growth shocks at horizon $h$ are given by%
\begin{equation}
\theta_{is}(h)=\frac{\omega_{s}^{2}\sum_{\ell=0}^{h}\mathbf{e}_{i}^{\prime
}\mathbf{b}_{\ell}\mathbf{b}_{\ell}^{\prime}\mathbf{e}_{i}}{\sum_{\ell=0}%
^{h}\mathbf{e}_{i}^{\prime}\mathbf{G}_{\ell}\mathbf{A}_{0}^{-1}%
\boldsymbol{\Sigma}\mathbf{A}_{0}^{\prime-1}\mathbf{G}_{\ell}^{\prime
}\mathbf{e}_{i}+\omega_{s}^{2}\sum_{\ell=0}^{h}\mathbf{e}_{i}^{\prime
}\mathbf{b}_{\ell}\mathbf{b}_{\ell}^{\prime}\mathbf{e}_{i}+\omega
_{\Delta\overline{y}}^{2}\sum_{\ell=0}^{h}\mathbf{e}_{i}^{\prime
}\boldsymbol{\kappa}_{\ell}\boldsymbol{\kappa}_{\ell}^{\prime}\mathbf{e}_{i}},
\label{FEVDsan}%
\end{equation}
and%
\begin{equation}
\theta_{i\Delta\overline{y}}(h)=\frac{\omega_{\Delta\overline{y}}^{2}%
\sum_{\ell=0}^{h}\mathbf{e}_{i}^{\prime}\boldsymbol{\kappa}_{\ell
}\boldsymbol{\kappa}_{\ell}^{\prime}\mathbf{e}_{i}}{\sum_{\ell=0}%
^{h}\mathbf{e}_{i}^{\prime}\mathbf{G}_{\ell}\mathbf{A}_{0}^{-1}%
\boldsymbol{\Sigma}\mathbf{A}_{0}^{\prime-1}\mathbf{G}_{\ell}^{\prime
}\mathbf{e}_{i}+\omega_{s}^{2}\sum_{\ell=0}^{h}\mathbf{e}_{i}^{\prime
}\mathbf{b}_{\ell}\mathbf{b}_{\ell}^{\prime}\mathbf{e}_{i}+\omega
_{\Delta\overline{y}}^{2}\sum_{\ell=0}^{h}\mathbf{e}_{i}^{\prime
}\boldsymbol{\kappa}_{\ell}\boldsymbol{\kappa}_{\ell}^{\prime}\mathbf{e}_{i}},
\label{FEVDzw}%
\end{equation}
respectively. Since all the shocks are assumed to be orthogonal, then it
follows that $\sum_{j=1}^{m}\theta_{ij}(h)+\theta_{is}(h)+\theta
_{i\Delta\overline{y}}(h)=1$.

\section{Measures of sanctions intensity\label{Sec: sanctions measure}}

Sanctions against Iran have been imposed with different degrees of intensity
over the past forty years. To account for both the prolonged nature of
sanctions and their time-varying intensity, we construct "sanctions on" and
"sanctions off" measures based on newspaper coverage of the imposition and the
occasional lifting of sanctions. Newspaper coverage has been used in the
literature and was initially formalized by \cite{baker2016} for measuring the
effects of economic uncertainty on macroeconomic aggregates. But, to our
knowledge, the idea of using newspaper coverage to quantify sanctions
intensity is new.

We consider six of the world's major daily newspapers, namely the New York
Times, the Washington Post, the Los Angeles Times and the Wall Street Journal
in the U.S., and the Guardian and the Financial Times in the U.K.. We then
count the number of articles published in these newspapers that deal with
sanctions and Iran.\footnote{The selected newspapers represent a balanced
sample of the most-read and well-informed articles over the past forty years,
and provide a good blend of both generalist press and those that focus on
economic-finance issues. Also, by including two different geographic regions,
we hope to cover a more diversified sample.} We are careful not to confound
our measures with articles that cover international sanctions against Iraq but
also mention Iran. Sources and details of how the searches were carried out
are provided in Section \ref{Sec. sanction index construction} of the online supplement.

One can think of our approach as measuring a proxy for an underlying latent
sanctions intensity process. The true process generates a signal, part of
which is captured in daily articles published in the six newspapers under
consideration. To be specific, let$\ n_{jdt}$ be the number of articles
published about sanctions on Iran in newspaper $j$ during day $d$ of month
$t$, and denote the true (latent) sanctions intensity variable during month
$t$ by $s_{t}^{\ast}$. The relationship between $n_{jdt}$ and $s_{t}^{\ast}$
is specified as%
\begin{equation}
n_{jdt}=\theta_{j}s_{t}^{\ast}+\zeta_{jdt}\text{,} \label{njdt}%
\end{equation}
where $\theta_{j}>0$ is loading of newspaper $j$ on the true signal,
$s_{t}^{\ast}$, and $\zeta_{jdt}\ $is an idiosyncratic serially uncorrelated
error term assumed to be distributed independently of the true signal,
$s_{dt}^{\ast}$, with zero means and finite variances. Equation (\ref{njdt})
could be viewed as a single factor model where $\theta_{j}$ is the
newspaper-specific factor loading. The number of articles published in
newspaper $j$ correlates with the true signal depending on the size of
$\theta_{j}$ and the variance of the idiosyncratic term. Clearly, not all
published articles capture the true signals, but by averaging across
newspapers and different days in a given month it is possible to reduce the
effects of the noise, $\zeta_{jdt}$, and obtain a consistent estimator of
$s_{t}^{\ast}$, up to a scalar constant. Both simple and weighted averages can
be used. Taking a simple average across the $J$ newspapers and the number of
days, $D_{t}$, in month $t$, we have $\bar{n}_{t}=\bar{\theta}_{J}s_{t}^{\ast
}+\overline{\zeta}_{t}$, where $\bar{n}_{t}=J^{-1}D_{t}^{-1}\sum_{j=1}^{J}%
\sum_{d=1}^{D_{t}}n_{jdt}$, $s_{t}^{\ast}=D_{t}^{-1}\sum_{d=1}^{D_{t}}%
s_{dt}^{\ast}$, and $\bar{\theta}_{J}=J^{-1}\sum_{j=1}^{J}\theta_{j}$. We
considered $6$ newspapers ($J=6$) over a number of publishing days per month
$D_{t}$, typically 26 days, resulting in about 156 data points over which to
average. This in turn ensures that the idiosyncratic errors get diversified,
and as a result the average error, $\overline{\zeta}_{t}$, becomes reasonably
small. Specifically%
\[
\overline{\zeta}_{t}=J^{-1}D_{t}^{-1}\sum_{j=1}^{J}\sum_{d=1}^{D_{t}}%
\zeta_{jdt}=O_{p}\left(  J^{-1}D_{t}^{-1}\right)  ,
\]
and we have $s_{t}^{\ast}=\bar{\theta}_{J}^{-1}\bar{n}_{t}+o_{p}(1).$These
monthly measures can then be time aggregated further to obtain quarterly or
annual series which are then used to identify the effects of $s_{t}^{\ast}$
(up to the scaling factor $\bar{\theta}_{J}^{-1}$) in our macro-econometric
model. We could also consider a weighted average version of $\bar{n}_{t}$
along the lines suggested in the literature, where the number of newspaper
articles (the raw count) is weighted by the inverse of their respective
standard deviations, $\hat{\sigma}_{jT}$, computed over the full data set,
using $\hat{\sigma}_{jT}=\sqrt{\left(  T-1\right)  ^{-1}\sum\nolimits_{t=1}%
^{T}\left(  \overline{n}_{jt}-\overline{n}_{j}\right)  ^{2}}$, $\overline
{n}_{jt}=D_{t}^{-1}\sum_{d=1}^{D_{t}}n_{jdt},$ and $\overline{n}_{j}%
=T^{-1}\sum_{t=1}^{T}\overline{n}_{jt}$. See \cite{baker2016} and
\cite{plante2019}. But, as reported in Figure
\ref{fig: sanction with and without weights} of the online supplement, the
simple and weighted averages, after being suitably scaled, are very close in
the case of our application.

Although most sanctions news has been about imposing new or tightening old
sanctions, there are some isolated periods where sanctions have been lifted,
as in 1981 after the release of the U.S. hostages, and over the period
2016q1--2018q2 after the implementation of JCPOA. Accordingly, we construct
two sanctions measures: an `on' measure, denoted by $s_{t,on}$, and an `off'
measure, denoted by $s_{t,off}$, and we normalize them such that they both lie
between $0$ and $1$, with $1$ representing the maximum sanctions intensity
over the full sample. We then define a net sanctions measure as $s_{t}%
=s_{t,on}-w\times s_{t,off}$, where $w>0$ represents the weight attached to
the sanctions off indicator compared to the sanctions on indicator. The
weight, $w,$ is estimated to be $\widehat{w}=0.4$ using a grid search method
over values of $w\in(0,1)$.\footnote{Section
\ref{Sec. sanction index construction} in the data appendix of the online
supplement provides further details on the estimation of $\widehat{w}.$}

\begin{sidewaysfigure}%
\caption{Sanctions intensity variable over the period 1989q1--2020q3}
\label{fig: s_graph}%

\[%
{\includegraphics[
height=4.6193in,
width=9.1177in
]%
{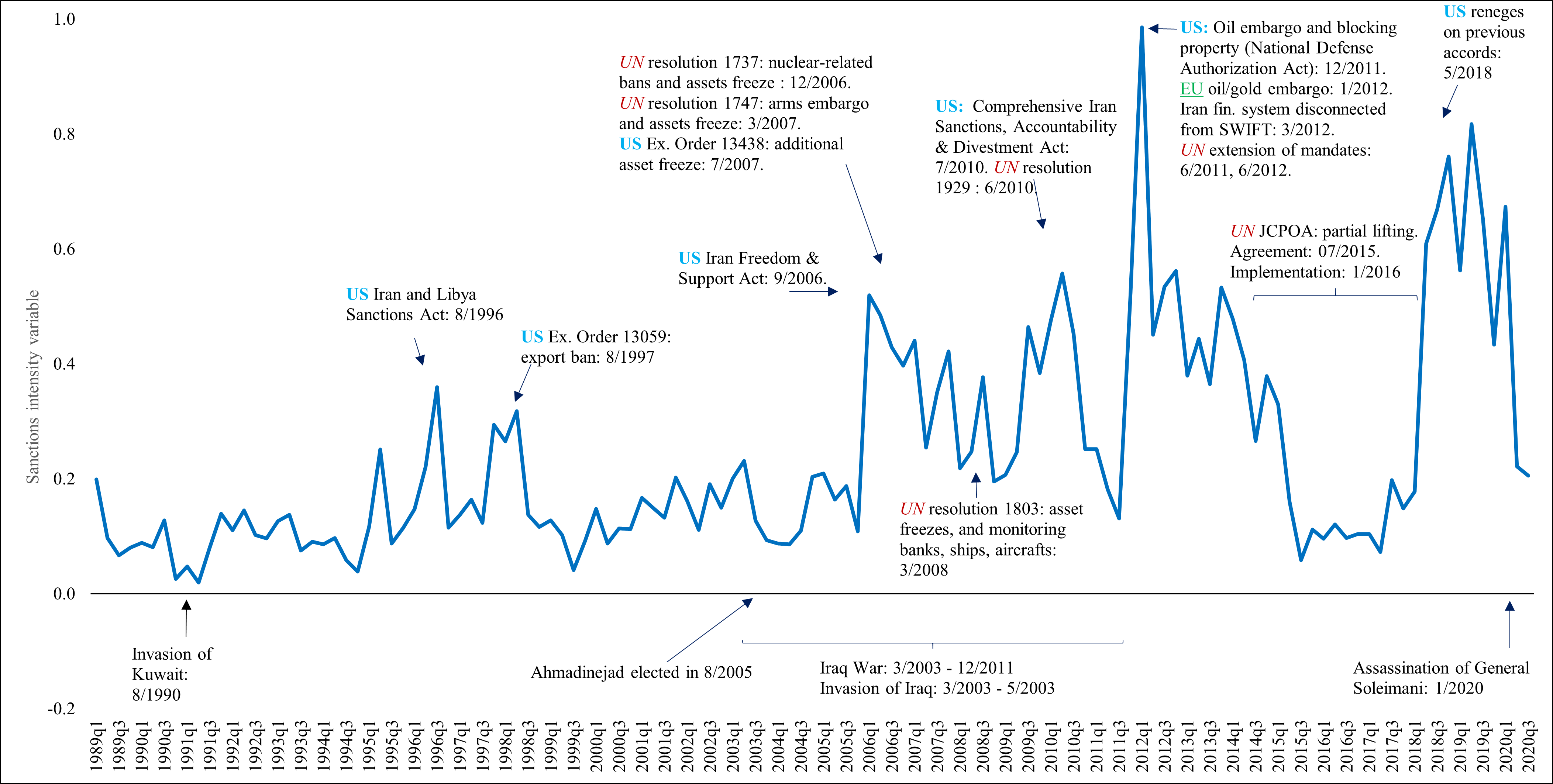}%
}
\]

\footnotesize
{}\textbf{Notes}: Major events related to the Middle East are indicated by
arrows below the $x$-axis. Major sanctions episodes related to Iran are
indicated by arrows above the $x$-axis. See Section
\ref{Sec. sanction index construction} in the data appendix of the online
supplement for details on the construction of the sanctions intensity variable.%

\end{sidewaysfigure}

Figure \ref{fig: s_graph} displays the quarterly estimates of $s_{t}$ over the
period from 1989q1 to 2020q3, which takes its maximum value at the end of 2011
when Iran was sanctioned simultaneously by the U.N., the U.S. and the E.U..
Important historical events are annotated in the lower part of the figure,
while specifics of particular sanctions are shown on the upper part of the figure.

The fact that intensity of sanctions against Iran has been quite varied can be
clearly seen from Figure \ref{fig: s_graph}. Most notably there are three
major spikes in sanctions intensity. The first is in 2006 after Ahmadinejad
was elected and Iran began its uranium enrichment program\textbf{, }when the
U.S. passed the "Iran Freedom and Support Act", which extended the coercive
measures against Iran -- most notably secondary sanctions on non-U.S.
corporations and institutions doing business with Iran and very strict
sanctions related to investments in the energy sector. An even larger spike
occurs between 2011 and 2012, when the Obama administration joined efforts
with the United Nations and the European Union to tighten the sanctions even
further with the aim of bringing Iran to negotiations over the nuclear
program. The U.S. passed stiff measures at the end of December 2011 under the
"National Defense and Authorization Act for Fiscal Year 2012", with Iran
threatening to block oil shipments through the Strait of Hormuz as a response.
At the same time the E.U. initiated a total disconnect of Iranian financial
institutions from the international payments system (\emph{SWIFT})\emph{ }in
March 2012,\footnote{SWIFT stands for the "Society for Worldwide Interbank
Financial Telecommunications", and it is a vast and secure network used by
banks and other financial institutions to operate financial transactions
across the globe.}\emph{ }while the U.N. proceeded to extend the mandates of
their previous resolutions between June 2011 and June 2012. The third, and
most recent, spike is registered in 2018q2 after Trump decided to unilaterally
withdraw the U.S. from the JCPOA accord and begin a strategy of "maximum
pressure". There are also minor spikes in 1996 when the Clinton administration
signed the "Iran and Libya Sanctions Act", and in 1997 when the U.S.
introduced an export ban to reduce the threat of potential weapons of mass
destruction being built, and in 2010 when the \emph{CISADA} ("Comprehensive
Iran Sanctions Accountability and Divestment Act") was signed into law and the
U.N. Security Council passed the fourth round of sanctions against Iran with
its 1929 resolution.%

Lows of the sanctions intensity variable are recorded during the
reconstruction period under President Rafsanjani and the pragmatic rule under
Khatami's administration, and more recently over the period between the JCPOA
agreement in August 2015 and January 2018, when the U.S. unilaterally withdrew
from the agreement. Table \ref{table: s_descriptives} provides summary
statistics (minimum, median, mean, maximum and standard deviations) of $s_{t}$
over a number of sub-periods. A number of interesting observations follow from
this table. First, the summary statistics for $s_{t}$ over the low sanctions
periods under Rafsanjani and Khatami are very close to those recorded for the
period 2015q1--2018q1 when sanctions were partially lifted after JCPOA.
Second, the peak of sanctions occurred during the internationally coordinated
efforts of 2011/2012 rather than after 2018, when the U.S. began their
"maximum pressure" strategy under Trump and Bolton. In the period after
2018q2, the degree of intensity of our indicator is 82 per cent of its peak in
2011. However, after 2018 the intensity of sanctions against Iran seems to
have been much more persistent: the mean and median are higher during the
2018q2--2020q3 period than during 2012q1--2014q4. Finally, we notice that
after the Iran-Iraq War, the median of the sanctions intensity has been only
two thirds of the mean: $0.16$ vs. $0.24$. This feature stems from the several
tail events that characterize the series of sanctions against Iran, and as an
overall measure the median is to be preferred to the mean.

For the analysis of the effects of sanctions on Iran, it is also important to
note that $s_{t}$ shows a considerable degree of persistence over time. When
sanctions are intensified they tend to remain high for some time before
subsiding. Table \ref{table: s-ar1-ar2} in the online supplement provides
estimates of first- and second-order autoregressive processes (AR) fitted to
$s_{t}$, and shows that an AR(1) model captures well the sanctions intensity
process, with a relatively large and highly significant AR coefficient, namely
$0.743$ $(0.059).$

\begin{table}[h!]
	\small\renewcommand{\arraystretch}{1.3}
\caption
{Descriptive statistics of the sanctions intensity variable over relevant time periods}
\vspace{-0.6cm}
\hspace{-0.2cm}%
\label{table: s_descriptives}%

\begin{center}
	\begin{tabular}{l|l|ccccc}
		\hline\hline\multicolumn{1}{r}{} & \multicolumn{1}{l}{Time period}
& Min   & Median & Mean  & Max   & St. Dev. \\
		\hline
Rafsanjani \& Khatami presidencies & 1989q3--2005q2 & 0.02  & 0.12  & 0.13  & 0.36  & 0.07 \\
		Ahmadinejad presidency & 2005q3--2013q2 & 0.11  & 0.39  & 0.38  & 1.0   & 0.17 \\
		U.N./U.S. max sanctions & 2012q1--2014q4 & 0.27  & 0.45  & 0.48  & 1.0   & 0.18 \\
		JCPOA agreement & 2015q1--2018q1 & 0.06  & 0.11  & 0.14  & 0.33  & 0.07 \\
		U.S. "maximum pressure" & 2018q2--2020q3 & 0.21  & 0.63  & 0.56  & 0.82  & 0.21 \\
		\hline\textit{Full sample}
(post Iran-Iraq War) & 1989q1--2020q3 & 0.02  & 0.16  & 0.24  & 1.0   & 0.19 \\
		\hline\hline\end{tabular}
\end{center}

\vspace{0.2cm}%
\footnotesize
{}\textbf{Notes: }See Section \ref{Sec. sanction index construction} in the
data appendix of the online supplement for details on the construction of the
sanctions intensity variable.%

\end{table}%

Finally, as a robustness check we also attempted to create an alternative
measure of sanctions intensity based on the number of Iranian entities being
sanctioned by the U.S.. We used the U.S. Treasury data set on entries and
exits of sanctioned companies, individuals and vessels. We were able to build
an indicator from 2006 to present. Although the two measures (newspaper
coverage and U.S. Treasury data) capture the sanctions phenomenon from
different perspectives, they correlate rather well at 38 per cent. For further
details see Section \ref{Sec. sanction index construction} of the online supplement.

\section{A time series structural model of
Iran\ \label{Sec: Structural model results}}

Equipped with the sanctions intensity measure, $s_{t}$, we now report the
results of estimating the SVAR model set out in Section
\ref{Sec: Channels and dynamics of sanctions} with the exchange rate variable,
$\Delta e_{ft}$, included first followed by money supply growth, $\Delta
m_{t}$, inflation, $\Delta p_{t}$, and output growth, $\Delta y_{t}$. We
estimated the four equations of the SVAR model including different sub-sets of
the control variables: world output growth, changes in international oil
prices, global realized volatility, world real equity returns, changes in long
term interest rates, and global real exchange rate changes against U.S.
dollar. The full set of regressions results are provided in Tables
\ref{table: svar_dfx} to \ref{table: svar_dy} in the online supplement. As can
be seen, none of the other control variables play a significant role in our
analysis with the exception of the world output growth. Accordingly, in Table
\ref{table: svar_main} we provide estimates of SVAR model including only the
world output growth ($\Delta\overline{y}_{wt}$) as a control variable.

The estimates of the effects of sanctions on the free market foreign exchange
rate variable, $\Delta e_{ft}$, are given under Column (1).\footnote{Note that
the exchange rate is expressed as the number of Iranian rials per one U.S.
dollar, and therefore a rise in the exchange rate variable corresponds to a
depreciation of the rial.} We first note that it is moderately persistent with
a persistent coefficient of $0.33$, partly reflecting inefficiencies in Iran's
foreign exchange market. Second, we observe that the rial depreciates strongly
in the same quarter in which sanctions are raised. The median fall in its
value is between $4.7$ and $5$ per cent per quarter.\footnote{To assess the
effect of sanctions, we need to multiply all coefficients related to $s_{t}$
and $s_{t-1}$ by the median $(0.16)$. See Table \ref{table: s_descriptives}.
In this way, we can assess the difference with respect to a case with no
sanctions \thinspace$(s_{t}=0)$. We consider the median rather than the
average intensity to maintain a conservative approach given that $s_{t}$
follows a non-Gaussian process with several outliers.} However, there is a
significant degree of overshooting, with the sanctions variable having the
opposite effect on exchange rate after one quarter. The rial appreciates by
about $3.8$ per cent in the following quarter, resulting in a less pronounced
overall impact of sanctions on the rial depreciation of around $2.5$ per cent
per quarter, or $10$ per cent per annum, which is still quite
substantial.\footnote{Such overshooting is well-known in the international
finance literature. See, for example,\emph{ }\cite{dornbusch1976}.} As can be
seen from Table \ref{table: svar_dfx} of the online supplement, these
estimates are remarkably stable and statistically significant\ at the $1$ per
cent level across all specifications regardless of the number of global
control variables included in the regression equation.\ In fact, none of the
lagged domestic variables (inflation, money supply growth, and output growth)
have a statistically significant effect on the exchange rate, and only foreign
output growth proves to be statistically significant at $10$ per cent level.
Most notably, changes in international oil prices do not have a statistically
significant impact on the exchange rate, which could be due to the fact that
once we condition on the sanctions variable, a rise in oil prices is less
likely to benefit Iran when oil exports are severely limited due to sanctions.
The adjusted $R^{2}$ of the exchange rate equations with world output growth
included is around $24$ per cent. This is high by the standard of exchange
rate equations, and is partly explained by the presence of the contemporaneous
sanctions variable in the regression. Its use for prediction requires
predicting the sanctions variable which adds another layer of uncertainty. It
is also noteworthy that there is no statistically significant evidence of
residual serial correlation in any of the exchange rate equations in Table
\ref{table: svar_dfx}. Lack of residual serial correlation is important for a
valid impulse response analysis and forecast error variance decompositions
that we shall consider below.

The estimates for the money supply growth ($\Delta m_{t}$) equation are
summarized in Column (2) of\textbf{ }Table \ref{table: svar_main}. As can be
seen, only lagged money supply growth is significant. It is also interesting
that the coefficient of lagged money supply growth has a negative sign
suggesting some overshooting of money supply growth which is then reversed in
the subsequent period. Notably, we do not find any feedback effects from
inflation to money supply growth.

The estimates for inflation ($\Delta p_{t}$) are summarized in Column (3) of
Table \ref{table: svar_main}. As discussed in Section
\ref{Sec: sanctions issues}, inflation in Iran has been persistently high over
the past forty years, and to capture its persistence it proved necessary to
include $\Delta p_{t-2}$, as well as $\Delta p_{t-1}$ in the regression
equation. It does not seem necessary to include second order lags of other
variables in the inflation equation.\footnote{See also Table
\ref{table: svar_dp} of the online supplement where different sub-sets of
control variables are also included in the regressions for the inflation
equation.} Perhaps not surprisingly, the estimates also show that exchange
rate depreciation is an important determinant of inflation in Iran, a factor
which is statistically significant and quantitatively important. The immediate
effect of one per cent depreciation of the free market exchange rate is to
raise prices by around $0.15$ to $0.16$ per cent, as many imported goods items
tend to rise with the fall in exchange rate. Sanctions affect inflation
indirectly through the exchange rate as well as directly, but the direct
effects of sanctions do not last and the net direct effects of sanctions on
inflation seem to be negligible. It is also interesting and quite surprising
that money supply growth or lagged output growth do not seem to have any
significant direct effects on inflation. But we do find some evidence of
global output growth positively affecting inflation, a kind of international
Phillips curve effect that leads to higher international prices that are in
turn reflected in Iran's import prices and hence domestic inflation.

Finally, Column (4) of Table \ref{table: svar_main} provides the results for
real output growth. Output growth in Iran is negatively autocorrelated, with a
coefficient estimated to be around $-0.195$ which is statistically
significant. This contrasts the positively autocorrelated output growth
observed for many other countries. The sanctions intensity variable affects
output growth with a lag, as it takes a few months for different sectors of
the economy to adjust to sanctions. After only one quarter, the effect of
sanctions on output growth is statistically highly significant.\footnote{Table
\ref{table: svar_dy} in the online supplement proves this to be a stable
finding under different specifications when we allow for a variety of control
variables.} Within two quarters the regression predicts Iran's output growth
to slow down by about $0.9$ per cent per quarter ($3.6$ per cent per annum).
In addition to this direct effect, sanctions also influence output growth
through exchange rate depreciation, which is also highly statistically
significant. This indirect effect amounts to around $0.125$ per cent per
quarter drop in output growth when the rial depreciates by one per cent.
Output growth is also negatively affected by lagged inflation, which
highlights the adverse effects of high and persistent inflation without any
short term Phillips curve type of trade off between inflation and output
growth. Interestingly enough, none of the global factors seem to have any
significant effects on Iran's output growth, partly due to Iran's relative
economic and financial isolation from the rest of the global economy.\ See
Table \ref{table: svar_dy} of the online supplement for further details.%

\begin{table}[h!]
\caption
{Quarterly estimates of the SVAR model of Iran with domestic variables ordered as: foreign exchange rate returns, money supply growth,
inflation, and output growth, estimated over the period 1989q1--2020q1}
\vspace*{-0.8cm}
\small\renewcommand{\arraystretch}{1.15}%
\label{table: svar_main}%

\begin{center}
		\begin{tabular}{@{\extracolsep{-15pt}}lD{.}{.}{-3} D{.}{.}{-3} D{.}{.}{-3}
D{.}{.}{-3} }
			\\[-1.8ex]\hline\hline\\[-2.7ex]
			\\[-1.8ex] & \multicolumn{1}{c}{$\Delta e_{ft}$} & \multicolumn{1}%
{c}{$\Delta m_{t}$} & \multicolumn{1}{c}{$\Delta p_{t}$} & \multicolumn{1}%
{c}{$\Delta y_{t}$} \\ \cline{2-5}
			\\[-1.8ex] & \multicolumn{1}{c}{(1)} & \multicolumn{1}{c}{(2)}
& \multicolumn{1}{c}{(3)} & \multicolumn{1}{c}{(4)}\\
			\hline\\[-1.8ex]
			$s_{t}$ & 0.303^{***} & -0.001 & -0.033^{***} & 0.021 \\
			& (0.061) & (0.024) & (0.012) & (0.025) \\
			$s_{t-1}$ & -0.245^{***} & 0.011 & 0.037^{***} & -0.058^{**} \\
			& (0.063) & (0.024) & (0.013) & (0.026) \\
			$\Delta e_{ft}$ &  & -0.015 & 0.162^{***} & -0.125^{***} \\
			&  & (0.033) & (0.017) & (0.044) \\
			$\Delta m_{t}$ &  &  & -0.032 & 0.097 \\
			&  &  & (0.048) & (0.094) \\
			$\Delta p_{t}$ &  &  &  & 0.341^{*} \\
			&  &  &  & (0.180) \\
			$\Delta\bar{y}_{wt}$ & -2.059^{*} & 0.135 & 0.721^{***} & -0.049 \\
			& (1.123) & (0.405) & (0.209) & (0.432) \\
			$\Delta e_{f,t-1}$ & 0.341^{***} & -0.040 & -0.010 & 0.033 \\
			& (0.090) & (0.034) & (0.019) & (0.035) \\
			$\Delta m_{t-1}$ & 0.350 & -0.289^{***} & -0.038 & -0.020 \\
			& (0.250) & (0.090) & (0.048) & (0.095) \\
			$\Delta p_{t-1}$ & -0.376 & 0.132 & 0.490^{***} & -0.496^{***} \\
			& (0.331) & (0.118) & (0.086) & (0.165) \\
			$\Delta y_{t-1}$ & -0.126 & -0.064 & 0.023 & -0.195^{**} \\
			& (0.242) & (0.086) & (0.046) & (0.088) \\
			$\Delta p_{t-2}$ &  &  & 0.174^{**} &  \\
			&  &  & (0.076) &  \\
			\hline\\[-1.8ex]
			Residual serial & 6.013 & 7.165 & 8.236 & 7.108 \\
			correlation test & [0.198] & [0.127] & [0.083] & [0.130] \\ \hline
Adjusted $R^{2}$ & 0.240 & 0.047 & 0.669 & 0.129 \\
			\hline\hline\\[-1.8ex]
		\end{tabular}
\end{center}%

\vspace{-0.4cm}%
\footnotesize
\textbf{Notes}: The variables are ordered as: $\Delta e_{ft},$ $\Delta m_{t},$
$\Delta p_{t},$\ and $\Delta y_{t},$ where: $\Delta e_{ft}=\ln(E_{ft}%
/E_{f,t-1}),$ $E_{ft}$ is the quarterly rial/U.S. dollar free market exchange
rate; $\Delta m_{t}=\ln(M_{2t}/M_{2,t-1}),$ $M_{2t}$ is obtained by summing
the aggregates $M1$ and "quasi-money"; $\Delta p_{t}=\ln(P_{t}/P_{t-1}),$
$P_{t}$ is the quarterly consumer price index of Iran; $\Delta y_{t}=\ln
(Y_{t}/Y_{t-1}),$ $Y_{t}$ is the quarterly real output of Iran. $s_{t}$ is the
quarterly sanctions intensity variable. $\Delta\overline{y}_{wt}$ is the
quarterly world output growth, computed as $\overline{y}_{wt}=\sum
\nolimits_{i=1}^{n}w_{i}y_{it},$ with $\left\{  y_{it}\right\}  _{i=1}^{n}$
being the natural log of real output for 33 major economies, and $\left\{
w_{i}\right\}  _{i=1}^{n}$ are GDP-PPP weights. Numbers in parentheses are
standard errors, and those in square brackets are p-values. ***$p<0.01$,
**$p<0.05$, *$p<0.1$. "Residual serial correlation test" is the
Breusch--Godfrey LM test of serially uncorrelated errors with lag order of the
test set to $4$.

See Sections \ref{Sec. sanction index construction},
\ref{Sec. calendar conversion}, and
\ref{Sec. socio-economic vars construction} in the data appendix of the online
supplement for details on the construction of the sanctions intensity
variable, calendar conversions, and sources of the data used. Regressions
results that include other global control variables (e.g. oil prices) are
provided in Tables \ref{table: svar_dfx}--\ref{table: svar_dy} in the online supplement.%

\end{table}%

It is also worth noting that our main findings are not much affected by
re-ordering of the domestic variables. In Section
\ref{Sec: online supplement reduced form results} of the online supplement we
provide results of estimating the SVAR model in (\ref{A0qt}), with two other
orderings of the domestic variables, namely ($\Delta p_{t},\Delta
e_{ft},\Delta m_{t},\Delta y_{t}$), and $\left(  \Delta m_{t},\Delta
e_{ft},\Delta p_{t},\Delta y_{t}\right)  $. The results are summarized in
Tables \ref{table: svar_dp order v2} to \ref{table: svar_dy order v2} and
\ref{table: svar_dm2 order v3} to \ref{table: svar_dy order v3}, respectively,
and clearly show that money supply growth plays a minimal role in
determination of inflation and exchange rate variations, and exchange rate
remains the primary\ driver of inflation and output growth.

Overall sanctions have affected Iran in a number of ways and through different
direct and indirect channels, the most important of which is the exchange rate
depreciation. The exchange rate depreciation itself could have its roots in
persistently high levels of inflation, coupled with a reduction in oil
revenues and anticipated decline in private sector activity. The currency
depreciation in turn leads to higher import prices and lower economic growth.
We also find that the \emph{direct} effect of sanctions on inflation is rather
small, compared to an average annual inflation norm of around $18$ per cent in
Iran (See Table \ref{table: descriptives by presidency}).

Money supply growth seems to follow patterns which are neither related to
sanctions nor to any of the domestic variables, notably inflation, which could
be due to the underdevelopment of capital and money markets in Iran, as
highlighted recently by \cite{mazarei2019}. These results seem quite robust to
other measures of liquidity such as M1 or private sector
credit.\footnote{Estimates based on these alternative measures of liquidity
are available upon request.}

\subsection{Impulse response analysis\label{Sec: IRF results}}

The estimates of the individual equations provided in Table
\ref{table: svar_main} provide a snap-shot of how sanctions interact with some
of the key macroeconomic variables. However, given the dynamic and
simultaneous nature of the model, to fully understand and evaluate the nature
and consequences of these interactions\ we need to resort to impulse response
functions (\emph{IRFs}) and forecast error variance decompositions
(\emph{FEVDs}) discussed in Sections \ref{Sec: irf description} and
\ref{FEVD}, respectively. We have seen that money supply growth does not play
much of a role in the determination of inflation and output growth, and is
hardly affected by sanctions. Also, amongst the control variables, only
foreign output growth seems to exert statistically significant effects on
inflation and output growth. For these reasons, to compute IRFs and FEVDs we
will be focussing on a parsimonious SVAR model without the money supply
growth, and only including $\Delta\overline{y}_{wt}$ as the control variable.
We also use AR(1) models for $s_{t}$ and $\Delta\overline{y}_{wt}$ to capture
the dynamics of these exogenous processes.\footnote{Time series evidence in
support of our choice of AR(1) specifications for $s_{t}$ and $\Delta
\overline{y}_{wt}$ are provided in Tables \ref{table: s-ar1-ar2} \ and
\ref{table: yw_ar1_ar2} of the online supplement. It is also worth noting that
the assumed AR(1) processes for $s_{t}$ and $\Delta\overline{y}_{wt}$ only
affect the IRFs and FEVDs, and do not affect the estimates of the SVAR model.}

\begin{figure}%
\caption
{Impulse responses of the effects of sanctions and domestic shocks on foreign exchange,
inflation, and output growth}%
\label{fig: IRF}%

\footnotesize
\vspace{-0.15cm}%
\[%
\begin{tabular}
[c]{c}%
$
\vspace{0.2cm}%
$\emph{Panel A}: One positive standard error shock to the sanctions intensity
variable\\
$%
\vspace{0.35cm}%
$%
{\includegraphics[
height=1.8458in,
width=5.6314in
]
{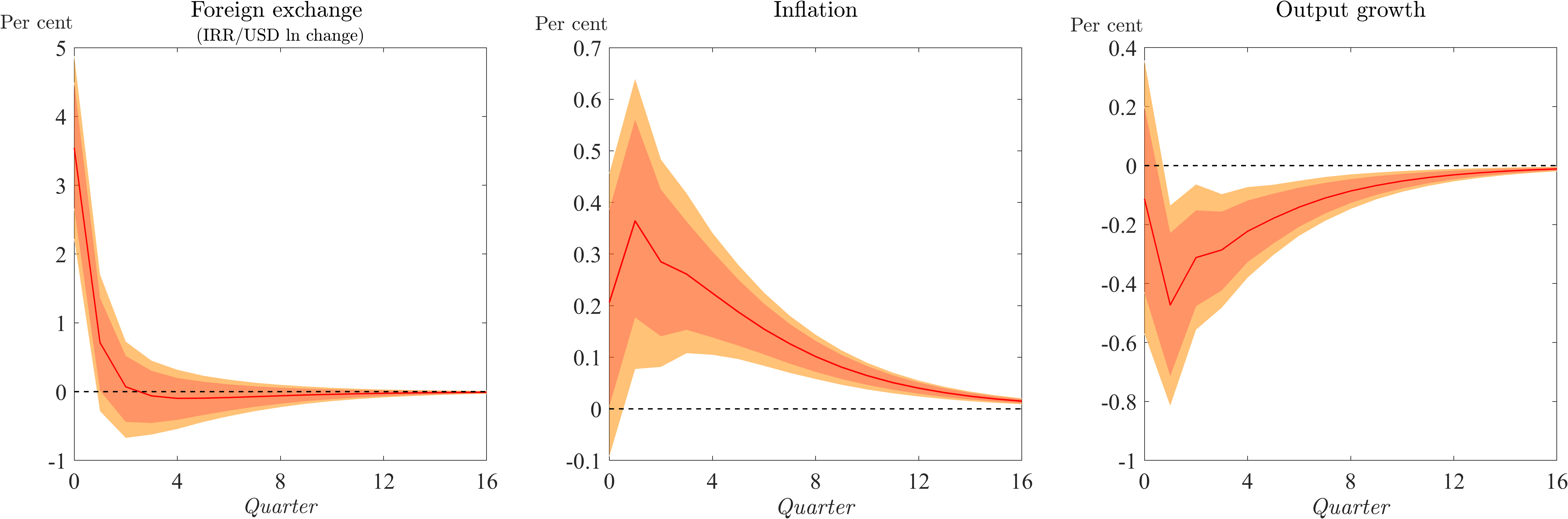}%
}
\\
$
\vspace{0.2cm}%
$\emph{Panel B: }One positive standard error shock to foreign exchange\\
$
\vspace{0.35cm}%
$%
{\includegraphics[
height=1.8458in,
width=5.6236in
]
{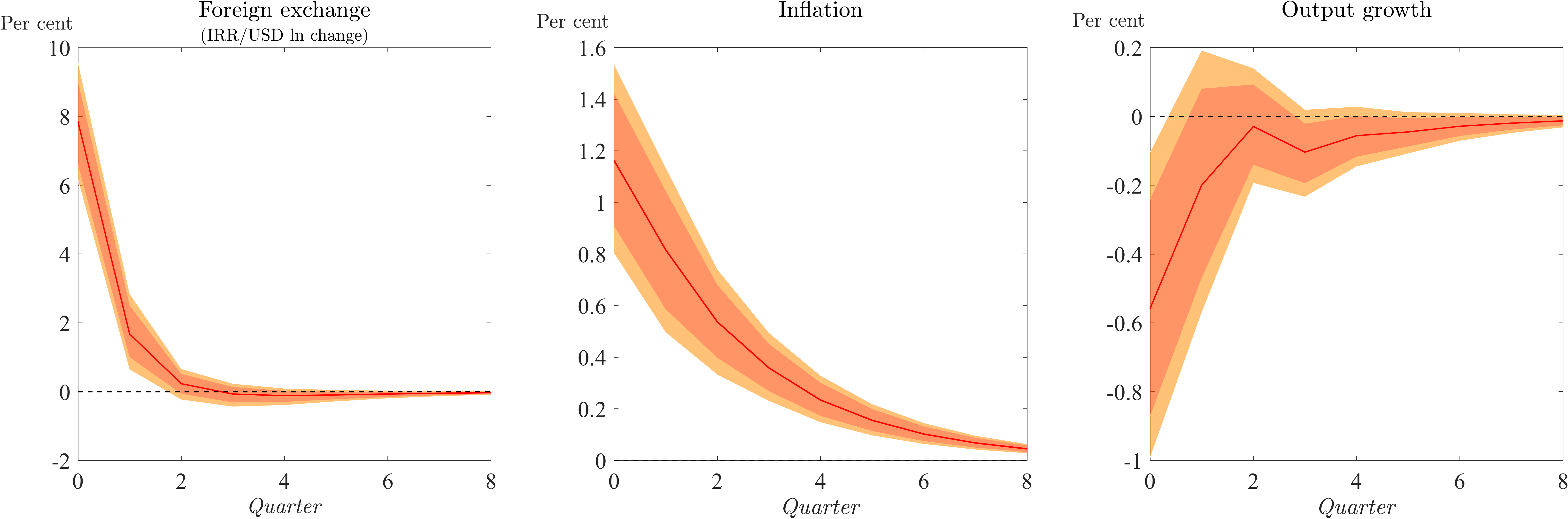}%
}
\\
$
\vspace{0.2cm}%
$\emph{Panel C: }One positive standard error shock to inflation\\
$
\vspace{0.35cm}%
$
{\includegraphics[
height=1.8458in,
width=5.6697in
]%
{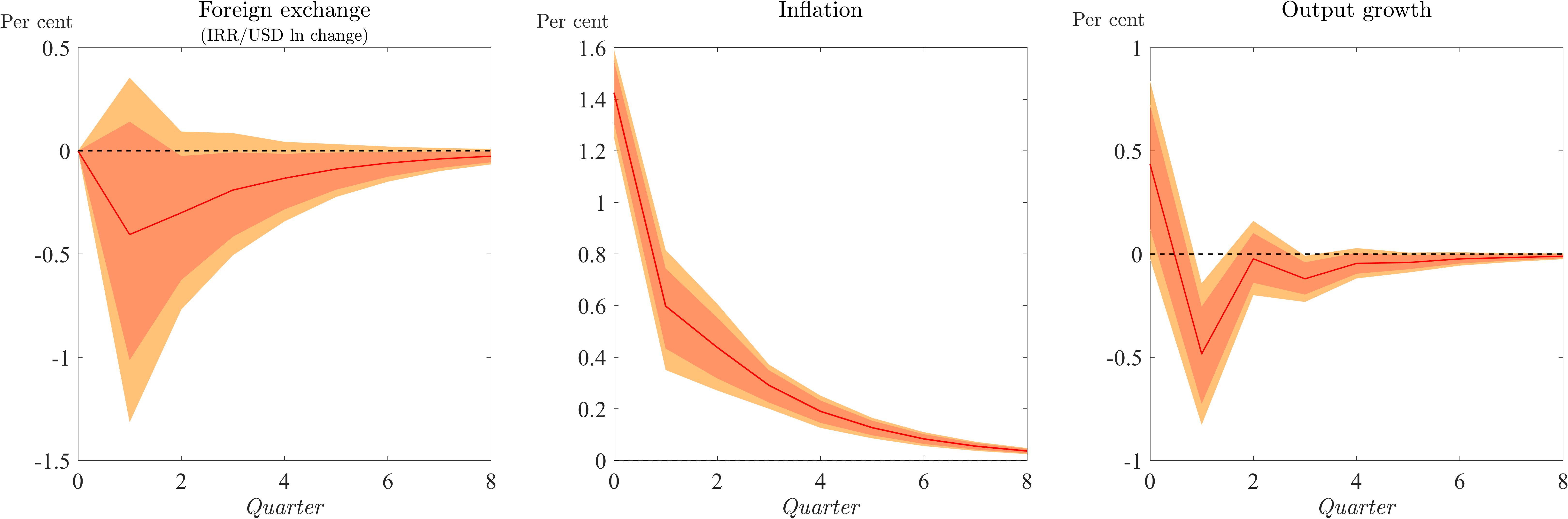}%
}
\\
$
\vspace{0.2cm}%
$\emph{Panel D: }One positive standard error shock to output growth\\%
{\includegraphics[
height=2.039in,
width=5.7028in
]%
{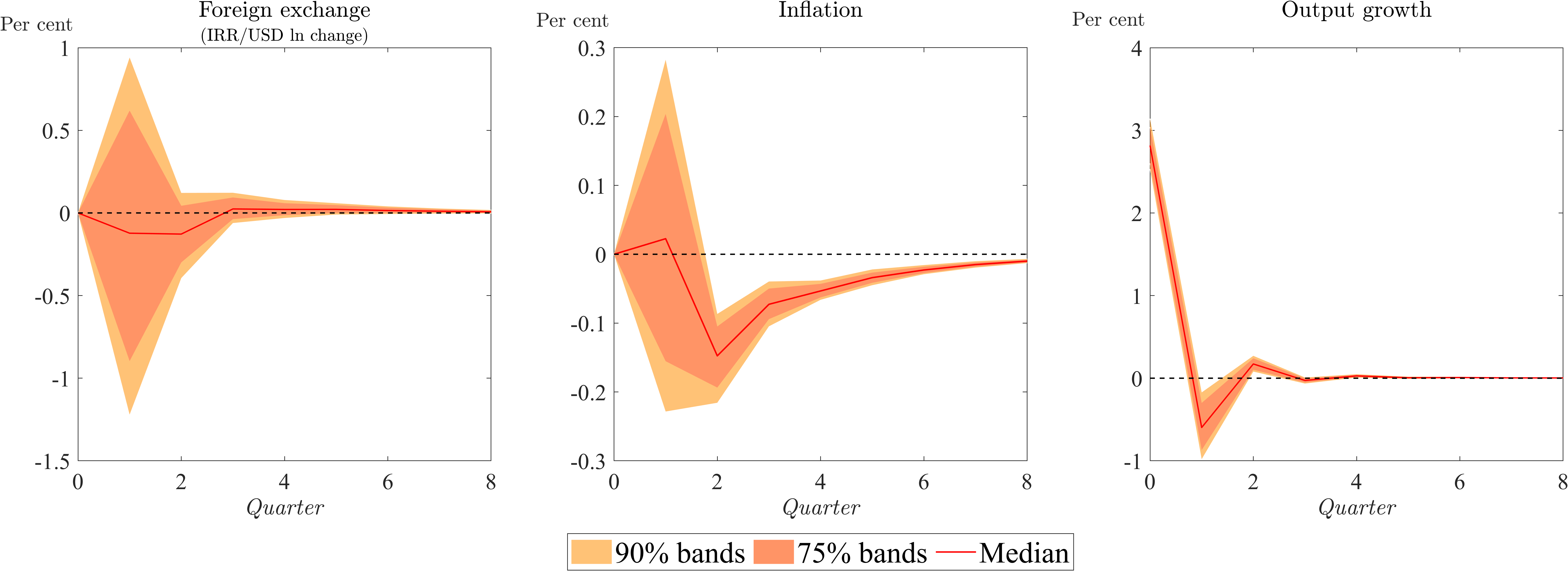}%
}%

\end{tabular}
\ \ \ \ \ \ \ \ \
\]
\bigskip%

\end{figure}

The IRFs for positive one standard error (s.e.) shocks to the three sanction
domestic shocks are displayed in Figure \ref{fig: IRF}. \emph{Panel A} of this
figure shows the results for the sanction shock.\footnote{One standard
deviation is equal to 0.125. See Table \ref{table: s-ar1-ar2}.} A single
quarter shock to sanctions intensity causes the foreign exchange rate to
depreciate by about $3$ per cent in the same quarter, but its effects are
rather short lived and become statistically insignificant two quarters after
the shock. For inflation and output growth the effects of the sanction shock
last much longer. Its effects on inflation are particularly persistent and
last at least for four years after the shock, although its magnitude is
relatively small: $0.3$ per cent increase per quarter in the first year. The
effects of sanction shock on output growth, on the other hand, are much larger
in size. A single period one standard error shock to sanctions causes output
growth to fall by more than $0.4$ per cent per quarter ($1.6$ per cent per
annum). The loss in output growth is still close to $0.2$ per cent per quarter
two years after the shock.

The results for the foreign exchange rate shock (independent of the sanction
shock) are given in \emph{Panel B} of Figure \ref{fig: IRF}. This shock
generates a sizeable and precisely estimated effect (of around $8$ per cent
per quarter) on exchange rate, but similar to the effects of the sanction
shock, it does not last long and its effects dissipate very quickly after two
quarters. The exchange rate shock raises inflation on impact by around $1.2$
per cent per quarter, and then starts to fall and vanishes completely after
about two years. The same is not true of real output growth. The direct
effects of foreign exchange shock on output growth are negative and
statistically significant but small in magnitude, around $-0.50$ per cent on
impact, which then moves towards zero very quickly.

\emph{Panel C} of Figure \ref{fig: IRF} gives the results for an inflation
shock (for example, due to a domestic expansionary policy). Again, because of
the highly persistent nature of inflation in Iran, the most pronounced effects
of the inflation shock is on inflation itself, raising inflation by $1.5$ per
cent per quarter on impact and then falling gradually to zero after two years.
Interestingly, the effect of inflation shock on exchange rate is not
statistically significant, suggesting that the causal link between them is
from exchange rate to inflation and not \textit{vice versa}. Compare the IRFs
for exchange rate and inflation in Panels B and C of Figure \ref{fig: IRF}.
The effects of inflation shock on output growth are positive on impact but
small in magnitude, and reverse quickly after one quarter, suggesting that it
might not be possible to increase output by expansionary policies.%

Finally, the IRFs of the effects of a positive shock to output growth are
displayed in \emph{Panel D} of Figure \ref{fig: IRF}. A positive output shock
could be due to technological advance or fundamental reforms that reduce
economic distortions and enhance production opportunities. The output shock
seems to have little impact (in short or medium term) on exchange rate, which
seems to be primarily driven by sanctions and its own dynamics. But the
positive output shock has a positive, but rather moderate, effect on
inflation, lowering inflation by $0.1$ per cent per quarter after two
quarters. The primary effects of the output shock are on output itself,
raising output by $2.8$ per cent per quarter on impact before losing momentum
in less than a year. The initial very large increase in output is somewhat of
an over-reaction which is then corrected slightly, yet providing a net $2$ per
cent rise in output within the year of the shock. Once again this result
highlights the importance of supply side policies for improving Iran's output
growth in the long run.\footnote{Figure \ref{fig: IRF global output growth} in
Section \ref{Sec: online supplement reduced form results}\ \ of the online
supplement provides an illustration of how the domestic variables respond to a
positive global output growth shock. It is found that the effects are
statistically insignificant, except for a small appreciation of the Iranian
rial in the first quarter.}

The impulse response analysis confirms some of the preliminary conclusions set
out in Section \ref{Sec: Structural model results}. Sanctions have their most
impact on free market exchange rate, and to a lesser extent on output growth.
Inflation has its own dynamics and is hardly affected by sanctions. The roots
of high and persistent inflation must be found in domestic economic
mismanagement. Also, sanctions do adversely affect output growth after one
quarter but such effects are short lived.

\subsection{Forecast error variance decompositions\label{Sec: FEVD results}}

We now turn to a quantification of the relative importance of sanctions as
compared to the three domestic shocks and the foreign output shock. Table
\ref{table: fevd single shock} presents the results.\footnote{FEVDs are
computed using equations (\ref{FEVDij}), (\ref{FEVDsan}), and (\ref{FEVDzw}).}
\emph{Panel A} of the table gives the results for the foreign exchange
variable. Not surprisingly, foreign exchange shocks are the most important,
and account for $82$ per cent of forecast error variance on impact and decline
only slightly, falling to $79$ per cent after $5$ quarters. Sanctions shock
accounts for $16$ per cent of the variance, with the other shocks contributing
very little. Therefore, isolated sanctions do not drive Iran's exchange rate,
and only become a dominant force if we consider prolonged periods over which
sanction shocks are in place with the same intensity.%

\begin{table}%
\caption
{Forecast error variance decomposition for domestic variables in the SVAR model with a single shock to sanctions}
	\renewcommand{\arraystretch}{1.05}
\small
\label{table: fevd single shock}%

\[
\text{\emph{Panel A:} Forecast error variance decomposition for foreign
exchange}%
\]

\vspace{-0.35cm}%
\begin{center}
	\begin{tabular}{c| p{0.1cm} p{1.8cm}p{1.5cm}p{1.5cm} p{1.5cm} p{1.5cm}}
		\hline\hline\multicolumn{1}{c}{\textit{Quarter }} &       & \multicolumn
{5}{c}{Proportion explained by a shock to:} \\
		\multicolumn{1}{c}{\textit{ahead}} &       & $s_{t}$ & $\Delta
e_{ft}$ & $\Delta p_{t}$ & $\Delta y_{t}$ & $\Delta\bar{y}_{wt}$ \\
		\hline0     &       & 0.16  & 0.82  & 0.00  & 0.00  & 0.02 \\
		1     &       & 0.16  & 0.80  & 0.00  & 0.00  & 0.03 \\
		2     &       & 0.16  & 0.80  & 0.01  & 0.00  & 0.03 \\
		3     &       & 0.16  & 0.80  & 0.01  & 0.00  & 0.03 \\
		4     &       & 0.16  & 0.80  & 0.01  & 0.00  & 0.03 \\
		5     &       & 0.16  & 0.79  & 0.01  & 0.00  & 0.03 \\
		6     &       & 0.16  & 0.79  & 0.01  & 0.00  & 0.03 \\
		7     &       & 0.16  & 0.79  & 0.01  & 0.00  & 0.03 \\
		8     &       & 0.16  & 0.79  & 0.01  & 0.00  & 0.03 \\
		\hline\hline\end{tabular}
\end{center}%

\[
\text{\emph{Panel B:} Forecast error variance decomposition for inflation}%
\]

\vspace{-0.35cm}%
\begin{center}
	\begin{tabular}{c|p{0.1cm} p{1.5cm}p{1.5cm}p{1.5cm} p{1.5cm} p{1.5cm}}
		\hline\hline\multicolumn{1}{c}{\textit{Quarter}} &       & \multicolumn
{5}{c}{Proportion explained by a shock to:} \\
		\multicolumn{1}{c}{\textit{ahead}} &       & $s_{t}$ & $\Delta
e_{ft}$ & $\Delta p_{t}$ & $\Delta y_{t}$ & $\Delta\bar{y}_{wt}$ \\
		\hline0     &       & 0.01  & 0.43  & 0.54  & 0.00  & 0.02 \\
		1     &       & 0.03  & 0.48  & 0.47  & 0.00  & 0.02 \\
		2     &       & 0.05  & 0.49  & 0.44  & 0.00  & 0.02 \\
		3     &       & 0.06  & 0.50  & 0.43  & 0.00  & 0.02 \\
		4     &       & 0.06  & 0.50  & 0.42  & 0.00  & 0.02 \\
		5     &       & 0.07  & 0.50  & 0.42  & 0.00  & 0.02 \\
		6     &       & 0.07  & 0.50  & 0.42  & 0.00  & 0.02 \\
		7     &       & 0.07  & 0.50  & 0.42  & 0.00  & 0.02 \\
		8     &       & 0.07  & 0.50  & 0.42  & 0.00  & 0.02 \\
		\hline\hline\end{tabular}
\end{center}%

\[
\text{\emph{Panel C:} Forecast error variance decomposition for output growth}%
\]

\vspace{-0.35cm}%
\begin{center}
	\begin{tabular}{c|p{0.1cm} p{1.5cm}p{1.5cm}p{1.5cm} p{1.5cm} p{1.5cm}}
		\hline\hline\multicolumn{1}{c}{\textit{Quarter}} &       & \multicolumn
{5}{c}{Proportion explained by a shock to:} \\
		\multicolumn{1}{c}{\textit{ahead}} &       & $s_{t}$ & $\Delta
e_{ft}$ & $\Delta p_{t}$ & $\Delta y_{t}$ & $\Delta\bar{y}_{wt}$ \\
		\hline0     &       & 0.00  & 0.04  & 0.03  & 0.93  & 0.00 \\
		1     &       & 0.02  & 0.04  & 0.05  & 0.88  & 0.00 \\
		2     &       & 0.04  & 0.04  & 0.05  & 0.87  & 0.00 \\
		3     &       & 0.04  & 0.04  & 0.05  & 0.86  & 0.00 \\
		4     &       & 0.05  & 0.04  & 0.05  & 0.85  & 0.00 \\
		5     &       & 0.05  & 0.04  & 0.05  & 0.85  & 0.00 \\
		6     &       & 0.05  & 0.04  & 0.05  & 0.85  & 0.00 \\
		7     &       & 0.05  & 0.04  & 0.05  & 0.85  & 0.00 \\
		8     &       & 0.05  & 0.04  & 0.05  & 0.85  & 0.00 \\
		\hline\hline\end{tabular}
\end{center}

\vspace{-0.15cm}%

\footnotesize

\textbf{Notes}: $s_{t}$ is the quarterly sanctions intensity variable. $\Delta
e_{ft}=\ln(E_{ft}/E_{f,t-1}),$ $E_{ft}$ is the Iran rial/U.S. dollar quarterly
free market exchange rate. $\ \Delta p_{t}=\ln(P_{t}/P_{t-1}),$ $P_{t}$ is the
quarterly consumer price index of Iran. $\Delta y_{t}=\ln(Y_{t}/Y_{t-1}),$
$Y_{t}$ is the quarterly real output of Iran. $\Delta\overline{y}_{wt}$ is the
quarterly world output growth: $\overline{y}_{wt}=\sum\nolimits_{i=1}^{n}%
w_{i}y_{it},$ with $\left\{  y_{it}\right\}  _{i=1}^{n}$ being the natural log
of real output for 33 major economies, and $w_{i}$ the GDP-PPP weights.

See Sections \ref{Sec. sanction index construction},
\ref{Sec. calendar conversion}, and
\ref{Sec. socio-economic vars construction} in the data appendix of the online
supplement for details on the construction of the sanctions intensity
variable, calendar conversions, and sources of the data used.%

\label{tab:addlabel}
\end{table}%

The FEVDs of inflation, reported in \emph{Panel B} of Table
\ref{table: fevd single shock}, show that foreign exchange and inflation
shocks account for the bulk of the variance, with sanction shocks accounting
for the remainder. Domestic and foreign output shocks make little
contribution. On impact, inflation shock accounts for $54$ per cent of the
variance, flattening out at $42$ per cent after $4$ quarters. In contrast, the
contribution of the foreign exchange shock rises from $43$ per cent on impact
to $50$ per cent after $3$ quarters. The contribution of the sanction shock is
not particularly large, and starts at $1$ per cent, but rises to $7$ per cent
after $4$ quarters. Once again we see that inflation and exchange rates in
Iran are mainly driven by domestic factors. But sanctions effects could
accumulate very quickly if we consider sanctions being in place over a
prolonged period of time.

Finally, the FEVDs of output growth are reported in \emph{Panel C} of Table
\ref{table: fevd single shock}. As can be seen, the output shock is by far the
most important shock and accounts for $93$ per cent of forecast error variance
of output growth on impact and falls only slightly to $85$ per cent after $4$
quarters. In line with our estimates, sanctions shocks do not affect output
growth on impact, and end up explaining only $5$ per cent of the variance
after $4$ quarters. Foreign output shocks do not have any explanatory power
for Iran's output growth. The other two domestic shocks (inflation and
exchange rate) together account for $9$ per cent of forecast error variance of
output growth after $1$ quarter.

These results show that without reducing the inherent volatility of output
growth (by reforms and better management), it is unlikely that sanctions
removal would enable the Iranian economy to return to single digit inflation
and sustained growth.

\subsection{Estimates of sanctions-induced output
losses\label{Sec: reduced form dy}}

So far we have focussed on the channels through which changes in the intensity
of sanctions affect the Iranian economy and the time profile of their
propagations. Here we consider possible output losses due to the direct and
indirect effects of the sanctions. However, the output growth equation
included in the SVAR model, and its estimates presented in Column (4) of Table
\ref{table: svar_main}, is not suitable for this purpose, since output growth
is specified to contemporaneously depend on exchange rate and inflation, and
therefore does not take account of output losses that could arise indirectly
through these variables. Instead, we consider the reduced form output growth
regressions set out in Equation (\ref{Dy2}), and focus on specifications with
$s_{t-1}$ as the intervention variable. We favor this specification over the
one that includes both $s_{t}$ and $s_{t-1}$, since "sanctions news" does not
contain anticipatory effects, and one would not expect contemporaneous changes
in $s_{t}$ to affect output growth, as time is required for the real economy
to adjust to sanctions news.\footnote{We are grateful to Nick Bloom for
drawing our attention to this point.} The estimates of the reduced form output
growth equations computed over the period 1989q1-2020q1 are summarized in
Table \ref{table: reduced form s(t-1) on dy} in the online supplement, where
we report both the short- and long-run effects of sanctions on output growth,
whilst allowing for a host of both domestic and foreign control
variables.\footnote{Amongst the domestic variables, only inflation has a
statistically significant impact on output growth. But, once again, we find
that global factors such as global volatility or output growth do not affect
Iran's output growth, largely due to Iran's relative economic isolation. The
only global factor which is statistically significant in the output growth
equation is the exchange rate. The negative effect of inflation on output
growth could be due to price distortions and allocation inefficiencies that
are often associated with high and persistent levels of inflation, as has been
the case in Iran. The negative effect of the global exchange rate variable on
output growth is more difficult to rationalize.} The parameter of interest is
the long run effect of sanctions on output growth reported at the bottom panel
of Table \ref{table: reduced form s(t-1) on dy}. It is estimated to be around
$-0.031$\ $(0.013)$, which is statistically significant and remarkably robust
across the seven different specifications reported.\footnote{Similar estimates
are obtained if both $s_{t}$ and $s_{t-1}$ are included in the regressions.
See Table \ref{table: reduced form s(t) and s(t-1) on dy} of the online
supplement.} This in turn suggests output growth losses of around $2$ per cent
per annum if we use the median value of $s_{t}$ over the sample under
consideration, or $3$ per cent if we use the mean value of $s_{t}%
$.\footnote{The median and mean values of $s_{t}$, are $0.16$ and $0.24$,
respectively, as summarized in Table \ref{table: s_descriptives}.} Due to the
large outliers in the sanctions intensity variable, we favor the lower
estimate of $2$ per cent based on the median value of $s_{t}$, which in turn
suggests that in the absence of sanctions and sanctions-induced mismanagement
Iran's average annual growth over 1989q1--2020q1 could have been around $4-5$
per cent, as compared to the $3$ per cent realized, a counterfactual outcome
which is close to the growth of emerging economies such as Indonesia, South
Korea, Thailand, and Turkey whose average annual growth rate over the same
sample period amounted to 4.8, 4.5, 4.2 and 4.0 per cent, respectively.

\section{Wider economic and socio-demographic effects of
sanctions\label{Sec: reduced form results}}

In this section we consider the effects of sanctions on the sectoral
composition of output, employment, labor force participation, education, and
consider if sanctions tend to have a gender bias. For this purpose we shall be
using annual data, as quarterly observations are not available for many
indicators of interest.

\subsection{Differential effects of sanctions on sectoral output
growths\label{Sec: reduced form results - structural transf}}

Since sanctions primarily affect oil revenues, foreign trade and international
payments, they are likely to have differential sectoral effects with traded
goods and financial sectors being more affected. In the case of Iran the least
traded goods sector is agricultural and parts of the service sector. It is,
therefore, plausible to expect sanctions to have less impact on the
agricultural sector as compared to the manufacturing and service sectors. To
investigate the sectoral effects of sanctions, we estimate VAR(1) models in
sectoral growth rates with the results summarized in Table
\ref{table: sector_var} of the online supplement. As can be seen, output
growths of manufacturing and service sectors are affected negatively by
sanctions and these adverse effects are statistically highly significant. But
the same can not be said about the output growth of the agriculture sector,
which seems to be hardly affected by sanctions. It is also interesting that
whilst there are significant inter-linkages between manufacturing and
services, there are no significant feedbacks between agriculture and the rest
of the economy. These findings are robust across the various specifications
that allow for a large combination of control variables.

We also note that sanctions have affected the manufacturing sector much harder
than their effects on the service sector. The median annual sanctions-induced
fall in manufacturing growth is about $2.4$ per cent, as compared to $1.2$ per
cent per for services. Even though industries such as banking and finance have
been seriously hampered by sanctions, services might have suffered less than
manufacturing as a result of the ongoing shift towards the knowledge-based
economy under way in Iran.

In the online supplement we report additional results for the shares of each
of these sectors out of the total value added.\footnote{See Tables
\ref{table: var_sector_share1} to \ref{table: var_sector_share3} of the online
supplement.} Our main finding is that sanctions targeting Iran have caused the
agricultural sector to become more important with respect to the overall
economy, while the manufacturing sector shrinks. Conversely, the share of
services out of the total is not significantly affected. Therefore, even
though sanctions have somewhat slowed down the growth of services value added,
the share of services out of the total value added has not declined in a
statistically significant way as a result of that.

These findings seem to confirm that sanctions are reflected not only in lower
output growth rates but also in a change of the sectoral composition. The
sectors that are more trade-intensive shrink the most, while a greater
reliance on agriculture and home-made innovations potentially comes into play.

\subsection{Effect of sanctions on labor
outcomes\label{Sec: reduced form results - labor}}

It is reasonable to expect that sanctions-induced disruptions and economic
slow downs have spill-over effects on the labor market, both in terms of
employment and labor force participation. To this end we run regressions of a
few labor market indicators on the sanctions intensity variable, with the
results summarized in Tables \ref{table: employment rate} to
\ref{table: labor force female} of the online supplement. We measure these
indicators relative to the trends in other MENA countries to take account of
common socio-economic factors of the region. We also include a number of
additional control variables in the regressions to reduce possible confounding effects.

Table \ref{table: employment rate} gives the regressions results for the rate
of change of employment in Iran relative to other countries in the MENA
region, and shows that Iran's employment rate contracts by about $1$ per cent
per year more than in other MENA countries as a reaction to
sanctions.\footnote{Recall that for assessing the effects of sanctions, we
need to multiply all coefficients related to $s_{t},$ and $s_{t-1}$, by their
median value of $0.16$, given in Table \ref{table: s_descriptives} of the
online supplement. In this way, we compare the effects of sanctions at their
median values with the no sanctions \thinspace$(s_{t}=0)$ outcomes.} No other
control variables have statistically significant effects except some mild
positive spill-overs from Turkey output growth, a feature that may relate to
the interconnections between the two economies.\footnote{Similar results are
obtained when using total labor force participation.\ See Table
\ref{table: labor force} of the online supplement.}

More interestingly, we find statistically significant evidence of a gender
bias. Sanctions, measured at their median value of $0.16$, resulted in male
labor force participation declining by $0.5$ per cent per year -- an impact
that is statistically significant but economically small. See Table
\ref{table: labor force male} of the online supplement.\textbf{ }However,
female labor participation was much more affected by sanctions, with the
coefficient of the sanctions variable being statistically highly significant
and much larger as compared to the results for males. See Table
\ref{table: labor force female} of the online supplement.\textbf{ }A median
level rise in sanctions intensity has caused the rate of female labor force
participation in Iran to fall by about $3.8$ per cent relative to other
MENA\ countries. To give a sense of the magnitude of such an effect, we notice
that after the period of highest international pressure 2006--2014, female
labor force participation rate declined by more than $30$ per cent (from its
peak of 19.5 per cent in 2005 to 13.4 per cent in 2014)\emph{.}

Overall, we confirm the economic intuition that links output declines to
employment losses. The worsening of labor market conditions is sizeable and
statistically significant. However, to the best of our knowledge, our study is
the first to provide quantitatively robust measures of gender bias in Iran
resulting from sanctions. There are many economic as well as social factors
behind this outcome. One plausible scenario points to resource allocation away
from females at times of reduced oil revenues and budget cuts, with males
given priority for jobs and wages over females.

\subsection{Effect of sanctions on education outcomes
\label{Sec: reduced form results - education}}

Finally, we ask if the education system has also been affected by sanctions.
If economic sanctions impair the government's budget then we would expect to
see some decline in expenditures on schools and teachers. We provide evidence
for the effects of sanctions on the number of lower secondary schools and high
schools in Tables \ref{table: school lower secondary} and
\ref{table: school high school} in the online supplement. For comparison we
also provide similar results for the number of primary schools in Table
\ref{table: school primary}. These results clearly show that indeed lower
secondary schools and high schools have been adversely affected by sanctions.
But, interestingly enough, the number of primary schools does not seem to have
been affected by sanctions, which could be due to the compulsory nature of
primary education in Iran, that obligates the government to provide schools
and teachers commensurate with the demographic rise in the number of
students.\footnote{The impact on the number of teachers is very similar to the
one experienced by the number of schools. See Tables
\ref{table: teachers primary}, \ref{table: teachers lower secondary} and
\ref{table: teachers high schools} in the online supplement.}

To investigate whether the gender bias documented in the case of labor force
participation is also present in the education sector, in Table
\ref{table: student gender ratio} of the online supplement we show regressions
of the rate of change of female-to-male student ratio on the sanctions
variable, as well as a number of control variables. We see that sanctions have
depressed this ratio, an effect which is sizeable as well as being
statistically highly significant.\textbf{ }Data from 1989 onwards confirm a
sanctions-induced reduction of about $0.5$ per cent per annum on average.

\section{Concluding remarks\label{Sec: conclusion}}

In this paper, using a novel measure of the intensity of sanctions based on
newspaper coverage, we have quantified the effects of sanctions on exchange
rate, inflation, output growth, employment, labor force participation, and
secondary and high school education in Iran. Our empirical analyses also show
that sanctions, perhaps unintentionally, accentuate rather than reduce gender
bias, and divert resources from education to other more pressing immediate
needs such as maintaining consumption at the expense of physical and human investments.

There is no doubt that sanctions have harmed the Iranian economy, but one
should not underestimate the damage done by years of economic mismanagement.
Iran's low output growth relative to its potential, high inflation and excess
output growth volatility cannot all be traced to sanctions and have domestic
roots stemming from prolonged periods of economic mismanagement, distorted
relative prices, rent seeking, a weak banking system and under-developed
financial institutions. Sanctions have accentuated some of these trends and
delayed the implementation of badly needed reforms.

A more comprehensive analysis of sanctions also requires detailed
investigation into how sanctions and their variability over the past forty
years have affected policy decisions at all levels, from monetary and fiscal
policies to industrial, regional and social policies. It is generally agreed
that, at times of increased sanctions intensity, governments fearful of
political consequences are reluctant to curtail distortionary policies, such
as large subsidies on food and energy, and they might even accentuate them, or
resort to multiple exchange rates to reduce the inflationary effects of sanctions.

In evaluating the direct and indirect costs of sanctions, we have followed the
literature and attempted to control for possible confounders, namely external
and domestic factors that affect the economy but are unrelated to sanctions,
such as advances in technology, world output growth, international prices, and
economic performance of some neighboring economies. Using a reduced form
regression of output growth on our sanctions intensity variable we estimate
Iran's output loss to be around 2 per cent per annum, which is considerable
when cumulated over time. There is, of course, a high degree of uncertainty
associated with such estimates which should be born in mind. But even if we
compare Iran's growth performance over the 1989q1-2021q1 period with that of
Turkey and other similar size emerging economies we find that Iran's realized
output growth of 3 percent still lies below the average growth of 4.4 per cent
experienced by Indonesia, Turkey, South Korea and Thailand over the same period.

We also recognize the importance of investigating the possible effects of
sanctions on price distortions and rising income inequality and poverty. Real
expenditure of households in Iran has been falling; more in urban than in
rural areas. The income gap is most pronounced between the top and the bottom
deciles of expenditure distribution. The Gini coefficient, a measure of
overall inequality, in Iran has been rising and is almost at the same level of
the Gini coefficient of the U.S., which is amongst the highest in advanced
economies. Another important area if the possible effects of sanction on price
distortions, and their implications for productivity and employment. However,
detailed studies of these important topics are beyond the scope of the present paper.

Sanctions have also led to some positive unintended effects. Non-oil exports
have risen from \$600 million before the Revolution to around \$40 billion,
resulting in greater foreign exchange diversification. The high-tech sector
has seen exponential growth over the past 10 years and is now one of the
regions' fastest growing sectors. Iran's major web-based companies have been
protected by potential competition from their U.S. counterparts shown in
brackets including: Digikala (Amazon), Aparat (YouTube), Cafe Bazaar (Google
Play), Snapp (Uber), Divar (Craigslist). It is estimated that over 65 per cent
of Iranian households are now connected to the internet. This rapid expansion
was facilitated by the government and security apparatus making affordable
high-speed internet a reality in Iran. The Mobile Telecommunication Company of
Iran, largely controlled by the Islamic Revolutionary Guard Corps now has over
43 million subscribers. Sanctions have also resulted in significant advances
in the areas of missiles and other military-related technologies. It is
estimated that IRGC control between 10-30 per cent of the economy, with large
stakes in the oil and gas sectors, construction, telecom, banking, and
tourism. One could argue that IRGC has been a major beneficiary of U.S. sanctions.

Our analysis also clearly shows that sanctions can only explain a relatively
small fraction of the variance of output growth, and the cause of Iran's low
and excessively volatile output growth should be found elsewhere. This in turn
suggests that lifting of sanctions will most likely have short-term positive
effects and for long term sustainability major reforms are required, in
addition to better management of the economy. This involves wide ranging
reforms of banking, the tax and subsidy system, unification of exchange rates,
and the development of a coherent system of social safety net to protect the
poor. Regional development policies should be initiated by giving priority to
remote regions that have been left behind. Government policies should become
more transparent with greater openness to private sector initiatives and
foreign investments. Insulating the economy against oil revenue volatility
will also become an urgent policy issue once sanctions are removed. In the
past, the Iranian economy has been very sensitive to oil price volatility and,
over the longer term, Iran needs a sovereign wealth fund to smooth the impact
of the volatility of oil revenues. The primary challenge is how to integrate
the Iranian economy within the regional and global economic system, so that
Iran's true economic potential can be realized.

Finally, it is important to bear in mind that Covid-19 could not have come at
a worse time for the Iranian economy. Our sample does not cover the period
from March 2020 when Covid-19 effects started to be felt in Iran. But it is
clear Covid-19 could have important medium term consequences, particularly for
the traditional service sector. The Covid shock has been truly global -- it
has hit almost 200 countries with different degrees of severity, with its
effects magnified through global trade and financial linkages. The full
economic impact of Covid-19 on the Iranian economy is unknown and requires
further investigation.

\begin{onehalfspacing}%
\small

\end{onehalfspacing}

\newpage%

\renewcommand{\thesection}{S.\arabic{section}}
\setcounter{table}{0}
\setcounter{figure}{0}
\setcounter{equation}{0} \setcounter{footnote}{0}
\setcounter{page}{1}
\setcounter{section}{0}
\renewcommand{\thetable}{S.\arabic{table}}
\renewcommand{\thefigure}{S.\arabic{figure}}
\renewcommand{\theequation}{S.\arabic{equation}}
\renewcommand{\thepage}{S.\arabic{page}}
\addcontentsline{toc}{section}{Supplement}
\onehalfspacing

\begin{center}
{\Large Online supplement to "Identifying the Effects of Sanctions on the
Iranian Economy using Newspaper Coverage"\bigskip}

{\large Dario Laudati}

{\small University of Southern California\bigskip}

{\large M. Hashem Pesaran}

{\small University of Southern California, USA, and Trinity College,
Cambridge, UK\bigskip}%

\date{August 2021}
{\Large \bigskip}
\end{center}

\section{Introduction}

This online supplement is composed of four main sections. Section
\ref{Sec: data appendix} gives details of data sources and construction of
some of the key variables used in our analysis. Sub-section
\ref{Sec. sanction index construction} provides details of how the proposed
sanctions intensity variables are constructed. Sub-section
\ref{Sec. calendar conversion} gives information on conversion of data from
the Iranian calender to the Gregorian calender. Sub-section
\ref{Sec. socio-economic vars construction} provides details of data sources
for the socio-economic variables. In Section \ref{Sec: additional methods} we
present details of the bootstrapping procedure used to compute error bands for
the impulse response functions (IRFs). Section
\ref{Sec: online supplement reduced form results} reports other empirical
results such as the AR specifications for the world output growth, additional
IRFs for a shock to the global output growth not presented in the paper, and
further estimates on the impact of sanctions on the Iranian economy and its
education system. Finally, a comprehensive list of all major sanctions against
Iran from November 1979 to January 2021 is provided in Table
\ref{table: chrono1}.

\section{Data appendix \label{Sec: data appendix}}

\subsection{Sanctions intensity variables
\label{Sec. sanction index construction}}

Our sanctions intensity variable, $s_{t}$, is based on newspapers coverage of
sanction events against Iran. Articles were retrieved from the platform
\emph{ProQuest}\textit{ } (www.proquest.com) which covers the whole period of
interest 1979q1--2020q3. ProQuest has detailed newspapers archives with good
search capabilities. The only exception to ProQuest was the \emph{Financial
Times Historical Archive} accessed through \emph{Gale} \emph{Historical
Newspapers }(www.gale.com/intl/primary-sources/historical-newspapers)\emph{,
}which helped to fill a gap left by ProQuest for articles published in the
Financial Times before 1996.

\bigskip

\noindent\textbf{Criteria of inclusion}

\noindent We focused on six major newspapers: the New York Times, the
Washington Post, the Los Angeles Times, the Wall Street Journal, the Guardian,
and the Financial Times. We only selected articles published in the newspapers
print version thus disregarding blogs, websites and other digital formats
which are only available more recently; however, we did allow for all types of
articles to be included, e.g. we included both editorials and main articles.

ProQuest has both a general \emph{ProQuest Central} database, holding
information for the relatively more recent publications, and several
historical newspaper-specific collections for the most highly printed world
outlets, \emph{ProQuest Historical Newspapers}, which proved useful in order
to extend our series back to 1979. Accordingly, we used the ProQuest Central
data for the maximum period available for each newspaper, and complemented
each series with the \emph{ad-hoc} historical data sets before such dates. See
Table \ref{Sec: newspapers sources} for details. As mentioned already, the
only exception was the articles published in the Financial Times before 1996,
for which there does not exist a historical archive on ProQuest, and instead
Gale Historical Newspapers were used.

To create the index of sanctions imposed on Iran ("sanctions on"), articles
were required to include the following terms: "economic*", "sanction*",
"against", "Iran*", with the additional feature of excluding articles in which
"lift*" was present. The star at the end of the previous words allowed the
search engine to pick words beginning with the same initial letters thus
including terms such as: "sanctioning", "Iranian", "lifting" etc.. Although
the number of potential synonyms and keywords to describe the phenomenon is
virtually very high, this set of words seemed to capture rather well the
extent to which Iran was mentioned as target of international measures. We
also found that further complicating the search did not produce sensible
results, as the new commands often could not be recognized by the search engine.

The search was carried out for each newspaper series separately by specifying
the name of the newspaper in the options list "Publication title -- PUB". For
some newspapers the search engine produced a handful of duplicates of the same
articles despite the option "Exclude duplicate documents"\ under "Result page
options"\ had being ticked. To address this issue, all articles were manually
checked before starting the download in order to avoid double-counting of
articles.\footnote{The extent of this technical hurdle varied considerably
amongst outlets. It was particularly severe for journals such as the Los
Angeles Times, while virtually non-existent for other newspapers such as the
New York Times.}

For the period 1990q3--1991q2, the search commands for sanctions against Iran
were updated to exclude also the word "Iraq". This adjustment was necessary in
order to avoid confounding noise due to the events of the Iraq invasion of
Kuwait in August 1990, and the subsequent Gulf War period, from January to
February 1991. These events received massive press coverage, which led Iran to
be mentioned for geopolitical reasons, not because of sanctions. Also, some
newspapers reported two additional small spikes not strictly related to Iran:
(\textit{i}) For the terrorist attacks happened between December 1985 (in Rome
and Vienna airports) and April 1986 (in a West Berlin discotheque);
(\textit{ii}) For the "1998 Coimbatore bombings"\ attacks in southern India.
In both cases, Iran was not the target of new sanctions therefore a manual
check deletion of these small number of occurrences had to be carried out.

The intensity variable to capture the partial lifting of sanctions ("sanctions
off") included the words beginning with "economic*", "sanction*", "against",
"Iran*"\ but now allowing also for \textit{at least} one of the following
words: "lift*", "waive*"\ and "accord*". An exception was made for the
Historical Database of the Financial Times, which does not support
sophisticated search structures. Therefore, a simple research allowing for
"sanctions against Iran"\ and "deal*"\ was conducted to capture the highest
number of articles, which were subsequently checked and skimmed manually to
meet our criteria of inclusion.

A detailed chronological study of economic sanctions against Iran allowed us
to restrict our search of "sanctions off" on two time periods only. First, in
1981 when the Algiers Accords were signed and the "Tehran hostage
crisis"\ ended; second,\ from 2016q1 to 2018q2, when the Joint Comprehensive
Plan of Action (\emph{JCPOA}) was enacted by all world major powers before
U.S. President Trump withdrew the country from the agreement. Accordingly, for
construction of the sanction-off index we focussed on the periods
1981q1--1981q4 and 2015q1--2018q2 in order to avoid unnecessary noise for the
time in between and after Trump's announcement. The "sanctions off" period of
our indicator was extended to one year before the actual implementation of the
JCPOA in order to allow for possible anticipatory effects.%

\begin{table}
\caption{Sources of newspaper articles over the period 1979m1--2020m9}
\renewcommand{\arraystretch}{1.1}
\label{Sec: newspapers sources}

\begin{center}
	\begin{tabular}{p{4cm} P{4cm} | P{4cm} }
		\hline\hline& \multicolumn{2}{c}{Period} \\
		\cline{2-3}          & Historical dataset & Modern dataset \\
		\cline{2-3}    New York Times & 1979m1--1980m12 & 1981m1--2020m9 \\
		Los Angeles Times & 1979m1--1984m12 & 1985m1--2020m9 \\
		Washington Post & 1979m1--2002m12 & 2003m1--2020m9 \\
		Wall Street Journal & 1979m1--1983m12 & 1984m1--2020m9 \\
		Guardian & 1979m1--1996m12 & 1997m1--2020m9 \\
		Financial Times & 1979m1--1995m12 & 1996m1--2020m9 \\
		\hline\hline\end{tabular}
\end{center}%

\footnotesize
\textbf{Notes}: "Historical data set" is the \emph{ProQuest Historical
Newspapers} data set for all newspapers except the Financial Times, for which
information have been retrieved from \emph{Gale Historical Newspapers.
}"Modern data set" is \emph{ProQuest Central} database for all newspapers considered.

\end{table}%

\bigskip

\noindent\textbf{Sanctions intensity variable construction}

\noindent Having obtained a number of daily articles related to the sanctions
imposed ("sanctions on") and lifted ("sanctions off"), we proceeded with the
following steps in order to build our estimator, $s_{t}(w)=s_{t,on}-w\times
s_{t,off}$. Here we focus on the construction of $s_{t,on}$. The same
procedure was used to construct $s_{t,off}.$

First, we computed a monthly series for each of our $J$ newspapers ($J=6$) by
averaging our daily series over the number of articles per month. In turn, we
carried out a simple average across newspapers, which led us to have a single
monthly series of "sanctions on" articles; subsequently, we averaged the
monthly observations over each quarter to obtain the quarterly series. The
"sanctions on" average was then divided by its maximum value over the period
1989q1--2020q3 in order to obtain the indicator $s_{t,on};$ so that $s_{t,on}$
index was\ defined on the $(0,1)$ range. We obtained a second variable
$s_{t,off\ }$ from our "sanctions off" raw count by following the same steps
just described. Finally, we estimated the weight, $w\in(0,1),$ with a grid
search in order to derive our final sanctions intensity variable
$s_{t}=s_{t,on}-w\times s_{t,off}$. The grid search was performed by running
the regressions:%
\[
\Delta y_{t}=\beta_{0}+\beta_{1}\Delta y_{t-1}+\beta_{2}s_{t-1}(w)+\varepsilon
_{t},
\]
\noindent over the period 1989q1--2020q1, with $\Delta y_{t}$\ being Iran's
quarterly real output growth,\ and with a step size of our grid equal to
$0.1$. The optimal weight was estimated as $\hat{w}=0.4,$ although the shape
of the likelihood was rather flat.\footnote{The \emph{annual} series for
$s_{t}$ was computed as a simple average of the quarterly intensity variable
over each year.}

\begin{figure}
\caption
{Sanctions intensity variable and standardized sanctions intensity variable over the period 1989q1--2020q3}
\label{fig: sanction with and without weights}

\[%
{\includegraphics[
height=3.0581in,
width=4.2608in
]%
{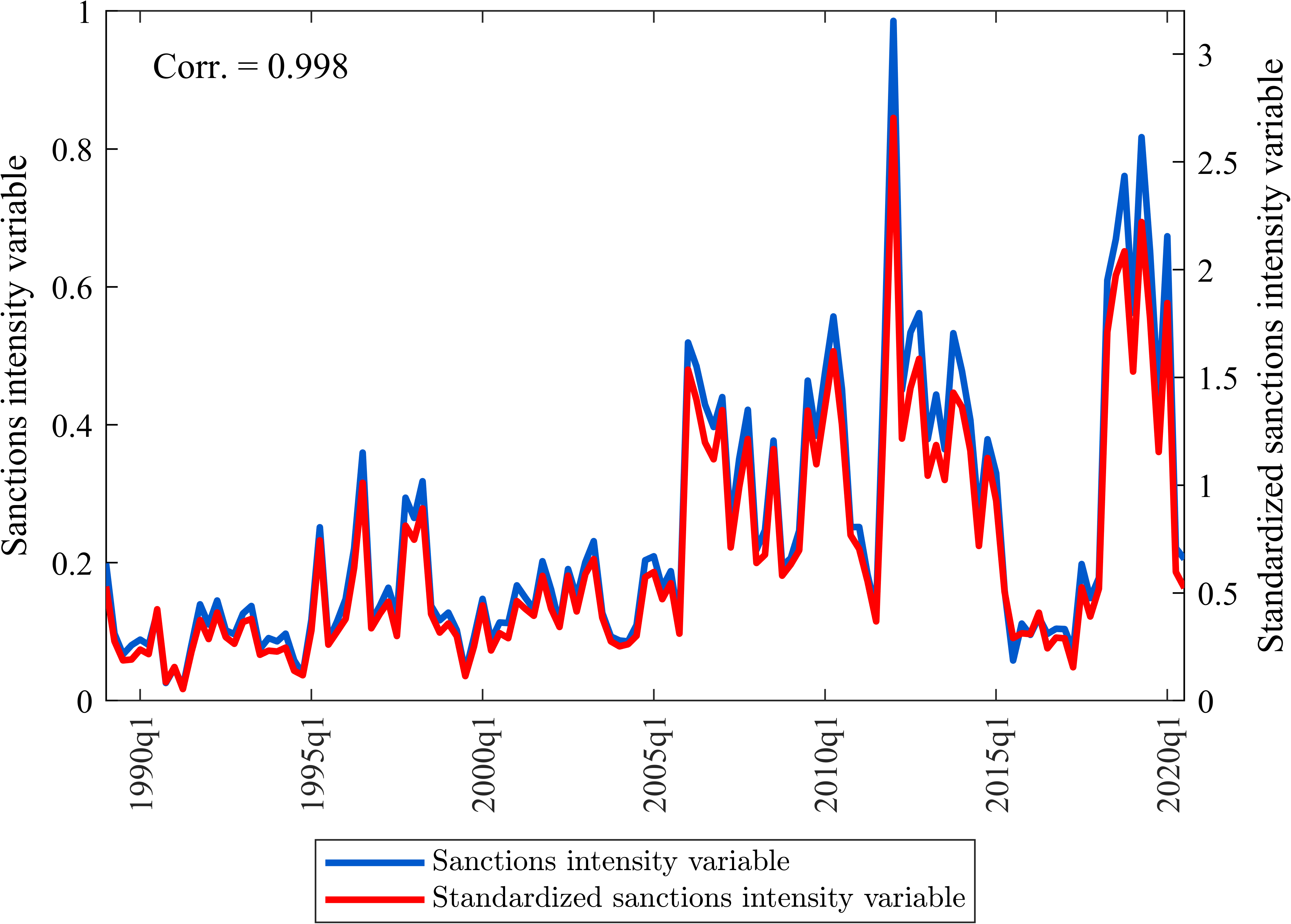}%
}
\]
\footnotesize
{}\textbf{Notes}: See Section \ref{Sec: sanctions measure} of the paper for
the sanctions intensity variable definition over the range (0,1). See Section
\ref{Sec. sanction index construction} in the data appendix of the online
supplement for details on construction of both the sanctions intensity variables.

\end{figure}

As a robustness check, we created a standardized version of our indicator by
following the approach advanced by \cite{baker2016}. We divided each of the
$J$ newspapers monthly \emph{raw} series by their respective standard
deviations.\footnote{For a measure of "sanctions on", we considered the
standard deviations over the entire period 1979m1--2020m9. For "sanctions
off", the monthly raw counts during 1981m1--1981m12 and 2015m1--2018m6 were
divided by the standard deviations over their respective periods.} The final
standardized intensity variable was obtained as before by averaging across
newspapers at monthly frequency, taking the simple mean for each quarter (for
both sanctions "on" and "off"), dividing each series by their respective
maxima over the period 1989q1--2020q3, and subtracting the "standardized
sanctions off" series from the "standardized sanctions on". We found these
weighted sanction on and sanction off series to be very close to the ones
based on simple averages, and as a result the grid search applied to the
weighted series also resulted in the estimate $\hat{w}$\thinspace$=0.4.$ Even
though this procedure was meant to avoid newspapers with a larger number of
articles per issue to carry unwarranted weight, the two series co-move almost
perfectly $(\rho=0.998)$. See Figure
\ref{fig: sanction with and without weights}. This finding is consistent with
\cite{plante2019}, who adjusts for the number of total articles per month and
finds that his two measures correlate at 0.97.%

\bigskip

\noindent\textbf{U.S. Treasury sanctions variable construction}

\noindent We also constructed a measure of sanctions intensity based on the
U.S. Treasury "Specially Designated Nationals And Blocked Persons List
(\emph{SDN})". The online database of the Treasury keeps track only of the
entities \textit{currently} sanctioned. To compile a complete time series list
of all Iranian entities, individuals, and vessels being sanctioned by the
U.S., we used yearly \emph{pdf} files available in the online archive of the
U.S. Treasury. In this way, we were able to follow over time each entity
entering and exiting the database.\footnote{The documents specify the exact
day in which entities enter/exit the list during the year considered.} The
list of sanctioned entities can be retrieved from 1994 onwards but the number
of entries for Iran up to 2005 is negligible. This is in line with the
historical record of U.S. sanctions against Iran. Therefore, we focussed on
building our entry-exit matrix from 2006 onwards.

To construct the \textit{U.S. Treasury sanctions variable,} we first summed
the total number of Iranian entities, individuals, and vessels being hit by
U.S. sanctions.\footnote{Notice that SDN lists specify to which sanctions
programs each entry belongs. In other words, according to whether the aim is
to hit entities related to Iran vis-\`{a}-vis other nations (say, North Korea)
different codes are attached to them.} In the SDN lists, entities refer to
companies (and institutions) of Iranian nationality, foreign companies having
offices in Iran, and -- in light of secondary sanctions -- all other foreign
companies doing business with sanctioned Iranian companies. Iranian
individuals, or foreigners doing business with sanctioned Iranians, were
tracked by First and Last Name, and Passport number or National ID -- when
available. For vessels, we did not confine ourselves to vessels name or
national flag given that these attributes were often changed. Instead, the
International Maritime Organization (\emph{IMO}) unique identification number
proved to be important and completely reliable to follow vessels history.

\begin{sidewaysfigure}%
\caption
{Sanctions intensity variable and the U.S. Treasury sanctions variable over the period 2006q1--2020q3}

\label{fig: s_SDN_graph}%

\[%
{\includegraphics[
height=4.6124in,
width=9.1177in
]
{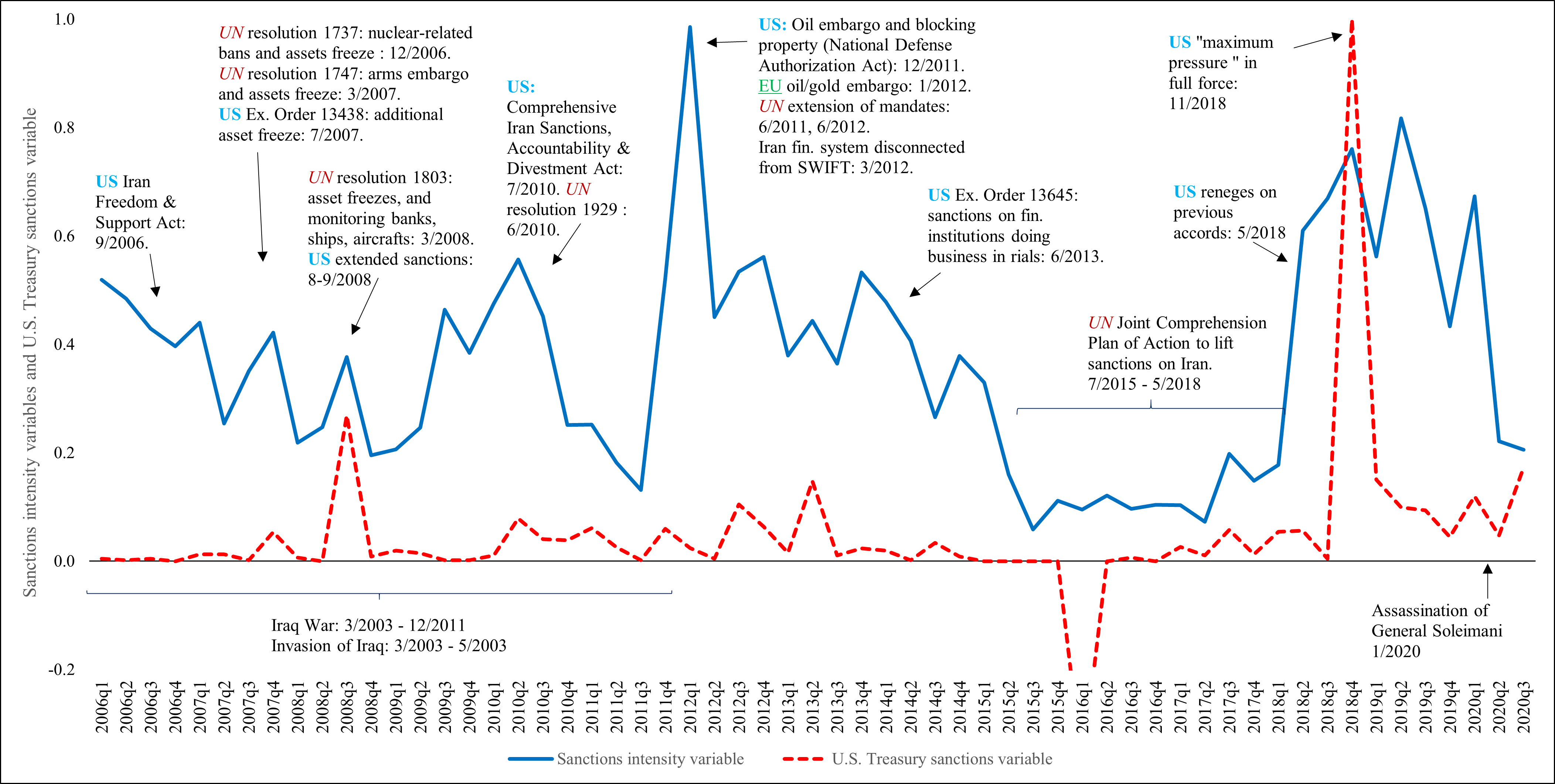}%
}
\]

\footnotesize
{}\textbf{Notes}: The U.S. Treasury sanctions variable is computed from the
number of newly introduced and removed entries in the "Specially Designated
Nationals And Blocked Persons List" (\emph{SDN}) of the U.S. Department of the
Treasury. Major sanctions-related historical events are indicated by arrows
and brackets. See Section \ref{Sec. sanction index construction} for details
of the construction of the sanctions variables.%

\end{sidewaysfigure}

The number of Iranian entries added to the SDN list allowed us to build an
"SDN sanctions on" time series; similarly, the entries removed from the list
provided the information for an "SDN sanctions off" index. We obtained our
final "U.S. Treasury sanctions variable" by attaching a weight to the "SDN
sanctions off" count equal to the newspaper-based indicator ($w=0.4$) and
subtracting it from the "SDN sanctions on" count. The final series was then
re-scaled by dividing it for its maximum value. See Figure
\ref{fig: s_SDN_graph}. The correlation between the U.S. Treasury measure and
$s_{t}$ is equal to 38 per cent over the period 2006q1--2020q3. Notice that
the series based on SDN has inevitably negative values over the JCPOA period
regardless of the weight one is willing to choose. This feature is due to the
fact that no new Iranian entities were added, while a large number of
previously sanctioned entities were removed.\footnote{This could also be
considered as a shortcoming of using such measure given that -- in our
framework -- a negative value of the sanctions intensity variable means an
attempt to subsidize the Iranian economy through transfers, something far from
the actual process happening over the period 2016q1--2018q2.}%

\subsection{Conversions from Iranian to Gregorian calendar
\label{Sec. calendar conversion}}

The data we use in our analysis are in Gregorian calendar. However, data
retrieved from Iranian sources, namely from the Central Bank of Iran and the
Statistical Center of Iran, follow the Iranian calendar format. The Iranian
year starts on March $21^{st}$ of the corresponding Gregorian year.
Accordingly, we carried out three calendar conversions in order for the
Iranian data to be in line with the ones in the Gregorian format. In the
following expressions, $G_{y},G_{q},$ and $G_{m}$ stand for the variables
transformed in the Gregorian calendar at yearly, quarterly, and monthly
frequencies, respectively, while $I_{y},I_{q},$ and $I_{m}$ are the data in
the original Iranian format. For annual statistics, the following formula was
applied:\footnote{Eighty days of the Gregorian year (from Jan. $1^{st}$ to
Mar. $21^{st}$) were to be attributed to the previous Iranian year.}
$G_{y}=\frac{80}{365}I_{y-1}+\frac{285}{365}I_{y}.$ For quarterly data, we
converted the\ Iranian series according to:\footnote{In the following
expression, $8/9$ represents the eighty days out of the approximately ninety
days within a given quarter.} $G_{q}=\frac{8}{9}I_{q-1}+\frac{1}{9}I_{q}.$
Finally, for the monthly time series -- we applied the following
transformation: $G_{m}=\frac{1}{3}I_{m-1}+\frac{2}{3}I_{m}.$

\subsection{Economic and socio-demographic
variables\label{Sec. socio-economic vars construction}}

In this section, we will refer to some of the Iranian data as being retrieved
from the "Quarterly Iran Data Set 2020". In this case, we extend and update
the data for Iran in the GVAR Data Set compiled by \cite{mohaddes_raissi2020}
until 2018q2 (and available upon request); more recent observations for Iran
were added by splicing forward the previously available series with new
observations from Iranian sources. In this respect, the conversions mentioned
in Section \ref{Sec. calendar conversion} were applied. All data from the
Central Bank of Iran (\emph{CBI}) were obtained from the \emph{Economic Time
Series Database}. For global factors we will refer to the "GVAR Data Set
2020". In this case, we use the latest version of the GVAR Data Set provided
by \cite{mohaddes_raissi2020}, which we extend to include observations for
2020q1.\footnote{\cite{mohaddes_raissi2020} data set ends in 2019q4. The
extended data set is available upon request.}

\bigskip

\noindent\textbf{Quarterly data}\textit{ }

\noindent The quarterly real output of Iran was obtained by splicing forward
the GVAR series in the Quarterly Iran Data Set 2020 available until 2018q2
with the "Iran's Quarterly National Accounts" released by the Statistical
Center of Iran until 2020q1.

Iran's inflation was computed as first difference of the natural logarithm of
Iran consumer price index (\emph{CPI}). CPI data from the GVAR series in the
Quarterly Iran Data Set 2020 available until 2018q2 were extended forward with
data from the Statistical Center of Iran, which provides Iranian monthly
inflation bulletins. After having converted the monthly series to the
Gregorian calendar, it was possible to compute the quarterly inflation rate,
and splice forward the Quarterly Iran Data Set until 2021q1. The CPI was then
rebased to have value equal to 100 in 1979q2.

The \emph{official} foreign exchange statistics from 1979q2 to 2020q3 were
retrieved in quarterly format from Bank Markazi (Iran's Central Bank), and
converted to the Gregorian calendar. The \emph{free market} foreign exchange
rate in quarterly format from 1979q2 to 2017q4 was also retrieved from Bank
Markazi. For 2018 onward the series were spliced forward with data from
\emph{bonbast.com} -- a highly cited website tracking the Iran's rial free
market rate against all major currencies. In this regard, \emph{bonbast.com}
presents information for "buy" and "sell" rates at daily frequency. We used
the average of buy and sell rates. In this way we were able to extend the
historical series from Bank Markazi until 2021q1.

Monetary statistics were also downloaded from the Bank Markazi website. The
monetary aggregate $M2$ was computed as the sum of $M1$ and "quasi-money".
Data were available at quarterly frequency, and -- before converting them to
the Gregorian calendar -- the observations from 2015q2 onwards had to be
multiplied by 1,000 given a change of format from billions to trillions of rials.

In order to account for global factors, we augmented our analyses with several
variables. $\Delta p_{t}^{0}$ is the rate of change of the oil price (first
difference of the natural logarithm). The oil price considered was the Brent
crude (U.S. dollars/barrel). Data at quarterly frequency until 2020q1 were
taken from the GVAR Data Set 2020. Observations for 2020q2 and 2020q3 were
obtained by splicing the series with data from the U.S. Energy Information
Administration (series name: "Europe Brent Spot Price FOB, Dollars per
Barrel"). The E.I.A. provided information at monthly frequency therefore we
first averaged the oil prices over each quarter, and then spliced forward our
GVAR time series.

The quarterly global realized volatility, $grv_{t},$ was taken directly from
the GVAR Data Set 2020 for the whole period 1979q2--2020q1; details about its
construction can be found in \cite{chudik_etal2020}.

We used the GVAR Data Set 2020 and followed the procedure indicated by
\cite{chudik_etal2020} also for the construction of the other global factors.
The factors we considered are: the world real output growth, $\Delta
\overline{y}_{wt};$ the rate of change of the world real exchange rate against
the U.S. dollar, $\Delta\overline{e}_{wt}$; the world real equity returns,
$\Delta\overline{req}_{wt}$; and the per cent change of the world nominal
long-term interest rate, $\Delta\overline{r}_{wt}$. These control variables
were obtained by taking the first difference of the following weighted
cross-sectional averages: $\overline{y}_{wt}=\sum_{i=0}^{n}w_{i}y_{it},$
$\overline{e}_{wt}=\sum_{i=0}^{n}w_{i}e_{it},$ $\overline{req}_{wt}=\sum
_{i=0}^{n}w_{i}eq_{it},$ $\overline{r}_{wt}=\sum_{i=0}^{n}w_{i}r_{it},$ where
$y_{it},e_{it},$ $eq_{it},$ $r_{it}$ \ are: the log of real output, the log of
the real exchange rate against the U.S. dollar, the log of real equity prices,
and the nominal long term interest rates of country $i$ in quarter $t$. The
sample included 33 of the world major economies, and the weights, $w_{i}$,
were computed as the GDP-PPP average by country $i$ out of the overall world
average output over the period 2014--2016:
\begin{equation}
w_{i}=\frac{\sum_{t=2014}^{2016}Y_{it}^{PPP}}{\sum_{i=0}^{n}\sum
_{t=2014}^{2016}Y_{it}^{PPP}}. \label{gdp-ppp weights}%
\end{equation}
The GDP-PPP measure allows for international comparisons, and it was retrieved
at yearly frequency from the \emph{World Bank Open Data }repository. The 33
countries are: Argentina, Australia, Austria, Belgium, Brazil, Canada, China,
Chile, Finland, France, Germany, India, Indonesia, Italy, Japan, South Korea,
Malaysia, Mexico, the Netherlands, Norway, New Zealand, Peru, the Philippines,
South Africa, Saudi Arabia, Singapore, Spain, Sweden, Switzerland, Thailand,
Turkey, the U.K.,\ and the U.S.A..

For some of the 33 countries, real equities returns, $eq_{it},$ and nominal
long term interest rates, $r_{it},$ were not available. As such, to compute
$\overline{req}_{wt}$ and $\overline{r}_{wt}$ \ we focussed on the countries
for which we had information, and rescaled the weights accordingly. In
particular, the historical real equity prices, $eq_{it},$ were available for
26 out of 33 countries (excluded were Brazil, China, Indonesia, Mexico, Peru,
Saudi Arabia, and Turkey). For the long run interest rates, $r_{it},$ data
were available for 18 of the 33 countries (excluded were Argentina, Brazil,
China, Chile, Finland, India, Indonesia, Malaysia, Mexico, Peru, the
Philippines, Saudi Arabia, Singapore, Thailand, and Turkey).

We also considered real output growth for oil exporters, $\Delta\overline
{y}_{t}^{0},$ and Turkey's output growth, $\Delta y_{t}^{Tur}$.\ The real
output for oil exporters is the cross-sectional weighted average $\overline
{y}_{t}^{0}=\sum_{i=0}^{n_{oil}}w_{i}^{0}y_{it}^{0}$, with the countries
composing $n_{oil}$ being: Brazil, Canada, Mexico, Norway, and Saudi Arabia.
The weights, $\left\{  w_{i}^{0}\right\}  _{i=1}^{n_{oil}}$, were computed
following the same procedure described in Equation (\ref{gdp-ppp weights}) but
restricting the sample from $n$ to $n_{oil}$. Turkey real output growth is
$\Delta y_{t}^{Tur}=\ln(Y_{t}^{Tur}/Y_{t-1}^{Tur}),$ with $Y_{t}^{Tur}$ being
Turkey real output in quarter $t$.\ The output series were retrieved from the
GVAR Data Set 2020.

We also retrieved information on the U.S. CPI from the OECD accessed through
the Federal Reserve Bank of St. Louis (\emph{FRED}) data set. Table
\ref{table: data_quarterly} provides an overview of the data sources.

\bigskip

\noindent\textbf{Annual data}

\noindent\emph{Economic variables }-- We cumulated quarterly series, whenever
available, to obtain annual series, such as Iran real output, $Y_{t}$; Iran
consumer price index, $P_{t};$ Brent crude oil price, $P_{t}^{0}$; world, oil
exporters, and Turkey real output data $\{\overline{y}_{wt},\overline{y}%
_{t}^{0},y_{t}^{Tur}\};$ and the global factors $\{\overline{e}_{wt},$
$\overline{req}_{wt},$ $\overline{r}_{wt},$ $grv_{t}\}$.

Data on value added by sectors (agriculture, manufacturing and mining, and
services) were retrieved at annual frequency from the Bank Markazi. The series
at constant 2004 prices was spliced forward in 2010 with the series at
constant 2011 prices. Before splicing, each data series was converted to the
Gregorian calendar format. For sectoral composition analyses: agriculture,
manufacturing, and services shares of value added were computed as a fraction
of the total value added represented by the sum of the three components.

Total exports and imports revenues in millions of dollars were retrieved from
the Bank Markazi at annual frequency, and converted to the Gregorian calendar
format. Oil and gas exports revenues, and non-oil exports revenues (in
millions of U.S. dollars) were also obtained from Bank Markazi website.
However, it is important to bear in mind that for both oil and gas, and
non-oil exports, Bank Markazi holds two data sets based on the "fourth
edition"\ and "fifth edition" directives of the annual IMF Balance of Payment
statistics.\footnote{Under the fourth edition oil exports category was defined
as \textquotedblleft Crude oil, oil products, liquefied gas and natural
gas\textquotedblright. In the fifth edition, the description is:
\textquotedblleft Value of crude oil, oil products, natural gas, natural gas
condensate and liquids (Tariff codes: 2709, 2710 and 2711) exported by
National Iranian Oil Company (NIOC), National Iranian Gas Company (NIGC),
National Iranian Oil Refining and Distribution Company (NIORDC), petrochemical
companies, and others (customs and non-customs)\textquotedblright. Tariff
codes refer to unique identifying numbers adopted by the UN Statistics
Division.} We followed the fourth edition, which was available for almost the
whole of the period under consideration (until 2018), and obtained data for
2019 by splicing forward the series with information from the fifth
edition.\footnote{The correlation of oil and non-oil exports growth under the
two systems of reference was 99.5 per cent and 92.8 per cent, respectively,
for the overlapping period 1997--2018 for which we had data for both series.}

The historical data series on annual oil exports in thousands of barrels per
day was also compiled from the Bank Markazi. But this series ended in 2018
therefore -- after having converted it to the Gregorian calendar -- we spliced
it forward with IMF data "Crude Oil Exports for Iran" series for the years
2019 and 2020, which we retrieved from FRED.

\bigskip

\noindent\emph{Education variables }-- Bank Markazi "Economic Time Series
Database"\emph{ }allows to retrieve further statistics on education such as
the number of schools (from 1979 to 2018), and the number of teachers (from
1979 to 2019). In each case, it was possible to obtain data on the number of
schools and teachers both at aggregate level and divided by grade: primary,
lower secondary, and high school. Furthermore, we retrieved data on the
overall number of students by gender. $\,$Information on the number of
students by gender for each school grade was discontinued in 2013. The total
number of schools was computed as the sum of primary, lower secondary, and
high schools. When using the series on the number of teacher employed, we
always divided it by the yearly Iran population in the 25--64 age group, in
order to control for population increases.

The "Female-to-male students enrollment ratio", $\Delta fm_{t},$ is the log
difference of the ratio between female and male students. Female and male
enrolled students were divided by the female and male population in the 5--19
age groups, respectively, to account for general population increases by
gender. Specifically, $\Delta fm_{t}=\ln(FM_{t}/FM_{t-1}),$ with
$FM_{t}=(S_{ft}/Pop_{ft})/(S_{mt}/Pop_{mt})$ where $S_{ft}$ and $S_{mt}$ are
the number of female and male students, respectively, and $Pop_{ft}$ and
$Pop_{mt}$ are Iran female and male population in the 5--19 age group.

\bigskip

\noindent\emph{Other socio-demographic variables}\textbf{ }-- Several other
yearly socio-economic statistics were obtained from the "World Development
Indicators" of the World Bank. More specifically, the male and female labor
force participation rate, the employment rate, total national population for a
number of world economies (listed below), and Iran male and female population
in the age brackets 5--9, 10--14, 15--19, 20--24, 15--64, as fraction of total
Iran's male and female population. The absolute values for the population by
sex and age bracket were computed by multiplying each fraction by the total
number of males and females in Iran each year, while the population in the
25--64 age group was obtained by subtracting the population in the 15--24 age
groups from the ones in the 15--64 age groups (which was available). The
information about age brackets was used as demographic control when estimating
the rate of change in the number of students enrolled in schools by gender.
Similarly, the 25-64 population was used as a demographic control for the
number of teachers employed in Iran.

We also constructed geographic regional aggregates for labor statistics, in
particular for Middle East and North Africa (\emph{MENA}). Each aggregate
annual labor series, say $\overline{x}_{wt}$, was built as a weighted average
of countries data series, $x_{it},$ i.e. $\overline{x}_{it}=\sum_{i=0}%
^{n}\widetilde{w}_{i}x_{it}$. Weights, $\widetilde{w}_{i},$ were based on the
country annual population average out of the total population for the specific
regional aggregate considered over the period 2014--2018:
\[
\widetilde{w}_{i}=\frac{\sum_{t=2014}^{2018}Pop_{it}}{\sum_{i=0}^{n}%
\sum_{t=2014}^{2018}Pop_{it}}.
\]
\noindent The MENA\textit{\ }region in our statistics includes the following
17 countries: Bahrain, Iraq, Israel, Jordan, Kuwait, Lebanon, Oman, Qatar,
Saudi Arabia, Syrian Arab Republic, United Arab Emirates, Yemen, Algeria,
Egypt, Libya, Morocco, and Tunisia.%

\begin{table}[H]
	\caption{Sources of quarterly data}
	\hspace*{-0.8cm}
	\footnotesize\renewcommand{\arraystretch}{1.15}%
\label{table: data_quarterly}%

\vspace{-0.3cm}%

\begin{tabular}{m{6.9cm} m{0.4cm} m{8.7cm}}
		\hline\hline\\ [-10pt]
		\normalsize\textbf{Data series} &       & \normalsize\textbf{Source} \\
		\hline&       &  \\
		\normalsize\textit{Iranian variables}$^{1}$ &       &  \\
\\ [-11pt]		
		Consumer price index &       & \multicolumn{1}{l}%
{Quarterly Iran Data Set 2020} \\
\\ [-11pt]
		Foreign exchange rate, Free Market &       & \multicolumn{1}{l}%
{Central Bank of Iran and \textit{bonbast.com}} \\
\\ [-11pt]
		Foreign exchange rate, Official rate &       & \multicolumn{1}%
{l}{Central Bank of Iran} \\
\\ [-11pt]
		Money supply: M1 and Quasi-money &       & \multicolumn{1}{l}%
{Central Bank of Iran} \\		
\\ [-11pt]
		Real output &       & \multicolumn{1}{l}{Quarterly Iran Data Set 2020} \\
		&       &  \\
		\normalsize\textit{Global and regional control variables}$^{2}%
$ &       & \multicolumn{1}{l}{GVAR Data Set 2020  and World Bank} \\
		\hline\hline\end{tabular}%

\vspace{0.3cm}

\footnotesize
{}\textbf{Notes}: \emph{1.} The Quarterly Iran Data Set 2020 extends and
updates the GVAR Data Set compiled by \cite{mohaddes_raissi2020}, whose
observations for Iran are available up to 2018q2. Such version of the data
base including Iran is available upon request. The most recent observations
for the consumer price index taken from the Statistical Center of Iran can be
retrieved from the monthly inflation bulletins available at
\emph{www.amar.org.ir.} The data provided by the Central Bank of Iran on
foreign exchange rates are available from the Economic Time Series Database:
\emph{tsd.cbi.ir, }under "External\ Sector/Value of Financial Assets (Exchange
Rate and Coin Price)"\emph{. }Recent data on free market foreign exchange data
can be retrieved from \emph{www.bonbast.com}. Money supply statistics are
available under "Monetary and Credit Aggregates" at \emph{tsd.cbi.ir.} The
data used to extend the Iran's real output series are taken from the
Statistical Center of Iran and can be retrieved under "Iran's Quarterly
National Accounts (base year = 1390)" from \emph{www.amar.org.ir. }

\emph{2. }Raw data for each country composing the global and regional averages
were retrieved from the GVAR Data Set compiled by \cite{mohaddes_raissi2020}
and available at \emph{www.mohaddes.org/gvar}. We extended the original data
set from 2019q4 to 2020q1 -- data are available upon request. The World Bank
data (\emph{data.worldbank.org}) have been used to construct the GDP-PPP
weights for each country (code indicator: "NY.GDP.MKTP.PP.CD"). The variables
included in this set of controls are: global nominal long term interest rate,
global real equity price, global real exchange rate, global real output,
global realized volatility, oil exporters real output, oil price (Brent
crude), and Turkey real output. For oil price, the observations for 2020q2 and
2020q3 were obtained from the U.S. Energy Information Administration (series
name: "Europe Brent Spot Price FOB, Dollars per Barrel") available at
\emph{www.eia.gov}. Information on the U.S. consumer price index was retrieved
from the FRED data base \emph{fred.stlouisfed.org} (series name: "Consumer
Price Index: Total All Items for the United States, Index 2015=100, Quarterly,
Seasonally Adjusted").

See Section \ref{Sec: data appendix} for further details on variables construction.%

\vskip0.1cm
\end{table}%

\begin{table}[H]
	\caption{Sources of annual data}
	\hspace*{-0.8cm}
	\footnotesize\renewcommand{\arraystretch}{1.15}%
\label{table: data_yearly}%

\begin{tabular}{m{6.9cm} m{0.4cm} m{8.7cm}}
		\hline\hline\\ [-10pt]
		\normalsize\textbf{Data series} &       & \normalsize\textbf{Source} \\
		\hline&       &  \\
		\normalsize\textit{Iranian variables} &       &  \\
\\ [-11pt]		
		Consumer price index &       & \multicolumn{1}{l}%
{Quarterly Iran Data Set 2020} \\
\\ [-11pt]
		Crude oil exports (th. barrels/day)$^{1}$ &       & \multicolumn{1}%
{l}{Central Bank of Iran and IMF} \\
\\ [-11pt]
		Education statistics: Schools, students by gender, teachers$^{1}
$ &       & \multicolumn{1}{l}{Central Bank of Iran} \\
\\ [-11pt]
		Exports revenues (non-oil, oil \& gas, and total)$^{1}
$ &       & \multicolumn{1}{l}{Central Bank of Iran} \\
\\ [-11pt]
		Foreign exchange rate, Free Market &       & \multicolumn{1}{l}%
{Central Bank of Iran and bonbast.com} \\
\\ [-11pt]
		Foreign exchange rate, Official rate &       & \multicolumn{1}%
{l}{Central Bank of Iran} \\
\\ [-11pt]
		Import revenues$^{1}$ &       & \multicolumn{1}{l}{Central Bank of Iran} \\
\\ [-11pt]
		Labor statistics: Employment rate, Labor force participation rates$^{2}%
$ &       & \multicolumn{1}{l}{World Bank, World Development Indicators} \\
\\ [-11pt]
		Money supply: M1 and Quasi-money &       & \multicolumn{1}{l}%
{Central Bank of Iran} \\
\\ [-11pt]
		Population statistics$^{2}$ &       & \multicolumn{1}{l}%
{World Bank, World Development Indicators} \\
\\ [-11pt]
		Real output &       & \multicolumn{1}{l}{Quarterly Iran Data Set 2020} \\
\\ [-11pt]
		Value added by sector$^{1}$ &       & \multicolumn{1}{l}%
{Central Bank of Iran} \\
		&       &  \\
		\normalsize\textit{Global and regional control variables}
&       & \multicolumn{1}{l}{GVAR Data Set 2020 and World Bank} \\
		\hline\hline\end{tabular}%

\vspace{0.3cm}

\footnotesize
{}\textbf{Notes}: \emph{1.} Crude oil exports (th. barrels/day) data from
\ the Central Bank of Iran are available from the Economic Time Series
Database: \emph{tsd.cbi.ir }under "Energy Sector" statistics\emph{. }The
observations from the IMF were retrieved through the FRED data base of the
Federal Reserve Bank of St. Louis: \emph{fred.stlouisfed.org} (series name:
"Crude Oil Exports for Iran"). Data on education were retrieved from "Human
resource and Employment" statistics available on the Central Bank of Iran
repository \emph{tsd.cbi.ir. }Export and import revenues are available from
the same website under "External Sector/Balance of Payments (Manual)"
statistics. \ Value added data are also taken from the same source under
"National Accounts (1383=100)" and "National Accounts (1390=100)" statistics.

\emph{2.} Labor and national population statistics are from the World
Development Indicators of the World Bank for both Iran and other MENA
countries accessed through \emph{data.worldbank.org. }Employment rate (ages
15+, total) indicator code: "SL.EMP.TOTL.SP.ZS". Female labor force
participation rate (per cent of female population ages 15+) indicator code:
"SL.TLF.CACT.FE.ZS". Male labor force participation rate (per cent of male
population ages 15+) indicator code: "SL.TLF.CACT.MA.ZS". National total
population statistics indicator code: "SP.POP.TOTL".

See notes in Table \ref{table: data_quarterly} for details on the sources of
the other variables. See Section \ref{Sec: data appendix} for further details
on variables construction.%

\end{table}%

\bigskip

\section{Computation of IRFs, FEVDs and their error bands by
bootstrap\label{Sec: additional methods}}

\subsection{IRFs and FEVDs alternative
computation\label{Sec: IRF-FEVD alternative computation}}

To compute the IRFs and FEVDs, we provide an alternative computation approach
with respect to the one described in the paper. We confirm that we obtained
the same numerical results as when we used the formulae in the paper, which we
had included for pedagogic reasons.

Re-write Equation (\ref{psi0zt_supplement}) as:%
\[
\widetilde{\mathbf{z}}_{t}=\widetilde{\boldsymbol{\Psi}}_{0}^{-1}\left(
\widetilde{\mathbf{a}}\mathbf{+}\widetilde{\boldsymbol{\Psi}}_{1}%
\widetilde{\mathbf{z}}_{t-1}+\widetilde{\boldsymbol{\Psi}}_{2}%
\widetilde{\mathbf{z}}_{t-2}+\widetilde{\mathbf{u}}_{t}\right)  ,
\]

\noindent with $\widetilde{\mathbf{z}}_{t}=\left(  \Delta e_{ft},\Delta
p_{t},\Delta y_{t},s_{t},\Delta\overline{y}_{wt}\right)  ^{\prime},$ and
$\widetilde{\mathbf{u}}_{t}=(\varepsilon_{\Delta e_{ft}},\varepsilon_{\Delta
p_{t}},\varepsilon_{\Delta y_{t}},\varepsilon_{s_{t}},\varepsilon
_{\Delta\overline{y}_{wt}})^{\prime}.$ The IRF can be computed by following
the approach described in the paper as:%
\[
IRF_{\mathbf{z}}\mathcal{(}h)=\sqrt{\sigma_{jj}}(\mathbf{F}_{h}%
\widetilde{\boldsymbol{\Psi}}_{0}^{-1}\mathbf{e}_{j}),
\]

\noindent where $\mathbf{e}_{j}$ is a $(m+2)\times1$ selection\ vector of
zeros except for its $j^{th}$ element, which is unity, and%
\[
\mathbf{F}_{\ell}=\widetilde{\boldsymbol{\Phi}}_{1}\mathbf{F}_{\ell
-1}+\widetilde{\boldsymbol{\Phi}}_{2}\mathbf{F}_{\ell-2},\text{ for }%
\ell=1,2,\dots
\]
\noindent where $\widetilde{\boldsymbol{\Phi}}_{1}=$
$\widetilde{\boldsymbol{\Psi}}_{0}^{-1}\widetilde{\boldsymbol{\Psi}}_{1},$
$\widetilde{\boldsymbol{\Phi}}_{2}=$ $\widetilde{\boldsymbol{\Psi}}_{0}%
^{-1}\widetilde{\boldsymbol{\Psi}}_{2},$ with $\mathbf{F}_{-1}=\mathbf{0}$,
and $\mathbf{F}_{0}=\mathbf{I}_{m+2}$. Consequently, the impulse response
effects of a positive one standard error change in the $j^{th}$ domestic
shock, $\varepsilon_{jt}$, on the $i^{th}$ variable (the $i^{th}$ element of
$\widetilde{\mathbf{z}}_{t}$) are given by:%
\[
IRF_{ij}(h)=\sqrt{\sigma_{jj}}(\mathbf{e}_{i}^{\prime}\mathbf{F}%
_{h}\widetilde{\boldsymbol{\Psi}}_{0}^{-1}\mathbf{e}_{j}),\text{ for
}h=0,1,...,H\text{, }i,j=\Delta e_{ft},\Delta p_{t},\Delta y_{t},s_{t}%
,\Delta\overline{y}_{wt}\text{ .}%
\]
\noindent The forecast errors can be now written more succinctly as:%
\[
\widetilde{\boldsymbol{\xi}}_{t}(n)=\sum_{\ell=0}^{n}\mathbf{F}_{\ell
}\widetilde{\boldsymbol{\Psi}}_{0}^{-1}\widetilde{\mathbf{u}}_{t+n-\ell},
\]
\noindent where, as before, $\widetilde{\mathbf{u}}_{t}$ is a vector of
$(m+2)\times1$ shocks. Similarly, the proportion of the forecast error
variance of the $i^{th}$ variable due to a shock to the $j^{th}$ variable at
horizon $h$ is given by:%
\[
\theta_{ij}(h)=\frac{\sigma_{jj}\sum_{\ell=0}^{h}\left(  \mathbf{e}%
_{i}^{\prime}\mathbf{F}_{\ell}\widetilde{\boldsymbol{\Psi}}_{0}^{-1}%
\mathbf{e}_{j}\right)  ^{2}}{\sum_{\ell=0}^{h}\mathbf{e}_{i}^{\prime
}\mathbf{F}_{\ell}\widetilde{\boldsymbol{\Psi}}_{0}^{-1}\boldsymbol{\Sigma
}\mathbf{\widetilde{\boldsymbol{\Psi}}}_{0}^{\prime-1}\mathbf{F}_{\ell
}^{\prime}\mathbf{e}_{i}}\text{, for }i,j=\Delta e_{ft},\Delta p_{t},\Delta
y_{t},s_{t},\Delta\overline{y}_{wt}\text{ },
\]
\noindent with $\boldsymbol{\Sigma}=Diag(\sigma_{\Delta e\Delta e}%
,\sigma_{\Delta p\Delta p},...,\sigma_{\Delta\overline{y}\Delta\overline{y}%
}).$ It can be proved that $\sum_{j=1}^{m}\theta_{ij}(h)+\theta_{is}%
(h)+\theta_{i\Delta\overline{y}_{w}}(h)=1.$

\subsection{Bootstrapping procedure\label{Sec: bootstrapping}}

In order to compute the impulse response functions (IRFs) and the associated
confidence bands, we followed a bootstrap procedure by simulating the
in-sample values of $\mathbf{z}_{t}$ in Equation (\ref{psi0zt}), which we
report here for convenience:%
\begin{equation}
\boldsymbol{\Psi}_{0}\mathbf{z}_{t}=\mathbf{a+}\boldsymbol{\Psi}_{1}%
\mathbf{z}_{t-1}+\boldsymbol{\Psi}_{2}\mathbf{z}_{t-2}+\mathbf{u}_{t}.
\label{psi0zt_supplement}%
\end{equation}
In Equation (\ref{psi0zt_supplement}), $\mathbf{z}_{t}=\left(  \mathbf{q}%
_{t},s_{t},\mathbf{\bar{z}}_{wt}\right)  ^{\prime}$ is a vector of $m$
domestic policy variables ($\mathbf{q}_{t}$), the sanctions intensity variable
($s_{t}$), and the $k$ global factors ($\mathbf{\bar{z}}_{wt}$); $\mathbf{a}$
is a $(m+k+1)\times1$ vector of constants, and $\mathbf{u}_{t}$\ are the
residuals of the system. In order to generate our bootstrap replications, we
proceed as follows:%

\begin{enumerate}%

\item
Generate the simulated residuals $\left\{  \mathbf{u}_{t}^{(r)},\text{
}r=1,2,...,R\right\}  $ by resampling with replacement from the estimated
residuals of each equation separately $\left\{  \widehat{\mathbf{u}}%
_{t},t=3,4,...,T\right\}  $, where $R=1,000$ is the number of random samples.%

\item
Let $\mathbf{z}_{1989q1}^{(r)}=\mathbf{z}_{1989q1}$, $\;\;\mathbf{z}%
_{1989q2}^{(r)}=\mathbf{z}_{1989q2}$ $\ \forall r$, and compute:%
\[
\mathbf{z}_{t}^{(r)}=\widehat{\boldsymbol{\Psi}}_{0}^{-1}\left(
\widehat{\mathbf{a}}\mathbf{+}\widehat{\boldsymbol{\Psi}}_{1}\mathbf{z}%
_{t-1}^{(r)}+\widehat{\boldsymbol{\Psi}}_{2}\mathbf{z}_{t-2}^{(r)}%
+\mathbf{u}_{t}^{(r)}\right)  \qquad t=1989q3,...,2020q1
\]%
\item
Use the data computed at point 2 to estimate the bootstrapped coefficients for
each replication:%
\[
\mathbf{z}_{t}^{(r)}=\widehat{\boldsymbol{\Psi}}_{0}^{-1,(r)}\left(
\widehat{\mathbf{a}}^{(r)}+\widehat{\boldsymbol{\Psi}}_{1}^{(r)}%
\mathbf{z}_{t-1}^{(r)}+\widehat{\boldsymbol{\Psi}}_{2}^{(r)}\mathbf{z}%
_{t-2}^{(r)}+\mathbf{u}_{t}^{(r)}\right)  .
\]%
\end{enumerate}%

The procedure just described can help evaluating $\mathbf{z}_{t}%
^{(r)}(baseline)$ vis-\`{a}-vis other scenarios $\mathbf{z}_{t}^{(r)}(Shock)$
with the major shock profiles derived in the paper.

\section{Additional empirical
results\label{Sec: online supplement reduced form results}}

In this section we provide additional empirical results in support of our
analyses. Table \ref{table: s-ar1-ar2} provides estimates of AR(1) and AR(2)
processes for the sanctions intensity index, $s_{t}.$ As explained in the
paper, the process is highly persistent, and a first-order specification
describes the process sufficiently well, with additional lags not being
statistically significant. Table \ref{table: yw_ar1_ar2} gives the estimates
of first- and second-order autoregressive processes (AR) for the world output
growth, and shows that the AR(1) specification used in the paper provides a
reasonable approximation.

In the main paper we presented estimates of SVAR model under our preferred
ordering, namely with the exchange rate variable ($\Delta e_{ft}$) included
first, followed by money supply growth ($\Delta m_{t}$), inflation ($\Delta
p_{t}$), and output growth ($\Delta y_{t}$), including the world output growth
as the control variables. Tables \ref{table: svar_dfx} to \ref{table: svar_dy}
display the regression results including all the control variables, and a
number of their sub-sets. As can be seen, the estimates of the effects of
sanctions on domestic variables are highly stable and consistent across all
specifications. It is also worth noting that changes in international oil
prices do not have a statistically significant impact on the exchange rate,
which could be due to the fact that once we condition on the sanctions
variable, a rise in oil prices is less likely to benefit Iran when oil exports
are severely limited due to sanctions. Also, none of the global factors seem
to have any significant impact on Iran's output growth, partly due to Iran's
relative economic and financial isolation from the rest of the global
economy\textbf{.}

Tables \ref{table: svar_dp order v2} to \ref{table: svar_dy order v2} and
\ref{table: svar_dm2 order v3} to \ref{table: svar_dy order v3} present
results of the SVAR model given by Equation (\ref{A0qt}) of the paper, but
with the different orderings of the domestic variables. Tables
\ref{table: svar_dp order v2} to \ref{table: svar_dy order v2} provide the
results when the domestic variables are ordered with inflation ($\Delta p_{t}%
$) first followed by the rate of change of the free market foreign exchange
rate ($\Delta e_{ft}$), money supply growth ($\Delta m_{t}$), and output
growth ($\Delta y_{t}$). Tables \ref{table: svar_dm2 order v3} to
\ref{table: svar_dy order v3} give the results when the variables are ordered
as $\Delta m_{t},$ $\Delta e_{ft},$ $\Delta p_{t},$ and $\Delta y_{t}$.$\,$As
can be seen, re-ordering of the variables do not alter our main findings
summarized in Section \ref{Sec: Structural model results}\ of the paper.

Figure \ref{fig: IRF global output growth} displays the impulse response
results for one positive standard error shock to the global output growth on
Iran's free market foreign exchange rate depreciation, inflation, and output
growth.\footnote{For a theoretical derivation, see Equation (\ref{IRF yw}) in
the main text.} As such, it complements the IRFs results from our SVAR model
presented in Figure \ref{fig: IRF} of the paper. Following one quarter shock
to global output growth, the Iran's rial appreciates by about 1.5 per cent in
the same quarter. However, the results are not particularly persistent, and
become quantitatively less important three quarters ahead, and lose
statistical significance about five quarters ahead. The effects of global
output growth on both inflation and Iran's output growth, on the other hand,
are not statistically significant. These results are in line with Iran's
relative economic isolation from the main advanced economies. Most of the
global shocks are reflected in the movements of the free market foreign
exchange rate, while the domestic economic mismanagement is a factor that can
not be underplayed to explain the dynamics of the Iranian economy.

Table \ref{table: reduced form s(t) and s(t-1) on dy}\ gives the estimates of
the reduced form output growth equation given by (\ref{Dy2}), where we include
both current lagged values of the sanctions intensity variable $s_{t}$. As can
be seen, the estimates are very close to the ones presented in the paper,
where only the lagged value of $s_{t}$ was included.

As reported in the paper, sanctions have adversely affected the growth of
manufacturing and services, but seem to have had little impact on the growth
of the agricultural sector. Here we also present results on the effects of
sanctions on the rate of change of the shares of agriculture, manufacturing
and mining, and services value added out of the total. To this end we estimate
VAR(1) models in two of the three share variables at a time, since the sum of
the shares add up to unity. The estimation results are summarized in Tables
\ref{table: var_sector_share1}, \ref{table: var_sector_share2} and
\ref{table: var_sector_share3}. As can be seen, sanctions have reduced the
share of the manufacturing sector, whilst resulting in an increase in the
share of the agricultural sector. The share of the services do not seem to be
much affected by the sanctions. Overall, it is the manufacturing sector which
is most affected by the sanctions.

In terms of labor statistics, we provide a robustness check for the effects of
sanctions on the labor force participation rate in Iran with respect to other
MENA countries. See Table \ref{table: labor force}. We notice that sanctions
induced a median decline which is perfectly consistent with the results
obtained for the employment rate shown in Table \ref{table: employment rate}
of the paper.

Finally, we provide additional robustness checks on sanctions-induced
education outcomes by considering changes in the number of teachers employed
across different education grades, and the total number of schools and
teachers. The estimates are summarized in Tables \ref{table: teachers primary}%
, \ref{table: teachers lower secondary}, \ref{table: teachers high schools},
\ref{table: teachers total} and \ref{table: school total}. These results
support our main conclusion that the primary effects of sanctions has been on
lower secondary and high school grades, with little impact on primary
education (which is compulsory in Iran). In terms of the total number of
schools and teachers, we find that sanctions have adversely affected the total
number of schools, but the effects of sanctions on the total number of
teachers is less clear cut.%

\begin{table}[H]%
\caption
{Quarterly estimates of the sanctions intensity variable AR(1) and AR(2) models over the period 1989q1--2020q3}
\small
\label{table: s-ar1-ar2}%

\begin{center}
	\begin{tabular}{@{\extracolsep{10pt}}lll}
		\\[-3.8ex]\hline\hline\\[-1.8ex]
		& \multicolumn{2}{c}{$s_{t}$} \\
		\cline{2-3}\\[-1.8ex]
		& (1) & (2)\\
		\hline\\[-1.8ex]
		$s_{t-1}$ & 0.743$^{***}$ & 0.639$^{***}$ \\
		& (0.059) & (0.089) \\
		$s_{t-2}$ &  & 0.139 \\
		&  & (0.089) \\
		Constant & 0.063$^{***}$ & 0.055$^{***}$ \\
		& (0.018) & (0.019) \\
		\hline\\[-1.8ex]
		Adjusted R$^{2}$ & 0.551 & 0.557 \\
		S.E. of regression ($\hat{\omega}_{s}$) & 0.125 & 0.125 \\
		\hline\hline\\
	\end{tabular}
\end{center}

\footnotesize
\textbf{Notes}: Standard errors in parentheses. ***$p<0.01$, **$p<0.05$,
*$p<0.1$. See Section \ref{Sec. sanction index construction} in the data
appendix of the online supplement for details on the construction of the
sanctions intensity variable.%

\end{table}

\begin{table}[H]%
\caption
{Quarterly estimates of the world real output growth AR(1) and AR(2) models over the period 1989q1--2020q1}
\small\renewcommand{\arraystretch}{1.0}%
\label{table: yw_ar1_ar2}%

\vspace{-0.7cm}%
\begin{center}
	\begin{tabular}{@{\extracolsep{5pt}}lD{.}{.}{-3} D{.}{.}{-3} }
		\\[-1.8ex]\hline\hline\\[-1.8ex]
		& \multicolumn{2}{c}{$ \Delta\bar{y}_{wt}$} \\
		\cline{2-3}
		\\[-1.8ex] & \multicolumn{1}{c}{(1)} & \multicolumn{1}{c}{(2)}\\
		\hline\\[-1.8ex]
		$\Delta\bar{y}_{w,t-1}$ & 0.468^{***} & 0.439^{***} \\
		& (0.108) & (0.118) \\
		$ \Delta\bar{y}_{w,t-2}$ &  & 0.073 \\
		&  & (0.118) \\
		Constant & 0.005^{***} & 0.005^{***} \\
		& (0.001) & (0.002) \\
		\hline\\[-1.8ex]
		Adjusted R$^{2}$ & 0.125 & 0.121 \\
		Residual Std. Error & 0.006 & 0.006 \\
		\hline\hline\\[-1.8ex]
	\end{tabular}
\end{center}%

\vspace{-0.45cm}%
\footnotesize
\textbf{Notes}: $\Delta\overline{y}_{wt}$ is the quarterly world output
growth: $\overline{y}_{wt}=\sum\nolimits_{i=1}^{n}w_{i}y_{it},$ with $\left\{
y_{it}\right\}  _{i=1}^{n}$ being the natural log of real output for 33 major
economies, and $w_{i}$ the GDP-PPP weights. See Section
\ref{Sec. socio-economic vars construction} in the data appendix of the online
supplement for details on the construction and sources of the data used.%

\end{table}

\bigskip

\begin{table}[H]
\renewcommand{\thetable}{S.\arabic{table}a}
\caption
{Quarterly estimates of the equation for the rate of change of the free market foreign exchange rate
in the SVAR model of Iran with domestic variables ordered as: foreign exchange rate returns, money supply growth,
inflation, and output growth, estimated over the period 1989q1--2020q1}
\vspace*{-0.8cm}
\small\renewcommand{\arraystretch}{1.15}%
\label{table: svar_dfx}%

\begin{center}
	\begin{tabular}{@{\extracolsep{-15pt}}lD{.}{.}{-3} D{.}{.}{-3} D{.}{.}{-3}
D{.}{.}{-3} D{.}{.}{-3} D{.}{.}{-3} D{.}{.}{-3} }
		\\[-1.8ex]\hline\hline\\[-1.8ex]
		& \multicolumn{7}{c}{$\Delta e_{ft}$} \\
		\cline{2-8}
		\\[-1.8ex] & \multicolumn{1}{c}{(1)} & \multicolumn{1}{c}{(2)}
& \multicolumn{1}{c}{(3)} & \multicolumn{1}{c}{(4)} & \multicolumn{1}{c}{(5)}
& \multicolumn{1}{c}{(6)} & \multicolumn{1}{c}{(7)}\\
		\hline\\[-1.8ex]
		$s_{t}$ & 0.311^{***} & 0.303^{***} & 0.295^{***} & 0.296^{***}
& 0.297^{***} & 0.304^{***} & 0.303^{***} \\
		& (0.061) & (0.061) & (0.061) & (0.061) & (0.060) & (0.061) & (0.061) \\
		$s_{t-1}$ & -0.243^{***} & -0.245^{***} & -0.238^{***} & -0.240^{***}
& -0.236^{***} & -0.242^{***} & -0.243^{***} \\
		& (0.064) & (0.063) & (0.064) & (0.063) & (0.063) & (0.063) & (0.063) \\
		$\Delta e_{f,t-1}$ & 0.335^{***} & 0.341^{***} & 0.345^{***} & 0.337^{***}
& 0.327^{***} & 0.328^{***} & 0.335^{***} \\
		& (0.091) & (0.090) & (0.090) & (0.090) & (0.089) & (0.089) & (0.090) \\
		$\Delta m_{t-1}$ & 0.298 & 0.350 & 0.366 & 0.321 & 0.379 & 0.385 & 0.406 \\
		& (0.251) & (0.250) & (0.251) & (0.251) & (0.251) & (0.251) & (0.254) \\
		$\Delta p_{t-1}%
$ & -0.304 & -0.376 & -0.366 & -0.413 & -0.357 & -0.391 & -0.411 \\
		& (0.332) & (0.331) & (0.331) & (0.330) & (0.329) & (0.332) & (0.334) \\
		$\Delta y_{t-1}%
$ & -0.135 & -0.126 & -0.133 & -0.146 & -0.113 & -0.114 & -0.108 \\
		& (0.245) & (0.242) & (0.242) & (0.241) & (0.239) & (0.240) & (0.241) \\
		$\Delta\bar{y}_{wt}$ &  & -2.059^{*} & -2.513^{**} & -3.467^{**}
& -3.211^{**} & -3.048^{**} & -3.105^{**} \\
		&  & (1.123) & (1.205) & (1.348) & (1.344) & (1.359) & (1.366) \\
		$\Delta p^{0}_{t}$ &  &  & 0.054 & 0.037 & 0.079 & 0.076 & 0.094 \\
		&  &  & (0.052) & (0.053) & (0.058) & (0.058) & (0.065) \\
		$grv_{t}$ &  &  &  & -0.365 & -0.387 & -0.542^{*} & -0.550^{*} \\
		&  &  &  & (0.237) & (0.235) & (0.299) & (0.300) \\
		$\Delta\bar{e}_{wt}$ &  &  &  &  & 0.748^{*} & 0.739^{*} & 0.799^{*} \\
		&  &  &  &  & (0.427) & (0.428) & (0.439) \\
		$\Delta\overline{req}_{wt}$ &  &  &  &  &  & -0.126 & -0.117 \\
		&  &  &  &  &  & (0.150) & (0.151) \\
		$\Delta\bar{r}_{wt}$ &  &  &  &  &  &  & -7.731 \\
		&  &  &  &  &  &  & (12.260) \\
		\hline\\[-2.3ex]
		Residual serial & 6.968 & 6.013 & 5.356 & 3.905 & 5.079 & 5.171 & 5.465 \\
		correlation test & [0.138] & [0.198] & [0.253] & [0.419] & [0.279] & [0.270] & [0.243] \\ \hline
Adjusted $R^{2}$ & 0.225 & 0.240 & 0.240 & 0.249 & 0.263 & 0.261 & 0.257 \\
		\hline\hline\\[-1.8ex]
	\end{tabular}
\end{center}

\vspace{-0.4cm}%
\footnotesize
\textbf{Notes}: The variables are ordered as: $\Delta e_{ft},$ $\Delta m_{t},$
$\Delta p_{t},$\ and $\Delta y_{t},$ where: $\Delta e_{ft}=\ln(E_{ft}%
/E_{f,t-1}),$ $E_{ft}$ is the quarterly rial/U.S. dollar free market exchange
rate; $\Delta m_{t}=\ln(M_{2t}/M_{2,t-1}),$ $M_{2t}$ is obtained by summing
the aggregates $M1$ and "quasi-money"; $\Delta p_{t}=\ln(P_{t}/P_{t-1}),$
$P_{t}$ is the quarterly consumer price index of Iran; $\Delta y_{t}=\ln
(Y_{t}/Y_{t-1}),$ $Y_{t}$ is the quarterly real output of Iran. $s_{t}$ is the
quarterly sanctions intensity variable. $\Delta\overline{y}_{wt}$ is the
quarterly world output growth, computed as $\overline{y}_{wt}=\sum
\nolimits_{i=1}^{n}w_{i}y_{it},$ with $\left\{  y_{it}\right\}  _{i=1}^{n}$
being the natural log of real output for 33 major economies, and $\left\{
w_{i}\right\}  _{i=1}^{n}$ are GDP-PPP weights. $\Delta p_{t}^{0}=\ln
(P_{t}^{0}/P_{t-1}^{0}),\ P_{t}^{0}$ is the quarterly oil price (Brent crude).
$grv_{t}$ is the quarterly global realized volatility. $\Delta\overline
{e}_{wt}$ is the quarterly rate of change of the global real exchange rate
vis-\`{a}-vis the U.S. dollar: $\overline{e}_{wt}=\sum\nolimits_{i=1}^{n}%
w_{i}e_{it}$, $e_{it}$ is the natural log of the real exchange rate of country
$i$ in quarter $t$. $\Delta\overline{req}_{wt}$ is the quarterly rate of
change of the global real equity price index: $\overline{req}_{wt}%
=\sum\nolimits_{i=1}^{n}w_{i}req_{it},$ $req_{it}$ is the natural log of the
real equity price of country $i$ in quarter $t$. $\Delta\overline{r}_{wt}$ is
the quarterly change of the global nominal long term interest rate:
$\overline{r}_{wt}=\sum\nolimits_{i=1}^{n}w_{i}r_{it},$ $r_{it}$ is the long
term nominal interest rate of country $i$ in quarter $t$. Numbers in
parentheses are standard errors, and those in square brackets are p-values.
***$p<0.01$, **$p<0.05$, *$p<0.1$. "Residual serial correlation test" is the
Breusch--Godfrey LM test of serially uncorrelated errors with lag order of the
test set to $4$.

See Sections \ref{Sec. sanction index construction},
\ref{Sec. calendar conversion}, and
\ref{Sec. socio-economic vars construction} in the data appendix of the online
supplement for details on the construction of the sanctions intensity
variable, calendar conversions, and sources of the data used.%

\end{table}%

\bigskip

\begin{table}[H]
\addtocounter{table}{-1}
\renewcommand{\thetable}{S.\arabic{table}b}
\caption{Quarterly estimates of the equation for the money supply growth
in the SVAR model of Iran with domestic variables ordered as: foreign exchange rate returns, money supply growth,
inflation, and output growth, estimated over the period 1989q1--2020q1}
\vspace*{-0.8cm}
\small\renewcommand{\arraystretch}{1.15}%
\label{table: svar_dm2}%

\begin{center}
	\begin{tabular}{@{\extracolsep{-15pt}}lD{.}{.}{-3} D{.}{.}{-3} D{.}{.}{-3}
D{.}{.}{-3} D{.}{.}{-3} D{.}{.}{-3} D{.}{.}{-3} }
		\\[-1.8ex]\hline\hline\\[-1.8ex]
		& \multicolumn{7}{c}{$\Delta m_{t}$} \\
		\cline{2-8}
		\\[-1.8ex] & \multicolumn{1}{c}{(1)} & \multicolumn{1}{c}{(2)}
& \multicolumn{1}{c}{(3)} & \multicolumn{1}{c}{(4)} & \multicolumn{1}{c}{(5)}
& \multicolumn{1}{c}{(6)} & \multicolumn{1}{c}{(7)}\\
		\hline\\[-1.8ex]
		$s_{t}$ & -0.001 & -0.001 & -0.001 & 0.002 & 0.002 & 0.002 & 0.002 \\
		& (0.024) & (0.024) & (0.024) & (0.024) & (0.024) & (0.024) & (0.025) \\
		$s_{t-1}$ & 0.011 & 0.011 & 0.012 & 0.009 & 0.009 & 0.008 & 0.008 \\
		& (0.024) & (0.024) & (0.024) & (0.024) & (0.024) & (0.024) & (0.025) \\
		$\Delta e_{ft}%
$ & -0.016 & -0.015 & -0.015 & -0.023 & -0.023 & -0.023 & -0.023 \\
		& (0.032) & (0.033) & (0.033) & (0.033) & (0.034) & (0.034) & (0.034) \\
		$\Delta e_{f,t-1}%
$ & -0.039 & -0.040 & -0.040 & -0.040 & -0.040 & -0.040 & -0.039 \\
		& (0.034) & (0.034) & (0.034) & (0.034) & (0.034) & (0.034) & (0.035) \\
		$\Delta m_{t-1}$ & -0.285^{***} & -0.289^{***} & -0.289^{***} & -0.304^{***}
& -0.304^{***} & -0.304^{***} & -0.302^{***} \\
		& (0.089) & (0.090) & (0.090) & (0.090) & (0.092) & (0.092) & (0.093) \\
		$\Delta p_{t-1}$ & 0.126 & 0.132 & 0.132 & 0.110 & 0.110 & 0.108 & 0.106 \\
		& (0.117) & (0.118) & (0.119) & (0.119) & (0.120) & (0.121) & (0.122) \\
		$\Delta y_{t-1}%
$ & -0.064 & -0.064 & -0.064 & -0.070 & -0.070 & -0.070 & -0.070 \\
		& (0.086) & (0.086) & (0.087) & (0.086) & (0.087) & (0.087) & (0.087) \\
		$\Delta\bar{y}_{wt}%
$ &  & 0.135 & 0.127 & -0.272 & -0.273 & -0.265 & -0.271 \\
		&  & (0.405) & (0.439) & (0.494) & (0.498) & (0.504) & (0.508) \\
		$\Delta p^{0}_{t}$ &  &  & 0.001 & -0.006 & -0.006 & -0.006 & -0.005 \\
		&  &  & (0.019) & (0.019) & (0.021) & (0.021) & (0.024) \\
		$grv_{t}$ &  &  &  & -0.145^{*} & -0.145^{*} & -0.154 & -0.155 \\
		&  &  &  & (0.085) & (0.086) & (0.110) & (0.111) \\
		$\Delta\bar{e}_{wt}$ &  &  &  &  & -0.006 & -0.006 & -0.001 \\
		&  &  &  &  & (0.156) & (0.157) & (0.162) \\
		$\Delta\overline{req}_{wt}$ &  &  &  &  &  & -0.007 & -0.006 \\
		&  &  &  &  &  & (0.055) & (0.055) \\
		$\Delta\bar{r}_{wt}$ &  &  &  &  &  &  & -0.606 \\
		&  &  &  &  &  &  & (4.461) \\
		\hline\\[-2.3ex]
		Residual serial & 7.051 & 7.165 & 7.174 & 6.363 & 6.411 & 6.653 & 6.650 \\
		correlation test & [0.133] & [0.127] & [0.127] & [0.174] & [0.171] & [0.155] & [0.156] \\ \hline
Adjusted $R^{2}$ & 0.054 & 0.047 & 0.038 & 0.054 & 0.045 & 0.037 & 0.029 \\
		\hline\hline\\[-1.8ex]
	\end{tabular}
\end{center}

\vspace{-0.4cm}%
\footnotesize
\textbf{Notes}: See the notes to Table \ref{table: svar_dfx} for further
details on the construction and sources of data used.%

\end{table}%

\bigskip

\begin{table}[H]
\addtocounter{table}{-1}
\renewcommand{\thetable}{S.\arabic{table}c}
\caption{Quarterly estimates of the equation for the inflation rate
in the SVAR model of Iran with domestic variables ordered as: foreign exchange rate returns, money supply growth,
inflation, and output growth, estimated over the period 1989q1--2020q1}
\vspace*{-0.8cm}
\small\renewcommand{\arraystretch}{1.05}%
\label{table: svar_dp}%

\begin{center}
	\begin{tabular}{@{\extracolsep{-15pt}}lD{.}{.}{-3} D{.}{.}{-3} D{.}{.}{-3}
D{.}{.}{-3} D{.}{.}{-3} D{.}{.}{-3} D{.}{.}{-3} }
		\\[-1.8ex]\hline\hline\\[-1.8ex]
		& \multicolumn{7}{c}{$\Delta p_{t}$} \\
		\cline{2-8}
		\\[-1.8ex] & \multicolumn{1}{c}{(1)} & \multicolumn{1}{c}{(2)}
& \multicolumn{1}{c}{(3)} & \multicolumn{1}{c}{(4)} & \multicolumn{1}{c}{(5)}
& \multicolumn{1}{c}{(6)} & \multicolumn{1}{c}{(7)}\\
		\hline\\[-1.8ex]
		$s_{t}$ & -0.033^{**} & -0.033^{***} & -0.032^{**} & -0.032^{**}
& -0.033^{***} & -0.030^{**} & -0.030^{**} \\
		& (0.013) & (0.012) & (0.013) & (0.013) & (0.012) & (0.013) & (0.013) \\
		$s_{t-1}$ & 0.034^{**} & 0.037^{***} & 0.036^{***} & 0.036^{***}
& 0.036^{***} & 0.033^{**} & 0.033^{**} \\
		& (0.013) & (0.013) & (0.013) & (0.013) & (0.013) & (0.013) & (0.013) \\
		$\Delta e_{ft}$ & 0.153^{***} & 0.162^{***} & 0.164^{***} & 0.163^{***}
& 0.168^{***} & 0.166^{***} & 0.167^{***} \\
		& (0.018) & (0.017) & (0.017) & (0.017) & (0.017) & (0.017) & (0.017) \\
		$\Delta m_{t}%
$ & -0.026 & -0.032 & -0.031 & -0.033 & -0.032 & -0.032 & -0.031 \\
		& (0.050) & (0.048) & (0.048) & (0.049) & (0.049) & (0.048) & (0.048) \\
		$\Delta e_{f,t-1}%
$ & -0.003 & -0.010 & -0.010 & -0.010 & -0.008 & -0.006 & -0.007 \\
		& (0.020) & (0.019) & (0.019) & (0.020) & (0.019) & (0.019) & (0.019) \\
		$\Delta m_{t-1}%
$ & -0.016 & -0.038 & -0.042 & -0.043 & -0.056 & -0.054 & -0.059 \\
		& (0.050) & (0.048) & (0.049) & (0.049) & (0.049) & (0.049) & (0.050) \\
		$\Delta p_{t-1}$ & 0.456^{***} & 0.490^{***} & 0.484^{***} & 0.484^{***}
& 0.460^{***} & 0.443^{***} & 0.441^{***} \\
		& (0.090) & (0.086) & (0.087) & (0.087) & (0.087) & (0.088) & (0.088) \\
		$\Delta y_{t-1}$ & 0.026 & 0.023 & 0.025 & 0.024 & 0.022 & 0.023 & 0.023 \\
		& (0.048) & (0.046) & (0.046) & (0.047) & (0.046) & (0.046) & (0.046) \\
		$\Delta p_{t-2}$ & 0.180^{**} & 0.174^{**} & 0.181^{**} & 0.179^{**}
& 0.199^{**} & 0.207^{***} & 0.216^{***} \\
		& (0.080) & (0.076) & (0.077) & (0.078) & (0.078) & (0.078) & (0.079) \\
		$\Delta\bar{y}_{wt}$ &  & 0.721^{***} & 0.801^{***} & 0.777^{***}
& 0.751^{***} & 0.795^{***} & 0.813^{***} \\
		&  & (0.209) & (0.225) & (0.258) & (0.256) & (0.257) & (0.259) \\
		$\Delta p^{0}_{t}$ &  &  & -0.009 & -0.010 & -0.018 & -0.019^{*}
& -0.023^{*} \\
		&  &  & (0.010) & (0.010) & (0.011) & (0.011) & (0.012) \\
		$grv_{t}$ &  &  &  & -0.009 & -0.002 & -0.049 & -0.046 \\
		&  &  &  & (0.045) & (0.045) & (0.057) & (0.057) \\
		$\Delta\bar{e}_{wt}$ &  &  &  &  & -0.138^{*} & -0.141^{*} & -0.156^{*} \\
		&  &  &  &  & (0.081) & (0.081) & (0.084) \\
		$\Delta\overline{req}_{wt}$ &  &  &  &  &  & -0.038 & -0.040 \\
		&  &  &  &  &  & (0.028) & (0.028) \\
		$\Delta\bar{r}_{wt}$ &  &  &  &  &  &  & 1.802 \\
		&  &  &  &  &  &  & (2.287) \\
		\hline\\[-2.3ex]
		Residual serial & 13.954 & 8.236 & 9.993 & 10.393 & 10.984 & 9.239 & 9.526 \\
		correlation test & [0.007] & [0.083] & [0.041] & [0.034] & [0.027] & [0.055] & [0.049] \\ \hline
Adjusted $R^{2}$ & 0.638 & 0.669 & 0.669 & 0.666 & 0.672 & 0.674 & 0.673 \\
		\hline\hline\\[-1.8ex]
	\end{tabular}
\end{center}

\vspace{-0.4cm}%

\footnotesize
\textbf{Notes}: See the notes to Table \ref{table: svar_dfx} for further
details on the construction and sources of data used.%

\end{table}%

\begin{table}[H]%
\addtocounter{table}{-1}
\renewcommand{\thetable}{S.\arabic{table}d}
\caption{Quarterly estimates of the equation for the output growth
in the SVAR model of Iran with domestic variables ordered as: foreign exchange rate returns, money supply growth,
inflation, and output growth, estimated over the period 1989q1--2020q1}
\vspace*{-0.8cm}
\small\renewcommand{\arraystretch}{1.05}%
\label{table: svar_dy}%

\begin{center}
	\begin{tabular}{@{\extracolsep{-15pt}}lD{.}{.}{-3} D{.}{.}{-3} D{.}{.}{-3}
D{.}{.}{-3} D{.}{.}{-3} D{.}{.}{-3} D{.}{.}{-3} }
		\\[-1.8ex]\hline\hline\\[-1.8ex]
		& \multicolumn{7}{c}{$\Delta y_{t}$} \\
		\cline{2-8}
		\\[-1.8ex] & \multicolumn{1}{c}{(1)} & \multicolumn{1}{c}{(2)}
& \multicolumn{1}{c}{(3)} & \multicolumn{1}{c}{(4)} & \multicolumn{1}{c}{(5)}
& \multicolumn{1}{c}{(6)} & \multicolumn{1}{c}{(7)}\\
		\hline\\[-1.8ex]
		$s_{t}$ & 0.021 & 0.021 & 0.020 & 0.021 & 0.016 & 0.017 & 0.017 \\
		& (0.025) & (0.025) & (0.025) & (0.025) & (0.025) & (0.025) & (0.026) \\
		$s_{t-1}$ & -0.057^{**} & -0.058^{**} & -0.057^{**} & -0.058^{**}
& -0.055^{**} & -0.056^{**} & -0.057^{**} \\
		& (0.025) & (0.026) & (0.026) & (0.026) & (0.026) & (0.026) & (0.026) \\
		$\Delta e_{ft}$ & -0.123^{***} & -0.125^{***} & -0.127^{***} & -0.130^{***}
& -0.113^{**} & -0.113^{**} & -0.116^{**} \\
		& (0.042) & (0.044) & (0.045) & (0.045) & (0.046) & (0.046) & (0.046) \\
		$\Delta m_{t}$ & 0.096 & 0.097 & 0.097 & 0.088 & 0.086 & 0.085 & 0.084 \\
		& (0.094) & (0.094) & (0.095) & (0.096) & (0.095) & (0.096) & (0.096) \\
		$\Delta p_{t}$ & 0.335^{*} & 0.341^{*} & 0.348^{*} & 0.344^{*} & 0.304^{*}
& 0.300 & 0.305^{*} \\
		& (0.171) & (0.180) & (0.181) & (0.182) & (0.181) & (0.183) & (0.183) \\
		$\Delta e_{f,t-1}%
$ & 0.033 & 0.033 & 0.035 & 0.034 & 0.034 & 0.034 & 0.038 \\
		& (0.034) & (0.035) & (0.035) & (0.035) & (0.035) & (0.035) & (0.035) \\
		$\Delta m_{t-1}%
$ & -0.022 & -0.020 & -0.016 & -0.024 & -0.051 & -0.051 & -0.040 \\
		& (0.094) & (0.095) & (0.096) & (0.097) & (0.097) & (0.098) & (0.098) \\
		$\Delta p_{t-1}$ & -0.490^{***} & -0.496^{***} & -0.499^{***} & -0.502^{***}
& -0.495^{***} & -0.495^{***} & -0.509^{***} \\
		& (0.157) & (0.165) & (0.166) & (0.167) & (0.165) & (0.166) & (0.167) \\
		$\Delta y_{t-1}$ & -0.195^{**} & -0.195^{**} & -0.196^{**} & -0.199^{**}
& -0.210^{**} & -0.210^{**} & -0.208^{**} \\
		& (0.087) & (0.088) & (0.088) & (0.088) & (0.088) & (0.088) & (0.088) \\
		$\Delta\bar{y}_{wt}%
$ &  & -0.049 & -0.145 & -0.278 & -0.313 & -0.298 & -0.335 \\
		&  & (0.432) & (0.467) & (0.524) & (0.519) & (0.527) & (0.529) \\
		$\Delta p^{0}_{t}$ &  &  & 0.010 & 0.008 & -0.009 & -0.009 & -0.001 \\
		&  &  & (0.019) & (0.019) & (0.021) & (0.021) & (0.024) \\
		$grv_{t}$ &  &  &  & -0.050 & -0.039 & -0.053 & -0.058 \\
		&  &  &  & (0.088) & (0.088) & (0.112) & (0.113) \\
		$\Delta\bar{e}_{wt}$ &  &  &  &  & -0.291^{*} & -0.291^{*} & -0.261 \\
		&  &  &  &  & (0.159) & (0.160) & (0.164) \\
		$\Delta\overline{req}_{wt}$ &  &  &  &  &  & -0.011 & -0.006 \\
		&  &  &  &  &  & (0.055) & (0.056) \\
		$\Delta\bar{r}_{wt}$ &  &  &  &  &  &  & -3.734 \\
		&  &  &  &  &  &  & (4.487) \\
		\hline\\[-2.3ex]
		Residual serial & 7.157 & 7.108 & 6.991 & 7.143 & 7.813 & 7.815 & 7.782 \\
		correlation test & [0.128] & [0.130] & [0.136] & [0.129] & [0.099] & [0.099] & [0.100] \\ \hline
Adjusted $R^{2}$ & 0.137 & 0.129 & 0.124 & 0.119 & 0.137 & 0.129 & 0.127 \\
		\hline\hline\\[-1.8ex]
	\end{tabular}
\end{center}

\vspace{-0.4cm}
\footnotesize
\textbf{Notes}: See the notes to Table \ref{table: svar_dfx} for further
details on the construction and sources of data used.%

\end{table}

\bigskip

\begin{table}[H]
\renewcommand{\thetable}{S.\arabic{table}a}
\caption{Quarterly estimates of the equation for the inflation rate
in the SVAR model of Iran with domestic variables ordered as:
inflation, foreign exchange rate returns, money supply growth, and output growth, estimated over the period 1989q1--2020q1
}
\vspace*{-0.8cm}
\small\renewcommand{\arraystretch}{1.05}

\label{table: svar_dp order v2}%

\begin{center}
\begin{tabular}{@{\extracolsep{-15pt}}lD{.}{.}{-3} D{.}{.}{-3} D{.}{.}{-3}
D{.}{.}{-3} D{.}{.}{-3} D{.}{.}{-3} D{.}{.}{-3} }
\\[-1.8ex]\hline\hline\\[-1.8ex]
& \multicolumn{7}{c}{$\Delta p_{t}$} \\
\cline{2-8}
\\[-1.8ex] & \multicolumn{1}{c}{(1)} & \multicolumn{1}{c}{(2)} & \multicolumn
{1}{c}{(3)} & \multicolumn{1}{c}{(4)} & \multicolumn{1}{c}{(5)} & \multicolumn
{1}{c}{(6)} & \multicolumn{1}{c}{(7)}\\
\hline\\[-1.8ex]
$s_{t}$ & 0.015 & 0.017 & 0.017 & 0.017 & 0.017 & 0.020 & 0.021 \\
& (0.015) & (0.015) & (0.015) & (0.015) & (0.015) & (0.015) & (0.016) \\
$s_{t-1}$ & -0.004 & -0.004 & -0.004 & -0.004 & -0.004 & -0.007 & -0.007 \\
& (0.016) & (0.016) & (0.016) & (0.016) & (0.016) & (0.016) & (0.016) \\
$\Delta e_{f,t-1}$ & 0.050^{**} & 0.049^{**} & 0.049^{**} & 0.046^{*}
& 0.047^{*} & 0.048^{*} & 0.048^{*} \\
& (0.024) & (0.024) & (0.024) & (0.025) & (0.025) & (0.025) & (0.025) \\
$\Delta m_{t-1}$ & 0.036 & 0.027 & 0.027 & 0.020 & 0.019 & 0.021 & 0.020 \\
& (0.061) & (0.062) & (0.062) & (0.062) & (0.063) & (0.063) & (0.064) \\
$\Delta p_{t-1}$ & 0.396^{***} & 0.411^{***} & 0.411^{***} & 0.411^{***}
& 0.410^{***} & 0.384^{***} & 0.384^{***} \\
& (0.114) & (0.114) & (0.115) & (0.115) & (0.117) & (0.117) & (0.118) \\
$\Delta y_{t-1}$ & 0.009 & 0.007 & 0.007 & 0.003 & 0.003 & 0.005 & 0.004 \\
& (0.061) & (0.061) & (0.062) & (0.062) & (0.062) & (0.062) & (0.062) \\
$\Delta p_{t-2}$ & 0.194^{*} & 0.192^{*} & 0.192^{*} & 0.181^{*} & 0.182^{*}
& 0.194^{*} & 0.196^{*} \\
& (0.102) & (0.102) & (0.102) & (0.103) & (0.105) & (0.104) & (0.106) \\
$\Delta\bar{y}_{wt}$ &  & 0.381 & 0.385 & 0.217 & 0.214 & 0.292 & 0.295 \\
&  & (0.275) & (0.296) & (0.334) & (0.337) & (0.338) & (0.341) \\
$\Delta p^{0}_{t}$ &  &  & -0.001 & -0.003 & -0.004 & -0.005 & -0.007 \\
&  &  & (0.013) & (0.013) & (0.015) & (0.015) & (0.016) \\
$grv_{t}$ &  &  &  & -0.064 & -0.064 & -0.135^{*} & -0.135^{*} \\
&  &  &  & (0.059) & (0.059) & (0.074) & (0.075) \\
$\Delta\bar{e}_{wt}$ &  &  &  &  & -0.009 & -0.015 & -0.018 \\
&  &  &  &  & (0.108) & (0.108) & (0.111) \\
$\Delta\overline{req}_{wt}$ &  &  &  &  &  & -0.059 & -0.059 \\
&  &  &  &  &  & (0.037) & (0.038) \\
$\Delta\bar{r}_{wt}$ &  &  &  &  &  &  & 0.443 \\
&  &  &  &  &  &  & (3.081) \\
\hline\\[-2.3ex]
Residual serial & 3.293 & 3.802 & 3.941 & 4.168 & 4.101 & 4.563 & 4.734 \\
correlation test & [0.510] & [0.433] & [0.414] & [0.384] & [0.392] & [0.335] & [0.316] \\ \hline
Adjusted $R^{2}$ & 0.406 & 0.411 & 0.406 & 0.407 & 0.401 & 0.409 & 0.404 \\
\hline\hline\\[-1.8ex]
\end{tabular}
\end{center}

\vspace{-0.4cm}%

\footnotesize

\textbf{Notes}: The variables are ordered as: $\Delta p_{t},$ $\Delta
e_{ft},\ \Delta m_{t},\ $\ and $\Delta y_{t},$ where: $\Delta p_{t}=\ln
(P_{t}/P_{t-1}),$ $P_{t}$ is the quarterly consumer price index of Iran;
$\Delta e_{ft}=\ln(E_{ft}/E_{f,t-1}),$ $E_{ft}$ is the quarterly rial/U.S.
dollar free market exchange rate; $\Delta m_{t}=\ln(M_{2t}/M_{2,t-1}),$
$M_{2t}$ is obtained by summing the aggregates $M1$ and "quasi-money";$\ $
$\Delta y_{t}=\ln(Y_{t}/Y_{t-1}),$ $Y_{t}$ is the quarterly real output of
Iran. See the notes to Table \ref{table: svar_dfx} for further details on the
construction and sources of data used.%

\end{table}

\begin{table}[H]
\addtocounter{table}{-1}
\renewcommand{\thetable}{S.\arabic{table}b}
\caption
{Quarterly estimates of the equation for the rate of change of the free market foreign exchange rate
in the SVAR model of Iran with domestic variables ordered as:
inflation, foreign exchange rate returns,
money supply growth, and output growth, estimated over the period 1989q1--2020q1}
\vspace*{-0.8cm}
\small\renewcommand{\arraystretch}{1.1}
\label{table: svar_dfx order v2}%

\begin{center}
\begin{tabular}{@{\extracolsep{-15pt}}lD{.}{.}{-3} D{.}{.}{-3} D{.}{.}{-3}
D{.}{.}{-3} D{.}{.}{-3} D{.}{.}{-3} D{.}{.}{-3} }
\\[-1.8ex]\hline\hline\\[-1.8ex]
& \multicolumn{7}{c}{$\Delta e_{ft}$} \\
\cline{2-8}
\\[-1.8ex] & \multicolumn{1}{c}{(1)} & \multicolumn{1}{c}{(2)} & \multicolumn
{1}{c}{(3)} & \multicolumn{1}{c}{(4)} & \multicolumn{1}{c}{(5)} & \multicolumn
{1}{c}{(6)} & \multicolumn{1}{c}{(7)}\\
\hline\\[-1.8ex]
$s_{t}$ & 0.286^{***} & 0.272^{***} & 0.265^{***} & 0.266^{***} & 0.267^{***}
& 0.266^{***} & 0.265^{***} \\
& (0.048) & (0.046) & (0.046) & (0.046) & (0.045) & (0.046) & (0.046) \\
$s_{t-1}$ & -0.250^{***} & -0.255^{***} & -0.248^{***} & -0.248^{***}
& -0.245^{***} & -0.244^{***} & -0.245^{***} \\
& (0.050) & (0.048) & (0.048) & (0.048) & (0.047) & (0.048) & (0.048) \\
$\Delta p_{t}$ & 2.541^{***} & 2.672^{***} & 2.668^{***} & 2.637^{***}
& 2.626^{***} & 2.631^{***} & 2.629^{***} \\
& (0.293) & (0.282) & (0.281) & (0.283) & (0.279) & (0.282) & (0.283) \\
$\Delta e_{f,t-1}$ & 0.257^{***} & 0.263^{***} & 0.267^{***} & 0.264^{***}
& 0.255^{***} & 0.254^{***} & 0.261^{***} \\
& (0.072) & (0.068) & (0.068) & (0.068) & (0.067) & (0.068) & (0.068) \\
$\Delta m_{t-1}$ & 0.175 & 0.246 & 0.261 & 0.241 & 0.295 & 0.294 & 0.312 \\
& (0.198) & (0.189) & (0.189) & (0.190) & (0.189) & (0.190) & (0.192) \\
$\Delta p_{t-1}$ & -1.696^{***} & -1.877^{***} & -1.865^{***} & -1.870^{***}
& -1.812^{***} & -1.811^{***} & -1.827^{***} \\
& (0.306) & (0.295) & (0.295) & (0.295) & (0.291) & (0.293) & (0.294) \\
$\Delta y_{t-1}%
$ & -0.082 & -0.066 & -0.073 & -0.079 & -0.049 & -0.049 & -0.044 \\
& (0.192) & (0.183) & (0.182) & (0.182) & (0.180) & (0.181) & (0.181) \\
$\Delta\bar{y}_{wt}$ &  & -3.102^{***} & -3.515^{***} & -3.945^{***}
& -3.704^{***} & -3.724^{***} & -3.774^{***} \\
&  & (0.854) & (0.912) & (1.021) & (1.011) & (1.027) & (1.032) \\
$\Delta p^{0}_{t}$ &  &  & 0.049 & 0.041 & 0.081^{*} & 0.081^{*} & 0.097^{**}
\\
&  &  & (0.039) & (0.040) & (0.043) & (0.044) & (0.049) \\
$grv_{t}$ &  &  &  & -0.169 & -0.190 & -0.172 & -0.180 \\
&  &  &  & (0.180) & (0.178) & (0.229) & (0.230) \\
$\Delta\bar{e}_{wt}$ &  &  &  &  & 0.696^{**} & 0.697^{**} & 0.750^{**} \\
&  &  &  &  & (0.321) & (0.322) & (0.331) \\
$\Delta\overline{req}_{wt}$ &  &  &  &  &  & 0.014 & 0.022 \\
&  &  &  &  &  & (0.114) & (0.115) \\
$\Delta\bar{r}_{wt}$ &  &  &  &  &  &  & -6.829 \\
&  &  &  &  &  &  & (9.235) \\
\hline\\[-2.3ex]
Residual serial & 11.560 & 8.001 & 8.665 & 7.670 & 8.223 & 8.175 & 7.930 \\
correlation test & [0.021] & [0.092] & [0.070] & [0.104] & [0.084] & [0.085] & [0.094] \\ \hline
Adjusted $R^{2}$ & 0.523 & 0.568 & 0.571 & 0.570 & 0.584 & 0.580 & 0.578 \\
\hline\hline\\[-1.8ex]
\end{tabular}
\end{center}

\vspace{-0.4cm}
\footnotesize
\textbf{Notes}: See the notes to Table \ref{table: svar_dfx} for further
details on the construction and sources of data used.%

\end{table}

\begin{table}[H]
\addtocounter{table}{-1}
\renewcommand{\thetable}{S.\arabic{table}c}
\caption{Quarterly estimates of the equation for the money supply growth
in the SVAR model of Iran with domestic variables ordered as:
inflation, foreign exchange rate returns,
money supply growth, and output growth, estimated over the period 1989q1--2020q1}
\vspace*{-0.8cm}
\small\renewcommand{\arraystretch}{1.15}
\label{table: svar_dm2 order v2}%

\begin{center}
\begin{tabular}{@{\extracolsep{-15pt}}lD{.}{.}{-3} D{.}{.}{-3} D{.}{.}{-3}
D{.}{.}{-3} D{.}{.}{-3} D{.}{.}{-3} D{.}{.}{-3} }
\\[-1.8ex]\hline\hline\\[-1.8ex]
& \multicolumn{7}{c}{$\Delta m_{t}$} \\
\cline{2-8}
\\[-1.8ex] & \multicolumn{1}{c}{(1)} & \multicolumn{1}{c}{(2)} & \multicolumn
{1}{c}{(3)} & \multicolumn{1}{c}{(4)} & \multicolumn{1}{c}{(5)} & \multicolumn
{1}{c}{(6)} & \multicolumn{1}{c}{(7)}\\
\hline\\[-1.8ex]
$s_{t}$ & -0.005 & -0.007 & -0.007 & -0.004 & -0.005 & -0.004 & -0.004 \\
& (0.025) & (0.025) & (0.025) & (0.025) & (0.025) & (0.025) & (0.025) \\
$s_{t-1}$ & 0.016 & 0.018 & 0.018 & 0.016 & 0.016 & 0.015 & 0.015 \\
& (0.025) & (0.025) & (0.025) & (0.025) & (0.025) & (0.025) & (0.026) \\
$\Delta p_{t}%
$ & -0.120 & -0.151 & -0.151 & -0.160 & -0.163 & -0.168 & -0.167 \\
& (0.169) & (0.177) & (0.179) & (0.177) & (0.179) & (0.181) & (0.182) \\
$\Delta e_{ft}$ & 0.002 & 0.010 & 0.010 & 0.003 & 0.005 & 0.005 & 0.005 \\
& (0.041) & (0.044) & (0.044) & (0.044) & (0.045) & (0.045) & (0.046) \\
$\Delta e_{f,t-1}%
$ & -0.041 & -0.044 & -0.044 & -0.044 & -0.044 & -0.044 & -0.044 \\
& (0.034) & (0.034) & (0.035) & (0.034) & (0.034) & (0.035) & (0.035) \\
$\Delta m_{t-1}$ & -0.285^{***} & -0.292^{***} & -0.292^{***} & -0.307^{***}
& -0.309^{***} & -0.308^{***} & -0.307^{***} \\
& (0.089) & (0.090) & (0.091) & (0.090) & (0.092) & (0.092) & (0.094) \\
$\Delta p_{t-1}$ & 0.198 & 0.226 & 0.226 & 0.209 & 0.210 & 0.209 & 0.207 \\
& (0.154) & (0.162) & (0.163) & (0.162) & (0.163) & (0.163) & (0.165) \\
$\Delta y_{t-1}%
$ & -0.064 & -0.064 & -0.064 & -0.070 & -0.071 & -0.071 & -0.071 \\
& (0.086) & (0.086) & (0.087) & (0.086) & (0.087) & (0.087) & (0.088) \\
$\Delta\bar{y}_{wt}$ &  & 0.245 & 0.246 & -0.152 & -0.155 & -0.137 & -0.142 \\
&  & (0.426) & (0.461) & (0.512) & (0.515) & (0.523) & (0.527) \\
$\Delta p^{0}_{t}$ &  &  & -0.0002 & -0.007 & -0.008 & -0.008 & -0.007 \\
&  &  & (0.019) & (0.019) & (0.021) & (0.021) & (0.024) \\
$grv_{t}$ &  &  &  & -0.147^{*} & -0.147^{*} & -0.162 & -0.163 \\
&  &  &  & (0.085) & (0.086) & (0.110) & (0.111) \\
$\Delta\bar{e}_{wt}$ &  &  &  &  & -0.023 & -0.024 & -0.020 \\
&  &  &  &  & (0.158) & (0.158) & (0.163) \\
$\Delta\overline{req}_{wt}$ &  &  &  &  &  & -0.013 & -0.012 \\
&  &  &  &  &  & (0.055) & (0.056) \\
$\Delta\bar{r}_{wt}$ &  &  &  &  &  &  & -0.449 \\
&  &  &  &  &  &  & (4.468) \\
\hline\\[-2.3ex]
Residual serial & 6.516 & 6.331 & 6.329 & 4.946 & 4.989 & 5.283 & 5.266 \\
correlation test & [0.164] & [0.176] & [0.176] & [0.293] & [0.288] & [0.259] & [0.261] \\ \hline
Adjusted $R^{2}$ & 0.050 & 0.044 & 0.036 & 0.052 & 0.044 & 0.036 & 0.027 \\
\hline\hline\\[-1.8ex]
\end{tabular}
\end{center}

\vspace{-0.4cm}
\footnotesize
\textbf{Notes}: See the notes to Table \ref{table: svar_dfx} for further
details on the construction and sources of data used.

\end{table}

\begin{table}[H]
\addtocounter{table}{-1}
\renewcommand{\thetable}{S.\arabic{table}d}
\caption{Quarterly estimates of the equation for the output growth
in the SVAR model of Iran with domestic variables ordered as:
inflation, foreign exchange rate returns,
money supply growth, and output growth, estimated over the period 1989q1--2020q1}
\vspace*{-0.8cm}
\small\renewcommand{\arraystretch}{1.05}
\label{table: svar_dy order v2}

\begin{center}
	\begin{tabular}{@{\extracolsep{-15pt}}lD{.}{.}{-3} D{.}{.}{-3} D{.}{.}{-3}
D{.}{.}{-3} D{.}{.}{-3} D{.}{.}{-3} D{.}{.}{-3} }
		\\[-1.8ex]\hline\hline\\[-1.8ex]
		& \multicolumn{7}{c}{$\Delta y_{t}$} \\
		\cline{2-8}
		\\[-1.8ex] & \multicolumn{1}{c}{(1)} & \multicolumn{1}{c}{(2)}
& \multicolumn{1}{c}{(3)} & \multicolumn{1}{c}{(4)} & \multicolumn{1}{c}{(5)}
& \multicolumn{1}{c}{(6)} & \multicolumn{1}{c}{(7)}\\
		\hline\\[-1.8ex]
		$s_{t}$ & 0.021 & 0.021 & 0.020 & 0.021 & 0.016 & 0.017 & 0.017 \\
		& (0.025) & (0.025) & (0.025) & (0.025) & (0.025) & (0.025) & (0.026) \\
		$s_{t-1}$ & -0.057^{**} & -0.058^{**} & -0.057^{**} & -0.058^{**}
& -0.055^{**} & -0.056^{**} & -0.057^{**} \\
		& (0.025) & (0.026) & (0.026) & (0.026) & (0.026) & (0.026) & (0.026) \\
		$\Delta e_{ft}$ & -0.123^{***} & -0.125^{***} & -0.127^{***} & -0.130^{***}
& -0.113^{**} & -0.113^{**} & -0.116^{**} \\
		& (0.042) & (0.044) & (0.045) & (0.045) & (0.046) & (0.046) & (0.046) \\
		$\Delta m_{t}$ & 0.096 & 0.097 & 0.097 & 0.088 & 0.086 & 0.085 & 0.084 \\
		& (0.094) & (0.094) & (0.095) & (0.096) & (0.095) & (0.096) & (0.096) \\
		$\Delta p_{t}$ & 0.335^{*} & 0.341^{*} & 0.348^{*} & 0.344^{*} & 0.304^{*}
& 0.300 & 0.305^{*} \\
		& (0.171) & (0.180) & (0.181) & (0.182) & (0.181) & (0.183) & (0.183) \\
		$\Delta e_{f,t-1}%
$ & 0.033 & 0.033 & 0.035 & 0.034 & 0.034 & 0.034 & 0.038 \\
		& (0.034) & (0.035) & (0.035) & (0.035) & (0.035) & (0.035) & (0.035) \\
		$\Delta m_{t-1}%
$ & -0.022 & -0.020 & -0.016 & -0.024 & -0.051 & -0.051 & -0.040 \\
		& (0.094) & (0.095) & (0.096) & (0.097) & (0.097) & (0.098) & (0.098) \\
		$\Delta p_{t-1}$ & -0.490^{***} & -0.496^{***} & -0.499^{***} & -0.502^{***}
& -0.495^{***} & -0.495^{***} & -0.509^{***} \\
		& (0.157) & (0.165) & (0.166) & (0.167) & (0.165) & (0.166) & (0.167) \\
		$\Delta y_{t-1}$ & -0.195^{**} & -0.195^{**} & -0.196^{**} & -0.199^{**}
& -0.210^{**} & -0.210^{**} & -0.208^{**} \\
		& (0.087) & (0.088) & (0.088) & (0.088) & (0.088) & (0.088) & (0.088) \\
		$\Delta\bar{y}_{wt}%
$ &  & -0.049 & -0.145 & -0.278 & -0.313 & -0.298 & -0.335 \\
		&  & (0.432) & (0.467) & (0.524) & (0.519) & (0.527) & (0.529) \\
		$\Delta p^{0}_{t}$ &  &  & 0.010 & 0.008 & -0.009 & -0.009 & -0.001 \\
		&  &  & (0.019) & (0.019) & (0.021) & (0.021) & (0.024) \\
		$grv_{t}$ &  &  &  & -0.050 & -0.039 & -0.053 & -0.058 \\
		&  &  &  & (0.088) & (0.088) & (0.112) & (0.113) \\
		$\Delta\bar{e}_{wt}$ &  &  &  &  & -0.291^{*} & -0.291^{*} & -0.261 \\
		&  &  &  &  & (0.159) & (0.160) & (0.164) \\
		$\Delta\overline{req}_{wt}$ &  &  &  &  &  & -0.011 & -0.006 \\
		&  &  &  &  &  & (0.055) & (0.056) \\
		$\Delta\bar{r}_{wt}$ &  &  &  &  &  &  & -3.734 \\
		&  &  &  &  &  &  & (4.487) \\
		\hline\\[-2.3ex]
		Residual serial & 7.157 & 7.108 & 6.991 & 7.143 & 7.813 & 7.815 & 7.782 \\
		correlation test & [0.128] & [0.130] & [0.136] & [0.129] & [0.099] & [0.099] & [0.100] \\ \hline
Adjusted $R^{2}$ & 0.137 & 0.129 & 0.124 & 0.119 & 0.137 & 0.129 & 0.127 \\
		\hline\hline\\[-1.8ex]
	\end{tabular}
\end{center}

\vspace{-0.4cm}
\footnotesize
\textbf{Notes}: See the notes to Table \ref{table: svar_dfx} for further
details on the construction and sources of data used.

\end{table}

\bigskip

\begin{table}[H]
\renewcommand{\thetable}{S.\arabic{table}a}
\caption{Quarterly estimates of the equation for the money supply growth
in the SVAR model of Iran with domestic variables ordered as: money supply growth,
foreign exchange rate returns, inflation, and output growth, estimated over the period 1989q1--2020q1}
\vspace*{-0.8cm}
\small\renewcommand{\arraystretch}{1.15}
\label{table: svar_dm2 order v3}

\begin{center}
	\begin{tabular}{@{\extracolsep{-15pt}}lD{.}{.}{-3} D{.}{.}{-3} D{.}{.}{-3}
D{.}{.}{-3} D{.}{.}{-3} D{.}{.}{-3} D{.}{.}{-3} }
		\\[-1.8ex]\hline\hline\\[-1.8ex]
		& \multicolumn{7}{c}{$\Delta m_{t}$} \\
		\cline{2-8}
		\\[-1.8ex] & \multicolumn{1}{c}{(1)} & \multicolumn{1}{c}{(2)}
& \multicolumn{1}{c}{(3)} & \multicolumn{1}{c}{(4)} & \multicolumn{1}{c}{(5)}
& \multicolumn{1}{c}{(6)} & \multicolumn{1}{c}{(7)}\\
		\hline\\[-1.8ex]
		$s_{t}$ & -0.006 & -0.005 & -0.005 & -0.005 & -0.005 & -0.005 & -0.005 \\
		& (0.021) & (0.022) & (0.022) & (0.022) & (0.022) & (0.022) & (0.022) \\
		$s_{t-1}$ & 0.015 & 0.015 & 0.015 & 0.014 & 0.014 & 0.014 & 0.014 \\
		& (0.022) & (0.022) & (0.023) & (0.023) & (0.023) & (0.023) & (0.023) \\
		$\Delta e_{f,t-1}%
$ & -0.044 & -0.045 & -0.045 & -0.048 & -0.047 & -0.047 & -0.047 \\
		& (0.032) & (0.032) & (0.032) & (0.032) & (0.032) & (0.032) & (0.033) \\
		$\Delta m_{t-1}$ & -0.290^{***} & -0.294^{***} & -0.294^{***} & -0.311^{***}
& -0.313^{***} & -0.312^{***} & -0.311^{***} \\
		& (0.088) & (0.089) & (0.089) & (0.089) & (0.090) & (0.091) & (0.092) \\
		$\Delta p_{t-1}$ & 0.131 & 0.137 & 0.137 & 0.120 & 0.118 & 0.117 & 0.116 \\
		& (0.116) & (0.117) & (0.118) & (0.118) & (0.119) & (0.120) & (0.121) \\
		$\Delta y_{t-1}%
$ & -0.061 & -0.062 & -0.062 & -0.067 & -0.068 & -0.068 & -0.067 \\
		& (0.086) & (0.086) & (0.086) & (0.086) & (0.086) & (0.087) & (0.087) \\
		$\Delta\bar{y}_{wt}%
$ &  & 0.165 & 0.164 & -0.193 & -0.201 & -0.195 & -0.198 \\
		&  & (0.398) & (0.429) & (0.480) & (0.485) & (0.492) & (0.495) \\
		$\Delta p^{0}_{t}$ &  &  & 0.0001 & -0.006 & -0.008 & -0.008 & -0.007 \\
		&  &  & (0.018) & (0.019) & (0.021) & (0.021) & (0.023) \\
		$grv_{t}$ &  &  &  & -0.137 & -0.136 & -0.141 & -0.142 \\
		&  &  &  & (0.084) & (0.085) & (0.108) & (0.109) \\
		$\Delta\bar{e}_{wt}$ &  &  &  &  & -0.023 & -0.023 & -0.020 \\
		&  &  &  &  & (0.154) & (0.155) & (0.159) \\
		$\Delta\overline{req}_{wt}$ &  &  &  &  &  & -0.004 & -0.004 \\
		&  &  &  &  &  & (0.054) & (0.055) \\
		$\Delta\bar{r}_{wt}$ &  &  &  &  &  &  & -0.427 \\
		&  &  &  &  &  &  & (4.442) \\
		\hline\\[-2.3ex]
		Residual serial & 7.577 & 7.715 & 7.720 & 7.444 & 7.503 & 7.676 & 7.718 \\
		correlation test & [0.108] & [0.103] & [0.102] & [0.114] & [0.112] & [0.104] & [0.102] \\ \hline
Adjusted $R^{2}$ & 0.060 & 0.053 & 0.045 & 0.058 & 0.050 & 0.042 & 0.033 \\
		\hline\hline\\[-1.8ex]
	\end{tabular}
\end{center}

\vspace{-0.4cm}
\footnotesize
\textbf{Notes}: The variables are ordered as: $\Delta m_{t},$ $\Delta e_{ft},$
$\Delta p_{t},$\ and $\Delta y_{t},$ where: $\Delta m_{t}=\ln(M_{2t}%
/M_{2,t-1}),$ $M_{2t}$ is obtained by summing the aggregates $M1$ and
"quasi-money"; $\Delta e_{ft}=\ln(E_{ft}/E_{f,t-1}),$ $E_{ft}$ is the
quarterly rial/U.S. dollar free market exchange rate; $\Delta p_{t}=\ln
(P_{t}/P_{t-1}),$ $P_{t}$ is the quarterly consumer price index of Iran;
$\Delta y_{t}=\ln(Y_{t}/Y_{t-1}),$ $Y_{t}$ is the quarterly real output of
Iran. See the notes to Table \ref{table: svar_dfx} for further details on the
construction and sources of data used.%

\end{table}

\begin{table}[H]
\addtocounter{table}{-1}
\renewcommand{\thetable}{S.\arabic{table}b}
\caption
{Quarterly estimates of the equation for the rate of change of the free market foreign exchange rate
in the SVAR model of Iran with domestic variables ordered as: money supply growth,
foreign exchange rate returns, inflation, and output growth, estimated over the period 1989q1--2020q1}
\vspace*{-0.8cm}
\small\renewcommand{\arraystretch}{1.1}
\label{table: svar_dfx order v3}

\begin{center}
	\begin{tabular}{@{\extracolsep{-15pt}}lD{.}{.}{-3} D{.}{.}{-3} D{.}{.}{-3}
D{.}{.}{-3} D{.}{.}{-3} D{.}{.}{-3} D{.}{.}{-3} }
		\\[-1.8ex]\hline\hline\\[-1.8ex]
		& \multicolumn{7}{c}{$\Delta e_{ft}$} \\
		\cline{2-8}
		\\[-1.8ex] & \multicolumn{1}{c}{(1)} & \multicolumn{1}{c}{(2)}
& \multicolumn{1}{c}{(3)} & \multicolumn{1}{c}{(4)} & \multicolumn{1}{c}{(5)}
& \multicolumn{1}{c}{(6)} & \multicolumn{1}{c}{(7)}\\
		\hline\\[-1.8ex]
		$s_{t}$ & 0.310^{***} & 0.302^{***} & 0.295^{***} & 0.295^{***}
& 0.296^{***} & 0.303^{***} & 0.302^{***} \\
		& (0.062) & (0.061) & (0.061) & (0.061) & (0.061) & (0.061) & (0.061) \\
		$s_{t-1}$ & -0.241^{***} & -0.244^{***} & -0.236^{***} & -0.237^{***}
& -0.234^{***} & -0.240^{***} & -0.241^{***} \\
		& (0.064) & (0.064) & (0.064) & (0.064) & (0.063) & (0.063) & (0.064) \\
		$\Delta m_{t}%
$ & -0.134 & -0.116 & -0.116 & -0.180 & -0.174 & -0.175 & -0.177 \\
		& (0.264) & (0.262) & (0.262) & (0.263) & (0.260) & (0.261) & (0.261) \\
		$\Delta e_{f,t-1}$ & 0.329^{***} & 0.336^{***} & 0.340^{***} & 0.329^{***}
& 0.319^{***} & 0.320^{***} & 0.327^{***} \\
		& (0.092) & (0.091) & (0.091) & (0.091) & (0.090) & (0.090) & (0.091) \\
		$\Delta m_{t-1}$ & 0.260 & 0.316 & 0.331 & 0.265 & 0.324 & 0.330 & 0.351 \\
		& (0.263) & (0.263) & (0.263) & (0.264) & (0.264) & (0.265) & (0.267) \\
		$\Delta p_{t-1}%
$ & -0.286 & -0.360 & -0.350 & -0.391 & -0.337 & -0.371 & -0.391 \\
		& (0.335) & (0.334) & (0.334) & (0.333) & (0.331) & (0.334) & (0.336) \\
		$\Delta y_{t-1}%
$ & -0.143 & -0.134 & -0.140 & -0.158 & -0.125 & -0.126 & -0.120 \\
		& (0.246) & (0.244) & (0.244) & (0.242) & (0.241) & (0.241) & (0.242) \\
		$\Delta\bar{y}_{wt}$ &  & -2.040^{*} & -2.494^{**} & -3.502^{**}
& -3.246^{**} & -3.082^{**} & -3.140^{**} \\
		&  & (1.128) & (1.210) & (1.352) & (1.348) & (1.364) & (1.370) \\
		$\Delta p^{0}_{t}$ &  &  & 0.054 & 0.035 & 0.078 & 0.075 & 0.093 \\
		&  &  & (0.052) & (0.053) & (0.058) & (0.058) & (0.065) \\
		$grv_{t}$ &  &  &  & -0.390 & -0.411^{*} & -0.567^{*} & -0.576^{*} \\
		&  &  &  & (0.240) & (0.238) & (0.302) & (0.303) \\
		$\Delta\bar{e}_{wt}$ &  &  &  &  & 0.744^{*} & 0.735^{*} & 0.795^{*} \\
		&  &  &  &  & (0.428) & (0.429) & (0.440) \\
		$\Delta\overline{req}_{wt}$ &  &  &  &  &  & -0.127 & -0.117 \\
		&  &  &  &  &  & (0.151) & (0.152) \\
		$\Delta\bar{r}_{wt}$ &  &  &  &  &  &  & -7.807 \\
		&  &  &  &  &  &  & (12.290) \\
		\hline\\[-2.3ex]
		Residual serial & 6.955 & 6.089 & 5.471 & 4.159 & 5.183 & 5.145 & 5.412 \\
		correlation test & [0.138] & [0.193] & [0.242] & [0.385] & [0.269] & [0.273] & [0.248] \\ \hline
Adjusted $R^{2}$ & 0.220 & 0.235 & 0.235 & 0.246 & 0.259 & 0.257 & 0.253 \\
		\hline\hline\\[-1.8ex]
	\end{tabular}
\end{center}

\vspace{-0.4cm}
\footnotesize
\textbf{Notes}: See the notes to Table \ref{table: svar_dfx} for further
details on the construction and sources of data used.%

\end{table}

\bigskip

\begin{table}[H]
\addtocounter{table}{-1}
\renewcommand{\thetable}{S.\arabic{table}c}
\caption{Quarterly estimates of the equation for the inflation rate
in the SVAR model of Iran with domestic variables ordered as: money supply growth,
foreign exchange rate returns, inflation, and output growth, estimated over the period 1989q1--2020q1}
\vspace*{-0.8cm}
\small\renewcommand{\arraystretch}{1.05}
\label{table: svar_dp order v3}

\begin{center}
	\begin{tabular}{@{\extracolsep{-15pt}}lD{.}{.}{-3} D{.}{.}{-3} D{.}{.}{-3}
D{.}{.}{-3} D{.}{.}{-3} D{.}{.}{-3} D{.}{.}{-3} }
		\\[-1.8ex]\hline\hline\\[-1.8ex]
		& \multicolumn{7}{c}{$\Delta p_{t}$} \\
		\cline{2-8}
		\\[-1.8ex] & \multicolumn{1}{c}{(1)} & \multicolumn{1}{c}{(2)}
& \multicolumn{1}{c}{(3)} & \multicolumn{1}{c}{(4)} & \multicolumn{1}{c}{(5)}
& \multicolumn{1}{c}{(6)} & \multicolumn{1}{c}{(7)}\\
		\hline\\[-1.8ex]
		$s_{t}$ & -0.033^{**} & -0.033^{***} & -0.032^{**} & -0.032^{**}
& -0.033^{***} & -0.030^{**} & -0.030^{**} \\
		& (0.013) & (0.012) & (0.013) & (0.013) & (0.012) & (0.013) & (0.013) \\
		$s_{t-1}$ & 0.034^{**} & 0.037^{***} & 0.036^{***} & 0.036^{***}
& 0.036^{***} & 0.033^{**} & 0.033^{**} \\
		& (0.013) & (0.013) & (0.013) & (0.013) & (0.013) & (0.013) & (0.013) \\
		$\Delta e_{ft}$ & 0.153^{***} & 0.162^{***} & 0.164^{***} & 0.163^{***}
& 0.168^{***} & 0.166^{***} & 0.167^{***} \\
		& (0.018) & (0.017) & (0.017) & (0.017) & (0.017) & (0.017) & (0.017) \\
		$\Delta m_{t}%
$ & -0.026 & -0.032 & -0.031 & -0.033 & -0.032 & -0.032 & -0.031 \\
		& (0.050) & (0.048) & (0.048) & (0.049) & (0.049) & (0.048) & (0.048) \\
		$\Delta e_{f,t-1}%
$ & -0.003 & -0.010 & -0.010 & -0.010 & -0.008 & -0.006 & -0.007 \\
		& (0.020) & (0.019) & (0.019) & (0.020) & (0.019) & (0.019) & (0.019) \\
		$\Delta m_{t-1}%
$ & -0.016 & -0.038 & -0.042 & -0.043 & -0.056 & -0.054 & -0.059 \\
		& (0.050) & (0.048) & (0.049) & (0.049) & (0.049) & (0.049) & (0.050) \\
		$\Delta p_{t-1}$ & 0.456^{***} & 0.490^{***} & 0.484^{***} & 0.484^{***}
& 0.460^{***} & 0.443^{***} & 0.441^{***} \\
		& (0.090) & (0.086) & (0.087) & (0.087) & (0.087) & (0.088) & (0.088) \\
		$\Delta y_{t-1}$ & 0.026 & 0.023 & 0.025 & 0.024 & 0.022 & 0.023 & 0.023 \\
		& (0.048) & (0.046) & (0.046) & (0.047) & (0.046) & (0.046) & (0.046) \\
		$\Delta p_{t-2}$ & 0.180^{**} & 0.174^{**} & 0.181^{**} & 0.179^{**}
& 0.199^{**} & 0.207^{***} & 0.216^{***} \\
		& (0.080) & (0.076) & (0.077) & (0.078) & (0.078) & (0.078) & (0.079) \\
		$\Delta\bar{y}_{wt}$ &  & 0.721^{***} & 0.801^{***} & 0.777^{***}
& 0.751^{***} & 0.795^{***} & 0.813^{***} \\
		&  & (0.209) & (0.225) & (0.258) & (0.256) & (0.257) & (0.259) \\
		$\Delta p^{0}_{t}$ &  &  & -0.009 & -0.010 & -0.018 & -0.019^{*}
& -0.023^{*} \\
		&  &  & (0.010) & (0.010) & (0.011) & (0.011) & (0.012) \\
		$grv_{t}$ &  &  &  & -0.009 & -0.002 & -0.049 & -0.046 \\
		&  &  &  & (0.045) & (0.045) & (0.057) & (0.057) \\
		$\Delta\bar{e}_{wt}$ &  &  &  &  & -0.138^{*} & -0.141^{*} & -0.156^{*} \\
		&  &  &  &  & (0.081) & (0.081) & (0.084) \\
		$\Delta\overline{req}_{wt}$ &  &  &  &  &  & -0.038 & -0.040 \\
		&  &  &  &  &  & (0.028) & (0.028) \\
		$\Delta\bar{r}_{wt}$ &  &  &  &  &  &  & 1.802 \\
		&  &  &  &  &  &  & (2.287) \\
		\hline\\[-2.3ex]
		Residual serial & 13.954 & 8.236 & 9.993 & 10.393 & 10.984 & 9.239 & 9.526 \\
		correlation test & [0.007] & [0.083] & [0.041] & [0.034] & [0.027] & [0.055] & [0.049] \\ \hline
Adjusted $R^{2}$ & 0.638 & 0.669 & 0.669 & 0.666 & 0.672 & 0.674 & 0.673 \\
		\hline\hline\\[-1.8ex]
	\end{tabular}
\end{center}

\vspace{-0.4cm}%
\footnotesize
\textbf{Notes}: See the notes to Table \ref{table: svar_dfx} for further
details on the construction and sources of data used.%

\end{table}

\bigskip

\begin{table}[H]
\addtocounter{table}{-1}
\renewcommand{\thetable}{S.\arabic{table}d}
\caption{Quarterly estimates of the equation for the output growth
in the SVAR model of Iran with domestic variables ordered as: money supply growth,
foreign exchange rate returns, inflation, and output growth, estimated over the period 1989q1--2020q1}
\vspace*{-0.8cm}
\small\renewcommand{\arraystretch}{1.05}%
\label{table: svar_dy order v3}%

\begin{center}
	\begin{tabular}{@{\extracolsep{-15pt}}lD{.}{.}{-3} D{.}{.}{-3} D{.}{.}{-3}
D{.}{.}{-3} D{.}{.}{-3} D{.}{.}{-3} D{.}{.}{-3} }
		\\[-1.8ex]\hline\hline\\[-1.8ex]
		& \multicolumn{7}{c}{$\Delta y_{t}$} \\
		\cline{2-8}
		\\[-1.8ex] & \multicolumn{1}{c}{(1)} & \multicolumn{1}{c}{(2)}
& \multicolumn{1}{c}{(3)} & \multicolumn{1}{c}{(4)} & \multicolumn{1}{c}{(5)}
& \multicolumn{1}{c}{(6)} & \multicolumn{1}{c}{(7)}\\
		\hline\\[-1.8ex]
		$s_{t}$ & 0.021 & 0.021 & 0.020 & 0.021 & 0.016 & 0.017 & 0.017 \\
		& (0.025) & (0.025) & (0.025) & (0.025) & (0.025) & (0.025) & (0.026) \\
		$s_{t-1}$ & -0.057^{**} & -0.058^{**} & -0.057^{**} & -0.058^{**}
& -0.055^{**} & -0.056^{**} & -0.057^{**} \\
		& (0.025) & (0.026) & (0.026) & (0.026) & (0.026) & (0.026) & (0.026) \\
		$\Delta e_{ft}$ & -0.123^{***} & -0.125^{***} & -0.127^{***} & -0.130^{***}
& -0.113^{**} & -0.113^{**} & -0.116^{**} \\
		& (0.042) & (0.044) & (0.045) & (0.045) & (0.046) & (0.046) & (0.046) \\
		$\Delta m_{t}$ & 0.096 & 0.097 & 0.097 & 0.088 & 0.086 & 0.085 & 0.084 \\
		& (0.094) & (0.094) & (0.095) & (0.096) & (0.095) & (0.096) & (0.096) \\
		$\Delta p_{t}$ & 0.335^{*} & 0.341^{*} & 0.348^{*} & 0.344^{*} & 0.304^{*}
& 0.300 & 0.305^{*} \\
		& (0.171) & (0.180) & (0.181) & (0.182) & (0.181) & (0.183) & (0.183) \\
		$\Delta e_{f,t-1}%
$ & 0.033 & 0.033 & 0.035 & 0.034 & 0.034 & 0.034 & 0.038 \\
		& (0.034) & (0.035) & (0.035) & (0.035) & (0.035) & (0.035) & (0.035) \\
		$\Delta m_{t-1}%
$ & -0.022 & -0.020 & -0.016 & -0.024 & -0.051 & -0.051 & -0.040 \\
		& (0.094) & (0.095) & (0.096) & (0.097) & (0.097) & (0.098) & (0.098) \\
		$\Delta p_{t-1}$ & -0.490^{***} & -0.496^{***} & -0.499^{***} & -0.502^{***}
& -0.495^{***} & -0.495^{***} & -0.509^{***} \\
		& (0.157) & (0.165) & (0.166) & (0.167) & (0.165) & (0.166) & (0.167) \\
		$\Delta y_{t-1}$ & -0.195^{**} & -0.195^{**} & -0.196^{**} & -0.199^{**}
& -0.210^{**} & -0.210^{**} & -0.208^{**} \\
		& (0.087) & (0.088) & (0.088) & (0.088) & (0.088) & (0.088) & (0.088) \\
		$\Delta\bar{y}_{wt}%
$ &  & -0.049 & -0.145 & -0.278 & -0.313 & -0.298 & -0.335 \\
		&  & (0.432) & (0.467) & (0.524) & (0.519) & (0.527) & (0.529) \\
		$\Delta p^{0}_{t}$ &  &  & 0.010 & 0.008 & -0.009 & -0.009 & -0.001 \\
		&  &  & (0.019) & (0.019) & (0.021) & (0.021) & (0.024) \\
		$grv_{t}$ &  &  &  & -0.050 & -0.039 & -0.053 & -0.058 \\
		&  &  &  & (0.088) & (0.088) & (0.112) & (0.113) \\
		$\Delta\bar{e}_{wt}$ &  &  &  &  & -0.291^{*} & -0.291^{*} & -0.261 \\
		&  &  &  &  & (0.159) & (0.160) & (0.164) \\
		$\Delta\overline{req}_{wt}$ &  &  &  &  &  & -0.011 & -0.006 \\
		&  &  &  &  &  & (0.055) & (0.056) \\
		$\Delta\bar{r}_{wt}$ &  &  &  &  &  &  & -3.734 \\
		&  &  &  &  &  &  & (4.487) \\
		\hline\\[-2.3ex]
		Residual serial & 7.157 & 7.108 & 6.991 & 7.143 & 7.813 & 7.815 & 7.782 \\
		correlation test & [0.128] & [0.130] & [0.136] & [0.129] & [0.099] & [0.099] & [0.100] \\ \hline
Adjusted $R^{2}$ & 0.137 & 0.129 & 0.124 & 0.119 & 0.137 & 0.129 & 0.127 \\
		\hline\hline\\[-1.8ex]
	\end{tabular}
\end{center}

\vspace{-0.4cm}
\footnotesize
\textbf{Notes}: See the notes to Table \ref{table: svar_dfx} for further
details on the construction and sources of data used.%

\end{table}

\begin{figure}[H]
\caption
{Impulse responses of the effects of world output growth shocks on foreign exchange,
inflation, and Iran output growth}
\label{fig: IRF global output growth}

\footnotesize
$%
\vspace{-0.15cm}%
$%
\[%
\begin{tabular}
[c]{c}%
$%
\vspace{0.2cm}%
$One positive standard error shock to the global output growth\\%
{\includegraphics[
height=2.3062in,
width=7.0978in
]%
{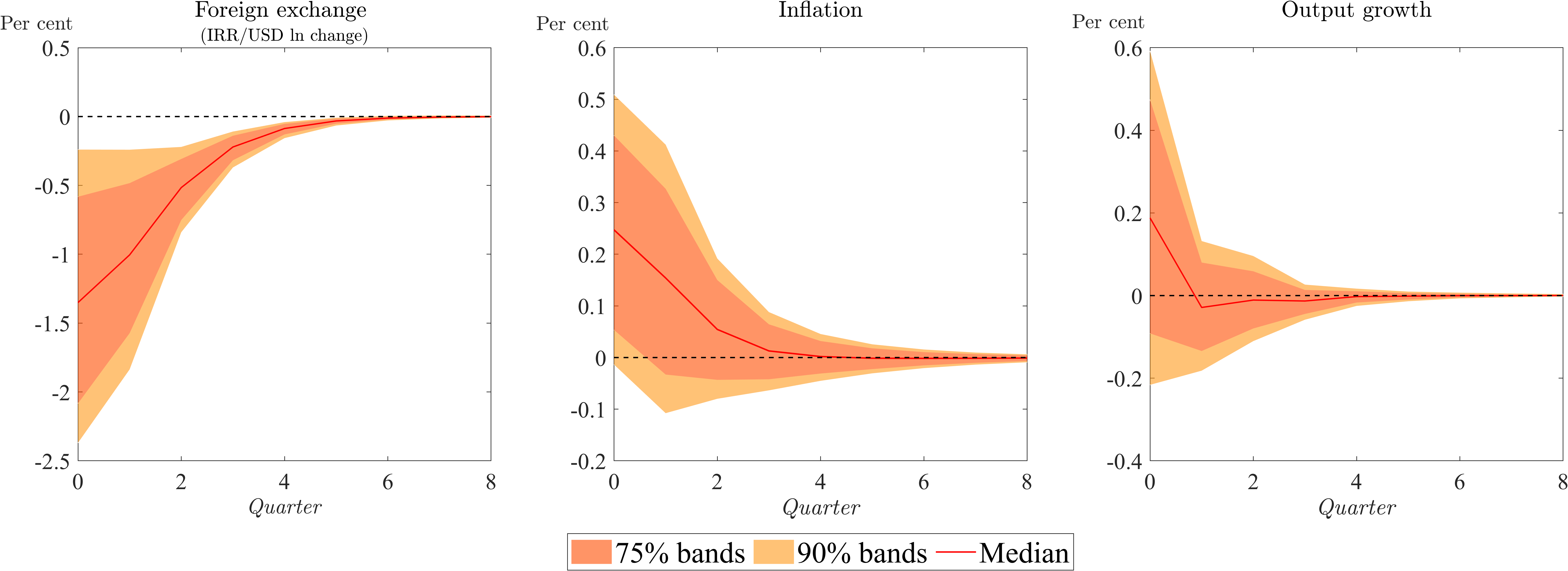}%
}
\end{tabular}
\]
\bigskip

\footnotesize
\textbf{Notes}: The structural VAR system studied in Equation (\ref{psi0zt}%
)\ is composed of five variables. The current figure supplements the four
panels available in Figure \ref{fig: IRF}.%

\end{figure}

\bigskip

\begin{table}[H]
\caption
{Estimates of reduced form Iran's output growth equation estimated over the period 1989q1--2020q1}
\vspace*{-0.8cm}
\small\renewcommand{\arraystretch}{1.15}
\label{table: reduced form s(t-1) on dy}

\begin{center}
	\begin{tabular}{@{\extracolsep{-15pt}}lD{.}{.}{-3} D{.}{.}{-3} D{.}{.}{-3}
D{.}{.}{-3} D{.}{.}{-3} D{.}{.}{-3} D{.}{.}{-3} }
		\\[-1.8ex]\hline\hline\\[-1.8ex]
		& \multicolumn{7}{c}{$\Delta y_{t}$} \\
		\cline{2-8}
		\\[-1.8ex] & \multicolumn{1}{c}{(1)} & \multicolumn{1}{c}{(2)}
& \multicolumn{1}{c}{(3)} & \multicolumn{1}{c}{(4)} & \multicolumn{1}{c}{(5)}
& \multicolumn{1}{c}{(6)} & \multicolumn{1}{c}{(7)}\\
		\hline\\[-1.8ex]
		$s_{t-1} (\beta_{s_{t-1}})$ & -0.037^{**} & -0.035^{**} & -0.035^{**}
& -0.035^{**} & -0.037^{**} & -0.037^{**} & -0.038^{**} \\
		& (0.016) & (0.016) & (0.016) & (0.016) & (0.016) & (0.016) & (0.016) \\
		$\Delta y_{t-1} (\lambda_{\Delta y_{t-1}})$ & -0.186^{**} & -0.188^{**}
& -0.188^{**} & -0.190^{**} & -0.206^{**} & -0.206^{**} & -0.204^{**} \\
		& (0.089) & (0.089) & (0.089) & (0.090) & (0.088) & (0.089) & (0.089) \\
		$\Delta e_{f,t-1}%
$ & -0.000 & -0.002 & -0.002 & -0.002 & 0.003 & 0.003 & 0.006 \\
		& (0.033) & (0.033) & (0.033) & (0.033) & (0.033) & (0.033) & (0.033) \\
		$\Delta m_{t-1}%
$ & -0.068 & -0.077 & -0.077 & -0.082 & -0.110 & -0.109 & -0.101 \\
		& (0.092) & (0.092) & (0.093) & (0.094) & (0.093) & (0.093) & (0.094) \\
		$\Delta p_{t-1}$ & -0.264^{**} & -0.250^{**} & -0.250^{**} & -0.255^{**}
& -0.283^{**} & -0.287^{**} & -0.295^{**} \\
		& (0.120) & (0.121) & (0.122) & (0.123) & (0.121) & (0.122) & (0.123) \\
		$\Delta\bar{y}_{wt}$ &  & 0.375 & 0.353 & 0.245 & 0.119 & 0.140 & 0.119 \\
		&  & (0.413) & (0.443) & (0.501) & (0.494) & (0.501) & (0.503) \\
		$\Delta p^{0}_{t}$ &  &  & 0.003 & 0.001 & -0.020 & -0.021 & -0.014 \\
		&  &  & (0.019) & (0.020) & (0.021) & (0.021) & (0.024) \\
		$grv_{t}$ &  &  &  & -0.041 & -0.030 & -0.052 & -0.055 \\
		&  &  &  & (0.088) & (0.087) & (0.110) & (0.111) \\
		$\Delta\bar{e}_{wt}$ &  &  &  &  & -0.370^{**} & -0.372^{**} & -0.349^{**}
\\
		&  &  &  &  & (0.158) & (0.158) & (0.163) \\
		$\Delta\overline{req}_{wt}$ &  &  &  &  &  & -0.018 & -0.014 \\
		&  &  &  &  &  & (0.055) & (0.056) \\
		$\Delta\bar{r}_{wt}$ &  &  &  &  &  &  & -2.932 \\
		&  &  &  &  &  &  & (4.542) \\
		\hline\\[-1.8ex]
		$\beta_{s_{t-1}} / (1-\lambda_{\Delta y_{t-1}})$ & -0.031^{**} & -0.029^{**}
& -0.029^{**} & -0.029^{**} & -0.031^{**} & -0.031^{**} & -0.031^{**} \\
		& (0.013) & (0.013) & (0.013) & (0.013) & (0.013) & (0.013) & (0.013) \\ \hline
Adjusted $R^{2}$ & 0.093 & 0.091 & 0.084 & 0.077 & 0.112 & 0.105 & 0.100 \\
		\hline\hline\\[-1.8ex]
	\end{tabular}
\end{center}

\vspace{-0.4cm}
\footnotesize
\textbf{Notes}: $\beta_{s_{t-1}}$ and $\lambda_{\Delta y_{t-1}}$ are the
coefficients of $s_{t-1}$ and $\Delta y_{t-1},$ respectively; $\beta_{s_{t-1}%
}$ $/(1-$ $\lambda_{\Delta y_{t-1}})$ represents the long run effect of
sanctions on output growth. See Chapter 6 of \cite{pesaran2015}. See the notes
to Table \ref{table: svar_dfx} for further details on the construction and
sources of data used.

\end{table}

\begin{table}[H]
\caption
{Estimates of reduced form Iran's output growth equation including contemporaneous sanction variable and estimated over the period 1989q1--2020q1}
\vspace*{-0.8cm}
\small\renewcommand{\arraystretch}{1.15}%
\label{table: reduced form s(t) and s(t-1) on dy}%

\begin{center}
	\begin{tabular}{@{\extracolsep{-15pt}}lD{.}{.}{-3} D{.}{.}{-3} D{.}{.}{-3}
D{.}{.}{-3} D{.}{.}{-3} D{.}{.}{-3} D{.}{.}{-3} }
		\\[-1.8ex]\hline\hline\\[-1.8ex]
		& \multicolumn{7}{c}{$\Delta y_{t}$} \\
		\cline{2-8}
		\\[-1.8ex] & \multicolumn{1}{c}{(1)} & \multicolumn{1}{c}{(2)}
& \multicolumn{1}{c}{(3)} & \multicolumn{1}{c}{(4)} & \multicolumn{1}{c}{(5)}
& \multicolumn{1}{c}{(6)} & \multicolumn{1}{c}{(7)}\\
		\hline\\[-1.8ex]
		$s_{t} (\beta_{s_{t}}%
)$ & -0.015 & -0.013 & -0.014 & -0.014 & -0.014 & -0.013 & -0.014 \\
			& (0.022) & (0.022) & (0.023) & (0.023) & (0.022) & (0.023) & (0.023) \\
			$s_{t-1} (\beta_{s_{t-1}}%
)$ & -0.025 & -0.025 & -0.024 & -0.024 & -0.026 & -0.027 & -0.027 \\
			& (0.023) & (0.023) & (0.024) & (0.024) & (0.023) & (0.023) & (0.024) \\
			$\Delta y_{t-1} (\lambda_{\Delta y_{t-1}})$ & -0.191^{**} & -0.192^{**}
& -0.193^{**} & -0.194^{**} & -0.211^{**} & -0.211^{**} & -0.208^{**} \\
			& (0.089) & (0.089) & (0.090) & (0.090) & (0.089) & (0.089) & (0.089) \\
			$\Delta e_{f,t-1}%
$ & -0.002 & -0.004 & -0.003 & -0.004 & 0.001 & 0.001 & 0.004 \\
			& (0.033) & (0.033) & (0.033) & (0.034) & (0.033) & (0.033) & (0.034) \\
			$\Delta m_{t-1}%
$ & -0.070 & -0.079 & -0.078 & -0.083 & -0.111 & -0.110 & -0.102 \\
			& (0.092) & (0.092) & (0.093) & (0.094) & (0.093) & (0.093) & (0.094) \\
			$\Delta p_{t-1}$ & -0.257^{**} & -0.244^{**} & -0.243^{**} & -0.249^{**}
& -0.276^{**} & -0.280^{**} & -0.287^{**} \\
			& (0.121) & (0.122) & (0.123) & (0.124) & (0.122) & (0.123) & (0.124) \\
			$\Delta\bar{y}_{wt}$ &  & 0.357 & 0.322 & 0.216 & 0.089 & 0.106 & 0.084 \\
			&  & (0.415) & (0.447) & (0.505) & (0.498) & (0.505) & (0.508) \\
			$\Delta p^{0}_{t}$ &  &  & 0.004 & 0.002 & -0.019 & -0.019 & -0.012 \\
			&  &  & (0.019) & (0.020) & (0.021) & (0.021) & (0.024) \\
			$grv_{t}$ &  &  &  & -0.041 & -0.030 & -0.046 & -0.049 \\
			&  &  &  & (0.089) & (0.087) & (0.111) & (0.112) \\
			$\Delta\bar{e}_{wt}$ &  &  &  &  & -0.371^{**} & -0.372^{**} & -0.349^{**}
\\
			&  &  &  &  & (0.158) & (0.159) & (0.163) \\
			$\Delta\overline{req}_{wt}$ &  &  &  &  &  & -0.013 & -0.009 \\
			&  &  &  &  &  & (0.056) & (0.056) \\
			$\Delta\bar{r}_{wt}$ &  &  &  &  &  &  & -2.982 \\
			&  &  &  &  &  &  & (4.556) \\
			\hline\\[-1.8ex]
			$\beta_{s_{t}} + \beta_{s_{t-1}}$ & -0.040^{**} & -0.038^{**} & -0.038^{**}
& -0.038^{**} & -0.040^{**} & -0.040^{**} & -0.041^{**} \\
			& (0.016) & (0.017) & (0.017) & (0.017) & (0.017) & (0.017) & (0.017) \\
			$(\beta_{s_{t}} + \beta_{s_{t-1}}) / (1-\lambda_{\Delta y_{t-1}%
})$ & -0.034^{**} & -0.032^{**} & -0.032^{**} & -0.032^{**} & -0.033^{**}
& -0.033^{**} & -0.034^{**} \\
			& (0.014) & (0.014) & (0.014) & (0.014) & (0.013) & (0.014) & (0.014) \\ \hline
Adjusted $R^{2}$ & 0.088 & 0.086 & 0.079 & 0.072 & 0.107 & 0.100 & 0.095 \\
			\hline\hline\\[-1.8ex]
		\end{tabular}
\end{center}

\vspace{-0.4cm}
\footnotesize
\textbf{Notes}: $\beta_{s_{t}},\beta_{s_{t-1}}$ and $\lambda_{\Delta y_{t-1}}$
are the coefficients of $s_{t},s_{t-1}$ and $\Delta y_{t-1},$ respectively;
$(\beta_{s_{t}}+\beta_{s_{t-1}})/(1-$ $\lambda_{\Delta y_{t-1}})$ represents
the long run effect of sanctions on output growth. See Chapter 6 of
\cite{pesaran2015}. See the notes to Table \ref{table: svar_dfx} for further
details on the construction and sources of data used.%

\end{table}

\bigskip

\begin{sidewaystable}
\small\renewcommand{\arraystretch}{1.1}
\caption
{Effects of sanctions on Iran's sectoral output growths estimated over the period 1989--2019}
\label{table: sector_var}%

\begin{tabular}{lllllllllllll}
		\hline\hline&       & \multicolumn{1}{c}{(1)} &       &       & \multicolumn
{1}{c}{(2)} &       &       & \multicolumn{1}{c}{(3)}
&       &       & \multicolumn{1}{c}{(4)} &  \\
		\cline{2-13}          & \multicolumn{1}{p{2em}}{$\Delta agr_{t}$}
& \multicolumn{1}{p{2em}}{$\Delta mfg_{t}$} & \multicolumn{1}{p{2em}|}{$\Delta
svcs_{t}$} & \multicolumn{1}{p{2em}}{$\Delta agr_{t}$}
& \multicolumn{1}{p{2em}}{$\Delta mfg_{t}$} & \multicolumn{1}{p{2em}|}{$\Delta
svcs_{t}$} & \multicolumn{1}{p{2em}}{$\Delta agr_{t}$}
& \multicolumn{1}{p{2em}}{$\Delta mfg_{t}$} & \multicolumn{1}{p{2em}|}{$\Delta
svcs_{t}$} & \multicolumn{1}{p{2em}}{$\Delta agr_{t}$}
& \multicolumn{1}{p{2em}}{$\Delta mfg_{t}$} & \multicolumn{1}{p{2em}}{$\Delta
svcs_{t}$} \\
		\hline&       &       & \multicolumn{1}{l|}{} &       &       & \multicolumn
{1}{l|}{} &       &       & \multicolumn{1}{l|}{} &       &       &  \\
		$s_{t}$ & -0.010 & -0.154*** & \multicolumn{1}{l|}{-0.074***}
& -0.008 & -0.158*** & \multicolumn{1}{l|}{-0.076***}
& 0.019 & -0.143*** & \multicolumn{1}{l|}{-0.067***}
& 0.038 & -0.154*** & -0.075*** \\
		& (0.058) & (0.057) & \multicolumn{1}{l|}{(0.028)}
& (0.051) & (0.054) & \multicolumn{1}{l|}{(0.025)}
& (0.050) & (0.055) & \multicolumn{1}{l|}{(0.025)}
& (0.054) & (0.059) & (0.027) \\
		\multicolumn{1}{p{5.5em}}{$\Delta agr_{t-1}$}
& 0.034 & -0.352* & \multicolumn{1}{l|}{-0.074}
& 0.113 & -0.284 & \multicolumn{1}{l|}{-0.034} & 0.189 & -0.242 & \multicolumn
{1}{l|}{-0.009} & 0.231 & -0.284 & -0.028 \\
		& (0.189) & (0.188) & \multicolumn{1}{l|}{(0.092)}
& (0.169) & (0.180) & \multicolumn{1}{l|}{(0.083)}
& (0.164) & (0.182) & \multicolumn{1}{l|}{(0.083)}
& (0.162) & (0.177) & (0.082) \\
		\multicolumn{1}{p{5.5em}}{$\Delta mfg_{t-1}$}
& 0.086 & -0.215 & \multicolumn{1}{l|}{-0.226**}
& -0.111 & -0.335* & \multicolumn{1}{l|}{-0.314***}
& -0.106 & -0.332* & \multicolumn{1}{l|}{-0.313***}
& -0.104 & -0.291 & -0.307*** \\
		& (0.187) & (0.186) & \multicolumn{1}{l|}{(0.091)}
& (0.178) & (0.189) & \multicolumn{1}{l|}{(0.087)}
& (0.167) & (0.186) & \multicolumn{1}{l|}{(0.084)}
& (0.173) & (0.190) & (0.087) \\
		\multicolumn{1}{p{5.5em}}{$\Delta svcs_{t-1}$}
& -0.048 & 1.271*** & \multicolumn{1}{l|}{0.878***}
& -0.084 & 1.235*** & \multicolumn{1}{l|}{0.866***}
& -0.244 & 1.145*** & \multicolumn{1}{l|}{0.812***}
& -0.348 & 1.238*** & 0.858*** \\
		& (0.302) & (0.300) & \multicolumn{1}{l|}{(0.147)}
& (0.266) & (0.283) & \multicolumn{1}{l|}{(0.130)}
& (0.262) & (0.292) & \multicolumn{1}{l|}{(0.132)}
& (0.270) & (0.296) & (0.136) \\
		\hline$\Delta p^{0}_{t}$ & yes   & yes   & \multicolumn{1}{l|}{yes}
& yes   & yes   & \multicolumn{1}{l|}{yes} & yes   & yes   & \multicolumn
{1}{l|}{yes} & yes   & yes   & yes \\
		$grv_{t}$ & yes   & yes   & \multicolumn{1}{l|}{yes}
& yes   & yes   & \multicolumn{1}{l|}{yes} & yes   & yes   & \multicolumn
{1}{l|}{yes} & yes   & yes   & yes \\
		$\Delta\bar{r}_{wt}$ &       &       & \multicolumn{1}{l|}{}
& yes   & yes   & \multicolumn{1}{l|}{yes} & yes   & yes   & \multicolumn
{1}{l|}{yes} & yes   & yes   & yes \\
		$\Delta\bar{e}_{wt}$ &       &       & \multicolumn{1}{l|}{}
& yes   & yes   & \multicolumn{1}{l|}{yes} & yes   & yes   & \multicolumn
{1}{l|}{yes} & yes   & yes   & yes \\
		$\Delta\overline{req}_{wt}$ &       &       & \multicolumn{1}{l|}{}
& yes   & yes   & \multicolumn{1}{l|}{yes} & yes   & yes   & \multicolumn
{1}{l|}{yes} & yes   & yes   & yes \\
		$\Delta\bar{y}_{wt}$ &       &       & \multicolumn{1}{l|}{}
&       &       & \multicolumn{1}{l|}{} & yes   & yes   & \multicolumn{1}%
{l|}{yes} & yes   & yes   & yes \\
		$\Delta\bar{y}^{0}_{t}$ &       &       & \multicolumn{1}{l|}{}
&       &       & \multicolumn{1}{l|}{} &       &       & \multicolumn{1}%
{r|}{} & yes   & yes   & yes \\
		$\Delta y^{Tur}_{t}$ &       &       & \multicolumn{1}{l|}{}
&       &       & \multicolumn{1}{l|}{} &       &       & \multicolumn{1}%
{l|}{} & yes   & yes   & yes \\
		\hline Adjusted R$^{2}%
$ & -0.020 & 0.385 & 0.545 & 0.102 & 0.378 & 0.598 & 0.167 & 0.368 & 0.601 & 0.133 & 0.365 & 0.585 \\
		\hline\hline\end{tabular}%

\vspace{0.3cm}

\footnotesize
\textbf{Notes}: $\Delta agr_{t},$ $\Delta mfg_{t},$ and $\Delta svcs_{t}$ are
the yearly rates of change of the value added of agriculture, manufacturing
and mining, and the services sector, respectively: $\Delta agr_{t}=\ln
(Agr_{t}/Agr_{t-1}),\Delta mfg_{t}=\ln(Mfg_{t}/Mfg_{t-1}),$ $\Delta
svcs_{t}=\ln(Svcs_{t}/Svcs_{t-1}),$ with $Agr_{t},Mfg_{t},Svcs_{t}$ being the
value added of the three sectors aforementioned in year $t$. See the notes to
Table \ref{table: svar_dfx} for further details on the construction and
sources of data used.

\end{sidewaystable}

\begin{sidewaystable}
\captionof{table}
{Effects of sanctions on the rate of change of value added of the agriculture and
manufacturing sectors as a share of total value added over the period 1989--2019}
\small\renewcommand{\arraystretch}{1.1}%
\label{table: var_sector_share1}%

\begin{center}
	\begin{tabular}{lllllllllll}
		\hline\hline& \multicolumn{2}{c}{(1)} & \multicolumn{2}{c}{(2)}
& \multicolumn{2}{c}{(3)} & \multicolumn{2}{c}{(4)} & \multicolumn{2}{c}{(5)}
\\
		\cline{2-11}     \rule{0cm}{15pt}     & \multicolumn{1}{p{4.43em}}%
{$\Delta\widetilde{agr}_{t}$} & \multicolumn{1}{p{4.43em}}{$\Delta
\widetilde{mfg}_{t}$} & \multicolumn{1}{p{4.43em}}{$\Delta\widetilde{agr}%
_{t}$} & \multicolumn{1}{p{4.43em}}{$\Delta\widetilde{mfg}_{t}$}
& \multicolumn{1}{p{4.43em}}{$\Delta\widetilde{agr}_{t}$} & \multicolumn
{1}{p{4.43em}}{$\Delta\widetilde{mfg}_{t}$} & \multicolumn{1}{p{4.43em}%
}{$\Delta\widetilde{agr}_{t}$} & \multicolumn{1}{p{4.43em}}{$\Delta
\widetilde{mfg}_{t}$} & \multicolumn{1}{p{4.43em}}{$\Delta\widetilde{agr}%
_{t}$} & \multicolumn{1}{p{4.43em}}{$\Delta\widetilde{mfg}_{t}$} \\
		\hline& \multicolumn{1}{c}{} & \multicolumn{1}{c|}{} & \multicolumn{1}{c}{}
& \multicolumn{1}{c|}{} & \multicolumn{1}{c}{} & \multicolumn{1}{c|}{}
& \multicolumn{1}{c}{} & \multicolumn{1}{c|}{} & \multicolumn{1}{c}{}
& \multicolumn{1}{c}{} \\
		$s_{t}$ & 0.105* & \multicolumn{1}{l|}{-0.066*} & 0.123** & \multicolumn
{1}{l|}{-0.074**} & 0.125** & \multicolumn{1}{l|}{-0.070**}
& 0.144*** & \multicolumn{1}{l|}{-0.076**} & 0.122** & -0.063* \\
		& (0.057) & \multicolumn{1}{l|}{(0.035)} & (0.054) & \multicolumn{1}%
{l|}{(0.035)} & (0.054) & \multicolumn{1}{l|}{(0.034)}
& (0.055) & \multicolumn{1}{l|}{(0.035)} & (0.056) & (0.037) \\
		\multicolumn{1}{p{7.215em}}{$\Delta\widetilde{agr}_{t-1}$}
& 0.598*** & \multicolumn{1}{l|}{-0.326***} & 0.662*** & \multicolumn{1}%
{l|}{-0.352***} & 0.671*** & \multicolumn{1}{l|}{-0.341***}
& 0.715*** & \multicolumn{1}{l|}{-0.364***} & 0.596*** & -0.274** \\
		& (0.188) & \multicolumn{1}{l|}{(0.115)} & (0.178) & \multicolumn{1}%
{l|}{(0.114)} & (0.178) & \multicolumn{1}{l|}{(0.112)}
& (0.175) & \multicolumn{1}{l|}{(0.111)} & (0.185) & (0.122) \\
		\multicolumn{1}{p{7.215em}}{$\Delta\widetilde{mfg}_{t-1}$}
& 0.708** & \multicolumn{1}{l|}{-0.176} & 0.711*** & \multicolumn{1}%
{l|}{-0.177} & 0.698*** & \multicolumn{1}{l|}{-0.193}
& 0.748*** & \multicolumn{1}{l|}{-0.197} & 0.608** & -0.165 \\
		& (0.289) & \multicolumn{1}{l|}{(0.178)} & (0.271) & \multicolumn{1}%
{l|}{(0.173)} & (0.270) & \multicolumn{1}{l|}{(0.171)}
& (0.266) & \multicolumn{1}{l|}{(0.169)} & (0.272) & (0.179) \\
		\hline$\Delta\bar{y}_{wt}$ &       & \multicolumn{1}{l|}{}
& yes   & \multicolumn{1}{l|}{yes} & yes   & \multicolumn{1}{l|}{yes}
& yes   & \multicolumn{1}{l|}{yes} & yes   & yes \\
		$\Delta p^{0}_{t}$ &       & \multicolumn{1}{l|}{} &       & \multicolumn
{1}{r|}{} & yes   & \multicolumn{1}{l|}{yes} & yes   & \multicolumn{1}%
{l|}{yes} & yes   & yes \\
		$\Delta\bar{y}^{0}_{t}$ &       & \multicolumn{1}{l|}{}
&       & \multicolumn{1}{l|}{} &       & \multicolumn{1}{l|}{}
& yes   & \multicolumn{1}{l|}{yes} & yes   & yes \\
		$\Delta y^{Tur}_{t}$ &       & \multicolumn{1}{l|}{} &       & \multicolumn
{1}{l|}{} &       & \multicolumn{1}{l|}{} & yes   & \multicolumn{1}{l|}{yes}
& yes   & yes \\
		$grv_{t}$ &       & \multicolumn{1}{l|}{} &       & \multicolumn{1}{l|}{}
&       & \multicolumn{1}{l|}{} &       & \multicolumn{1}{l|}{}
& yes   & yes \\
		$\Delta\bar{r}_{wt}$ &       & \multicolumn{1}{l|}{} &       & \multicolumn
{1}{l|}{} &       & \multicolumn{1}{l|}{} &       & \multicolumn{1}{l|}{}
& yes   & yes \\
		$\Delta\bar{e}_{wt}$ &       & \multicolumn{1}{l|}{} &       & \multicolumn
{1}{l|}{} &       & \multicolumn{1}{l|}{} &       & \multicolumn{1}{l|}{}
& yes   & yes \\
		$\Delta\overline{req}_{wt}$ &       & \multicolumn{1}{l|}{}
&       & \multicolumn{1}{l|}{} &       & \multicolumn{1}{l|}{}
&       & \multicolumn{1}{l|}{} & yes   & yes \\
		\hline
Adjusted R$^2$ & 0.182 & 0.147 & 0.256 & 0.162 & 0.235 & 0.159 & 0.218 & 0.139 & 0.207 & 0.048  \\
		\hline\hline\end{tabular}
\end{center}

\footnotesize
{}\textbf{Notes}: $\Delta\widetilde{agr}_{t}$ and $\Delta\widetilde{mfg}_{t}$
are the yearly rates of change of the value added of the agriculture, and
manufacturing and mining sectors as a fraction of total value added,
respectively: $\Delta\widetilde{agr}_{t}=\ln(\widetilde{Agr}_{t}%
/\widetilde{Agr}_{t-1}),\Delta\widetilde{mfg}_{t}=\ln(\widetilde{Mfg}%
_{t}/\widetilde{Mfg}_{t-1}),$ with $\widetilde{Agr}_{t},\widetilde{Mfg}_{t},$
being the value added of the two sectors aforementioned divided by the sum of
agriculture, manufacturing and mining and services value added in year $t$.
See the notes to Table \ref{table: svar_dfx} for further details on the
construction and sources of data used.%

\end{sidewaystable}
\clearpage

\bigskip

\clearpage\begin{sidewaystable}
\captionof{table}
{Effects of sanctions on the rate of change of value added of the agriculture and services
sectors as a share of total value added over the period 1989--2019}
\small\renewcommand{\arraystretch}{1.1}
\label{table: var_sector_share2}%

\begin{center}
	\begin{tabular}{lllllllllll}
		\hline\hline& \multicolumn{2}{c}{(1)} & \multicolumn{2}{c}{(2)}
& \multicolumn{2}{c}{(3)} & \multicolumn{2}{c}{(4)} & \multicolumn{2}{c}{(5)}
\\
		\cline{2-11}      \rule{0pt}{10pt}    & \multicolumn{1}{p{4.43em}}%
{$\Delta\widetilde{agr}_{t}$} & \multicolumn{1}{p{4.43em}}{$\Delta
\widetilde{svcs}_{t}$} & \multicolumn{1}{p{4.43em}}{$\Delta\widetilde{agr}%
_{t}$} & \multicolumn{1}{p{4.43em}}{$\Delta\widetilde{svcs}_{t}$}
& \multicolumn{1}{p{4.43em}}{$\Delta\widetilde{agr}_{t}$} & \multicolumn
{1}{p{4.43em}}{$\Delta\widetilde{svcs}_{t}$} & \multicolumn{1}{p{4.43em}%
}{$\Delta\widetilde{agr}_{t}$} & \multicolumn{1}{p{4.43em}}{$\Delta
\widetilde{svcs}_{t}$} & \multicolumn{1}{p{4.43em}}{$\Delta\widetilde{agr}%
_{t}$} & \multicolumn{1}{p{4.43em}}{$\Delta\widetilde{svcs}_{t}$} \\
		\hline&       & \multicolumn{1}{l|}{} &       & \multicolumn{1}{l|}{}
&       & \multicolumn{1}{l|}{} &       & \multicolumn{1}{l|}{} &       &  \\
		$s_{t}$ & 0.111* & \multicolumn{1}{l|}{0.017} & 0.125** & \multicolumn
{1}{l|}{0.018} & 0.129** & \multicolumn{1}{l|}{0.016}
& 0.152*** & \multicolumn{1}{l|}{0.016} & 0.130** & 0.014 \\
		& (0.058) & \multicolumn{1}{l|}{(0.012)} & (0.055) & \multicolumn{1}%
{l|}{(0.012)} & (0.055) & \multicolumn{1}{l|}{(0.012)}
& (0.055) & \multicolumn{1}{l|}{(0.012)} & (0.057) & (0.013) \\
		\multicolumn{1}{p{7.215em}}{$\Delta\widetilde{agr}_{t-1}$}
& 0.387** & \multicolumn{1}{l|}{0.077**} & 0.443*** & \multicolumn{1}%
{l|}{0.082**} & 0.459*** & \multicolumn{1}{l|}{0.075**}
& 0.492*** & \multicolumn{1}{l|}{0.080**} & 0.423*** & 0.064* \\
		& (0.157) & \multicolumn{1}{l|}{(0.033)} & (0.151) & \multicolumn{1}%
{l|}{(0.034)} & (0.151) & \multicolumn{1}{l|}{(0.033)}
& (0.148) & \multicolumn{1}{l|}{(0.033)} & (0.161) & (0.038) \\
		\multicolumn{1}{p{7.215em}}{$\Delta\widetilde{svcs}_{t-1}$}
& -1.703** & \multicolumn{1}{l|}{0.084} & -1.633** & \multicolumn{1}%
{l|}{0.090} & -1.612** & \multicolumn{1}{l|}{0.081} & -1.794*** & \multicolumn
{1}{l|}{0.092} & -1.488** & 0.085 \\
		& (0.680) & \multicolumn{1}{l|}{(0.144)} & (0.644) & \multicolumn{1}%
{l|}{(0.143)} & (0.639) & \multicolumn{1}{l|}{(0.139)}
& (0.631) & \multicolumn{1}{l|}{(0.139)} & (0.638) & (0.151) \\
		\hline$\Delta\bar{y}_{wt}$ &       & \multicolumn{1}{r|}{}
& yes   & \multicolumn{1}{l|}{yes} & yes   & \multicolumn{1}{l|}{yes}
& yes   & \multicolumn{1}{l|}{yes} & yes   & yes \\
		$\Delta p^{0}_{t}$ &       & \multicolumn{1}{l|}{} &       & \multicolumn
{1}{r|}{} & yes   & \multicolumn{1}{l|}{yes} & yes   & \multicolumn{1}%
{l|}{yes} & yes   & yes \\
		$\Delta\bar{y}^{0}_{t} $ &       & \multicolumn{1}{l|}{}
&       & \multicolumn{1}{l|}{} &       & \multicolumn{1}{r|}{}
& yes   & \multicolumn{1}{l|}{yes} & yes   & yes \\
		$\Delta y^{Tur}_{t}$ &       & \multicolumn{1}{l|}{} &       & \multicolumn
{1}{l|}{} &       & \multicolumn{1}{l|}{} & yes   & \multicolumn{1}{l|}{yes}
& yes   & yes \\
		$grv_{t}$ &       & \multicolumn{1}{r|}{} &       & \multicolumn{1}{r|}{}
&       & \multicolumn{1}{r|}{} &       & \multicolumn{1}{r|}{}
& yes   & yes \\
		$\Delta\bar{r}_{wt}$ &       & \multicolumn{1}{r|}{} &       & \multicolumn
{1}{r|}{} &       & \multicolumn{1}{r|}{} &       & \multicolumn{1}{r|}{}
& yes   & yes \\
		$\Delta\bar{e}_{wt}$ &       & \multicolumn{1}{r|}{} &       & \multicolumn
{1}{r|}{} &       & \multicolumn{1}{r|}{} &       & \multicolumn{1}{r|}{}
& yes   & yes \\
		$\Delta\overline{req}_{wt}$ &       & \multicolumn{1}{r|}{}
&       & \multicolumn{1}{r|}{} &       & \multicolumn{1}{r|}{}
&       & \multicolumn{1}{r|}{} & yes   & yes \\
		\hline Adjusted R$^{2}%
$ & 0.188 & 0.101 & 0.247 & 0.085 & 0.228 & 0.103 & 0.222 & 0.070 & 0.216 & -0.076 \\
		\hline\hline\end{tabular}
\end{center}

\footnotesize
{}\textbf{Notes}: $\Delta\widetilde{agr}_{t}$ and $\Delta\widetilde{svcs}_{t}$
are the rates of change of the value added of the agriculture and services
sectors as a fraction of total value added, respectively: $\Delta
\widetilde{agr}_{t}=\ln(\widetilde{Agr}_{t}/\widetilde{Agr}_{t-1}%
),\Delta\widetilde{svcs}_{t}=\ln(\widetilde{Svcs}_{t}/\widetilde{Svcs}%
_{t-1}),$ with $\widetilde{Agr}_{t},\widetilde{Svcs}_{t},$ being the value
added of the two sectors aforementioned divided by the sum of agriculture,
manufacturing and mining and services value added in year $t$. See the notes
to Table \ref{table: svar_dfx} for further details on the construction and
sources of data used.%

\end{sidewaystable}
\clearpage

\clearpage\begin{sidewaystable}
\captionof{table}
{Effects of sanctions on the rate of change of value added of manufacturing and services
sectors as a share of total value added over the period 1989--2019}
\small\renewcommand{\arraystretch}{1.1}%
\label{table: var_sector_share3}%

\begin{center}
	\begin{tabular}{lllllllllll}
		\hline\hline& \multicolumn{2}{c}{(1)} & \multicolumn{2}{c}{(2)}
& \multicolumn{2}{c}{(3)} & \multicolumn{2}{c}{(4)} & \multicolumn{2}{c}{(5)}
\\
		\cline{2-11}    \rule{0pt}{15pt}      & \multicolumn{1}{p{4.43em}}%
{$\Delta\widetilde{mfg}_{t}$} & \multicolumn{1}{p{4.43em}}{$\Delta
\widetilde{svcs}_{t}$} & \multicolumn{1}{p{4.43em}}{$\Delta\widetilde{mfg}%
_{t}$} & \multicolumn{1}{p{4.43em}}{$\Delta\widetilde{svcs}_{t}$}
& \multicolumn{1}{p{4.43em}}{$\Delta\widetilde{mfg}_{t}$} & \multicolumn
{1}{p{4.43em}}{$\Delta\widetilde{svcs}_{t}$} & \multicolumn{1}{p{4.43em}%
}{$\Delta\widetilde{mfg}_{t}$} & \multicolumn{1}{p{4.43em}}{$\Delta
\widetilde{svcs}_{t}$} & \multicolumn{1}{p{4.43em}}{$\Delta\widetilde{mfg}%
_{t}$} & \multicolumn{1}{p{4.43em}}{$\Delta\widetilde{svcs}_{t}$} \\
		\hline& \multicolumn{1}{c}{} & \multicolumn{1}{c|}{} & \multicolumn{1}{c}{}
& \multicolumn{1}{c|}{} & \multicolumn{1}{c}{} & \multicolumn{1}{c|}{}
& \multicolumn{1}{c}{} & \multicolumn{1}{c|}{} & \multicolumn{1}{c}{}
& \multicolumn{1}{c}{} \\
		$s_{t}$ & -0.072** & \multicolumn{1}{l|}{0.018} & -0.075** & \multicolumn
{1}{l|}{0.018} & -0.072** & \multicolumn{1}{l|}{0.017}
& -0.082** & \multicolumn{1}{l|}{0.017} & -0.067* & 0.014 \\
		& (0.036) & \multicolumn{1}{l|}{(0.012)} & (0.036) & \multicolumn{1}%
{l|}{(0.012)} & (0.036) & \multicolumn{1}{l|}{(0.012)}
& (0.036) & \multicolumn{1}{l|}{(0.013)} & (0.039) & (0.014) \\
		\multicolumn{1}{p{7.215em}}{$\Delta\widetilde{mfg}_{t-1}$}
& 0.859*** & \multicolumn{1}{l|}{-0.241**} & 0.881*** & \multicolumn{1}%
{l|}{-0.245**} & 0.824*** & \multicolumn{1}{l|}{-0.218**}
& 0.945*** & \multicolumn{1}{l|}{-0.242**} & 0.698* & -0.183 \\
		& (0.306) & \multicolumn{1}{l|}{(0.105)} & (0.304) & \multicolumn{1}%
{l|}{(0.105)} & (0.307) & \multicolumn{1}{l|}{(0.105)}
& (0.304) & \multicolumn{1}{l|}{(0.106)} & (0.357) & (0.127) \\
		\multicolumn{1}{p{7.215em}}{$\Delta\widetilde{svcs}_{t-1}$}
& 2.436*** & \multicolumn{1}{l|}{-0.474} & 2.460*** & \multicolumn{1}%
{l|}{-0.479} & 2.343*** & \multicolumn{1}{l|}{-0.421}
& 2.686*** & \multicolumn{1}{l|}{-0.473} & 2.007** & -0.337 \\
		& (0.863) & \multicolumn{1}{l|}{(0.296)} & (0.852) & \multicolumn{1}%
{l|}{(0.295)} & (0.853) & \multicolumn{1}{l|}{(0.292)}
& (0.853) & \multicolumn{1}{l|}{(0.296)} & (0.963) & (0.343) \\
		\hline$\Delta\bar{y}_{wt}$ &       & \multicolumn{1}{l|}{}
& yes   & \multicolumn{1}{l|}{yes} & yes   & \multicolumn{1}{l|}{yes}
& yes   & \multicolumn{1}{l|}{yes} & yes   & yes \\
		$\Delta p^{0}_{t}$ &       & \multicolumn{1}{l|}{} &       & \multicolumn
{1}{r|}{} & yes   & \multicolumn{1}{l|}{yes} & yes   & \multicolumn{1}%
{l|}{yes} & yes   & yes \\
		$\Delta\bar{y}^{0}_{t}$ &       & \multicolumn{1}{l|}{}
&       & \multicolumn{1}{l|}{} &       & \multicolumn{1}{l|}{}
& yes   & \multicolumn{1}{l|}{yes} & yes   & yes \\
		$\Delta y^{Tur}_{t}$ &       & \multicolumn{1}{l|}{} &       & \multicolumn
{1}{l|}{} &       & \multicolumn{1}{l|}{} & yes   & \multicolumn{1}{l|}{yes}
& yes   & yes \\
		$grv_{t}$ &       & \multicolumn{1}{l|}{} &       & \multicolumn{1}{l|}{}
&       & \multicolumn{1}{l|}{} &       & \multicolumn{1}{l|}{}
& yes   & yes \\
		$\Delta\bar{r}_{wt}$ &       & \multicolumn{1}{l|}{} &       & \multicolumn
{1}{l|}{} &       & \multicolumn{1}{l|}{} &       & \multicolumn{1}{l|}{}
& yes   & yes \\
		$\Delta\bar{e}_{wt}$ &       & \multicolumn{1}{l|}{} &       & \multicolumn
{1}{l|}{} &       & \multicolumn{1}{l|}{} &       & \multicolumn{1}{l|}{}
& yes   & yes \\
		$\Delta\overline{req}_{wt}$ &       & \multicolumn{1}{l|}{}
&       & \multicolumn{1}{l|}{} &       & \multicolumn{1}{l|}{}
&       & \multicolumn{1}{l|}{} & yes   & yes \\
		\hline Adjusted R$^{2}%
$ & 0.147 & 0.098 & 0.136 & 0.071 & 0.123 & 0.080 & 0.121 & 0.051 & 0.029 & -0.101 \\
		\hline\hline\end{tabular}
\end{center}

\footnotesize
{}\textbf{Notes}: $\Delta\widetilde{mfg}_{t}$ and $\Delta\widetilde{svcs}_{t}$
are the rates of change of the value added of the manufacturing and mining,
and services sectors as a fraction of total value added, respectively:
$\Delta\widetilde{mfg}_{t}=\ln(\widetilde{Mfg}_{t}/\widetilde{Mfg}%
_{t-1}),\Delta\widetilde{svcs}_{t}=\ln(\widetilde{Svcs}_{t}/\widetilde{Svcs}%
_{t-1}),$ with $\widetilde{Mfg}_{t},\widetilde{Svcs}_{t},$ being the value
added of the two sectors aforementioned divided by the sum of agriculture,
manufacturing and mining and services value added in year $t$. See the notes
to Table \ref{table: svar_dfx} for further details on the construction and
sources of data used.%

\end{sidewaystable}
\clearpage
\bigskip

\begin{table}[H]
\caption
{Effects of sanctions on the rate of change of employment rate in Iran relative to other MENA
countries estimated over the period 1995--2019}
	\small\renewcommand{\arraystretch}{0.92}%
\label{table: employment rate}%

\vspace{-0.73cm}

\begin{center}
	\begin{tabular}{@{\extracolsep{-10pt}}lD{.}{.}{-3} D{.}{.}{-3} D{.}{.}{-3}
D{.}{.}{-3} D{.}{.}{-3} }
		\\[-1.8ex]\hline\hline\\[-1.8ex]
		& \multicolumn{5}{c}{$\Delta\tilde{e}_{t}$} \\
		\cline{2-6}
		\\[-1.8ex] & \multicolumn{1}{c}{(1)} & \multicolumn{1}{c}{(2)}
& \multicolumn{1}{c}{(3)} & \multicolumn{1}{c}{(4)} & \multicolumn{1}{c}%
{(5)}\\
		\hline\\[-1.8ex]
		$s_{t}$ & -0.067^{**} & -0.061^{**} & -0.057^{*} & -0.055^{*} & -0.054^{*}
\\
		& (0.028) & (0.027) & (0.028) & (0.029) & (0.026) \\
		$\Delta\tilde{e}_{t-1}$ & 0.408^{**} & 0.443^{**} & 0.439^{**} & 0.420^{**}
& 0.373^{**} \\
		& (0.180) & (0.176) & (0.177) & (0.184) & (0.168) \\
		$\Delta y_{t}$ & -0.170 & -0.119 & -0.081 & -0.087 & -0.103 \\
		& (0.111) & (0.112) & (0.123) & (0.125) & (0.114) \\
		$\Delta p^{0}_{t}$ &  & -0.022 & -0.013 & -0.011 & -0.002 \\
		&  & (0.014) & (0.017) & (0.019) & (0.017) \\
		$\Delta\bar{y}_{wt}$ &  &  & -0.297 & -0.079 & -0.536 \\
		&  &  & (0.368) & (0.554) & (0.542) \\
		$\Delta\bar{y}^{0}_{t}$ &  &  &  & -0.217 & -0.335 \\
		&  &  &  & (0.408) & (0.373) \\
		$\Delta y^{Tur}_{t}$ &  &  &  &  & 0.235^{**} \\
		&  &  &  &  & (0.106) \\
		\hline\\[-1.8ex]
Adjusted R$^{2}$ & 0.236 & 0.287 & 0.274 & 0.245 & 0.381 \\
		\hline\hline\\[-1.8ex]
	\end{tabular}
\end{center}

\vspace{-0.33cm}%

\footnotesize
\textbf{Notes}: The dependent variable is the rate of change of yearly
employment rate in Iran relative to that of other MENA countries, namely
$\Delta\widetilde{e}_{t}=\Delta e_{t}-\Delta e_{t}^{MENA},$ where $\Delta
e_{t}=\ln(E_{t}/E_{t-1})$ and $\Delta e_{t}^{MENA}=\ln(E_{t}^{MENA}%
/E_{t-1}^{MENA})$. $E_{t}$ is Iran's employment rate in year $t$,
$E_{t}^{MENA}=\sum\nolimits_{j=1}^{n_{MENA}}w_{j}E_{jt},$ where $\left\{
E_{jt}\right\}  _{j=1}^{n_{MENA}}$ are the employment rates of 17 MENA
countries (excluding Iran) in year $t$, and $w_{j}$ the population weights for
each country. See the notes to Table \ref{table: svar_dfx} for further details
on the construction and sources of data used. See Section
\ref{Sec. socio-economic vars construction}\ in the data appendix of the
online supplement for details on labor statistics.%

\end{table}

\begin{table}[H]
\caption
{Effects of sanctions on the rate of change of male labor force participation in Iran relative to
other MENA countries estimated  over the period 1995--2019}
	\small\renewcommand{\arraystretch}{0.92}%
\label{table: labor force male}%

\vspace{-0.73cm}%

\begin{center}
	\begin{tabular}{@{\extracolsep{-10pt}}lD{.}{.}{-3} D{.}{.}{-3} D{.}{.}{-3}
D{.}{.}{-3} D{.}{.}{-3} }
		\\[-1.8ex]\hline\hline\\[-1.8ex]
		& \multicolumn{5}{c}{$\Delta\widetilde{lf}_{mt}$} \\
		\cline{2-6}
		\\[-1.8ex] & \multicolumn{1}{c}{(1)} & \multicolumn{1}{c}{(2)}
& \multicolumn{1}{c}{(3)} & \multicolumn{1}{c}{(4)} & \multicolumn{1}{c}%
{(5)}\\
		\hline\\[-1.8ex]
		$s_{t}$ & -0.034^{**} & -0.028^{**} & -0.026^{*} & -0.025^{*} & -0.024^{*}
\\
		& (0.015) & (0.013) & (0.013) & (0.013) & (0.013) \\
		$\Delta\widetilde{lf}_{m, t-1}$ & 0.522^{***} & 0.557^{***} & 0.554^{***}
& 0.535^{***} & 0.512^{***} \\
		& (0.170) & (0.141) & (0.140) & (0.145) & (0.143) \\
		$\Delta y_{t}$ & -0.106^{*} & -0.057 & -0.034 & -0.037 & -0.041 \\
		& (0.061) & (0.053) & (0.057) & (0.058) & (0.057) \\
		$\Delta p^{0}_{t}$ &  & -0.021^{***} & -0.016^{*} & -0.014 & -0.012 \\
		&  & (0.006) & (0.008) & (0.008) & (0.008) \\
		$\Delta\bar{y}_{wt}$ &  &  & -0.178 & -0.049 & -0.183 \\
		&  &  & (0.169) & (0.253) & (0.268) \\
		$\Delta\bar{y}^{0}_{t}$ &  &  &  & -0.129 & -0.163 \\
		&  &  &  & (0.185) & (0.184) \\
		$\Delta y^{Tur}_{t}$ &  &  &  &  & 0.069 \\
		&  &  &  &  & (0.052) \\
		\hline\\[-1.8ex]
Adjusted R$^{2}$ & 0.299 & 0.520 & 0.523 & 0.510 & 0.529 \\
		\hline\hline\\[-1.8ex]
	\end{tabular}
\end{center}

\vspace{-0.33cm}

\footnotesize
\textbf{Notes}: The dependent variable is the rate of change of yearly male
labor force participation rate in Iran relative to that of other MENA
countries, namely $\Delta\widetilde{lf}_{mt}=\Delta lf_{mt}-\Delta
lf_{mt}^{MENA},$ where $\Delta lf_{mt}=\ln(LF_{mt}/LF_{m,t-1})$ and $\Delta
lf_{mt}^{MENA}=\ln(LF_{mt}^{MENA}/LF_{m,t-1}^{MENA})$. $LF_{mt}$ is Iran's
male labor force participation rate in year $t$, $LF_{mt}^{MENA}%
=\sum\nolimits_{j=1}^{n_{MENA}}w_{j}LF_{jmt},$ where $\left\{  LF_{jmt}%
\right\}  _{j=1}^{n_{MENA}}$ are the male labor force participation rates of
17 MENA countries (excluding Iran) in year $t$, and $w_{j}$ the population
weights for each country. \ See the notes to Table \ref{table: svar_dfx} for
further details on the construction and sources of data used. See Section
\ref{Sec. socio-economic vars construction}\ in the data appendix of the
online supplement for details on labor statistics.%

\end{table}

\begin{table}[H]
\caption
{Effects of sanctions on the rate of change of female labor force participation in Iran relative to
other MENA countries estimated  over the period 1995--2019}
\small\renewcommand{\arraystretch}{0.95}%
\label{table: labor force female}%

\vspace{-0.7cm}%

\begin{center}
	\begin{tabular}{@{\extracolsep{-10pt}}lD{.}{.}{-3} D{.}{.}{-3} D{.}{.}{-3}
D{.}{.}{-3} D{.}{.}{-3} }
		\\[-1.8ex]\hline\hline\\[-1.8ex]
		& \multicolumn{5}{c}{$\Delta\widetilde{lf}_{ft}$} \\
		\cline{2-6}
		\\[-1.8ex] & \multicolumn{1}{c}{(1)} & \multicolumn{1}{c}{(2)}
& \multicolumn{1}{c}{(3)} & \multicolumn{1}{c}{(4)} & \multicolumn{1}{c}%
{(5)}\\
		\hline\\[-1.8ex]
		$s_{t}$ & -0.269^{***} & -0.238^{***} & -0.232^{***} & -0.229^{***}
& -0.229^{***} \\
		& (0.062) & (0.048) & (0.049) & (0.050) & (0.046) \\
		$\Delta\widetilde{lf}_{f, t-1}$ & 0.535^{***} & 0.624^{***} & 0.617^{***}
& 0.607^{***} & 0.588^{***} \\
		& (0.143) & (0.111) & (0.112) & (0.116) & (0.107) \\
		$\Delta y_{t}$ & -0.583^{**} & -0.393^{*} & -0.327 & -0.332 & -0.353 \\
		& (0.256) & (0.201) & (0.220) & (0.225) & (0.207) \\
		$\Delta p^{0}_{t}$ &  & -0.098^{***} & -0.084^{**} & -0.079^{**}
& -0.065^{*} \\
		&  & (0.024) & (0.031) & (0.033) & (0.031) \\
		$\Delta\bar{y}_{wt}$ &  &  & -0.491 & -0.150 & -0.920 \\
		&  &  & (0.641) & (0.960) & (0.957) \\
		$\Delta\bar{y}^{0}_{t}$ &  &  &  & -0.344 & -0.525 \\
		&  &  &  & (0.708) & (0.656) \\
		$\Delta y^{Tur}_{t}$ &  &  &  &  & 0.387^{*} \\
		&  &  &  &  & (0.186) \\
		\hline\\[-1.8ex]
Adjusted R$^{2}$ & 0.602 & 0.770 & 0.765 & 0.755 & 0.793 \\
		\hline\hline\\[-1.8ex]
	\end{tabular}
\end{center}

\vspace{-0.3cm}

\footnotesize
\textbf{Notes}: The dependent variable is the rate of change of yearly female
labor force participation rate in Iran relative to that of other MENA
countries, namely $\Delta\widetilde{lf}_{ft}=\Delta lf_{ft}-\Delta
lf_{ft}^{MENA},$ where $\Delta lf_{ft}=\ln(LF_{ft}/LF_{f,t-1})$ and $\Delta
lf_{ft}^{MENA}=\ln(LF_{ft}^{MENA}/LF_{f,t-1}^{MENA})$. $LF_{ft}$ is Iran's
female labor force participation rate in year $t$, $LF_{ft}^{MENA}%
=\sum\nolimits_{j=1}^{n_{MENA}}w_{j}LF_{jft},$ where $\left\{  LF_{jft}%
\right\}  _{j=1}^{n_{MENA}}$ are the female labor force participation rates of
17 MENA countries (excluding Iran) in year $t$, and $w_{j}$ the population
weights for each country. \ See the notes to Table \ref{table: svar_dfx} for
further details on the construction and sources of data used. See Section
\ref{Sec. socio-economic vars construction}\ in the data appendix of the
online supplement for details on labor statistics.

\end{table}
\bigskip

\begin{table}[H]
\caption
{Effects of sanctions on the rate of change of labor force participation in Iran relative to
other MENA countries estimated  over the period 1995--2019}
	\small\renewcommand{\arraystretch}{1.0}
\label{table: labor force}%

\vspace{-0.7cm}

\begin{center}
\begin{tabular}{@{\extracolsep{-10pt}}lD{.}{.}{-3} D{.}{.}{-3} D{.}{.}{-3}
D{.}{.}{-3} D{.}{.}{-3} }
\\[-1.8ex]\hline\hline\\[-1.8ex]
& \multicolumn{5}{c}{$\Delta\widetilde{lf}_{t}$} \\
\cline{2-6}
\\[-1.8ex] & \multicolumn{1}{c}{(1)} & \multicolumn{1}{c}{(2)} & \multicolumn
{1}{c}{(3)} & \multicolumn{1}{c}{(4)} & \multicolumn{1}{c}{(5)}\\
\hline\\[-1.8ex]
$s_{t}$ & -0.083^{***} & -0.073^{***} & -0.070^{***} & -0.069^{***}
& -0.068^{***} \\
& (0.023) & (0.018) & (0.018) & (0.018) & (0.017) \\
$\Delta\widetilde{lf}_{t-1}$ & 0.518^{***} & 0.578^{***} & 0.571^{***}
& 0.561^{***} & 0.529^{***} \\
& (0.152) & (0.118) & (0.119) & (0.124) & (0.116) \\
$\Delta y_{t}$ & -0.179^{*} & -0.104 & -0.074 & -0.076 & -0.082 \\
& (0.093) & (0.074) & (0.080) & (0.082) & (0.077) \\
$\Delta p^{0}_{t}$ &  & -0.035^{***} & -0.028^{**} & -0.027^{**} & -0.022^{*}
\\
&  & (0.009) & (0.011) & (0.012) & (0.011) \\
$\Delta\bar{y}_{w,t}$ &  &  & -0.223 & -0.106 & -0.370 \\
&  &  & (0.234) & (0.353) & (0.355) \\
$\Delta\bar{y}^{0}_{t}$ &  &  &  & -0.118 & -0.187 \\
&  &  &  & (0.260) & (0.244) \\
$\Delta y^{Tur}_{t}$ &  &  &  &  & 0.135^{*} \\
&  &  &  &  & (0.069) \\
\hline\\[-1.8ex]
Adjusted R$^{2}$ & \multicolumn{1}{c}{0.495} & \multicolumn{1}{c}{0.700}
& \multicolumn{1}{c}{0.698} & \multicolumn{1}{c}{0.685} & \multicolumn{1}%
{c}{0.728} \\
\hline\hline\\[-1.8ex]
\end{tabular}
\end{center}

\footnotesize
\textbf{Notes}: The dependent variable is the difference between the rate of
change of yearly labor force participation rate in Iran and other MENA
countries: $\Delta\widetilde{lf}_{t}=\Delta lf_{t}-\Delta lf_{t}^{MENA},$ with
$\Delta lf_{t}=\ln(LF_{t}/LF_{t-1})$ and $\Delta lf_{t}^{MENA}=\ln
(LF_{t}^{MENA}/LF_{t-1}^{MENA})$. $LF_{t}$ is Iran's labor force participation
rate in year $t$, $LF_{t}^{MENA}=\sum\nolimits_{j=1}^{n_{MENA}}w_{j}LF_{jt},$
with $\left\{  LF_{jt}\right\}  _{j=1}^{n_{MENA}}$ being the labor force
participation rate of 17 MENA countries (Iran not included) in year $t$, and
$w_{j}$ the population weights for each country. See the notes to Table
\ref{table: svar_dfx} for further details on the construction and sources of
data used. See Section \ref{Sec. socio-economic vars construction}\ in the
data appendix of the online supplement for details on labor statistics.%

\end{table}
\bigskip

\begin{table}[H]
\caption
{Effects of sanctions on the rate of change of the number of lower secondary schools in Iran estimated
over the period 1989--2018}
\small\renewcommand{\arraystretch}{1.0}
\label{table: school lower secondary}

\vspace{-0.7cm}
\begin{center}
	\begin{tabular}{@{\extracolsep{-10pt}}lD{.}{.}{-3} D{.}{.}{-3} D{.}{.}{-3}
D{.}{.}{-3} D{.}{.}{-3} }
		\\[-1.8ex]\hline\hline\\[-1.8ex]
		& \multicolumn{5}{c}{Lower secondary schools} \\
		\cline{2-6}
		\\[-1.8ex] & \multicolumn{1}{c}{(1)} & \multicolumn{1}{c}{(2)}
& \multicolumn{1}{c}{(3)} & \multicolumn{1}{c}{(4)} & \multicolumn{1}{c}%
{(5)}\\
		\hline\\[-1.8ex]
		$s_{t-1}$ & -0.217^{***} & -0.221^{***} & -0.227^{***} & -0.222^{***}
& -0.223^{***} \\
		& (0.052) & (0.052) & (0.050) & (0.054) & (0.055) \\
		$\Delta y_{t}$ & 0.124 & 0.166 & 0.130 & 0.117 & 0.114 \\
		& (0.167) & (0.167) & (0.163) & (0.172) & (0.181) \\
		$\Delta p^{0}_{t}$ &  & -0.036 & -0.066^{**} & -0.062^{*} & -0.062 \\
		&  & (0.027) & (0.032) & (0.035) & (0.036) \\
		$\Delta\bar{y}_{wt}$ &  &  & 1.072 & 1.288 & 1.265 \\
		&  &  & (0.637) & (0.979) & (1.068) \\
		$\Delta\bar{y}^{0}_{t}$ &  &  &  & -0.233 & -0.233 \\
		&  &  &  & (0.791) & (0.808) \\
		$\Delta y^{Tur}_{t}$ &  &  &  &  & 0.011 \\
		&  &  &  &  & (0.186) \\
		\hline\\[-1.8ex]
		Adjusted R$^{2}$ & 0.413 & 0.429 & 0.467 & 0.447 & 0.423 \\
		\hline\hline\\[-1.8ex]
	\end{tabular}
\end{center}

\vspace{-0.3cm}

\footnotesize
\textbf{Notes}: The dependent variable is the rate of change of the number of
lower secondary schools in Iran. See the notes to Table \ref{table: svar_dfx}
for further details on the construction and sources of data used. See Section
\ref{Sec. socio-economic vars construction}\ in the data appendix of the
online supplement for details on education statistics.

\end{table}

\begin{table}[H]
\caption
{Effects of sanctions on the rate of change of the number of high schools in Iran estimated
over the period 1989--2018}
\small\renewcommand{\arraystretch}{1.0}%
\label{table: school high school}%

\vspace{-0.7cm}%

\begin{center}
	\begin{tabular}{@{\extracolsep{-10pt}}lD{.}{.}{-3} D{.}{.}{-3} D{.}{.}{-3}
D{.}{.}{-3} D{.}{.}{-3} }
		\\[-1.8ex]\hline\hline\\[-1.8ex]
		& \multicolumn{5}{c}{High schools} \\
		\cline{2-6}
		\\[-1.8ex] & \multicolumn{1}{c}{(1)} & \multicolumn{1}{c}{(2)}
& \multicolumn{1}{c}{(3)} & \multicolumn{1}{c}{(4)} & \multicolumn{1}{c}%
{(5)}\\
		\hline\\[-1.8ex]
		$s_{t-1}$ & -0.253^{***} & -0.248^{***} & -0.254^{***} & -0.272^{***}
& -0.266^{***} \\
		& (0.074) & (0.075) & (0.075) & (0.080) & (0.081) \\
		$\Delta y_{t}$ & 0.031 & -0.010 & -0.044 & 0.004 & 0.047 \\
		& (0.238) & (0.243) & (0.244) & (0.256) & (0.266) \\
		$\Delta p^{0}_{t}$ &  & 0.035 & 0.006 & -0.010 & -0.012 \\
		&  & (0.039) & (0.047) & (0.053) & (0.053) \\
		$\Delta\bar{y}_{wt}$ &  &  & 1.025 & 0.256 & 0.647 \\
		&  &  & (0.955) & (1.456) & (1.571) \\
		$\Delta\bar{y}^{0}_{t}$ &  &  &  & 0.830 & 0.842 \\
		&  &  &  & (1.177) & (1.189) \\
		$\Delta y^{Tur}_{t}$ &  &  &  &  & -0.195 \\
		&  &  &  &  & (0.274) \\
		\hline\\[-1.8ex]
		Adjusted R$^{2}$ & 0.281 & 0.275 & 0.280 & 0.265 & 0.249 \\
		\hline\hline\\[-1.8ex]
	\end{tabular}
\end{center}

\vspace{-0.3cm}

\footnotesize
\textbf{Notes}: The dependent variable is the rate of change of the number of
high schools in Iran. See the notes to Table \ref{table: svar_dfx} for further
details on the construction and sources of data used. See Section
\ref{Sec. socio-economic vars construction}\ in the data appendix of the
online supplement for details on education statistics.%

\end{table}
\bigskip

\begin{table}[H]
\caption
{Effects of sanctions on the rate of change of the number of primary schools in Iran estimated
over the period 1989--2018}
\small\renewcommand{\arraystretch}{1.0}%
\label{table: school primary}%

\vspace{-0.7cm}
\begin{center}
	\begin{tabular}{@{\extracolsep{-10pt}}lD{.}{.}{-3} D{.}{.}{-3} D{.}{.}{-3}
D{.}{.}{-3} D{.}{.}{-3} }
		\\[-1.8ex]\hline\hline\\[-1.8ex]
		& \multicolumn{5}{c}{Primary schools} \\
		\cline{2-6}
		\\[-1.8ex] & \multicolumn{1}{c}{(1)} & \multicolumn{1}{c}{(2)}
& \multicolumn{1}{c}{(3)} & \multicolumn{1}{c}{(4)} & \multicolumn{1}{c}%
{(5)}\\
		\hline\\[-1.8ex]
		$s_{t-1}$ & -0.041 & -0.041 & -0.040 & -0.047 & -0.051 \\
		& (0.033) & (0.034) & (0.034) & (0.037) & (0.037) \\
		$\Delta y_{t}$ & 0.040 & 0.040 & 0.045 & 0.064 & 0.037 \\
		& (0.105) & (0.109) & (0.112) & (0.118) & (0.121) \\
		$\Delta p^{0}_{t}$ &  & -0.001 & 0.003 & -0.003 & -0.002 \\
		&  & (0.018) & (0.022) & (0.024) & (0.024) \\
		$\Delta\bar{y}_{wt}$ &  &  & -0.130 & -0.441 & -0.686 \\
		&  &  & (0.438) & (0.669) & (0.715) \\
		$\Delta\bar{y}^{0}_{t}$ &  &  &  & 0.336 & 0.329 \\
		&  &  &  & (0.541) & (0.542) \\
		$\Delta y^{Tur}_{t}$ &  &  &  &  & 0.122 \\
		&  &  &  &  & (0.125) \\
		\hline\\[-1.8ex]
		Adjusted R$^{2}$ & 0.007 & -0.031 & -0.068 & -0.095 & -0.097 \\
		\hline\hline\\[-1.8ex]
	\end{tabular}
\end{center}

\vspace{-0.3cm}

\footnotesize
\textbf{Notes}: The dependent variable is the rate of change of the number of
primary schools in Iran. See the notes to Table \ref{table: svar_dfx} for
further details on the construction and sources of data used. See Section
\ref{Sec. socio-economic vars construction}\ in the data appendix of the
online supplement for details on education statistics.

\end{table}

\begin{table}[H]
\caption
{Effects of sanctions on the number of female to male students across all grades
(rate of change) estimated over the period 1989--2019}
\small\renewcommand{\arraystretch}{1.0}%
\label{table: student gender ratio}%

\vspace{-0.7cm}%
\begin{center}
	\begin{tabular}{@{\extracolsep{-10pt}}lD{.}{.}{-3} D{.}{.}{-3} D{.}{.}{-3}
D{.}{.}{-3} D{.}{.}{-3} }
		\\[-1.8ex]\hline\hline\\[-1.8ex]
		& \multicolumn{5}{c}{Female-to-male  student  ratio} \\
		\cline{2-6}
		\\[-1.8ex] & \multicolumn{1}{c}{(1)} & \multicolumn{1}{c}{(2)}
& \multicolumn{1}{c}{(3)} & \multicolumn{1}{c}{(4)} & \multicolumn{1}{c}%
{(5)}\\
		\hline\\[-1.8ex]
		$s_{t}$ & -0.028^{***} & -0.028^{***} & -0.028^{***} & -0.028^{***}
& -0.028^{***} \\
		& (0.008) & (0.008) & (0.008) & (0.008) & (0.009) \\
		$\Delta y_{t}$ & 0.005 & 0.004 & 0.003 & -0.0001 & -0.004 \\
		& (0.027) & (0.029) & (0.029) & (0.031) & (0.032) \\
		$\Delta p^{0}_{t}$ &  & 0.001 & -0.001 & 0.0002 & 0.0005 \\
		&  & (0.005) & (0.006) & (0.006) & (0.006) \\
		$\Delta\bar{y}_{wt}$ &  &  & 0.042 & 0.093 & 0.063 \\
		&  &  & (0.111) & (0.170) & (0.184) \\
		$\Delta\bar{y}^{0}_{t}$ &  &  &  & -0.052 & -0.054 \\
		&  &  &  & (0.130) & (0.132) \\
		$\Delta y^{Tur}_{t}$ &  &  &  &  & 0.015 \\
		&  &  &  &  & (0.033) \\
		\hline\\[-1.8ex]
		Adjusted R$^{2}$ & 0.358 & 0.335 & 0.313 & 0.290 & 0.267 \\
		\hline\hline\\[-1.8ex]
	\end{tabular}
\end{center}

\vspace{-0.3cm}

\footnotesize
\textbf{Notes}: The dependent variable is the rate of change of the ratio of
Iran's females students (as a proportion of the Iran's female population ages
5--19) to the males students (as a proportion of the Iran's male population
ages 5--19). See the notes to Table \ref{table: svar_dfx} for further details
on the construction and sources of data used. See Section
\ref{Sec. socio-economic vars construction}\ in the data appendix of the
online supplement for details on education statistics.%

\end{table}

\begin{table}[H]
\caption
{Effects of sanctions on  the rate of change of the number of primary school teachers  in Iran estimated
over the period 1989--2019}
\small\renewcommand{\arraystretch}{1.0}%
\label{table: teachers primary}%

\vspace{-0.7cm}%

\begin{center}
	\begin{tabular}{@{\extracolsep{-10pt}}lD{.}{.}{-3} D{.}{.}{-3} D{.}{.}{-3}
D{.}{.}{-3} D{.}{.}{-3} }
		\\[-1.8ex]\hline\hline\\[-1.8ex]
		& \multicolumn{5}{c}{Primary school teachers} \\
		\cline{2-6}
		\\[-1.8ex] & \multicolumn{1}{c}{(1)} & \multicolumn{1}{c}{(2)}
& \multicolumn{1}{c}{(3)} & \multicolumn{1}{c}{(4)} & \multicolumn{1}{c}%
{(5)}\\
		\hline\\[-1.8ex]
		$s_{t-1}$ & 0.026 & 0.025 & 0.024 & -0.005 & -0.013 \\
		& (0.069) & (0.070) & (0.071) & (0.071) & (0.072) \\
		$\Delta y_{t}$ & 0.186 & 0.195 & 0.181 & 0.252 & 0.205 \\
		& (0.217) & (0.225) & (0.232) & (0.229) & (0.236) \\
		$\Delta p^{0}_{t}$ &  & -0.008 & -0.017 & -0.047 & -0.045 \\
		&  & (0.038) & (0.047) & (0.049) & (0.049) \\
		$\Delta\bar{y}_{wt}$ &  &  & 0.323 & -1.350 & -1.814 \\
		&  &  & (0.934) & (1.368) & (1.471) \\
		$\Delta\bar{y}^{0}_{t}$ &  &  &  & 1.744 & 1.743 \\
		&  &  &  & (1.069) & (1.073) \\
		$\Delta y^{Tur}_{t}$ &  &  &  &  & 0.228 \\
		&  &  &  &  & (0.258) \\
		\hline\\[-1.8ex]
		Adjusted R$^{2}$ & -0.044 & -0.081 & -0.117 & -0.050 & -0.059 \\
		\hline\hline\\[-1.8ex]
	\end{tabular}
\end{center}
\footnotesize
\textbf{Notes}: The dependent variable is the rate of change of the number of
primary school teachers in Iran as a proportion of the Iran 25--64 population.
See the notes to Table \ref{table: svar_dfx} for further details on the
construction and sources of data used. See Section
\ref{Sec. socio-economic vars construction} in the data appendix of the online
supplement for education statistics.%

\end{table}

\begin{table}[H]
\caption
{Effects of sanctions on  the rate of change of the number of lower secondary school teachers  in Iran estimated
over the period 1989--2019}
\small\renewcommand{\arraystretch}{1.0}%
\label{table: teachers lower secondary}%

\vspace{-0.7cm}

\begin{center}
	\begin{tabular}{@{\extracolsep{-10pt}}lD{.}{.}{-3} D{.}{.}{-3} D{.}{.}{-3}
D{.}{.}{-3} D{.}{.}{-3} }
		\\[-1.8ex]\hline\hline\\[-1.8ex]
		& \multicolumn{5}{c}{Lower secondary school teachers} \\
		\cline{2-6}
		\\[-1.8ex] & \multicolumn{1}{c}{(1)} & \multicolumn{1}{c}{(2)}
& \multicolumn{1}{c}{(3)} & \multicolumn{1}{c}{(4)} & \multicolumn{1}{c}%
{(5)}\\
		\hline\\[-1.8ex]
		$s_{t-1}$ & -0.222^{***} & -0.228^{***} & -0.233^{***} & -0.221^{***}
& -0.229^{***} \\
		& (0.069) & (0.068) & (0.065) & (0.068) & (0.069) \\
		$\Delta y_{t}$ & 0.054 & 0.114 & 0.047 & 0.019 & -0.023 \\
		& (0.218) & (0.218) & (0.212) & (0.219) & (0.226) \\
		$\Delta p^{0}_{t}$ &  & -0.052 & -0.096^{**} & -0.084^{*} & -0.082^{*} \\
		&  & (0.037) & (0.043) & (0.047) & (0.047) \\
		$\Delta\bar{y}_{wt}$ &  &  & 1.545^{*} & 2.215 & 1.801 \\
		&  &  & (0.856) & (1.306) & (1.407) \\
		$\Delta\bar{y}^{0}_{t}$ &  &  &  & -0.699 & -0.700 \\
		&  &  &  & (1.020) & (1.027) \\
		$\Delta y^{Tur}_{t}$ &  &  &  &  & 0.203 \\
		&  &  &  &  & (0.247) \\
		\hline\\[-1.8ex]
		Adjusted R$^{2}$ & 0.286 & 0.312 & 0.365 & 0.352 & 0.344 \\
		\hline\hline\\[-1.8ex]
	\end{tabular}
\end{center}%

\footnotesize
\textbf{Notes}: The dependent variable is the rate of change of the number of
lower secondary school teachers in Iran as a proportion of the Iran 25--64
population. See the notes to Table \ref{table: svar_dfx} for further details
on the construction and sources of data used. See Section
\ref{Sec. socio-economic vars construction} in the data appendix of the online
supplement for education statistics.%

\end{table}

\begin{table}[H]
\caption
{Effects of sanctions on  the rate of change of the number of high school teachers  in Iran estimated
over the period 1989--2019}
\small\renewcommand{\arraystretch}{1.0}%
\label{table: teachers high schools}%

\vspace{-0.7cm}

\begin{center}
	\begin{tabular}{@{\extracolsep{-10pt}}lD{.}{.}{-3} D{.}{.}{-3} D{.}{.}{-3}
D{.}{.}{-3} D{.}{.}{-3} }
		\\[-1.8ex]\hline\hline\\[-1.8ex]
		& \multicolumn{5}{c}{High school teachers} \\
		\cline{2-6}
		\\[-1.8ex] & \multicolumn{1}{c}{(1)} & \multicolumn{1}{c}{(2)}
& \multicolumn{1}{c}{(3)} & \multicolumn{1}{c}{(4)} & \multicolumn{1}{c}%
{(5)}\\
		\hline\\[-1.8ex]
		$s_{t-1}$ & -0.170^{**} & -0.163^{*} & -0.164^{*} & -0.190^{**}
& -0.185^{**} \\
		& (0.081) & (0.081) & (0.082) & (0.084) & (0.086) \\
		$\Delta y_{t}$ & 0.185 & 0.121 & 0.106 & 0.170 & 0.199 \\
		& (0.257) & (0.259) & (0.267) & (0.269) & (0.281) \\
		$\Delta p^{0}_{t}$ &  & 0.056 & 0.047 & 0.019 & 0.018 \\
		&  & (0.044) & (0.054) & (0.057) & (0.058) \\
		$\Delta\bar{y}_{wt}$ &  &  & 0.337 & -1.189 & -0.912 \\
		&  &  & (1.077) & (1.608) & (1.751) \\
		$\Delta\bar{y}^{0}_{t}$ &  &  &  & 1.591 & 1.592 \\
		&  &  &  & (1.257) & (1.277) \\
		$\Delta y^{Tur}_{t}$ &  &  &  &  & -0.136 \\
		&  &  &  &  & (0.307) \\
		\hline\\[-1.8ex]
		Adjusted R$^{2}$ & 0.163 & 0.182 & 0.154 & 0.173 & 0.145 \\
		\hline\hline\\[-1.8ex]
	\end{tabular}
\end{center}

\footnotesize
\textbf{Notes}: The dependent variable is the rate of change of the number of
high school teachers in Iran as a proportion of the Iran 25--64 population.
See the notes to Table \ref{table: svar_dfx} for further details on the
construction and sources of data used. See Section
\ref{Sec. socio-economic vars construction} in the data appendix of the online
supplement for education statistics.%

\end{table}

\begin{table}[H]
\caption
{Effects of sanctions on  the rate of change of the total number of teachers  in Iran estimated
over the period 1989--2019}
\small\renewcommand{\arraystretch}{1.0}
\label{table: teachers total}%

\vspace{-0.7cm}

\begin{center}
	\begin{tabular}{@{\extracolsep{-10pt}}lD{.}{.}{-3} D{.}{.}{-3} D{.}{.}{-3}
D{.}{.}{-3} D{.}{.}{-3} }
		\\[-1.8ex]\hline\hline\\[-1.8ex]
		& \multicolumn{5}{c}{Total number of teachers} \\
		\cline{2-6}
		\\[-1.8ex] & \multicolumn{1}{c}{(1)} & \multicolumn{1}{c}{(2)}
& \multicolumn{1}{c}{(3)} & \multicolumn{1}{c}{(4)} & \multicolumn{1}{c}%
{(5)}\\
		\hline\\[-1.8ex]
		$s_{t-1}$ & -0.089 & -0.089 & -0.090 & -0.108^{*} & -0.112^{*} \\
		& (0.054) & (0.056) & (0.056) & (0.057) & (0.058) \\
		$\Delta y_{t}$ & 0.163 & 0.158 & 0.133 & 0.177 & 0.154 \\
		& (0.172) & (0.178) & (0.182) & (0.184) & (0.191) \\
		$\Delta p^{0}_{t}$ &  & 0.004 & -0.013 & -0.031 & -0.030 \\
		&  & (0.030) & (0.037) & (0.039) & (0.040) \\
		$\Delta\bar{y}_{wt}$ &  &  & 0.575 & -0.457 & -0.680 \\
		&  &  & (0.735) & (1.098) & (1.193) \\
		$\Delta\bar{y}^{0}_{t}$ &  &  &  & 1.077 & 1.076 \\
		&  &  &  & (0.858) & (0.870) \\
		$\Delta y^{Tur}_{t}$ &  &  &  &  & 0.109 \\
		&  &  &  &  & (0.209) \\
		\hline\\[-1.8ex]
		Adjusted R$^{2}$ & 0.124 & 0.093 & 0.079 & 0.099 & 0.072 \\
		\hline\hline\\[-1.8ex]
	\end{tabular}
\end{center}%

\footnotesize
\textbf{Notes}: The dependent variable is the rate of change of the total
number of teachers in Iran. The total number of teachers is computed as the
sum of primary, lower secondary, and high schools teachers, as a proportion of
the 25--64 Iran population. See the notes to Table \ref{table: svar_dfx} for
further details on the construction and sources of data used. See Section
\ref{Sec. socio-economic vars construction} in the data appendix of the online
supplement for education statistics.%

\end{table}

\begin{table}[H]
\caption
{Effects of sanctions on  the rate of change of the total number of schools in Iran estimated
over the period 1989--2018}
\small\renewcommand{\arraystretch}{1.0}
\label{table: school total}

\vspace{-0.7cm}

\begin{center}
	\begin{tabular}{@{\extracolsep{-10pt}}lD{.}{.}{-3} D{.}{.}{-3} D{.}{.}{-3}
D{.}{.}{-3} D{.}{.}{-3} }
		\\[-1.8ex]\hline\hline\\[-1.8ex]
		& \multicolumn{5}{c}{Total number of schools} \\
		\cline{2-6}
		\\[-1.8ex] & \multicolumn{1}{c}{(1)} & \multicolumn{1}{c}{(2)}
& \multicolumn{1}{c}{(3)} & \multicolumn{1}{c}{(4)} & \multicolumn{1}{c}%
{(5)}\\
		\hline\\[-1.8ex]
		$s_{t-1}$ & -0.102^{***} & -0.102^{***} & -0.104^{***} & -0.111^{***}
& -0.112^{***} \\
		& (0.030) & (0.031) & (0.031) & (0.033) & (0.034) \\
		$\Delta y_{t}$ & 0.059 & 0.056 & 0.047 & 0.065 & 0.055 \\
		& (0.097) & (0.100) & (0.102) & (0.107) & (0.112) \\
		$\Delta p^{0}_{t}$ &  & 0.002 & -0.006 & -0.012 & -0.011 \\
		&  & (0.016) & (0.020) & (0.022) & (0.022) \\
		$\Delta\bar{y}_{wt}$ &  &  & 0.280 & -0.019 & -0.115 \\
		&  &  & (0.399) & (0.609) & (0.661) \\
		$\Delta\bar{y}^{0}_{t}$ &  &  &  & 0.323 & 0.320 \\
		&  &  &  & (0.492) & (0.501) \\
		$\Delta y^{Tur}_{t}$ &  &  &  &  & 0.048 \\
		&  &  &  &  & (0.115) \\
		\hline\\[-1.8ex]
		Adjusted R$^{2}$ & 0.307 & 0.281 & 0.266 & 0.249 & 0.222 \\
		\hline\hline\\[-1.8ex]
	\end{tabular}
\end{center}

\footnotesize
\textbf{Notes}: The dependent variable is the rate of change of the total
number of schools. The total number of schools is computed as the sum of
primary, lower secondary, and high schools. See the notes to Table
\ref{table: svar_dfx} for further details on the construction and sources of
data used. See Section \ref{Sec. socio-economic vars construction} in the data
appendix of the online supplement for education statistics.%

\end{table}

\bigskip

\bigskip

\begin{sidewaysfigure}
\captionof{table}{Chronology of major sanctions events against Iran
over the period from November 1979 to January 2021}
\ContinuedFloat
\label{table: chrono1}

\[%
{\includegraphics[
height=5.2424in,
width=8.2057in
]%
{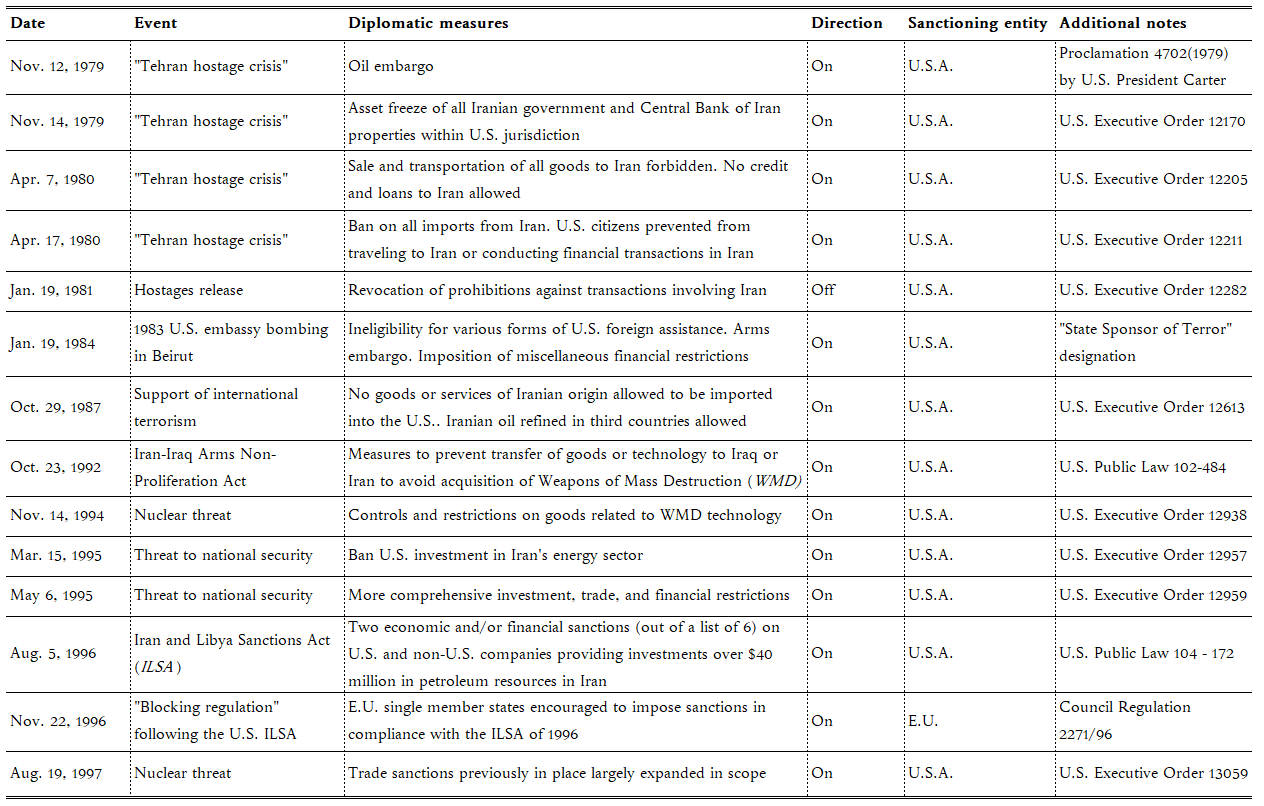}%
}
\]

\end{sidewaysfigure}

\bigskip

\begin{sidewaysfigure}
\captionof{table}{Chronology of major sanctions events against Iran
over the period from November 1979 to January 2021}
\ContinuedFloat
\label{table: chrono2}%
\[%
{\includegraphics[
height=5.2424in,
width=8.2091in
]%
{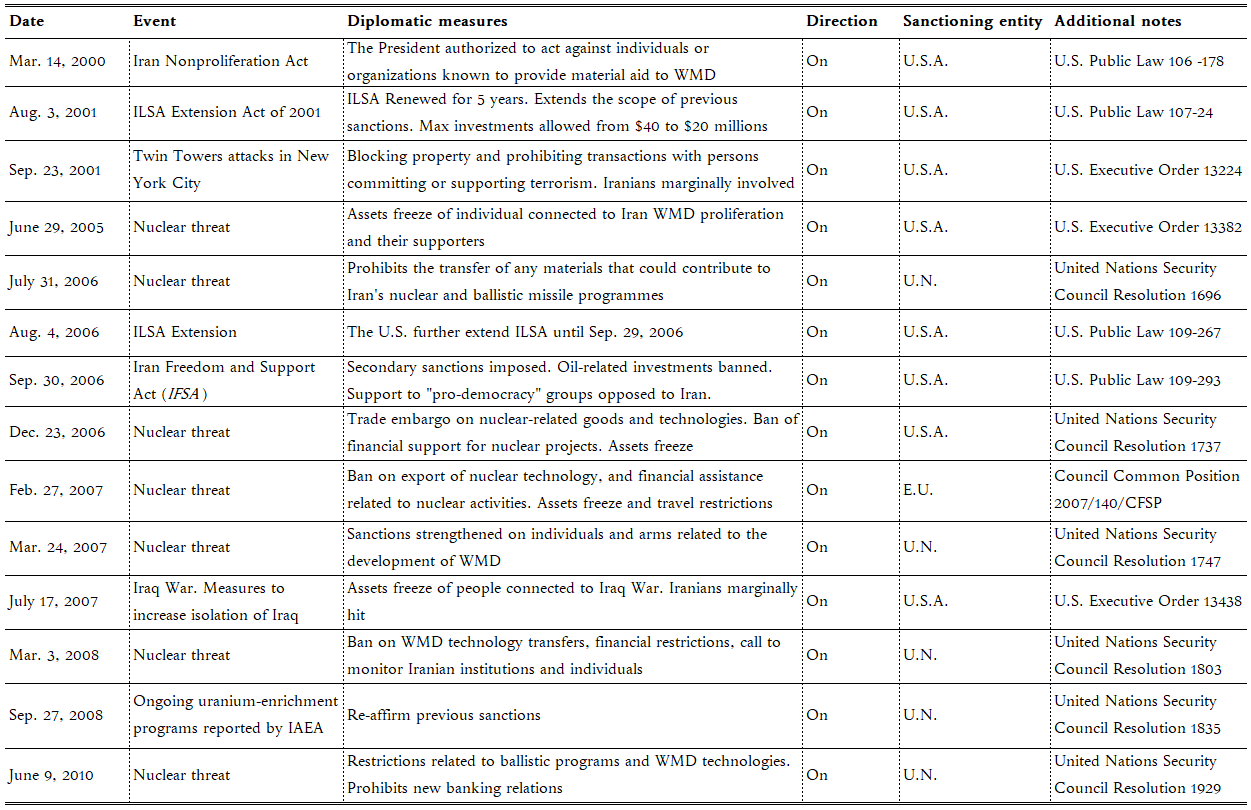}%
}%
\]

\end{sidewaysfigure}

\bigskip

\begin{sidewaysfigure}
\captionof{table}{Chronology of major sanctions events against Iran
over the period from November 1979 to January 2021}
\ContinuedFloat
\label{table: chrono3}
\[%
{\includegraphics[
height=5.245in,
width=8.2091in
]%
{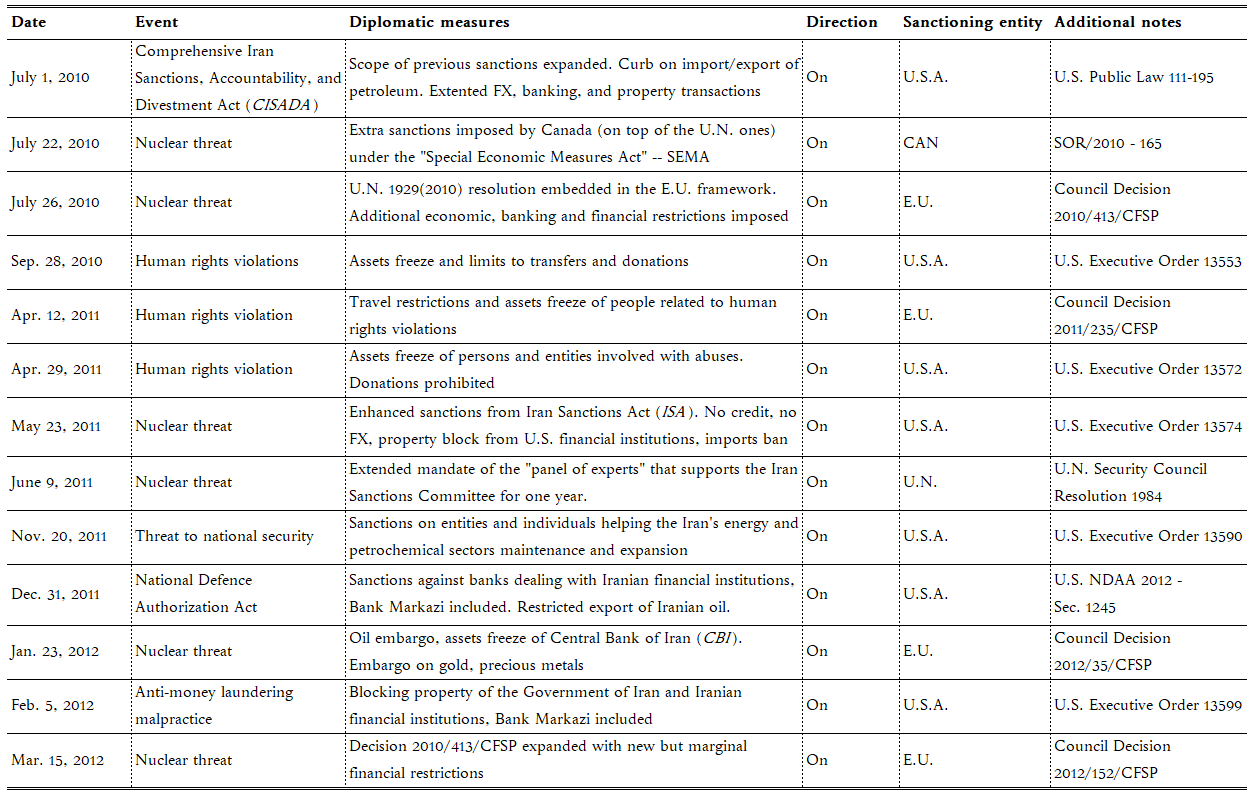}%
}
\]

\end{sidewaysfigure}

\bigskip

\begin{sidewaysfigure}
\captionof{table}{Chronology of major sanctions events against Iran
over the period from November 1979 to January 2021}
\ContinuedFloat
\label{table: chrono4}%
\[%
{\includegraphics[
height=5.2442in,
width=8.2091in
]%
{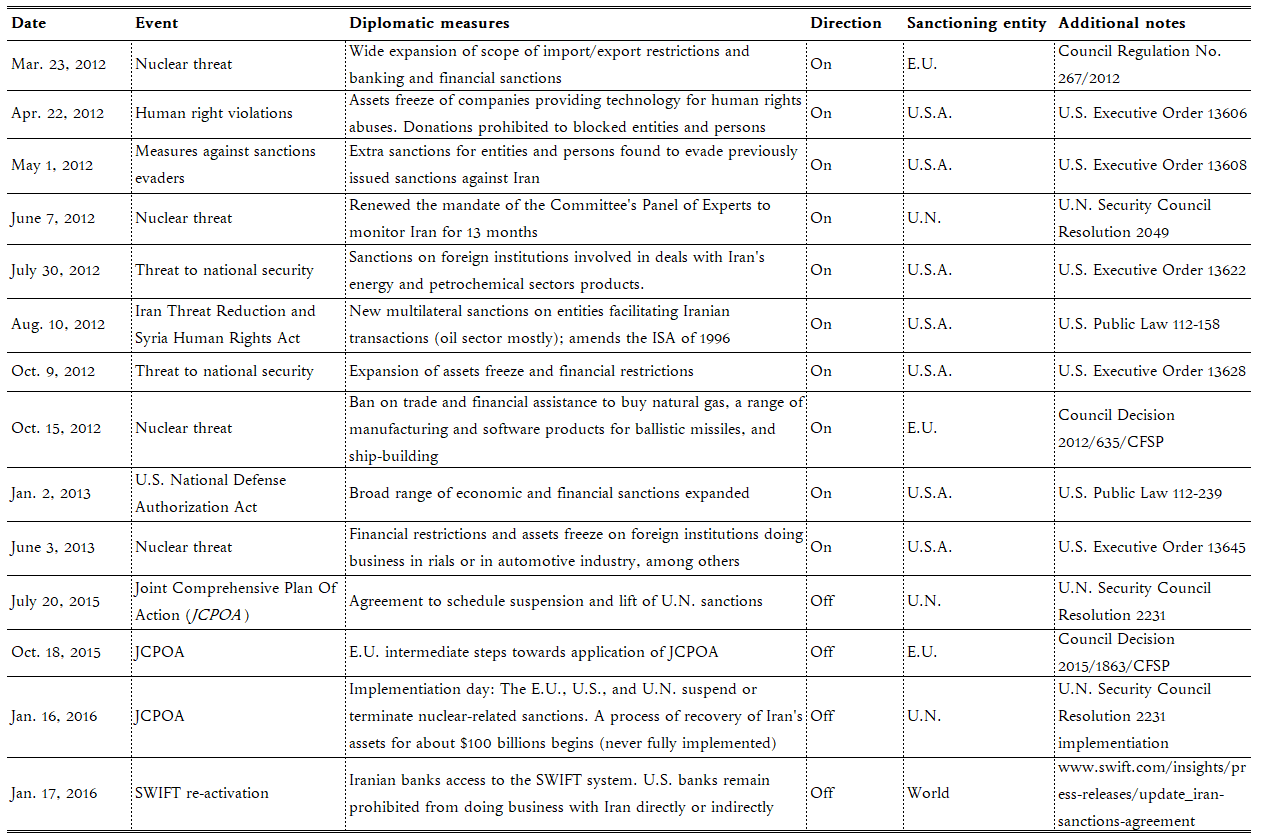}%
}%
\]
\end{sidewaysfigure}

\bigskip

\begin{sidewaysfigure}
\captionof{table}{Chronology of major sanctions events against Iran
over the period from November 1979 to January 2021}
\ContinuedFloat
\label{table: chrono5}
\[%
{\includegraphics[
height=5.2424in,
width=8.2091in
]
{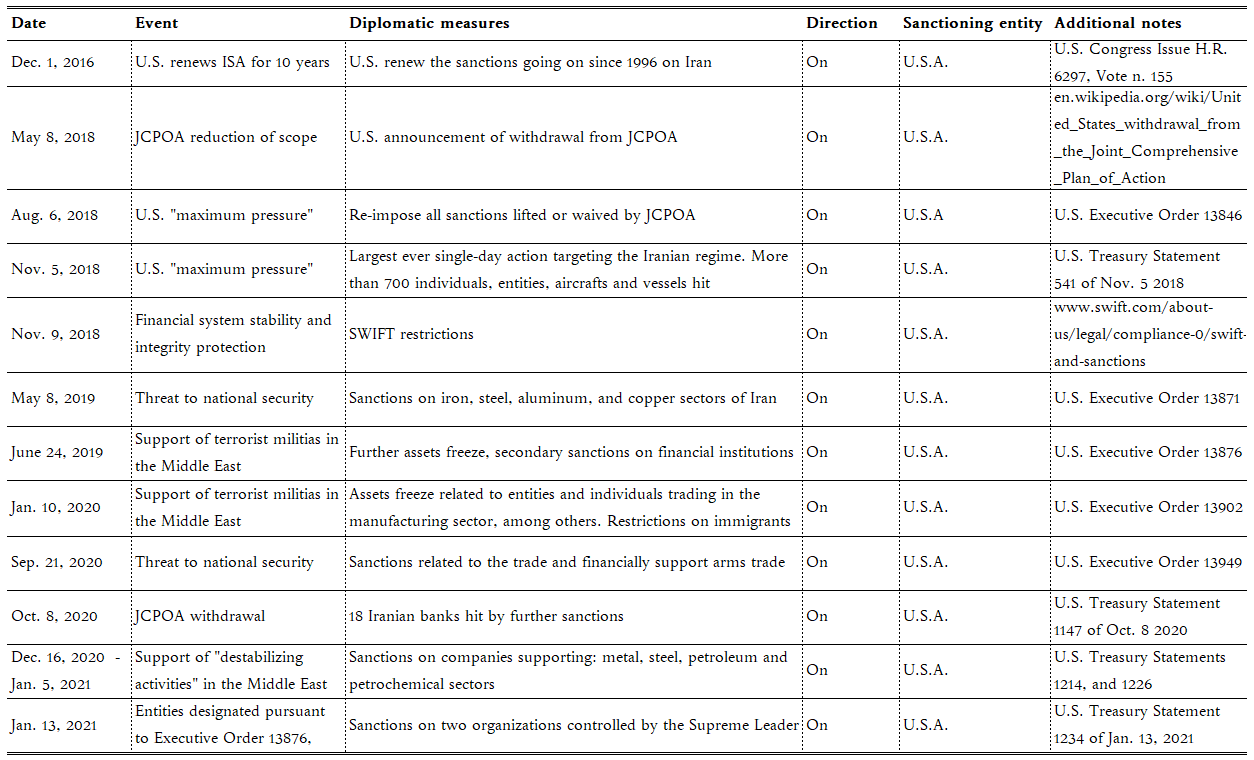}
}
\]

\end{sidewaysfigure}


\begin{thebibliography}{}
	
	\bibitem[\protect\citeauthoryear{Abadie, Diamond, and Hainmueller}{Abadie
		et~al.}{2010}]{abadie2010}
	Abadie, A., A.~Diamond, and J.~Hainmueller (2010).
	\newblock Synthetic {C}ontrol {M}ethods for {C}omparative {C}ase {S}tudies:
	{E}stimating the {E}ffect of {C}alifornia's {T}obacco {C}ontrol {P}rogram.
	\newblock {\em Journal of the American Statistical Association\/}~{\em 105},
	493--505.
	
	\bibitem[\protect\citeauthoryear{Abadie and Gardeazabal}{Abadie and
		Gardeazabal}{2003}]{abadie_gardezabal2003}
	Abadie, A. and J.~Gardeazabal (2003).
	\newblock The {E}conomic {C}osts of {C}onflict: A {C}ase {S}tudy of the
	{B}asque {C}ountry.
	\newblock {\em American Economic Review\/}~{\em 93}, 113--132.
	
	\bibitem[\protect\citeauthoryear{Alizadeh}{Alizadeh}{2017}]{alizadeh2017}
	Alizadeh, P. (2017).
	\newblock Female labour-force participation in {I}ran: A review of the
	literature.
	\newblock {\em Country Briefing Paper No: 06.17.4 prepared for the project
		"Dynamics of Gender Inequality in the Middle East, North Africa and South
		Asia"\/}.
	
	\bibitem[\protect\citeauthoryear{Amuzegar}{Amuzegar}{1997a}]{amuzegar1997b}
	Amuzegar, J. (1997a).
	\newblock Adjusting to {S}anctions.
	\newblock {\em Foreign Affairs\/}~{\em 76}, 31--41.
	
	\bibitem[\protect\citeauthoryear{Amuzegar}{Amuzegar}{1997b}]{amuzegar1997a}
	Amuzegar, J. (1997b).
	\newblock Iran's economy and the {US} sanctions.
	\newblock {\em Middle East Journal\/}~{\em 51}, 185--199.
	
	\bibitem[\protect\citeauthoryear{Andreas}{Andreas}{2005}]{andreas2005}
	Andreas, P. (2005).
	\newblock Criminalizing {C}onsequences of {S}anctions: {E}mbargo {B}usting and
	{I}ts {L}egacy.
	\newblock {\em International Studies Quarterly\/}~{\em 49}, 335--360.
	
	\bibitem[\protect\citeauthoryear{Bahmani-Oskooee}{Bahmani-Oskooee}{1996}]{bahmani1996}
	Bahmani-Oskooee, M. (1996).
	\newblock The {B}lack {M}arket {E}xchange {R}ate and {D}emand for {M}oney in
	{I}ran.
	\newblock {\em Journal of Macroeconomics\/}~{\em 18}, 171--176.
	
	\bibitem[\protect\citeauthoryear{Baker, Bloom, and Davis}{Baker
		et~al.}{2016}]{baker2016}
	Baker, S.~R., N.~Bloom, and S.~J. Davis (2016).
	\newblock Measuring economic policy uncertainty.
	\newblock {\em The Quarterly Journal of Economics\/}~{\em 131}, 1593--1636.
	
	\bibitem[\protect\citeauthoryear{Baumeister and Hamilton}{Baumeister and
		Hamilton}{2015}]{baumeister_hamilton2015}
	Baumeister, C. and J.~D. Hamilton (2015).
	\newblock Sign {R}estrictions, {S}tructural {V}ector {A}utoregressions, and
	{U}seful {P}rior {I}nformation.
	\newblock {\em Econometrica\/}~{\em 83}, 1963--1999.
	
	\bibitem[\protect\citeauthoryear{Bonato}{Bonato}{2008}]{bonato2008}
	Bonato, L. (2008).
	\newblock Money and {I}nflation in the {I}slamic {R}epublic of {I}ran.
	\newblock {\em Review of Middle East Economics and Finance\/}~{\em 4}, Article
	3.
	
	\bibitem[\protect\citeauthoryear{Borszik}{Borszik}{2016}]{borszik2016}
	Borszik, O. (2016).
	\newblock International sanctions against {I}ran and {T}ehran's responses:
	political effects on the targeted regime.
	\newblock {\em Contemporary {P}olitics\/}~{\em 22}, 20--39.
	
	\bibitem[\protect\citeauthoryear{Carswell}{Carswell}{1981}]{carswell1981}
	Carswell, R. (1981).
	\newblock Economic sanctions and the {I}ran experience.
	\newblock {\em Foreign Affairs\/}~{\em 60}, 247--265.
	
	\bibitem[\protect\citeauthoryear{Celasun and Goswami}{Celasun and
		Goswami}{2002}]{celasun_goswami2002}
	Celasun, O. and M.~Goswami (2002).
	\newblock An {A}nalysis of {M}oney {D}emand and {I}nflation in the {I}slamic
	{R}epublic of {I}ran.
	\newblock {\em IMF Working Paper No. 02/205\/}.
	
	\bibitem[\protect\citeauthoryear{Chudik, Mohaddes, Pesaran, Raissi, and
		Rebucci}{Chudik et~al.}{2020}]{chudik_etal2020}
	Chudik, A., K.~Mohaddes, M.~H. Pesaran, M.~Raissi, and A.~Rebucci (2020).
	\newblock A {C}ounterfactual {E}conomic {A}nalysis of {C}ovid-19 {U}sing a
	{T}hreshold {A}ugmented {M}ulti-{C}ountry {M}odel.
	\newblock {\em National Bureau of Economic Research Working Paper No. 27855\/}.
	
	\bibitem[\protect\citeauthoryear{Chudik, Pesaran, and Tosetti}{Chudik
		et~al.}{2011}]{chudik_etal2011}
	Chudik, A., M.~H. Pesaran, and E.~Tosetti (2011).
	\newblock Weak and strong cross-section dependence and estimation of large
	panels.
	\newblock {\em The Econometrics Journal\/}~{\em 14}, C45--C90.
	
	\bibitem[\protect\citeauthoryear{Dadkhah and Zangeneh}{Dadkhah and
		Zangeneh}{1998}]{dadkhah1998}
	Dadkhah, K. and H.~Zangeneh (1998).
	\newblock International {E}conomic {S}anctions {A}re {N}ot {Z}ero-{S}um
	{G}ames: {T}here {A}re {O}nly {L}osers.
	\newblock {\em Iranian Journal of Trade Studies Quarterly\/}~{\em 1}, 1--14.
	
	\bibitem[\protect\citeauthoryear{Dizaji}{Dizaji}{2014}]{dizaji2014}
	Dizaji, S.~F. (2014).
	\newblock The effects of oil shocks on government expenditures and government
	revenues nexus (with an application to {I}ran's sanctions).
	\newblock {\em Economic Modelling\/}~{\em 40}, 299--313.
	
	\bibitem[\protect\citeauthoryear{Dizaji and Farzanegan}{Dizaji and
		Farzanegan}{2021}]{dizaji2021}
	Dizaji, S.~F. and M.~R. Farzanegan (2021).
	\newblock Do {S}anctions {C}onstrain {M}ilitary {S}pending of {I}ran?
	\newblock {\em Defence and Peace Economics\/}~{\em 32}, 125--150.
	
	\bibitem[\protect\citeauthoryear{Dizaji, Nasab, van Bergeijk, and
		Assari}{Dizaji et~al.}{2014}]{dizaji_et_al_2014}
	Dizaji, S.~F., E.~H. Nasab, P.~A. van Bergeijk, and A.~Assari (2014).
	\newblock Exports, {G}overnment {S}ize and {E}conomic growth: {E}vidence from
	{I}ran as a {D}eveloping {O}il-based {E}conomy.
	\newblock {\em International Journal of Humanities\/}~{\em 21}, 45--86.
	
	\bibitem[\protect\citeauthoryear{Dizaji and van Bergeijk}{Dizaji and van
		Bergeijk}{2013}]{dizaji2013}
	Dizaji, S.~F. and P.~A.~G. van Bergeijk (2013).
	\newblock Potential early phase success and ultimate failure of economic
	sanctions: A {VAR} approach with an application to {I}ran.
	\newblock {\em Journal of Peace Research\/}~{\em 50}, 721--736.
	
	\bibitem[\protect\citeauthoryear{Dornbusch}{Dornbusch}{1976}]{dornbusch1976}
	Dornbusch, R. (1976).
	\newblock Expectations and {E}xchange {R}ate {D}ynamics.
	\newblock {\em Journal of Political Economy\/}~{\em 84}, 1161--1176.
	
	\bibitem[\protect\citeauthoryear{Doudchenko and Imbens}{Doudchenko and
		Imbens}{2016}]{doudchenko_imbens2016}
	Doudchenko, N. and G.~W. Imbens (2016).
	\newblock Balancing, {R}egression, {D}ifference-in-{D}ifferences and
	{S}ynthetic {C}ontrol {M}ethods: A {S}ynthesis.
	\newblock {\em National Bureau of Economic Research Working Paper No. 22791\/}.
	
	\bibitem[\protect\citeauthoryear{Downs and Maloney}{Downs and
		Maloney}{2011}]{downs2011}
	Downs, E. and S.~Maloney (2011).
	\newblock Getting {C}hina to {S}anction {I}ran: {T}he {C}hinese-{I}ranian {O}il
	{C}onnection.
	\newblock {\em Foreign Affairs\/}~{\em 90}, 15--21.
	
	\bibitem[\protect\citeauthoryear{Doxey}{Doxey}{1996}]{doxey1996}
	Doxey, M.~P. (1996).
	\newblock {\em International {S}anctions in {C}ontemporary {P}erspective\/}
	(2$^{nd}$ ed.).
	\newblock Palgrave Macmillan UK, London.
	
	\bibitem[\protect\citeauthoryear{Esfahani, Mohaddes, and Pesaran}{Esfahani
		et~al.}{2013}]{esfahani_etal2013}
	Esfahani, H.~S., K.~Mohaddes, and M.~H. Pesaran (2013).
	\newblock Oil exports and the {I}ranian economy.
	\newblock {\em The {Q}uarterly {R}eview of {E}conomics and {F}inance\/}~{\em
		53}, 221--237.
	
	\bibitem[\protect\citeauthoryear{Esfahani and Shajari}{Esfahani and
		Shajari}{2012}]{esfahani_etal2012}
	Esfahani, H.~S. and P.~Shajari (2012).
	\newblock Gender, education, family structure, and the allocation of labor in
	{I}ran.
	\newblock {\em Middle East Development Journal\/}~{\em 4},
	1250008/1--1250008/40.
	
	\bibitem[\protect\citeauthoryear{Farzanegan}{Farzanegan}{2011}]{farzanegan2011oil}
	Farzanegan, M.~R. (2011).
	\newblock Oil revenue shocks and government spending behavior in {I}ran.
	\newblock {\em Energy Economics\/}~{\em 33}, 1055--1069.
	
	\bibitem[\protect\citeauthoryear{Farzanegan}{Farzanegan}{2013}]{farzanegan2013}
	Farzanegan, M.~R. (2013).
	\newblock Effects of {I}nternational {F}inancial and {E}nergy {S}anctions on
	{I}ran's {I}nformal {E}conomy.
	\newblock {\em SAIS Review of International Affairs\/}~{\em 33}, 13--36.
	
	\bibitem[\protect\citeauthoryear{Farzanegan}{Farzanegan}{2014}]{farzanegan2014military}
	Farzanegan, M.~R. (2014).
	\newblock Military spending and economic growth: The case of {I}ran.
	\newblock {\em Defence and Peace Economics\/}~{\em 25}, 247--269.
	
	\bibitem[\protect\citeauthoryear{Farzanegan}{Farzanegan}{2019}]{farzanegan2019scm}
	Farzanegan, M.~R. (2019).
	\newblock The {E}ffects of {I}nternational {S}anctions on {M}ilitary {S}pending
	of {I}ran: A {S}ynthetic {C}ontrol {A}nalysis.
	\newblock {\em CESifo Working Paper No. 7937\/}.
	
	\bibitem[\protect\citeauthoryear{Farzanegan and Alaedini}{Farzanegan and
		Alaedini}{2016}]{farzanegan2016}
	Farzanegan, M.~R. and P.~Alaedini (2016).
	\newblock {\em Economic {W}elfare and {I}nequality in Iran: {D}evelopments
		since the {R}evolution}.
	\newblock Palgrave Macmillan US, New York.
	
	\bibitem[\protect\citeauthoryear{Farzanegan and Hayo}{Farzanegan and
		Hayo}{2019}]{farzanegan_hayo2019}
	Farzanegan, M.~R. and B.~Hayo (2019).
	\newblock Sanctions and the shadow economy: empirical evidence from {I}ranian
	provinces.
	\newblock {\em Applied Economics Letters\/}~{\em 26}, 501--505.
	
	\bibitem[\protect\citeauthoryear{Farzanegan and Markwardt}{Farzanegan and
		Markwardt}{2009}]{farzanegan_markwardt2009}
	Farzanegan, M.~R. and G.~Markwardt (2009).
	\newblock The effects of oil price shocks on the {I}ranian economy.
	\newblock {\em Energy Economics\/}~{\em 31}, 134--151.
	
	\bibitem[\protect\citeauthoryear{Gardeazabal and Vega-Bayo}{Gardeazabal and
		Vega-Bayo}{2017}]{gardeazabal_et_al2017}
	Gardeazabal, J. and A.~Vega-Bayo (2017).
	\newblock An empirical comparison between the synthetic control method and
	{H}siao et al.'s panel data approach to program evaluation.
	\newblock {\em Journal of Applied Econometrics\/}~{\em 32}, 983--1002.
	
	\bibitem[\protect\citeauthoryear{Gharehgozli}{Gharehgozli}{2017}]{gharehgozli2017}
	Gharehgozli, O. (2017).
	\newblock An estimation of the economic cost of recent sanctions on {I}ran
	using the synthetic control method.
	\newblock {\em Economics Letters\/}~{\em 157}, 141--144.
	
	\bibitem[\protect\citeauthoryear{Greenwald}{Greenwald}{2020}]{jhu2020}
	Greenwald, J. (Ed.) (2020).
	\newblock {\em Iran {U}nder {S}anctions}.
	\newblock John Hopkins University School of Advanced International Studies
	report.
	
	\bibitem[\protect\citeauthoryear{Haidar}{Haidar}{2017}]{haidar2017}
	Haidar, J.~I. (2017).
	\newblock Sanctions and export deflection: evidence from {I}ran.
	\newblock {\em Economic Policy\/}~{\em 32}, 319--355.
	
	\bibitem[\protect\citeauthoryear{Hsiao, Ching, and Ki~Wan}{Hsiao
		et~al.}{2012}]{hsiao_et_al2012}
	Hsiao, C., H.~S. Ching, and S.~Ki~Wan (2012).
	\newblock A panel data approach for program evaluation: Measuring the benefits
	of political and economic integration of {H}ong {K}ong with mainland {C}hina.
	\newblock {\em Journal of Applied Econometrics\/}~{\em 27}, 705--740.
	
	\bibitem[\protect\citeauthoryear{Hufbauer, Schott, and Elliott}{Hufbauer
		et~al.}{1990}]{hufbauer1990}
	Hufbauer, G.~C., J.~J. Schott, and K.~A. Elliott (1990).
	\newblock {\em Economic {S}anctions {R}econsidered: {H}istory and {C}urrent
		{P}olicy\/} (2$^{nd}$ ed.), Volume~1.
	\newblock Peterson Institute for International Economics, Washington, D.C.
	
	\bibitem[\protect\citeauthoryear{Karshenas and Pesaran}{Karshenas and
		Pesaran}{1995}]{karshenas_pesaran1995}
	Karshenas, M. and M.~H. Pesaran (1995).
	\newblock Economic reform and the reconstruction of the {I}ranian economy.
	\newblock {\em Middle East Journal\/}~{\em 49}, 89--111.
	
	\bibitem[\protect\citeauthoryear{Kokabisaghi}{Kokabisaghi}{2018}]{kokabisaghi2018}
	Kokabisaghi, F. (2018).
	\newblock Assessment of the {E}ffects of {E}conomic {S}anctions on {I}ranians'
	{R}ight to {H}ealth by {U}sing {H}uman {R}ights {I}mpact {A}ssessment {T}ool:
	{A} {S}ystematic {R}eview.
	\newblock {\em International Journal of Health Policy and Management\/}~{\em
		7}, 374--393.
	
	\bibitem[\protect\citeauthoryear{Liu and Adedeji}{Liu and
		Adedeji}{2000}]{liu_adedeji2000}
	Liu, O. and O.~Adedeji (2000).
	\newblock Determinants of {I}nflation in the {I}slamic {R}epublic of {I}ran: A
	{M}acroeconomic {A}nalysis.
	\newblock {\em IMF Working Paper No. 00/127\/}.
	
	\bibitem[\protect\citeauthoryear{Majbouri}{Majbouri}{2015}]{majbouri2015}
	Majbouri, M. (2015).
	\newblock Female {L}abor {F}orce {P}articipation in {I}ran: A {S}tructural
	{A}nalysis.
	\newblock {\em Review of Middle East Economics and Finance\/}~{\em 11}, 1--23.
	
	\bibitem[\protect\citeauthoryear{Majidpour}{Majidpour}{2013}]{majidpour2013}
	Majidpour, M. (2013).
	\newblock The {U}nintended {C}onsequences of {US}-led {S}anctions on {I}ranian
	{I}ndustries.
	\newblock {\em Iranian Studies\/}~{\em 46}, 1--15.
	
	\bibitem[\protect\citeauthoryear{Maloney}{Maloney}{2015}]{maloney2015}
	Maloney, S. (2015).
	\newblock {\em Iran's {P}olitical {E}conomy since the {R}evolution}.
	\newblock Cambridge University Press, Cambridge.
	
	\bibitem[\protect\citeauthoryear{Mazarei}{Mazarei}{2019}]{mazarei2019}
	Mazarei, A. (2019).
	\newblock Iran {H}as a {S}low {M}otion {B}anking {C}risis.
	\newblock {\em Peterson Institute for International Economics Policy Brief No.
		19-8\/}.
	
	\bibitem[\protect\citeauthoryear{Mazarei}{Mazarei}{2020}]{mazarei2020}
	Mazarei, A. (2020).
	\newblock Inflation {T}argeting in the {T}ime of {S}anctions and {P}andemic.
	\newblock In {\em J. Greenwald (ed.) "Iran {U}nder {S}anctions", John Hopkins
		University School of Advanced International Studies report}.
	
	\bibitem[\protect\citeauthoryear{Mohaddes and Pesaran}{Mohaddes and
		Pesaran}{2013}]{mohaddes_pesaran2013}
	Mohaddes, K. and M.~H. Pesaran (2013).
	\newblock One hundred years of oil income and the {I}ranian economy: A curse or
	a blessing?
	\newblock In {\em Alizadeh, P. and H. Hakimian, (eds.), "Iran and the Global
		Economy: Petro populism, Islam and economic sanctions"}, Chapter 1, 12--45.
	Routledge, New York.
	
	\bibitem[\protect\citeauthoryear{Mohaddes and Raissi}{Mohaddes and
		Raissi}{2020}]{mohaddes_raissi2020}
	Mohaddes, K. and M.~Raissi (2020).
	\newblock Compilation, {R}evision and {U}pdating of the {G}lobal {VAR} ({GVAR})
	{D}atabase, 1979{Q}2-2019{Q}4.
	\newblock {\em University of {C}ambridge: Judge {B}usiness {S}chool (mimeo)\/}.
	
	\bibitem[\protect\citeauthoryear{Morgan, Bapat, and Kobayashi}{Morgan
		et~al.}{2014}]{morgan2014}
	Morgan, T.~C., N.~Bapat, and Y.~Kobayashi (2014).
	\newblock Threat and imposition of economic sanctions 1945--2005: Updating the
	{TIES} dataset.
	\newblock {\em Conflict Management and Peace Science\/}~{\em 31}, 541--558.
	
	\bibitem[\protect\citeauthoryear{Naghavi and Pignataro}{Naghavi and
		Pignataro}{2015}]{naghavi2015}
	Naghavi, A. and G.~Pignataro (2015).
	\newblock Theocracy and resilience against economic sanctions.
	\newblock {\em Journal of Economic Behavior \& Organization\/}~{\em 111},
	1--12.
	
	\bibitem[\protect\citeauthoryear{Pape}{Pape}{1997}]{pape1997}
	Pape, R.~A. (1997).
	\newblock Why {E}conomic {S}anctions {D}o {N}ot {W}ork.
	\newblock {\em International {S}ecurity\/}~{\em 22}, 90--136.
	
	\bibitem[\protect\citeauthoryear{Pape}{Pape}{1998}]{pape1998}
	Pape, R.~A. (1998).
	\newblock Why {E}conomic {S}anctions \textit{{S}till} {D}o {N}ot {W}ork.
	\newblock {\em International {S}ecurity\/}~{\em 23}, 66--77.
	
	\bibitem[\protect\citeauthoryear{Peksen}{Peksen}{2009}]{peksen2009}
	Peksen, D. (2009).
	\newblock Better or {W}orse? {T}he {E}ffect of {E}conomic {S}anctions on
	{H}uman {R}ights.
	\newblock {\em Journal of Peace Research\/}~{\em 46}, 59--77.
	
	\bibitem[\protect\citeauthoryear{Peksen and Drury}{Peksen and
		Drury}{2010}]{peksen2010}
	Peksen, D. and A.~C. Drury (2010).
	\newblock Coercive or {C}orrosive: The {N}egative {I}mpact of {E}conomic
	{S}anctions on {D}emocracy.
	\newblock {\em International Interactions\/}~{\em 36}, 240--264.
	
	\bibitem[\protect\citeauthoryear{Pesaran}{Pesaran}{1992}]{pesaran1992}
	Pesaran, M.~H. (1992).
	\newblock The {I}ranian foreign exchange policy and the black market for
	dollars.
	\newblock {\em International Journal of Middle East Studies\/}~{\em 24},
	101--125.
	
	\bibitem[\protect\citeauthoryear{Pesaran}{Pesaran}{2000}]{pesaran2000}
	Pesaran, M.~H. (2000).
	\newblock Economic {T}rends and {M}acroeconomic {P}olicies in
	{P}ost-{R}evolutionary {I}ran.
	\newblock In {\em P. Alizadeh (ed.), "The Economy of {I}ran: Dilemmas of an
		Islamic State"}, Chapter 2, 63-99. I.B. Tauris, London.
	
	\bibitem[\protect\citeauthoryear{Pesaran}{Pesaran}{2006}]{pesaran2006}
	Pesaran, M.~H. (2006).
	\newblock Estimation and inference in large heterogeneous panels with a
	multifactor error structure.
	\newblock {\em Econometrica\/}~{\em 74}, 967--1012.
	
	\bibitem[\protect\citeauthoryear{Pesaran}{Pesaran}{2015}]{pesaran2015}
	Pesaran, M.~H. (2015).
	\newblock {\em Time {S}eries and {P}anel {D}ata {E}conometrics}.
	\newblock Oxford University Press, Oxford.
	
	\bibitem[\protect\citeauthoryear{Plante}{Plante}{2019}]{plante2019}
	Plante, M. (2019).
	\newblock {OPEC} in the news.
	\newblock {\em Energy Economics\/}~{\em 80}, 163--172.
	
	\bibitem[\protect\citeauthoryear{Popova and Rasoulinezhad}{Popova and
		Rasoulinezhad}{2016}]{popova2016}
	Popova, L. and E.~Rasoulinezhad (2016).
	\newblock Have {S}anctions {M}odified {I}ran's {T}rade {P}olicy? {A}n
	{E}vidence of {A}sianization and {D}e-{E}uropeanization through the {G}ravity
	{M}odel.
	\newblock {\em Economies\/}~{\em 4}, Article 24.
	
	\bibitem[\protect\citeauthoryear{Wan, Xie, and Hsiao}{Wan
		et~al.}{2018}]{wan_et_al2018}
	Wan, S.-K., Y.~Xie, and C.~Hsiao (2018).
	\newblock Panel data approach vs synthetic control method.
	\newblock {\em Economics Letters\/}~{\em 164}, 121--123.
	
	\bibitem[\protect\citeauthoryear{Weiss, Cortright, Lopez, and Minear}{Weiss
		et~al.}{1997}]{cortright1997}
	Weiss, T.~G., D.~Cortright, G.~A. Lopez, and L.~Minear (Eds.) (1997).
	\newblock {\em Political {G}ain and {C}ivilian {P}ain: {H}umanitarian {I}mpacts
		of {E}conomic {S}anctions}.
	\newblock Rowman \& Littlefield, Lanham: MD.
	
\end{thebibliography}
\end{document}